\documentstyle[12pt,mythes,epsfig,amssymb]{report}
\begin{document}
\bibliographystyle{apsrev}
\setcounter{secnumdepth}{4}
\def\lsim{\mathrel{
\def\arraystretch{.4}
\begin{array}{c}
$$<$$\\
$$\sim$$
\end{array}}}
\def\gsim{\mathrel{
\def\arraystretch{.4}
\begin{array}{c}
>$$\\
$$\sim$$
\end{array}}}
\title{\bf STUDY OF HEAVY ION INDUCED FISSION FRAGMENT ANGULAR AND MASS 
DISTRIBUTION AT NEAR AND SUB-COULOMB BARRIER ENERGIES} 
\author{\sl \bf  TILAK KUMAR GHOSH}
\dept{\bf JADAVPUR UNIVERSITY}

\newpage

\copyrightfalse
\figurespagefalse
\tablespagefalse
\beforepreface
\prefacesection{\bf CERTIFICATE}

   This is to certify that the thesis entitled {\bf Study of heavy ion induced 
fission fragment angular and mass distribution at near and sub-coulomb barrier 
energies} submitted by Sri {\bf Tilak Kumar Ghosh} who got his name registered
on 31.01.03 for the award of {\bf Ph.D.(Science) degree} 
of Jadavpur University, is absolutely based upon his own research
 work under the supervision of {\bf Prof. Pratap Bhattacharya, Saha Institute 
of Nuclear Physics, Kolkata}, and that neither this thesis nor any part of it 
has been submitted for any degree/diploma or any other academic award anywhere 
before.
 
\vspace{2in}
\begin{center}
${\overline {Prof.\  Pratap\ Bhattacharya}}$\\
\end{center}

\newpage
\vspace*{3in}
\begin{center}
{\it To my sister-in-law}\\
{\it \bf{Rekha Ghosh}}\\
{\it for her motherly love, affection and inspiration}
\end{center}
\newpage
\prefacesection{Acknowledgments} 

{\it \indent Just after finishing my Post M.Sc course when I had opted for 
the experimental nuclear physics as my subject of research, I was advised that
 one experimentalist should learn at least few things first: cooking, 
driving, spending sleepless night and ....., let it be secret, but it seemed
 to me that doing experiment is a fun. Now after spending more than five years 
when I am in the juncture of a major step toward my research carrier, I 
believe that choosing this field of research has allowed me to get to be 
acquainted with some outstanding feelings of life and existence.

\indent With gleeful heart I at the very outset offer my sincerest regard 
to my thesis supervisor, Prof. Pratap Bhattacharya who has allowed me to work 
independently and I have been blessed with his guidance whenever needed. This 
work couldn't be flourished without his enormous support. The close 
association of his family member made my tenure enjoyable here. While Puchi 
was a great critique (which turns out to be the words of appreciation after 
going out to ice-cream parlor with her!) of mine, Radha-di 
(Prof. Radha Bhattacharya) was always caring. I am obliged to them.

\indent With deep respect I remember the inspiration of Dr. Nanda Dasgupta 
and Prof Bhupal Chandra Samanta who taught me Nuclear Physics at the Burdwan 
University. I remember the encouragement of Nanda-madam who motivated me to 
join in research. I reagrdfully acknowledge her best wishes for my research 
carrier.

\indent All the experimental work of this thesis was carried out using the 
15 UD Pelletron of the Nuclear Science Centre (NSC), New Delhi. I thank the 
Accelerator User Committee for providing me a beam time account. Dr. S.K. Datta, who was my local collaborator at NSC, always extended his helping hand during
 our experiments. I am grateful to him. I am particularly obliged to Dr.
 Ranjan Bhowmik not only for his valuable and expert comment during the 
experiment but also for  many illuminating discussions with him prior to and 
following the experiments. Dr. Amit Roy, Director of Nuclear Science Centre 
showed interest in our work and I am thankful to him for sanctioning me from 
Director's quota an extra beam time which was quite useful to finish this 
thesis work. I am also thankful to all the operating staff members of NSC 
Pelletron, especially to Abhijit Sarkar for delivering good quality of 
pulsed beam ; Mr Debashis Sen for his scientific coordination in the official 
side , Avilash for helping target preparation and Bivash who helped a lot 
during one of my initial experiments.I would like to acknowledge Ms. K.S.Golda, my collaborator at NSC, whose immense help at every step was more than
 responsibility.

\indent I am thankful to all my collaborators from Saha Institute of Nuclear 
Physics, Prof. S. Chattopadhayay, Dr. N. Majumdar, Dr. T. Sinha and Mr. S. Pal. Thanks are due to Dr Manoj Sharan who took part one of the initial experiments
 at NSC. I acknowledge the help of Mr. Dipankar Das and Mrs Lipy Paul during 
the design and fabrication of the gas detectors.

\indent I would like to express my gratitude to my collaborators Dr D.C.Biswas, Dr. A.Saxena, Mr. P.K.Sahu of Bhaba Atomic Research Centre, Mumbai for their 
sincere involvement in the experiments. I would also thank Dr L.M.Pant and Mr. 
R.G.Thomas who took part in one of the preliminary experiment at TIFR Pelletron.

\indent I am indebted to Prof. B.C. Sinha, the Director of Saha Institute of
 Nuclear Physics for extending the different facilities to work in the 
institute. In this context I don't want to lose the opportunity to express my 
deepest gratitude to Prof Prasanta Sen, whose interest about my research work, 
despite his busy schedule of work, encouraged a lot to me.

\indent Special gratefulness is expressed to the staff of the workshop for 
their support, which helped a lot for the successful completion of this work. 
Thanks are due to the library staff of our institute.

\indent I consider myself lucky to met many people whom I consider as fountain 
of knowledge, during the course of my thesis work.  Discussions with Prof 
Binayak Dutta Roy (BDR-sir), one of the best teachers I saw in my student
 life, were always enjoyable and not necessarily restricted within the 
classroom or laboratory. He illustrated me the magnificence to resolve  problem 
from the very first principle. I deeply 
appreciate help of  Prof Susanta Dattagupta, Director of S. N. Bose  National 
Centre for Basic Science, for understanding basic ideas of nuclear dissipation. I am indebted to Dr. R. K. Choudhury, Director of Institute of Physics, 
Bhubaneswar for his keen interest in our work and suggestions regarding 
the work.

\indent I am grateful to all my friends who supported and inspired me during
 the awful phases of my life and shared joy and happiness during the good times. Immediately I remember to my friend Mishreyee who deserves special mention. 
Dr Dipak Goswami, one of my best friends from College days, has always extended 
a constant support and encouragement. I also cherish to share the hostel room
 with Sanjib (popularly name as Maji) during my Post M.Sc. days. My sincere 
thankfulness should be recorded to Shakti-da, Suresh-da and Madhu-da of the 
hostel canteen, for their care that made my hostel life homely. My thanks also 
to all my well-wishers, who have given their generous help and made my stay at 
Kolkata enjoyable.

\indent Finally my boundless gratitude are due to my mother and family members 
for their love and care and constant support without which I could not reach 
this stage.}\\

%\bigskip
\noindent Tilak Kumar Ghosh\\
%\noindent Kolkata,\\
%\today 

\prefacesection{List of publication}
\noindent{\bf\underline{Published and communicated in journals}}\\
%\noindent The thesis is covered by the follwing publications in international 
%journals. 
%\noindent{\bf\underline{PUBLISHED IN JOURNALS:}} 
\footnotesep=5.in
\footnotetext[1]{\vskip -5in {\large *} Indicates papers included in this 
thesis.}
\begin{description} 
\item [{{\large*}[1]~}] {\it Anomalous increase in width of fission fragment 
mass distribution in $^{19}$F+$^{232}$Th}.\\
{\bf  {T. K. Ghosh}}, S. Pal, T.Sinha, N. Majumdar, S. Chattopadhyay,   
P. Bhattacharya, A. Saxena, P. K. Sahu, K.S. Golda and S. K. Datta.\\ 
{\it Phys. Rev.} {\bf C69}, (2004) 031603. ({\bf Rapid Communication}). 

\item [{{\large*}[2]~}] {\it Anomalous increase in width of fission fragment 
mass distribution as a probe for onset of quasi fission reactions in deformed 
target-projectile system at near and sub-barrier energies}.\\
{\bf  {T. K. Ghosh}}, S. Pal, T.Sinha, S. Chattopadhyay,   
P. Bhattacharya, D.C. Biswas and K.S. Golda.\\ 
{\it Phys. Rev.} {\bf C70}, (2004) 011604. ({\bf Rapid Communication}).

\item [{{\large*}[3]~}] {\it Time of Flight (TOF) spectrometer for accurate 
measurement of mass and angular distribution of fission fragments in heavy 
ion induced fission reaction}.\\  
{\bf  {T. K. Ghosh}}, S. Pal, T.Sinha, S. Chattopadhyay and P. Bhattacharya.\\ 
{\it Nucl. Instrum. Phys. Res.} {\bf B540}, (2005) 285.

\item [{{\large*}[4]~}] {\it Evidence of microscopic effects in fragment mass 
distribution in fusion-fission reactions of light projectiles with heavy 
targets }.\\  
{\bf  {T. K. Ghosh}}, S. Pal, P. Bhattacharya and K.G. Golda.\\ 
{\it Accepted for publication in Phys. Lett. B.} ({\bf nucl-ex/0502013}). 

\item [{{\large*}[5]~}] {\it Comment on "Fission mass widths in $^{19}$F + 
$^{232}$Th, $^{16}$O + $^{235,238}$U reactions at near barrier energies"}.\\  
{\bf  {T. K. Ghosh}} and P. Bhattacharya \\ 
{\it Submitted to Phys. Rev. C.} ({\bf nucl-ex/0505019}).

\item[{[6]~}] {\it Observation of antimagnetic rotation 
in $^{108}$Cd} .\\
P. Datta, S. Chattopadhyay, S. Bhattacharya, {\bf {T. K. Ghosh}}, A. Goswami,
 S. Pal, M. Saha Sarkar, H. C. Jain, P. K. Joshi ,R. K. Bhowmik, R. Kumar, 
N. Madhaban, S.Muralithar, M. Rao, and R. P. Singh \\   
{\it Phys. Rev. {\bf C71,}} 041305 (2005) ({\bf Rapid Communication}). 

\item[{[7]~}] {\it Observation of magnetic rotation in odd-odd $^{104}$Ag}.\\
P. Datta , S. Chattopadhyay, P. Banerjee, S. Bhattacharya, B. Dasmahapatra, 
{\bf {T. K. Ghosh}}, A. Goswami, S. Pal, M. SahaSarkar, S. Sen , H.C. Jain,
P.K. Joshi, Amita\\  
{\it Phys. Rev. {\bf C 69}}, 044317 (2004). 
 
\item[{[8]~}] {\it Possible coexistence of principal and tilted  axis  
rotation  in $^{103}$Ag}.\\
P. Datta, S. Chattopadhyay, P. Banerjee, S. Bhattacharya, J. Chatterjee, 
B. Dasmahapatra,C. C. Dey, {\bf {T. K. Ghosh}}, A. Goswami, S. Pal, I. Ray, 
M. Saha Sarkar, S. Sen, H. C. Jain, P. K. Joshi , Amita\\   
{\it Phys. Rev. {\bf C 67}},014325 (2003). \\
\medskip
\end{description}
\clearpage

\noindent{ \bf\underline{Conference Proceedings:}} 
\begin{description}

\item [{{\large*}[1]~}] Performance characteristics of a MWPC detector for time of 
flight measurement\\
{\bf{T.K. Ghosh}}, T. Sinha, N. Majumdar, S. Chattopadhyay, P. Bhattacharya, 
D.C. Biswas, L.M. Pant, A. Saxena and R.K. Choudhury\\
 {\it Proc. Int. Nucl. Phys. Symp. (Mumbai, India)} {\bf 43B} (2000) 486.

\item [{{\large*}[2]~}] Mass distribution study of $^{16}$O + $^{209}$Bi\\
{\bf{T.K. Ghosh}}, S. Pal, N. Majumdar, D.C. Biswas, S. Chattopadhyay,
M.K. Sharan, S. Mukhopadhyay, S.K. Datta and P. Bhattacharya\\
{\it Proc. of DAE. Symp. (Tirunelvelli, India) } {\bf 45B }(2002) 166.

\item [{{\large*}[3]~}] Near barrier fragment mass and angular distribution in 
$^{19}$F + $^{209}$Bi\\
{\bf{T.K. Ghosh}}, S. Pal, N. Majumdar, S. Chattopadhyay,S. Mukhopadhyay, 
 P. Bhattacharya,P.K. Sahu, A. Saxena, K.S. Golda  and S.K. Datta\\
{\it Proc Int. Nucl. Phys. Symp. (Mumbai, India)} {\bf 46B }(2003).

\item[{[4]~}] Multi-particle response of a Cathode Pad Chamber (CPC) \\
P. Bhattacharya, S. Bose, S. Chattopadhyay, D. Das, P. Datta, {\bf{T.K. Ghosh}}, N. Majumdar, S. Mukhopadhyay, S. Pal, L. Paul, P. Roy, A. Sanyal, S. Sarkar, 
P. Sen,  M. Sharan, S.K. Sen, T. Sinha and B.C. Sinha\\
{\it Proc Int. Nucl. Phys. Symp.(Mumbai, India)} {\bf 43B}(2000) 504.

\item[{[5]~}] Design and fabrication of an ultra-thin Cathode Pad Chamber 
(CPC)\\
P. Bhattacharya, S. Bose, S. Chattopadhyay, D. Das, P. Datta, {\bf{T.K. Ghosh}}, N. Majumdar, S. Mukhopadhyay, S. Pal, L. Paul, P. Roy, A. Sanyal, S. Sarkar, 
P. Sen,  M. Sharan, S.K. Sen, T. Sinha and B.C. Sinha\\
{\it Proc. Int. Nucl. Phys. Symp.(Mumbai, India)} {\bf 43B}(2000) 506.

\item[{[6]~}] Simulation studies on the spatial resolution and reconstruction 
efficiency of cathode pad chamber (CPC)\\
P. Bhattacharya, S. Bose, S. Chattopadhyay, D. Das, P. Datta, {\bf{T. Ghosh}}, 
N. Majumdar, S. Mukhopadhyay, S. Pal, L. Paul, P. Roy, A. Sanyal, S. Sarkar, 
P. Sen, S. Sen, M.K. Sharan and B.C. Sinha\\
{\it Proc. DAE-BRNS Symp. Nucl. Phys.} {\bf 44B}(2001) 308.

\item[{[7]~}] Numerical simulation of gas detectors\\
S Pal, P. Datta, {\bf{T.K. Ghosh}}, S. Mukhopadhyay\\
{\it Proc Int. Nucl. Phys. Symp.} {\bf 45B }(2002) 456. \\

\item [{[8]~}]Study of the characteristics of Cathode Pad Chambers with new 
generation of MANAS 1.2-3.1\\
P. Bhattacharya, S. Bose, S. Chattopadhyay, D. Das, P. Datta, {\bf{T. Ghosh}}, 
 N. Majumdar, S. Mukhopadhyay, S. Pal, L. Paul, P. Roy, A. Sanyal, S. Sarkar, 
P. Sen, S. Sen, M. Sharan and B. Sinha\\
{\it Proc. DAE-BRNS Symp. Nucl. Phys.}{\bf 45B}(2002) 484.\\

\end{description}

\prefacesection{Summary}

\indent The thesis presents investigations on the angular and mass 
distribution of fission fragments on heavy ion induced fission reactions. 
The present investigations address current issues in heavy ion induced 
fission reactions. For the last few years, the focus on the research in this 
field was on the formation of compound nucleus close and below the Coulomb 
barrier, as such studies have a direct bearing on the synthesis of super heavy 
nuclei, an island of stable or quasi-stable nuclei far from the beta stability 
line far out in neutron and proton numbers from the known trans-
uranium elements. Since the principal decay mode of the super heavy nuclei 
would be fission reaction, the studies of the competition between the fusion 
and fission and the factors hindering the yields of super heavy elements are
 intensely followed. 
           
\indent One of the findings in the nineties in this field was the anomalous 
angular anisotropy observed in the near and sub-Coulomb barrier energies,
 particularly in reactions of heavy ions on heavy, deformed, actinide nuclei. 
The explanation of the anomaly was in terms of fission from a non-equilibrated 
compound nucleus or alternately, a quasi-fission mechanism signaling a 
possible hindrance to the production of super heavy nuclei.

\indent The present investigations were carried out to measure precisely the 
distribution of fragment mass in the same reactions, which showed departure 
from production of an equilibrated compound nucleus. We have used a double arm 
time-of flight spectrometer over a long flight path to measure 
the precise masses of complementary fission fragments. Necessary large 
area position sensitive gas detectors, the method of experiments and data
 analysis were developed for measurement of angular distribution and mass 
distribution of fission fragments, exclusively for events in which the 
incoming projectile momentum was fully transferred to the fissioning 
composite or compound nucleus. The experiments were done using pulsed heavy 
ions from the 15UD Pelletron at the Nuclear Science Centre, New Delhi. 

\indent	The first string of measurements were for a spherical target, 
$^{209}$Bi, with oxygen and fluorine projectiles. The angular distribution 
measurements in the same experiments supplemented the existing angular 
anisotropy measurements to establish beyond doubt the systems scrupulously 
followed the predictions based on the macroscopic theories of the production 
of equilibrated compound nucleus. The  mass distributions were symmetric 
around the average of the target and projectile mass, and the width of mass 
distribution varied smoothly with the beam energy, fully conforming to a 
statistical binary split of the compound nucleus.      

\indent The next series of experiments were done using a deformed target 
of $^{232}$Th and projectiles of carbon, oxygen and fluorine. The angular 
anisotropy data existing in these systems showed an anomalous increase of the 
anisotropy as the beam energy decreased through the Coulomb barrier. The mass 
distributions were measured for these systems. In case of all the systems with 
deformed target, at all energies, the mass distributions were symmetric, 
peaked around the average of the target and projectile masses, as in the case 
of systems with spherical target, viz, $^{209}Bi$. However, the width of the 
mass distribution, $\sigma_m^2$, showed completely anomalous behaviour. 
For $^{19}$F + $^{232}Th$, the  $\sigma_m^2$ decreased monotonically as the 
energy is decreased, but near the Coulomb barrier, value of  $\sigma_m^2$
 started to rise  and reached a value which is more than 150$\%$ of the 
extrapolated value at that energy. Thereafter, $\sigma_m^2$ value again 
started to decrease. Exactly similar trend of the $\sigma_m^2$ values were 
observed for oxygen and carbon projectile, although the sharp increase of  
$\sigma_m^2$ values progressively got smaller for oxygen and carbon projectiles.

\indent The close similarity of the trends of the fragment angular anisotropy 
and the width of mass distributions immediately suggested a common explanation
 of the observed  anomalous rise in both the observable for deformed target 
and the same energy regions. A review of the possible explanations of 
anomalous anisotropies, in terms of a pre-equilibrium model and orientation 
dependent quasi-fission prompted the probability of the latter explanation 
being applicable to explain the observed anomalous rise of the  $\sigma_m^2$
 values with decreasing energy.
	
\indent In orientation dependent quasi-fission formalism, it has been 
postulated that due to microscopic effects of the relative elongated 
configuration of the projectile fusion with the deformed target through the 
polar region, the fusion barriers are lowered and simultaneously, the system 
prefers to reach directly a saddle shape (which may be mass asymmetric) in a 
quasi-fission reaction, in contrast to the initial compact configuration 
leading preferentially to a formation of compound nucleus when the projectile 
hits the equatorial region of the deformed target. Hence, it was 
conjectured that up to a critical angle on the relative orientation of the 
symmetry axis of target with respect to the projectile trajectory, normal 
compound nuclear fission and the quasi fission mechanisms can be mixed. 
With cross section weighted admixture of the two fission modes, the observed 
anomalous rise in  $\sigma_m^2$ could be phenomenologically explained for 
all the systems.

\indent In last few years, considerable progress has been achieved in
 numerical simulation to calculate the path followed by fusing nuclei, 
through a multidimensional energy landscape. The present investigations 
show a possible scenario in the paths followed by the systems for a deformed
 target–projectile combination . In addition to the normal fusion over a 
barrier, followed by fission over a unconditional mass symmetric fission 
barrier, the system can hit a ridge for certain orientations, when the 
normal route is blocked, and transit to fission over a conditional saddle 
with zero or small mass asymmetry in a quasi fission mechanism. The present 
investigations re-emphasize the need of dynamical calculations of fusion paths
 of ions in a dissipative medium.\\

\noindent{\large{\bf\underline{Outline of the thesis:}}}\\
 
\indent The work to be presented in this thesis is divided into five chapters 
with a appendix. {\it Chapter 1} gives an overview of the subject, briefly 
mentioning the investigations of fission fragment mass and angular 
distribution study reported by other experimental groups. Necessity to 
introduce a new probe to study the fusion-fission dynamics, which is the main 
motivation of this work has been discussed in this very first chapter of 
the thesis. {\it Chapter 2} deals with design and fabrication of a large area
 multi-wire proportional counter that was developed in our laboratory at Saha 
Institute of Nuclear Physics. {\it Chapter 3} describe the details of the 
experimental technique and the data analysis procedure. Different methods 
to determine the mass distributions of fission fragments are compared and
 the performance of a dual time of flight (TOF) 
spectrometer using the two position sensitive  MWPCs used by us are discussed.
In {\it Chapter 4}, the results of the measurements of mass distributions for 
both spherical and deformed targets are presented.{\it Chapter 5} is devoted 
for the discussions on the findings of this thesis work. Conclusion 
of the thesis work is given at the end of this chapter. Some details of 
the formulations of angular and mass distributions are  derived at the 
appendix. Cumulative references are given at the end of the each chapter.

\afterpreface
\setcounter{equation}{0}
\setcounter{figure}{0}
\chapter{Introduction}
\markboth{nothing}{\it Introduction}

\newpage

\indent Nuclear fission was discovered sixty five years ago in which a heavy 
nucleus breaks up into two nuclei. The general features of fission reactions 
were well understood both experimentally and theoretically, still it continues 
to throw up new challenges from time to time consistently in last six decades. 
The reason for that is the involvement of a very large number of nucleons 
and the macroscopic and microscopic forces that influence the fission phenomena.

\indent The macroscopic forces largely determines the gross features of the 
fission phenomena. Liquid drop model of Bohr and Wheeler \cite{BohrPR39*1} 
explained fission, particularly tunneling of the fragments through a fission 
barrier in spontaneous or low energy fission reactions. Subsequently, the 
inclusion of shell effects \cite{StruNPA67*1} defined the finer effects of 
modulating the fission barrier as function of nuclear deformations and
 modifying the mass and kinetic energy distributions in the fission fragments.

\indent New observables were introduced as the experimental techniques  
advanced. Fission isomers, super-deformed shapes, pre-scission particle 
emission and fission times scales were introduced in neutron, light and 
heavy ion induced fission reaction studies. As the excitation energy and the 
angular momentum brought-in, in the heavy ion induced fission reaction, interest was renewed in the fission studies as the new probes of the distributions 
of the fragments with respect to the angle, in and out of plane of the reaction 
plane, total kinetic energy, mass, and the number and time of emissions of 
neutrons or charged particles. For the first time the probes presented the 
opportunity of studying the dynamics of the fission reactions. In other words, 
definite clues could be found about the entire history of the fission of two 
accelerated ions to a composite system and resepartion into two fragments 
with damping of radical motions and 
alterations in mass and excitations compared to the initial reactants. 

\indent The first insight into the dynamics of fission reaction was the 
observation that the reaction proceeds with complete oblivion of the 
initial experimental parameters of the mass and kinetic energies of the 
reactants, the target and projectile. It was observed that the angular 
distribution of the fission fragments do not follow the expected $1/sin\theta$ 
behaviour observed in reactions where the ejectiles are formed in one step, 
direct reactions. A statistical theory of fission \cite{HSGeneva58*1} based on 
the equilibrium properties of a hot, rotating nuclei successfully explained 
the observed ratio of yields of fission fragments, parallel and perpendicular 
to the beam axis. A very large number of experimental reports fully complied 
with the picture that fission takes place from a fully equilibrated compound 
nucleus undergoing shape changes to reach a saddle configuration following 
statistical rules, and undergo fission. Hence the first step in the heavy ion 
induced fission is clearly established - the target and projectile fuse 
together and then the composite system equilibrates before fission takes 
place. The damping of the incoming radial motion 
relaxes the excitation energies by inducing spinning of the composite system and also a statistical 
evaporation of particles (mostly neutron). Hence the average number of
 particles and their energy defined a time scale of the equilibration of the 
compound nucleus. A fission time scales of a few tens of $10^{-20}$ sec 
was inferred for the systematics of neutron evaporation before scission 
occurs.

\indent The main feature of the fusion-fission reactions of the compound 
nuclear fission was established, but by early nineties, it became quite 
apparent that considerable departures were possible and new reaction paths or 
mechanism were needed to explain {\it anomalous} properties in fission 
observables. As the total 
angular momentum $J$ of the rotating CN increased, the fission barrier began 
to drop and ultimately reached zero where the compound nucleus was spontaneously unstable against fission. Such prompt fission reaction was called fast 
fission \cite{GrePLB81*1}. The angular distributions of the fragments were 
forward peaked and the mass distributions were extremely wide. However,
 the mechanism did not point to any basic change of the reaction dynamics but a mere breakdown of the compound nucleus before it had chance of formation. 

\indent In energies intermediate between the Coulomb barrier and the 
onset of fast fission, and systems which are  mass symmetric, it was 
increasingly becoming evident that mass distributions are transforming from
symmetric shape peaking around average of target and projectile masses, 
to that of an asymmetric mass distribution. It was appropriate that such 
processes were named as "quasi-fission" as it is apparent that the systems 
are not proceeding along the 
fusion-fission path and following an entirely different path. This is an
important new clue about the dynamics of the heavy ion induced fusion-
fission reactions.

\indent Departures from the SSPM predictions of the fragment angular 
distributions were observed in many systems, particularly with highly 
deformed targets, in energies close to any below Coulomb barriers in 
ejectiles in 1980's and 1990's and quickly became subject of intense 
experimental and theoretical investigations. It had been of interest 
because of the excitation energies and the total 
angular momentum of the systems were within reasonable ranges and the systems 
were fully expected to follow the compound 
nuclear fusion-fission paths. Spectacular rise in the angular anisotropies 
of the fission fragments [defined as the ratio of yields, $A = W(0^\circ) or 
 W(180^\circ)/W(90^\circ)$] as the energy is lowered through the Coulomb
 barrier were observed \cite{ZhangPRC94*1,HindePRL95*1,NMPRL96*1}. As the 
angular 
anisotropy can be related to the macroscopic properties $A= 1 + <l^2>/4K_0^2$ 
where $<l^2>$ is the mean square angular momentum and $K_0^2$ is the width of 
the distribution of K-values; K being the projection of total angular 
momentum $J$ on the nuclear symmetry axis at saddle point, both 
$<l^2>$ or $K_0^2$ may be the reason for enhanced A values.

\indent The average value of  $<l^2>$ is determined from the fitting of 
the fusion-fission excitation functions and dependent on the reaction 
mechanism and optical model parameters. The average width of the 
K-distribution, is in turn, related to the temperature and the moment 
of inertia of the nucleus about the symmetry axis. The value of $<K_0^2>$ 
could be calculated from the macroscopic properties  of a 
finite rotating liquid drop in a heat bath and considered to be reliable. 
Hence, the initial explanation of anomalous angular anisotropies pointed to an 
anomalous widening of spin distribution i.e., $<l^2>_{exp}$ $>$ $<l^2>_{theo}$.

\indent The uncertainties in determining $<l^2>$ was slowly resolved, both 
experimentally and theoretically. In the systems where targets were heavy 
deformed actinides, like thorium or uranium, the fission of target like 
nuclei following excitation by transfer of few nucleons from the projectile, 
the so called transfer fission channel, was significantly populated. The 
mixture of transfer fission was argued to be the cause of the broadening of 
the spin distribution. The thrust of experimental investigations were 
directed to isolate the pure fusion-fission events.Folding angle between 
complementary fission fragments was found to be dependent on the recoil 
velocity of the CN and a precise measurement of the ($\theta,\phi$) 
distribution of the fission fragments easily separates exclusive 
fusion-fission events following full transfer of the incident momentum 
\cite{PBneuvo*1,ZhangPRC94*1}. However, exclusion of transfer fission from the 
total fission events increased the angular anisotropy compared to inclusive measurements. The angular anisotropy 
of the exclusive fusion-fission reaction became more anomalous \cite{NMPRC95*1}
 compared to that predicted by macroscopic theories.

\indent Smaller corrective steps, viz, taking into consideration of cooling of 
compound nucleus by pre-saddle 
neutron emissions, or the appropriate values of the level density parameter 
did not improve the fit between the experimental data and theoretical 
predictions. An important break-through, out of the impasse, was achieved 
when the calculation of the $<l^2>$  could be made more reliable. The 
basic uncertainty in $<l^2>$ determined from fission excitation function came 
from the optical model parameters of the coupled channel calculations.
[One-dimensional barrier penetration model was inadequate to explain the large 
magnitude of enhancement in fusion (fission) cross sections observed in 
below barrier energies]. However, another check of the model parameters could 
be done by comparing the predicted distribution of fusion barriers (which 
is a function of $d^2\sigma/dE^2$) with a precise experimental determination 
of the same quantity \cite{HindePRL95*1}. Hence a precise and 
unambiguous determination of $<l^2>$ and subsequent calculations of the 
angular anisotropy proved beyond doubt that width of the K-distribution, 
$K_0^2$ must have been smaller than that calculated from finite rotating 
liquid drop model to explain the observed anomalously enhanced angular 
anisotropies \cite{NMPRC95*1}.

\indent The above experimental observation of the anomalous fragment angular 
anisotropy and narrowing down of the cause to a diminished width of the 
K-distribution over that expected from complete equilibration, prompted 
theoretical assumptions regarding the fusion and fission reaction mechanisms. 
   
\indent In order to explain the angular distributions of fission fragments 
in energies much lower in which quasi-fission prevails, 
Kapoor and Ramamurthy (KR)  \cite{RamPRL85*1} postulated a pre-equilibrium 
model. According to KR, for fission time smaller than 8 $\times$ $10^{-20}$ sec, the K-quantum number do not reach equilibrium and the width of the 
K-distribution $<K_0^2> < <K_0^2>_{eq}$. Hence the smaller value of $K_0^2$ 
explained larger fragment angular anisotropies. However it was difficult to 
imagine that as energy is decreased through the fission barrier, the fission 
time scale becomes smaller than 8 $\times$ $10^{-20}$ sec and pre-equilibrium
 mode becomes more dominating reaction mode, particularly for deformed target 
projectile system. 

\indent An argument circumventing the above to apply the pre-equilibrium 
model at near and below Coulomb barrier was put forward by Vorkapic and 
Ivanisevic \cite{VorPRC95*1}. According to their idea, the width of the 
K-distribution,varies with the orientation of the nuclear symmetry 
axis. In lower energies, where the fusion cross sections are 
primarily due to reactions on the polar region of the deformed target, the 
calculated fission time scales following the macroscopic properties of 
liquid drop model are smaller than the K-equilibration times and boosts 
the fragment angular anisotropy.

\indent The pre-equilibrium model of Kapoor and Ramamurthy and the later 
modified version of Vorkapic and Ivanisevic do not change the basic path 
of the system in a multi-dimensional energy landscapes, but only indicates 
that the fission following fusion can be faster than the equilibration time 
of some quantum numbers (namely the projection of spin on symmetry axis). 
However, an entirely new postulation was made by Hinde {\it et al.,} 
\cite{HindePRL95*1} to introduce a quasi-fission mechanism, analogous to the 
one observed at much higher excitation energies and nearly symmetric nuclei 
pair, 
to near and below barrier energies, but applicable to only specific deformed 
target-projectile combinations. The postulated quasi-fission mechanism, which 
points to the system going over the initial two-nucleus configuration in 
the entrance channel to a final binary fragments in a fission like reaction, 
skipping the intermediate equilibrated compound nuclear state altogether,  
resulting in enhanced fragment angular anisotropy. The onset of the 
quasi-fission depends on the relative compactness of the target and projectile
 and is assumed to be probable for the projectile hitting the "polar" region 
of the deformed target rather than the equatorial region, as the former 
configuration favours a more elongated intermediate mono-nucleus prone
to reach the saddle shape on a asymmetric mass ridge in the energy 
landscape while for the later configuration, a compact mono-nucleus 
preferentially ends up to nearly spherical compound nucleus. As the energy is 
decreased, the reaction proceeds only through the polar region and the 
orientation dependent quasi-fission dominates. 

\indent Such a reaction mode composed of two fusion-fission paths - the normal 
path over a fusion barrier to CN and then riding over a mass symmetric 
fusion barrier, and the quasi-fission path over a ridge along the fusion 
barrier for deformed target and hitting a mass asymmetric barrier, with 
characteristic widths for K-distribution. The mixture of two modes could 
successfully explain the observed angular anisotropies 
\cite{HindePRL95*1,NMPRL96*1}.

 \indent Around late nineties, experiments on the angular distribution of 
fission fragment in heavy ion reactions could be summarized as to lead to 
two view points - in large number of systems the systems follow the statistical fusion-fission path, but in deformed target projectile systems, particularly 
around and below the Coulomb barrier, notable deviations are evident, 
presumably due to pre-equilibrium effects or due to a quasi-fission mechanism 
dependent on the relative orientation of the symmetry axis of the deformed 
target with respect to the fusing projectile. To choose between the two 
explanations, the focus shifted to other experimental probes.

\indent Hinde {\it et al.,}  tried to measure \cite{HindePRC96*1} the mass 
distribution of the fission 
fragments in $^{16}$O + $^{238}$U (ground state deformation $\beta = 0.275$) 
in energies close to the Coulomb barrier. The authors {\it a priori} assumed 
that mass 
distribution for quasi fission reaction would be asymmetrical and they 
analyzed the experimental distribution of fragment masses in terms of a 
mixture of three normal distributions - one symmetric for the normal 
fusion-fission path, the other two Gaussians due to mass asymmetric 
quasi-fission. Although the authors claimed a systematic change in the 
{\it ratio} of the asymmetric to symmetric fraction, absence of an evidence 
of a discernible separation of symmetric and asymmetric mass distributions in 
the experimental data rendered the probe  doubtful. 

\indent If quasi-fission is present, the fusion process would be hindered, 
particularly at energies where the cross section for the fission reactions 
are primarily enhanced due to reaction through the polar region of the 
deformed target. Berriman {\it et al.,} \cite{NatureBerri01*1} studied the  
relative fusion  cross sections with different target-projectile combinations
 leading to the same 
compound nucleus of $^{216}$Ra. They reported inhibition of the production 
of evaporation residues (ER) in more symmetric target-projectile combinations 
of $^{19}$F + $^{197}$Au and $^{30}$Si +$^{186}$W, but no inhibition of ER  
in asymmetric $^{12}$C + $^{204}$Pb system. Later experiments on a host 
of systems on production of $^{220}Th$ , the (HI, xn) cross-sections got 
hindered as the entrance channel mass asymmetry got lowered in several systems. However, it was also noted that microscopic effects such as the binding energies of the systems also played a significant role \cite{NatureMollerNV03*1}.

\indent The above findings clearly points to the possible quasi-fission 
reactions hindering the fusion process. However, the nice scenario achieved 
in the above experiments have become hazy once again, in the report by Nishio 
{\it et al.,} \cite{NishioPRL04*1}, of their inability to confirm any 
inhibition of production of the ER's in $^{16}$O + $^{238}$U system.

\indent	The present investigations started in early 2000 and it had been 
planned that we will concentrate on investigating the fusion-fission process
 with the help of an accurate determination of the masses of fission fragments.

\indent The determination of  precise masses itself was a challenging task, particularly for fission reactions with very low cross sections. In a series of initial 
experiments on fission  fragment mass measurements, which are not reported here,
 we realised that a double arm time of flight spectrometer for simultaneous
 detection of complementary fragments over a large flight path of at least 
50 cms offers a decent method, with  mass resolutions of a few atomic mass. 
We also noted that the complementary fragments detected with a position 
sensitive detectors also enable separation of the complete fusion reactions 
with incomplete momentum transfer events. 

\indent We took up the fabrication of large position sensitive gas detectors 
with high gas gains to have a good position and timing properties. Operating 
the detectors with small gas pressures, we could virtually make the detectors 
transparent to the projectile-like particles. 

\indent	We chose the target and projectiles with the option of studying the 
reactions on spherical and deformed target. The projectiles were chosen to
 have the entrance channel mass asymmetry larger and smaller than the 
Businaro-Gallone  ridge. Simultaneous measurement of mass and angular 
distributions were done in systems where reported data on the angular 
distribution data were not available or not extensive.

\indent The experiments were done at the 15 MV Pelletron accelerator at
 the Nuclear Science Centre, New Delhi. Pulsed and bunched beams of carbon, 
oxygen and fluorine beams were used. Typically 96 hour beam slots were utilized for the experiments with 2-4 pnA.  The center also offered for scattering
 experiments, a  1.5 m diameter scattering chamber with provisions of putting 
gas detectors inside the chamber. The analysis of signals from the detectors 
and the supply of high voltage supplies, along with slow control of the chamber parameters could be handled from remote locations. Data stored on optical 
disks could be analysed off line with data analysis softwares developed 
indigenously at the laboratory at SINP, Kolkata.

\indent We studied the systems of  oxygen and fluorine on a self supporting 
bismuth target. The angular distributions and mass distributions were studied. 
The angular distributions were found out from the yields of complementary
 fragments as a function of the angle of the fragment in the {\it forward
 detector}. The mass distributions were determined from the {\it difference} of 
time of flight between the two detectors. We concluded that the systems with 
spherical  target and projectile, fusion-fission paths follow the expected 
compound nuclear fission reaction. The mass distributions show a smooth 
variation of the width of the distribution with energy. In the experiments, 
we could eliminate any effect of elastic, quasi-elastic or transfer fission 
channels and precise measurements of the masses were possible with a mass 
resolution of 3-4 a.m.u. The systematic effects for elimination of 
experimental errors were finalized in this set of experiments.

\indent In the next series of experiments, projectiles of carbon, oxygen and 
fluorine were used on the deformed target of thorium. Self supporting rolled 
foil of thorium was used as target. Since angular distributions, excitation 
functions, barrier distributions were experimentally measured earlier, only 
precise measurement of the mass distributions of fragments were attempted. 
The experiments were repeated to confirm the results. In the case of $^{19}$F+ 
$^{232}$Th, the first experiments showed decisive departures of the width of 
the mass distributions from a smooth behaviour observed in the variation of  
width with energy in $^{19}$F + $^{209}$Bi system. Similar experimental 
observations on O+Th system established that width of the mass distribution 
is also an important tool to pick up the deviations from the normal compound
 nuclear fusion-fission reactions, as effectively as in the study of angular 
anisotropy of fission fragments. The effect of the direction of flow of 
nucleons were studied in the $^{12}$C+ $^{232}$Th system, with an entrance 
channel mass asymmetry greater than the Businaro-Gallone value.   

\indent The results on the variation of the widths of the mass distribution 
in deformed target and projectile system for the first time showed a clear 
effect due to non-compound fusion-fission channel. The variations of the 
widths could be phenomenologically explained with an assumed mixture of 
normal fusion-fission and quasi fission channels. We could successfully use 
fragment mass distributions as a probe and establish that postulation of the 
quasi-fission dependent on orientation of the nuclear symmetry axes can be a
 viable explanation of the observed departure of the systems from a  purely 
macroscopic picture.

\indent	During the course of work, detailed computer simulations of the paths 
of the fusion-fission  and other probable channels through a multidimensional 
energy landscape were reported. The present investigations strengthens the 
need for more detailed calculations and simulations to get an insight into 
the dynamics of the damping  and motion of the nucleons in a quantum 
dissipative medium.

\setcounter{equation}{0}
\setcounter{figure}{0}
\chapter{Position sensitive detector}
\markboth{nothing}{\it Position sensitive detector}

\newpage

\section{Introduction}

\indent In heavy ion induced reactions at bombarding energies below 
10 MeV/{\it amu} the reaction mechanisms which contribute to the total reaction cross sections are elastic and inelastic scattering, fusion like process (fast 
fission, quasi fission and pre-equilibrium fission) and compound nucleus 
formation followed by its statistical decay. If the compound nucleus is heavy 
and fissility is higher than unity, the most probable statistical decay 
mode is fission. A significant part of the information available 
on fission has been provided by the mass-energy-velocity analysis of fission 
fragment (FF) distributions. The measurement of the mass, energy and angular 
distributions for FF gives the possibility to study the {\it dynamic aspects 
of the fission process} when the compound system moves through the saddle point to scission.  However, in such studies, it is essential to {\it separate} the 
fission fragments from a compound nuclear reaction from those 
following elastic, quasi-elastic and non-compound fission reactions. 
The separation of different reaction channels can be obtained from 
precise measurement of the linear momentum 
transferred in the reaction. A signature of the linear 
momentum transferred in the heavy ion induced fission reaction is the folding 
angle between the complementary fission fragments. So, simultaneous 
measurements  of velocity, energy and angular distributions of the two 
correlated reaction products can give information about the contributions 
of the different types of reactions as well as dynamics of fission. 
 
\indent Based on the correlation method,  we designed a spectrometer 
 for investigating the binary fission process. Instead of determining 
the fragment-mass distributions by a combined velocity-energy measurement 
\cite{SchPRB65*2} of the reaction products (which requires high individual 
resolving powers of the time and energy detectors \cite{OedNIM84*2}) we 
preferred the accurate measurement of the TOF difference of the two correlated 
fragments. The fission fragments have to be isolated in presence of large 
back ground of undesired events, such as gamma quanta, electrons, light 
charged particles and neutrons. The task becomes more difficult as the cross 
sections for fusion fission reactions falls of steeply with decreasing 
projectile energy below the Coulomb barrier. Therefore, the experimental 
arrangement should be based on large-area position sensitive gas detectors.

\indent It is from the early 1970s when the new available accelerators 
could deliver good quality beams, sophisticated low pressure gas detectors 
started to take over - covering large solid angles. These detectors efficiently 
measure the TOF, position, energy loss and are capable of handling high particle rates without any radiation damage. The development of gas detectors was 
truly revolutionized by the invention of the Multi-Wire Proportional Chambers 
(MWPC) \cite{ChaNIM68*2}  by Georges Charpak in 1968. MWPCs are suitable to 
detect heavy ionising particle selectively when operated at very low gas 
pressure. These detectors offer very high gain, fast rise time, good position 
resolution and excellent detection efficiency for fission fragments 
\cite{ChaNIM78*2,ChaNIM79*2,BreNIM79*2,BreNIM83*2,MouNIM99*2}. With judicious choice of 
gas and gas pressure, heavy ions can easily be discriminated from a 
background of light ions. In early 1980s, Amos Breskin in The Weizmann 
Institute of Science, Israel, developed a detector (known as Breskin detector) 
 \cite{BreNIM82*2,BreNIM84*2} which consisted of a pre-amplification stage 
operating as a parallel plate chamber (PPAC) directly coupled to a MWPC. 
Breskin detectors proved to be the most suitable in heavy ion induced fission 
studies at low energies. These detectors has high gain (10$^5$ - 10$^6$; 
100 times more than MWPCs), good time resolutions ($\sim$ 200 ps fwhm) and 
position resolution ($\sim$ 200 {$\mu$}m fwhm). A number of Breskin detectors 
were designed and fabricated in our laboratory at Saha Institute of 
Nuclear Physics and was used in the TOF spectrometer which was set up inside 
the 1.5 meter diameter General Purpose Scattering Chamber (GPSC) at Nuclear 
Science Centre, New Delhi.     

\indent Design and construction details of the detectors are 
described in subsection 2. Operational principle of the detector
 has been discussed in subsection 3. Subsection 4 is devoted to the 
electronics and offline test of the detector at our laboratory.

\section{Design and Construction of Detector}

\indent For the detection of low energy heavy ions, the thickness of the 
entrance window of a detector should be very thin. This demands that, 
the detector be operated at low gas pressure (less than 5 Torr) . 
But at such low gas densities, heavy particles of 
low energies have comparatively low electronic energy loss, and therefore 
produce only a few primary electrons in the gas. For example, only few 
tens of electrons are produced per mm at 1 Torr of isobutane. Hence, 
Parallel Plate Avalanche Chamber (PPAC) with typical amplification 
of 10$^4$-10$^5$ are excluded. Even low pressure MWPC with typical 
gains of 10$^5$-10$^6$ do not have sufficient amplification to provide 
good timing and position information. In PPACs and 
MWPCs, the electrons are exponentially multiplied according to their distance 
from the anode, i.e., the avalanche basically generates from the primary 
electrons produced close to the cathode. So, under the low ionization 
conditions, poor response is obtained. 

\indent To solve this problem, the idea \cite{BreNIM84*2} was to increase 
the number of primary electrons by drifting the electrons produced along 
the ion path in the collection stage into a MWPC and thereby multiplying 
each of them with an equal gain. This structure improves the primary 
ionization statistics, but, due to the drift time of the electrons, rise 
time of the anode signals become larger and hence the time resolution 
goes poorer. However, this problem can be overcome by operating the 
first stage, not as a slow collection space, but as a pre-amplification 
stage. This gives an efficient multi-step operation mode, which provides 
high total gain, first time response and good position resolution.

%\pagebreak
\begin{figure}[ht]
\begin{center}
\includegraphics[height=14.0cm]{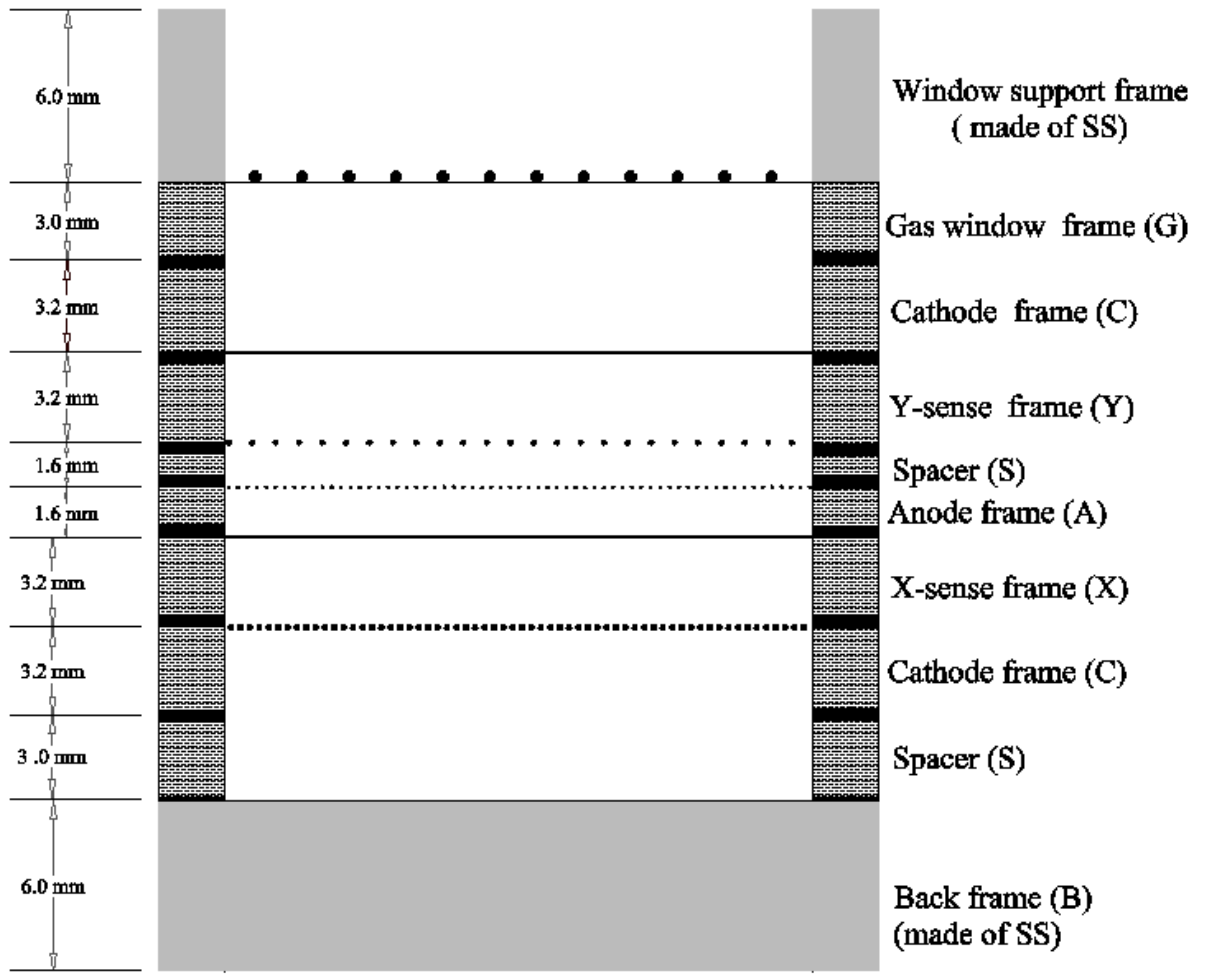}
\end{center}
\caption{\label{crosssectioanlview} Vertical cross sectional view of detector}
\end{figure}
\clearpage

\pagebreak
\begin{figure}[ht]
\begin{center}
\includegraphics[height=18.0cm]{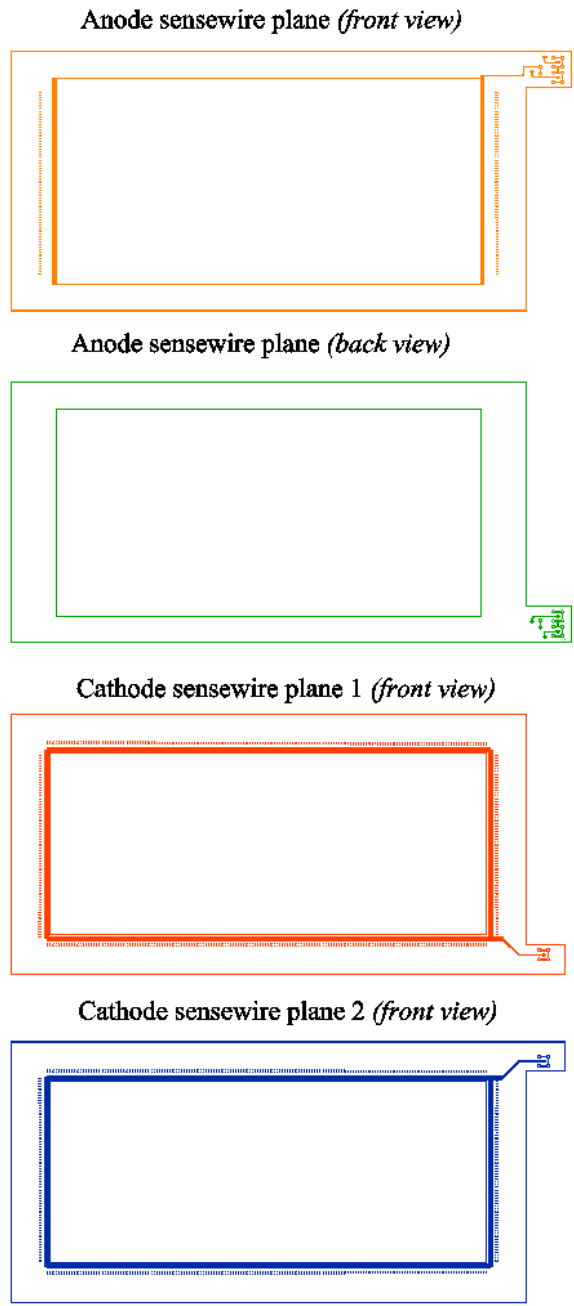}
\end{center}
\caption{\label{anodecathodethesis} Design of the PCB's 
of anode and cathode wire planes}
\end{figure}
\clearpage

\pagebreak
\begin{figure}[ht]
\begin{center}
\includegraphics[height=18.0cm]{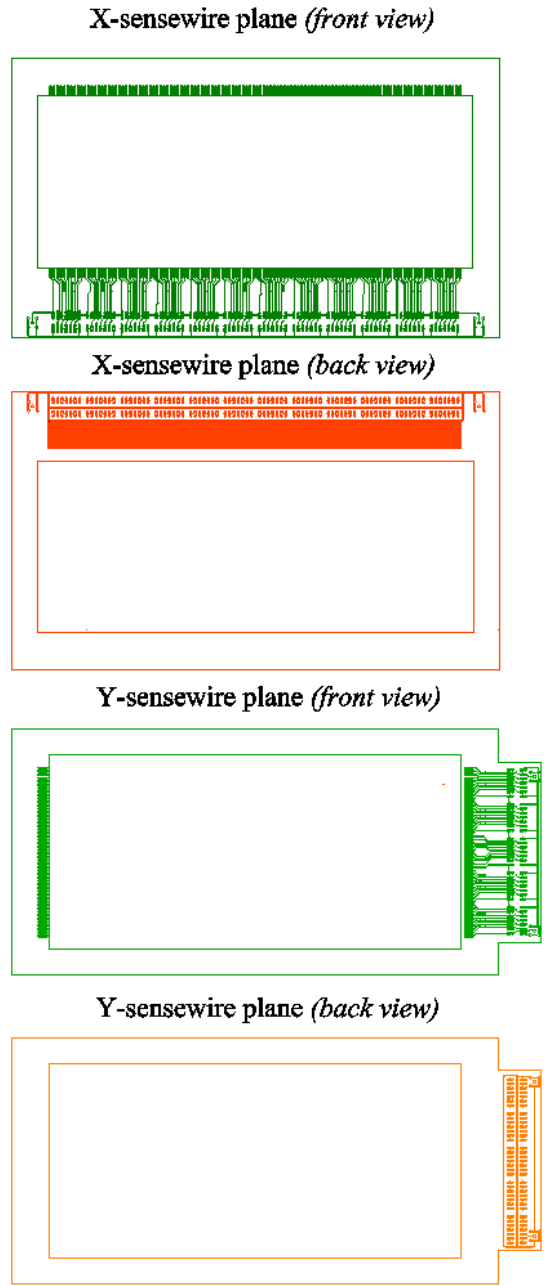}
\end{center}
\caption{\label{XYplanethesis} Design of the PCB's of 
X-sense and Y-sense wire planes}
\end{figure}
%\clearpage

%\pagebreak

\begin{figure}[h]
\begin{center}
\includegraphics[height=12.0cm]{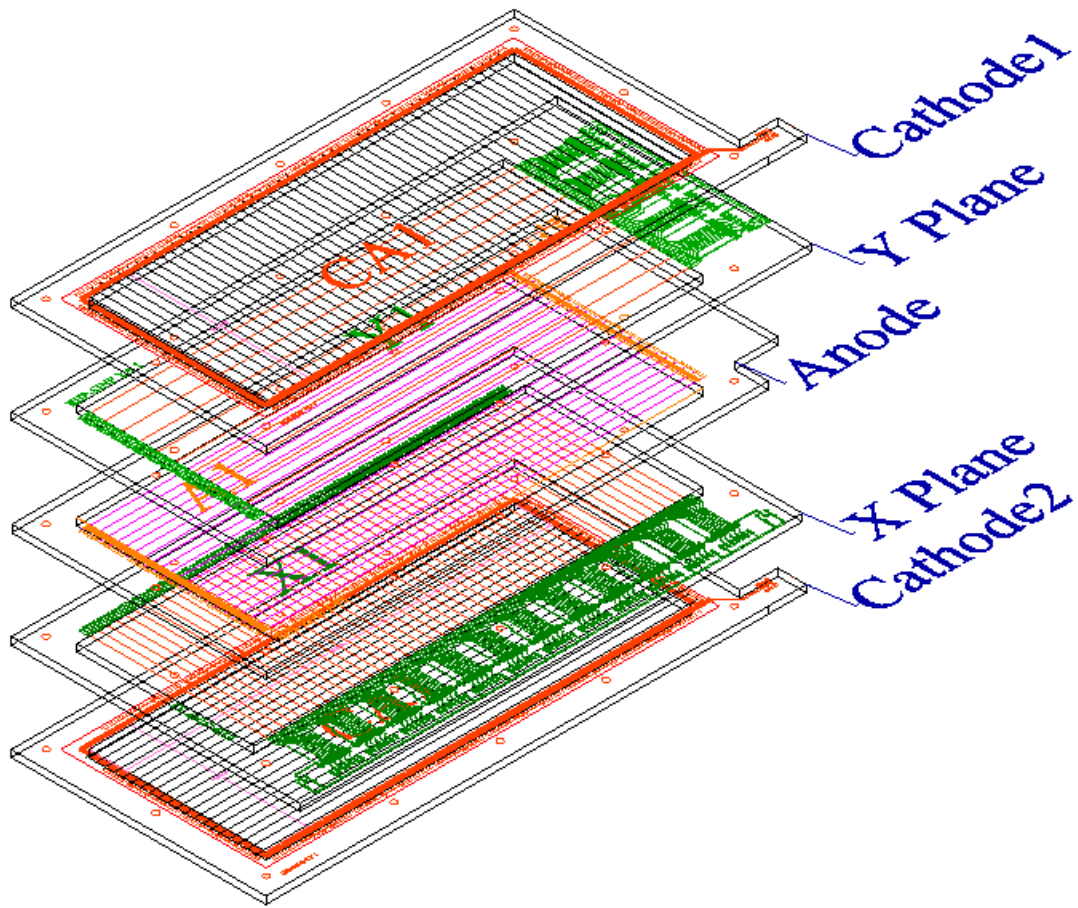}
\end{center}
\caption{\label{det3dview} View of the assembly of detector wire planes.}
\end{figure}
\clearpage

\indent The detectors we made, consisted of two PPAC stage coupled to a 
low pressure MWPC. The active area of the detector was 24 cm {$\times$} 10 cm. 
 A schematic diagram of the cross sectional view of the detector is shown in 
Fig.\ref{crosssectioanlview}. There was five wire planes, one anode (A), two 
sense wire planes (X, Y) and two cathode (C) wire planes. Design of the 
anode, cathode and sense wire  planes are shown in Fig.\ref{anodecathodethesis}
and Fig.\ref{XYplanethesis}. The anode wire planes consisted of 12.5 {$\mu$}m 
diameter gold plated tungsten wires (manufactured by LUMA, Sweden), soldered 
1 mm apart. The X and Y sense wire planes were perpendicular to each other and 
were made of 50 {$\mu$}m diameter gold coated tungsten wire, placed 2 mm apart. 
The cathode wire planes were also made of 50 {$\mu$}m (or 20 {$\mu$}m)  
diameter gold coated tungsten wire, placed 1 mm apart. The separation between 
anode and X (or Y) planes was 1.6 mm while separation between X (or Y) and a 
cathode plane was 3.2 mm. All the wire planes were made of G - 10 quality 
double sided epoxy, copper plated boards (PCB). A 3 dimensional view of the 
detector wire planes is shown in Fig.\ref{det3dview}

\indent  The position informations were derived from the X and Y sense wire 
planes with delay line read out. The X sense wire plane consisted of 
120 wires at a pitch of 2 mm and 50 wires at 2 mm pitch was used as Y sense 
wires. The position signals were read by tapped delay lines. The delay between 
successive X-sense wires was 2 ns, while that between Y-sense wires was 5 ns. 
The delay line chips, made of Rhombus Industries, USA  provided ten delay 
lines per chip and twelve such  (TZB12-5) delay lines chips were used for 
the X-sense plane where as five chips (TZB36-5) were used for Y-sense plane. 
The chips had characteristics impedance of 50 {$\Omega$} and fast rise 
time ($\sim$ 7 ns). One end of the delay line chain was terminated on board 
through 50 {$\Omega$} and signal was taken from the other end.

\indent The anode wires were soldered on to  conducting strips. 
 The cathode wires were similarly 
soldered to a conducting pad. The two cathode wire planes were shorted outside 
and connected to a power supply through a charge sensitive pre-amplifier 
(ORTEC 142 IH). This gives the provision to get the energy loss signal from 
the cathode. In X and Y sense wire planes, the wires were connected to the 
individual pad which were connected to successive pads of delay line chips. 
  
\begin{figure}[h]
\begin{center}
\includegraphics[height=9.0cm]{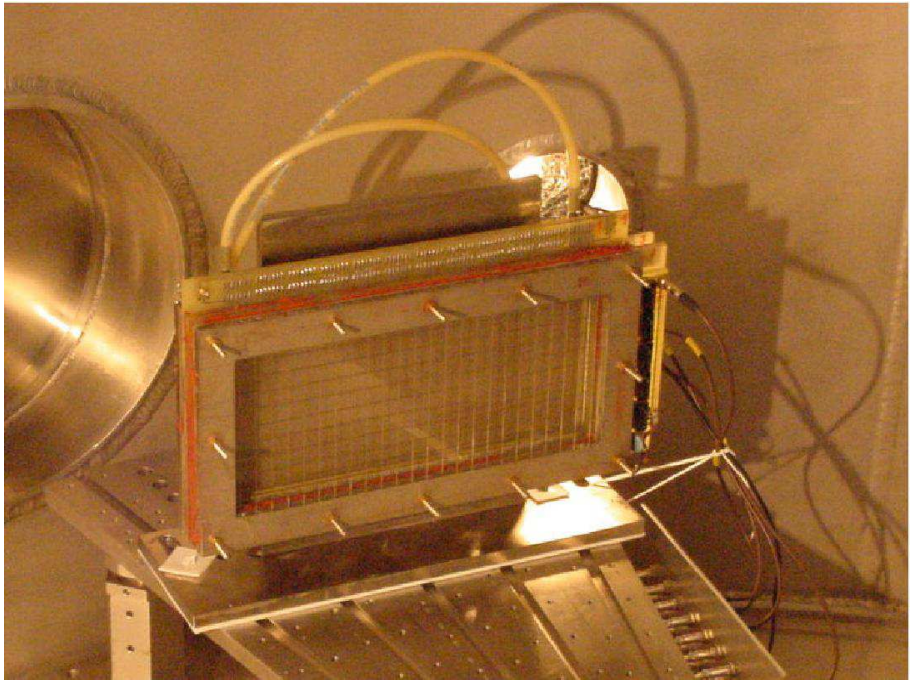}
\end{center}
\caption{\label{detphoto1} Photograph of a detector.}
\end{figure}

\indent Successive layers of PCB boards with spacers(S), also of G-10 boards 
were vacuum sealed with RTV88 sealant (General Electric, USA). Stretched 
polypropylene films of thickness 50 - 100 {$\mu$}g/ cm $^2$ were used as 
the entrance windows (G) of the detector. A 1 cm {$\times$} 1 cm wire mesh 
of stainless steel wires of diameter 0.4 mm was used as a support to the 
polypropylene film. Two gas feed-throughs were connected to the back support 
frame (B) which is made of stainless steel. Isobutane gas is continuously 
sent through the detector at a constant pressure flow mode with baratron 
feed-back closed loop flow control system (made of MKS, USA). Typical 
operating pressure was about 2 Torr. A photograph of the detector is shown 
in Fig.\ref{detphoto1}

%================================================
%COMMENTS:
%TO INCLUDE FIGURE USE THE FOLLOWING FEW LINES:
%NOTE: HERE CHAPII IS A DIRECTORY WHERE growth-mode.eps FIGURE FILE IS THERE.
%PUT AAL THE FIGURES FOR DIFFERENT CHAPTER IN DIFFERENT DIRECTORY .. PUT THE LINK
%AS BELOW.. FIGURE FILES SHOULD BE EITHER .eps or .ps FORMAT.. YOU DON'T NEED TO 
%PUT EXTENSION (.eps or .ps) IN \includegraphics LINE AS BELOW.

%===================================================

\section{Operational principle of the detector}

\noindent The properties of the multi-wire proportional counters was 
extensively investigated \cite{ChaNIM79*2,ChaRev70*2} since their introduction 
by Charpak in 1968. First measurements with MWPCs at pressure as low as 3 Torr 
showed \cite{BinonNIM71*2} that a time resolution of 2.5 ns (fwhm) could be 
reached with 5.5 MeV ${\alpha}$ particles. However, more careful study by 
 Breskin et al. \cite{BresNIM77*2} reported a time resolution about 0.8 ns at 
a pressure of 5 Torr of ethylene. This rather astonishing time performance 
was two orders of magnitude better than that usually achieved at normal gas 
pressure and was first attributed to a faster collection of electrons rapidly 
drifting in the presence of high electric field. Some further investigations
\cite{BreNIM79*2,BreIEEE80*2} had led to consider an entirely different 
operation process.

\indent In order to achieve the multi-step avalanche of electrons, the 
detectors were operated with a high reduced electric field for substantial gas 
multiplication. Typical voltages applied on anode and cathode planes were 
+280 Volt and -250 Volt respectively. The sense wire planes were grounded 
through delay lines. The operating pressure was about 2 Torr. The reduced 
electric field E/p, where E is the electric field between the cathode and 
and sense wire plane, and p is the gas pressure, was high enough (${\sim}$ 450 
Volt cm$^{-1}$/Torr) to produce secondary multiplication of the primary 
electrons produced in the region between cathode and sense wires. However,
the electric field in the cathode to anode region is a constant accelerating
field. The constant filed in this region is not qualitatively changed by the 
introduction of the grounded X, Y sense wire planes at 1.6 mm distance.
The intense field region around the central anode wire extends roughly
twenty times the diameter, {\it{ i.e.}}, in this case about 0.25 mm.  

\indent When an ionising particle passes through the detector volume 
primary ionisation are produced. In the gap between cathodes and sense 
wire planes, the detector works as a parallel plate avalanche counter 
which means the reduced field in this region is high enough to produce 
secondary electrons. The secondary electrons are accelerated by the field 
and produce a swarm of electrons which pass through the sense wire grid. In 
the sense wire to anode gap, the operation of the detector is similar to 
that of a MWPC. The swarm of electrons continue to grow and expand till 
those experience the intense attractive field around the anode. A large 
avalanche of electrons produces a fast localised current pulse on the 
concerned anode wire. The production of secondary electrons, all along the 
electron trajectories, combined with the intense avalanche near the anode 
produces very high gas amplification ($\sim$ 10$^8$). A large fast 
negative signal on the anode wire is produced which induces positive 
voltage signals in the nearby X, Y sense wire. Typical anode signal is 
$\sim$ 5 mV on 50 $\Omega$ and rise time $\sim$ 1 ns. The positive ion 
sheath is collected with in a few microseconds and the detector can withstand 
large counting rates. The position signals develop over a few adjacent 
wires for every event. The average position of the avalanche in the detector 
can be determined from the centroid of the avalanche triggering pulses in 
sense wires. These pulses are delayed by delay lines and average time 
delay measured with respect to the anode pulse is the measure of X, Y 
positions of an event.

\indent The detector was operated at pressure $\sim$ 2 torr. Hydrocarbons, 
like isobutane, n-heptane and propane are popularly used as operating gas. 
 The factors which govern the signal amplitude and pulse rise times are the 
inelastic electron scattering cross section and mean free path between 
collision and the mobility of the electron in the gas. Quick neutralisation 
of the positive ion sheath is also important. A slight admixture (1\%) of 
electronegative gas like freon may improve the performance of the detector. 
We used isobutane as a operating gas because the energy loss density 
{\it dE}/{\it dx}, is higher and so the thickness of the detector can be 
reduced. Also for this gas, the inelastic cross section for the interaction 
of electrons with the gas molecule is high. The detectors were operated 
in constant flow mode with precise electronic regulation of gas pressure.

\section{Offline test at laboratory} 

\noindent After the fabrication, the detectors were tested in the laboratory 
for uniformity of the position read outs and correspondence between the timing 
(anode pulse) and position (X-Y delay line signals). A $^{252}$Cf source was 
mounted in front of a detector which was placed inside a evacuated chamber. 

\begin{figure}[ht]
\begin{center}
\includegraphics[height=12.0cm, angle=90]{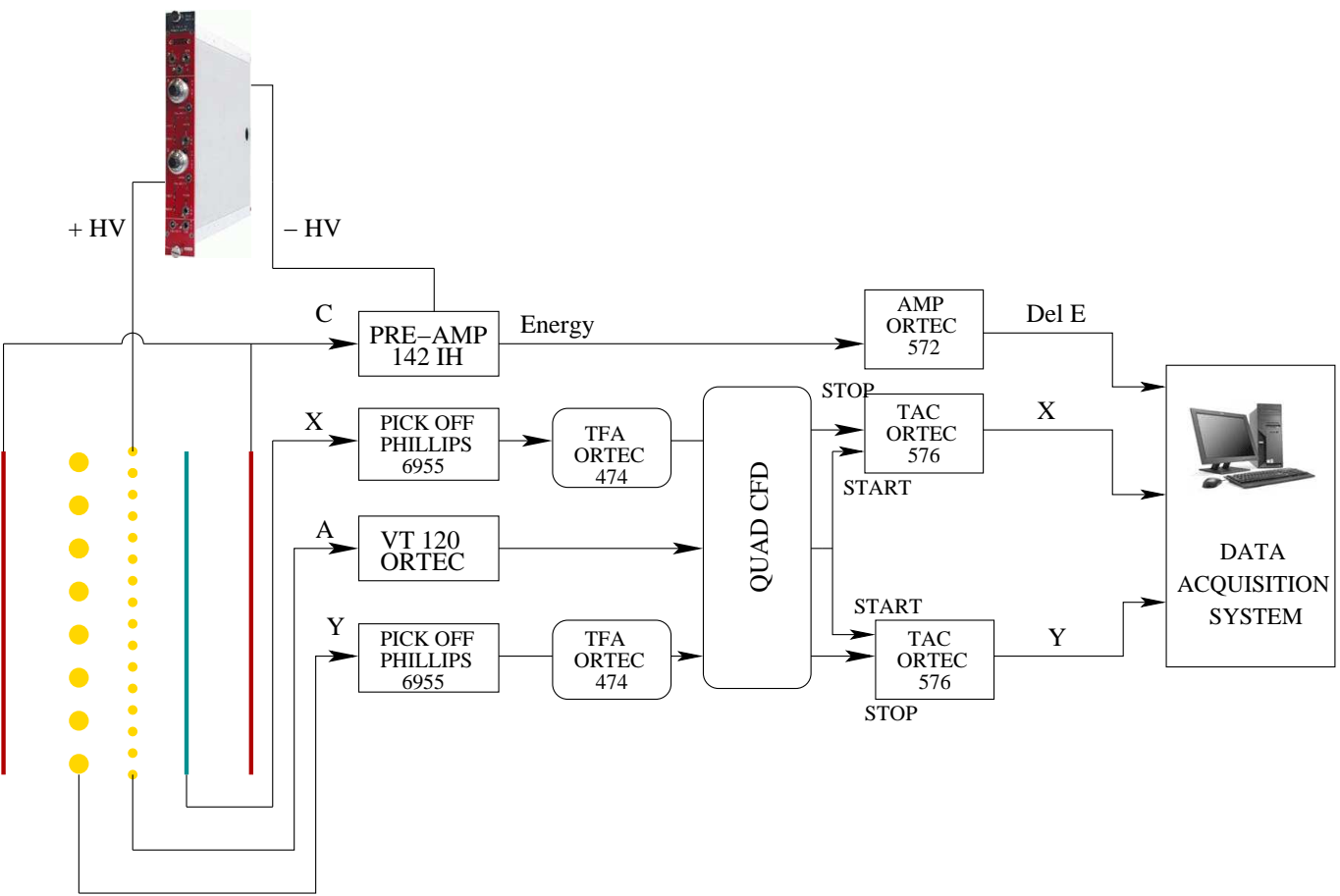}
\end{center}
\caption{\label{cirlab} Block diagram of electronics set up for operating 
a detector offline in the laboratory.}
\end{figure}

\indent The electronic set up for operating a single detector is shown in 
Fig.\ref{cirlab}. Positive and negative high voltages, required for anode 
and cathode of the detector can be supplied from the 2 fold over current 
protected high voltage supply module (N471A,CAEN). The current limit is 
set to 500 nA. 

\begin{table}[h]
\begin{center}
\caption{\label{tab:table1}~Typical characteristics of pulses from the detector
 using $^{252}$Cf source}
%\begin{ruledtabular}
\begin{tabular}[t]{||c|c||}
\hline\hline
Gas & Isobutane  \\
\hline
Operating pressure & 2.5 torr  \\
\hline
Operating bias: &  \\
Anode &  +280 Volt\\
Cathode & -250 Volt \\
\hline
Anode pulse height: &  \\
Fission fragment &  1-2 Volt\\
Alpha & 50 mV \\
Noise & 20 mV \\
\hline
Position pulse height: &  \\
Fission fragment &  400-600 mV\\
Alpha & Not observed \\
Noise & 30 mV \\

\hline\hline
\end{tabular}
\end{center}
\end{table}

The fast 
negative anode pulse is boosted by a fast current sensitive ORTEC VT120A 
preamplifier with high gain of 200 and large bandwidth of 1.2 GHz. The 
positive X, Y sense pulses from the detector are picked off  by PHILLIPS 
6955B preamplifier and are amplified and inverted by ORTEC 474 timing filter 
amplifiers (TFA) before deriving the time information. The characteristics of the test pulses of a
%The photographs of 
%the anode pulses for fission fragments and the associated X,Y sense wire 
%pulses, amplified by TFAs are shown in Pho.\ref{anodepulse, Xpulse,Ypulse}.
 
%\pagebreak
\clearpage 

\noindent  typical detector for fission 
fragments is tabulated in table \ref{tab:table1}.

\indent The negative outputs from PA and TFA are connected to TC 454 TENNELEC quad 
constant fraction discriminators (CFD). The timing informations for the 
anode and X,Y sense pulses of the detectors from each CFD are connected to 
ORTEC 567 time to amplitude converters (TAC). Appropriate delays are given 
between start (anode) and stop (X,Y) signals. The outputs from the TAC's 
measure the time differences between the arrival times of the anode and 
sense wire signals, thus representing the positions of the ionising particles 
hitting the detector. The position resolutions can be found by taking 
the image of a mask put on the detector and illuminating the detector 
with the fission fragments. The images of the stainless steel support 
wires for the thin window foils also serve as measure for position 
resolution.

\setcounter{equation}{0}
\setcounter{figure}{0}
\chapter{Experimental procedure and data analysis}
\markboth{nothing}{\it Experimental procedure and data analysis}

%\newpage
%\vspace*{5cm}
%\raisebox{12cm}{\fbox{\fbox{\parbox[b]{14cm}{
%\indent {\it  The experimental methods to measure 
%the angular and mass distributions of fission fragments following complete 
%fusion (FFCF) events for the fissioning systems under study are discussed, 
%along with the detail descriptions of the time of flight (TOF) spectrometer 
%set-up and electronics. Time calibration of the TOF spectrometer using specific%detector independence of mass or kinetic energy distributions are shown to 
%improve  mass resolution compared to the conventional calibration 
%technique using elastically scattered projectiles. Detail procedure of 
%off-line analysis of the experimental data are also discussed.}
%}}}}
\newpage

\indent The general reaction mechanism in the heavy ion induced reaction 
is quite well known. The dominant reaction is the damping of the radial 
motion and fission of the projectile and target. The combine system
 usually equilibrates in various kinematic and macroscopic degrees 
of freedom to a compound nucleus. If the fissility is large, the 
compound nucleus may also undergo binary compound nuclear fission. 

\indent The compound nuclear fission reaction is not the only reaction 
channel and the non-compound nuclear fission process like fission following 
transfer of few nucleons in several fissioning system is quite significant 
at energies near and below Coulomb barrier \cite{BiswasPRC97*3}. This mode 
of fission of a target like fragment, produced due to the transfer of a few 
nucleons, is of different characteristics vis-a-vis the direction and the
 velocities of the recoiling composite system prior to fission compared to 
the FFCF. The velocity and direction of recoil of the target like fissioning 
system depends on the momentum and the direction of the ejectiles and generally differs from the recoil of the compound nucleus in the beam direction with 
full transfer of the incoming projectile momentum. The energy, direction and 
intensity of the elastic and quasi-elastic particles may also be significant 
compared to the fission fragments and be serious contaminant of the spectrum 
of fission fragments. So, in the study of the dynamics of fusion-fission
 reaction, it is essential to separate the fission fragments (FF) from a 
compound nuclear reaction from elastics, quasi-elastics and transfer fission 
channels.

\indent The separation of fusion fission and transfer fission reaction can 
be obtained from precise measurement of the linear momentum transferred 
in the reaction. A signature of the linear momentum transferred in the heavy 
ion induced fission reaction is the folding angle between the complimentary 
fission fragments in the predominantly binary fission reaction.

\subsection{Folding angle for fusion-fission:}

\begin{figure}[h]
\begin{center}
\includegraphics[height=11.0cm,angle=0]{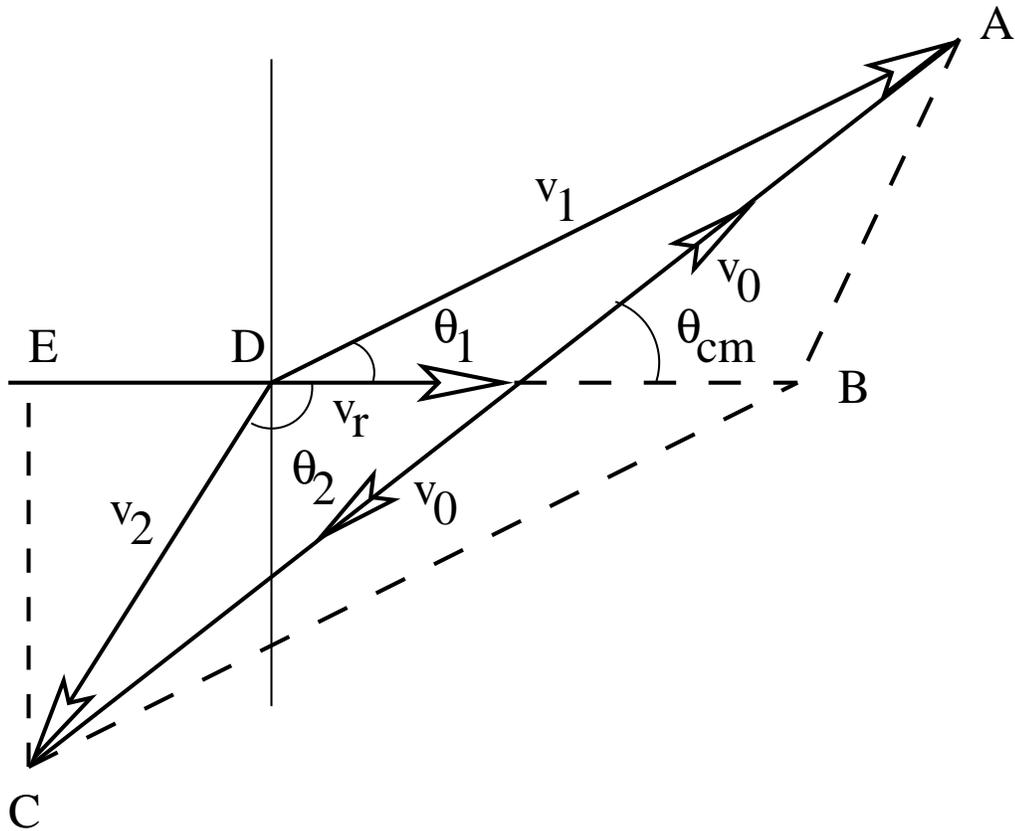}
\end{center}
\caption{\label{foldkine} Kinematics of fission following complete fusion }
\end{figure}

In compound 
nuclear reactions full projectile momentum is transferred to the fused 
fissioning system and the fragment folding angle depends on vector sums of 
the velocity of the fission fragments and the recoil velocity of the 
fissioning nucleus as shown in Fig. \ref{foldkine}.

\indent The first detector to detect the fragment F1 is kept at forward 
hemisphere (0$^\circ < \theta < 90^\circ$). The fragments F1 and F2 are 
emitted with velocities $\vec{v_1}$ and $\vec{v_2}$ in the laboratory 
frame and with velocity $\vec{v_0}$ in c.m. frame. $\vec{v_r}$ is the 
recoiling velocity of the compound nucleus. The folding angle can be 
written according to the diagram as,

\begin{equation}
\theta_{fold}=\theta_1 + \theta_2 
\end{equation}

\indent The angle $\theta_1$ is known and to calculate the folding angle, 
$\theta_2$ must be evaluated in terms of known quantities. Considering 
the geometry of the diagram,

\begin{displaymath}
tan\theta_2=-tan(\pi-\theta_2)=-\frac{EC}{DE}=-\frac{EC}{BE-BD}
\end{displaymath}

\noindent which can be written as,

\begin{equation}
tan\theta_2=-\frac{v_1sin\theta_1}{v_1cos\theta_1 - 2v_r}
=\frac{v_1sin\theta_1}{ 2v_r - v_1cos\theta_1 }
\end{equation}

\noindent Then finally, the folding angle can be written according to 
eqn.(3.1) as,

\begin{equation}
\theta_{fold}=\theta_1 + tan^{-1}[\frac{v_1sin\theta_1}{2v_r - 
v_1cos\theta_1}]
\end{equation}

\noindent From the figure, following the vectorial relationship,

\begin{displaymath}
\vec{v_0}=\vec{v_1}-\vec{v_r}
\end{displaymath}

\noindent Then,

\begin{displaymath}
v_0^2=v_1^2 + v_r^2 - 2v_1v_rcos\theta_1
\end{displaymath}

\noindent where $\theta_1$ is the angle where one detector is kept to detect 
the fragment F1. The above expression can be written as a quadratic equation of $v_1$.

\begin{equation}
v_1^2 - 2v_1v_rcos\theta_1 - (v_0^2 - v_r^2)= 0
\end{equation}

\noindent Then , solving equation 3.4, the roots can be  written as,

\begin{equation}
v_1=\frac{1}{2}\left[ 2v_rcos\theta_1 \pm \sqrt{\{ 4v_r^2cos^2\theta_1 + 4(v_0^2 - v_r^2) \}} \right]
\end{equation}

\noindent The positive root is considered only, as for negative root, for 
$\theta_1$ increasing $v_1$ becomes negative. The velocity of the fragment F1 
can be written as,

\begin{equation}
v_1=v_rcos\theta_1 + \sqrt{\left(v_r^2cos^2\theta_1 + v_0^2 - v_r^2 \right)}
\end{equation} 
 
\noindent The recoil velocity can be calculated as,

\begin{equation}
v_r=\sqrt{\frac{2E_r}{A_{pt}}}
\end{equation} 

\noindent where $E_r$ is the recoil energy. The mass of the fissioning system 
is $A_{pt}$. The recoil energy can be calculated from the incident momentum 
$\vec{p_i}$ because

\begin{displaymath}
\vec{p_i}=\vec{p_r}
\end{displaymath}

\noindent where $p_r$ is the recoil momentum. Then the recoil energy is,

\begin{equation}
E_r=\frac{p_i^2}{2A_{pt}}
\end{equation}

\noindent The fragment velocity in c.m. frame can be calculated from the
average total kinetic energy released in fission process.
 
\begin{equation}
v_0=\sqrt{\frac{2<E_k>}{A_{pt}}}
\end{equation}

\indent The energy $<E_k>$ can be determined from the mass $A_{pt}$ and 
the charge $Z_{pt}$ of the fissioning nucleus assuming symmetric mass split, 
following Viola's systematics \cite{Viola85*3}

\begin{equation}
<E_k> = \left[ 0.1189 \frac{Z_{pt}^2}{A_{pt}^{\frac{1}{3}}} 
+7.3  \right] MeV
\end{equation}

For the fissioning system $^{19}$F + $^{232}$Th the value of $<E_k>$ and 
$v_0$ are 192.04 MeV and 1.21 cm/ns respectively. With the quantities 
involved, calculated in this way, the folding angle can be determined for 
fission following complete fusion reactions.

\indent To convert the values of $\theta_1$ in lab. frame to c.m. frame, 
a relation must be established between $\theta_1$ and $\theta_{cm}$. From the 
Fig \ref{foldkine}, it can be written that

\begin{displaymath}
tan\theta_{cm}= \frac{v_1sin\theta_1}{v_1cos\theta_1 - v_r}
\end{displaymath}

\noindent So the relation between the angle of two frames of reference is

\begin{equation}
\theta_{cm}=tan^{-1}\left[\frac{v_1sin\theta_1}{v_1cos\theta_1 - v_r}\right]
\end{equation}

\noindent The angle in c.m. frame $\theta_{cm}$ can be evaluated in terms 
of known quantities, $v_1$ and $v_r$.

%%%%%%%%%%%%%%%%%%%%%%%%%%%%%%%%%%%%%%%%%%%%%%%%%%%%%%%%%%%%%%%%
%\begin{figure}[h]
%\begin{center}
%\includegraphics[height=10.0cm]{CHAPIII/kine1}
%\end{center}
%\caption{\label{kine1} Kinetic diagram for FFCF reaction with the 
%compound nucleus recoiling in the beam direction with a momentum, 
%$p_r=p_i$ where $p_i$ being the incident momentum. }
%\end{figure}
%%%%%%%%%%%%%%%%%%%%%%%%%%%%%%%%%%%%%%%%%%%%%%%%%%%%%%%%%%%%%%%%%

\begin{figure}[h]
\begin{center}
\includegraphics[height=10.0cm]{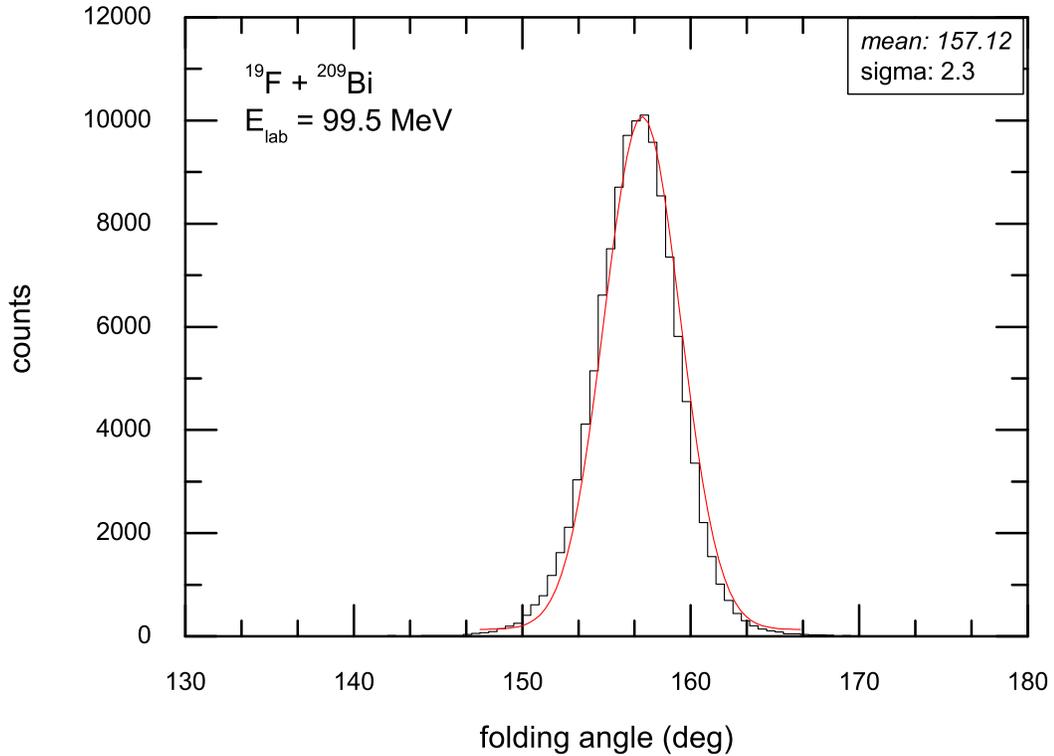}
\end{center}
\caption{\label{fold_100} Folding angle distributions for the all fission 
fragments in the reaction $^{19}$F + $^{209}$Bi reaction at 99.5 MeV. The 
Gaussian fit is shown by solid red line. }
\end{figure}

The folding angle and the width of the folding function for complementary 
fragments is a slowly 
varying function of the angle of emission of the fragment in laboratory
frame. The width of the folding function (angular range of the folding 
angle between complementary fragment pairs) depends on the kinematic 
factors due to asymmetric mass splitting and variation of total fragment 
kinetic energy on asymmetric mass splitting, and to a large extent due to 
pre and post saddle neutron emission. The velocity of the fission fragments 
can be calculated assuming symmetric mass splitting from the 
phenomenological rules (Viola's systematic \cite{Viola85*3}) and the 
velocity of the recoil. The distribution of folding angle is experimentally 
found to of Gaussian shape to even very forward angles of the detector. 
A typical distributions of folding angles for complementary pairs for 
the reaction $^{19}$F + $^{209}$Bi reaction at 99.5 MeV is shown in 
Fig \ref{fold_100}. The reaction cross section is almost purely 
FFCF ($>$ 98$\%$) and the folding angle distribution is well fitted with 
a single Gaussian.

%%%%%%%%%%%%%%%%%%%%%%%%%%%%%%%%%%%%%%%%%%%%%%%%%%%%%%%%%%%%%%%%%
%%%%%%%%%%%%%%%%%%%%%%%%%%%%%%%%%%%%%%%%%%%%%%%%%%%%%%%%%%%%%%%%

\begin{figure}[h]
\begin{center}
\includegraphics[height=12.0cm]{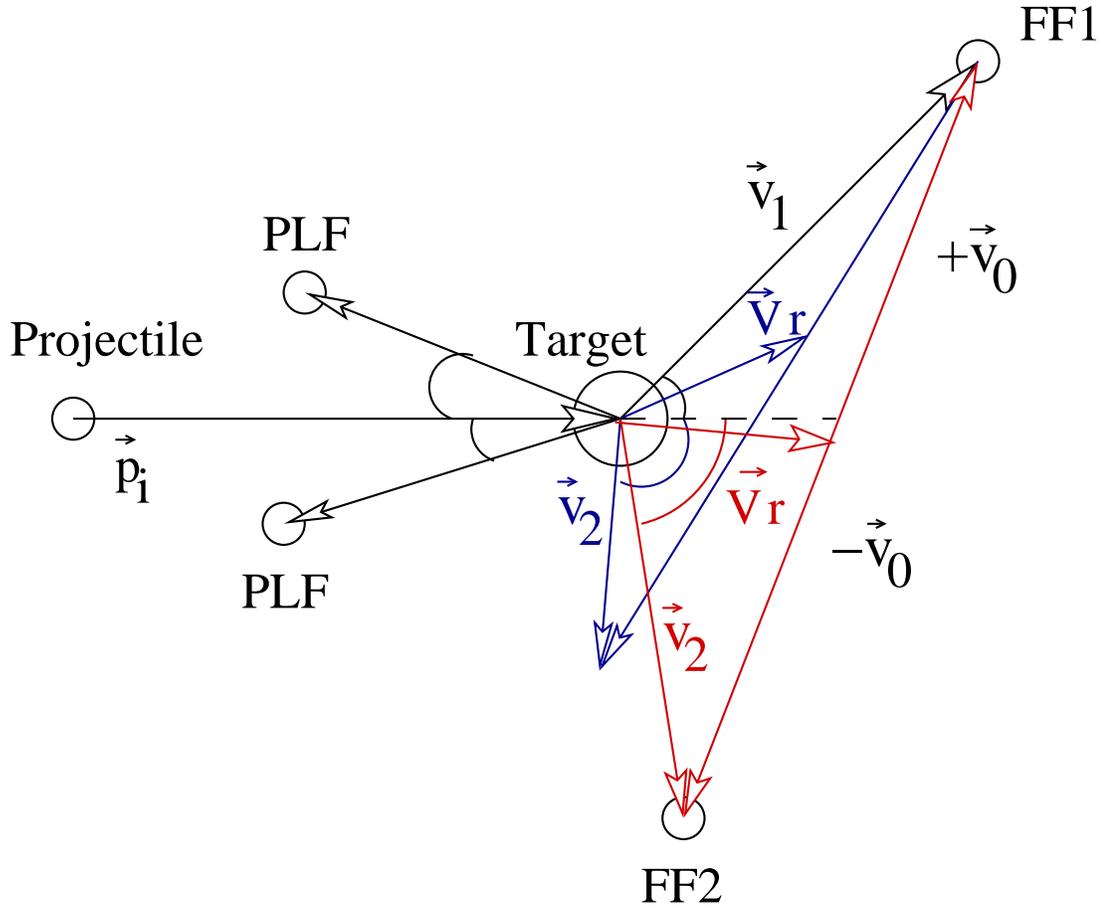}
\end{center}
\caption{\label{kine2} Kinetic diagram for target like fragment fission 
with target like fragment recoiling with a momentum, $p_r > p_i$ (red line) 
 and $p_r < p_i$ (blue line). }
\end{figure}

\indent At beam energies above Coulomb barrier, in transfer reactions, 
typically a few nucleons are transferred to the target and the erectile is 
emitted mostly in the forward directions and takes away a large portion 
of the incident momentum. As a result, fission takes place from a slower 
recoiling nucleus compared to that to the case of the compound nucleus 
in FFCF reaction. Thus the fragment folding angle in the case of TF is 
larger than that in FFCF (shown in Fig \ref{kine2}).

%%%%%%%%%%%%%%%%%%%%%%%%%%%%%%%%%%%%%%%%%%%%%%%%%%%%%%%%%%%
%%%%%%%%%%%%%%%%%%%%%%%%%%%%%%%%%%%%%%%%%%%%%%%%%%%%%%%%%%%%%

\begin{figure}[h]
\begin{center}
\includegraphics[height=11.0cm]{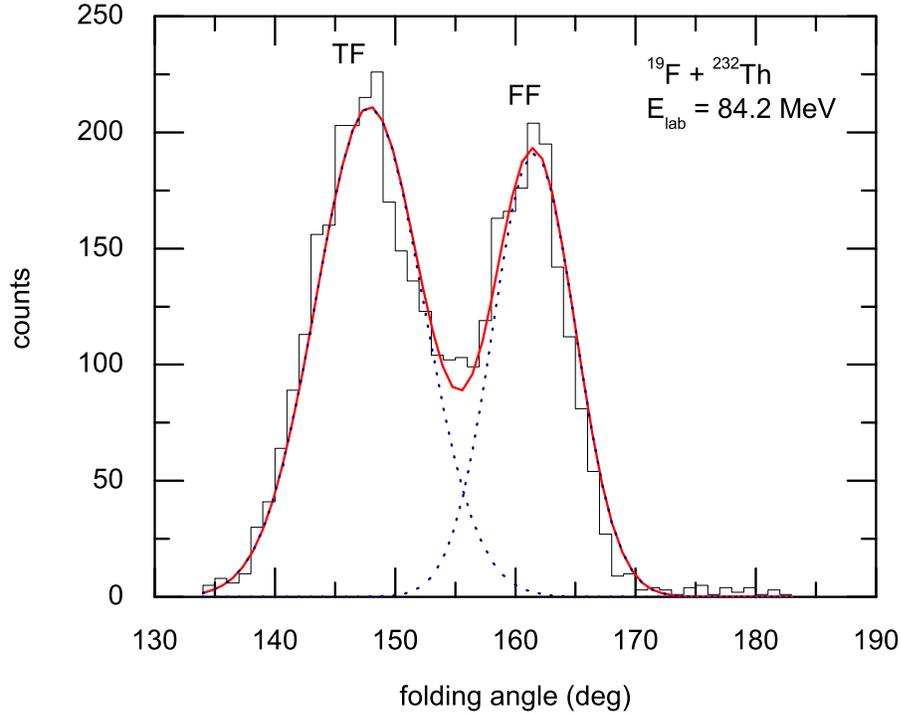}
\end{center}
\caption{\label{fold_fth86} Folding angle distributions for the all fission 
fragments in the  $^{19}$F + $^{232}$Th reaction at 84.2 MeV. 
Fusion fission (FF) and transfer fission (TF) component are fitted by 
two Gaussians ( dotted blue lines). The overall fitting is shown by   
solid red line. }
\end{figure}

However at sub-barrier energies, heavy ion transfer cross-sections peak in 
the backward angles, and the backward moving ejectiles imparts more momentum 
to the target than even in compound nuclear reaction. Consequently the 
fragment folding angle is smaller than that in FFCF (shown in Fig 
\ref{kine2}). The folding angle distributions for FFCF are thus 
separable from that of TF at much above and below Coulomb energies more 
efficiently. The width of the folding angle distributions for TF is much 
wider than that for compound nuclear fission. The kinematic broadening is 
larger due to large variations in the recoils and is a complicated function 
of the angle of the detectors. The folding angle distributions are also 
increasingly non-Gaussian in forward backward orientations of the 
complementary fragment detectors. 
Typical mixtures of folding angles for TF and FFCF reactions are shown 
in Fig \ref{fold_fth86}  for the $^{19}$F + $^{232}$Th at 84.2 MeV energy, 
where TF reaction cross section is significant. 

%%%%%%%%%%%%%%%%%%%%%%%%%%%%%%%%%%%%%%%%%%%%%%%%%%%%%%%%%%%
%%%%%%%%%%%%%%%%%%%%%%%%%%%%%%%%%%%%%%%%%%%%%%%%%%%%%%%%%%%%%%%%

\section{Conventional experimental techniques}

\begin{figure}[h]
\begin{center}
\includegraphics[height=7.0cm]{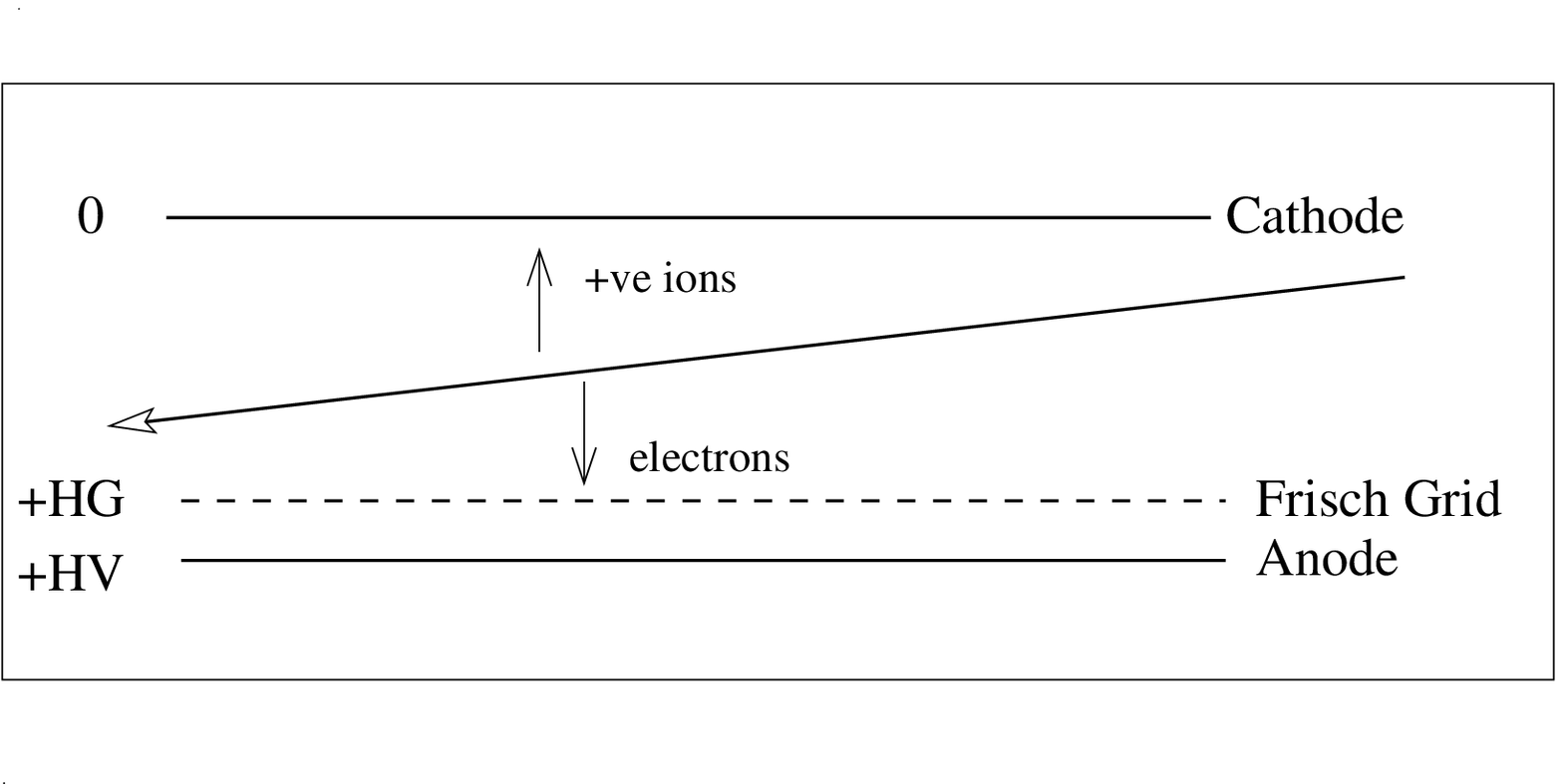}
\end{center}
\caption{\label{ic} Schematic diagram of an ionisation chamber.}
\end{figure}

\indent Probably the most simple device to detect individual fission 
events is an ionisation chamber (IC). A schematic diagram of IC is shown 
in Fig. \ref{ic}. IC is a gas filled parallel plate 
chamber with a voltage applied across the electrodes. When an ionizing 
particle like fission fragment travels through the gas volume, the 
electrons and positive ions will drift in the electric filed of the 
chamber toward the anode and cathode respectively. Usually the anode 
is shielded by a high transparent wire mesh, a so called Frisch grid, 
placed just in front of the anode. The Frisch grid has the effect that 
the movement of electrons and ions within the grid-anode gap are 
sensed by an electronic amplifier coupled to the anode, thus eliminating 
the spatial effect on the charge collection.
To a good approximation the total charge accumulated 
on the anode is a measure for the kinetic energy E of fission fragment 
being stopped in the gas volume. For binary fission events the two 
complementary fragments are recorded in coincidence in the two chambers.

%%%%%%%%%%%%%%%%%%%%%%%%%%%%%%%%%%%%%%%%%%%%%%%%%%%%%%%%%%%
%%%%%%%%%%%%%%%%%%%%%%%%%%%%%%%%%%%%%%%%%%%%%%%%%%%%%%%%%%%%%%%%

 Provided the detector has been properly calibrated, the total kinetic 
energy $E_{K}$ can be calculated from the energies $E_{L}$ and $E_{H}$ of the 
individual fragment as 

\begin{equation}
E_{K}=E_{L}+E_{H}
\end{equation}

\noindent where the indices L and H stand for the light and heavy fragment, 
respectively. From the correlated energies obtained from in the double 
energy  experiment, the fragment masses may also be found. Neglecting 
the prompt neutron evaporation, mass conservation for the primary 
fragments give,
 
\begin{equation}
M_{F}=M_{L}+M_{H}
\end{equation}
 
\noindent with $M_{L}$ and $M_{H}$ the masses of the primary fragments 
and $M_{F}$ the mass of the fissioning nucleus. From momentum conservation 
in the centre of mass system of the fissioning nucleus one gets in the 
non-relativistic limits,

\begin{equation}
M_{L}V_{L}=M_{H}V_{H}
\end{equation}

\noindent where $V_{L}$ and $V_{H}$ are the velocities of the primary fragments.
The above equations gives

\begin{equation}
\frac{E_{L}}{E_{H}}=\frac{M_{H}}{M_{L}}
\end{equation}
 
\noindent with $E_{L}$ and $E_{H}$ the kinetic energies of the primary 
fragments. Once these energies are known, the fragment masses are readily 
calculated from equations 3.13 and 3.14.

\indent Unfortunately, the determination of fragment masses from a
double energy (2E) measurement experiment is accurate  if the 
energy losses in the target is negligible and  no neutrons are evaporated. 
 In the large majority 
of events the evolution scheme laid down in equations 3.13 and 3.14 has to 
be modified and fragments after particle and neutron emission are 
detected. It is known that a number of prompt neutrons are emitted from 
the fully accelerated fragments. Assuming $\nu$ neutrons were evaporated 
isotropically relative to the fragment, the energy E would simply be reduced 
by a factor $(M-\nu)/M$, i.e., a mere shift in energy. A further complication 
arises with the neutron emission number $\nu$ being not constant but showing 
a distribution. The average emission numbers depend furthermore on the fragment  mass. Finally, any finite velocity of the neutron relative to the fragment 
will impart a momentum to the residual fragment upon neutron evaporation.
Therefore, starting with a primary fragment at fixed energy (say $E^\star$), 
 the observed energy E will not only be shifted, but also broadened into a 
distribution around an average energy $<E>$.

\indent At the time solid-state devices came into widespread use in 
electronics, the gas filled ionisation chambers were replaced by solid state 
detectors. The solid state devices are rugged and easy to use and show 
reasonably good energy and timing resolutions. For fission fragments the 
energy resolution (FWHM) $\delta$E achieved $=$ 1 to 2 MeV. However, there 
is one drawback of using solid state detectors. They suffer from a size-able
 pulse-height defect, i.e., for fixed incoming energy, heavy ions  produce 
a smaller integrated charge at the electrodes of the junction than light 
ions (e.g., $\alpha$ particles).

\indent The main tool to investigate the kinematics of fission fragments is 
the {\it time of flight} (TOF) method. In this approach the time of flight 
for a given flight path and thus velocity of the fragment is determined. 
Both timing 
devices, at the start and the stop of the flight path should of course 
have the best feasible time resolution. An additional requirement for 
the start detector is, however, that the least possible amount of material 
should be placed in the fragment path, in order to avoid excessive 
velocity and angular straggling being introduced by the measurement.
Energy loss in the start detector also affects the ultimate mass resolution. 
Start time is deduced from detector, usually a PPAC \cite{SwanNIM94*3,  
ArefNIM94*3, KumaNIM01*3} or a micro-channel plate picking secondary 
electrons 
produced in passage of the particles through a foil. The start time can also 
be picked off from the  pulsing 
of a bunched beam \cite{AhmNIM85*3}. For a pulsed beam, time spread of 
the bunched beam limits the achievable mass resolution.  The stopping time
 can be derived from a PPAC or  from a silicon detector. The TOF has to be 
calibrated by elastically scattered particles and the mass resolution 
critically depends on the flight path and the accuracy of flight time 
measurement.

\indent Thus, it is straightforward to perform a double velocity (2V)
 experiment where the complementary fragments are detected in coincidence.
 Mass of the fragments can be calculated from the 2V data using the mass 
and momentum conservation for the primary fragments prior to neutron emission.
As already discussed for a 2E-experiment, one again has to be aware of neutron 
emission.  Unlike kinetic energy, however, the fragment velocity is on 
the average not shifted by the isotropic evaporation of neutrons. However, 
for a fixed initial velocity V$^\star$, neutron emission will introduce a 
spread in the measured final velocity V of the secondary fragments. The 
variance of V leads in turn to a variance of the calculated mass M$^\star$.
It can be shown that \cite{TerrelPR62*3} that the mass variance of a 2V 
experiment is only about $1/4$ that of a 2E experiment. In this 
sense 2V data are superior in quality compared to 2E data.

\indent The measurements of angular distribution require separation of 
elastic and quasi-elastic reaction channels and the conventional energy 
measurements \cite{KaiPRC91*3, ZhaNPA92*3} with silicon or ionization chambers 
are not efficient to completely separate the the contaminants due to 
large energy straggling of fission fragments. Moreover, fragments due to 
different reaction mechanism (e.g., compound and non-compound channels) can 
not be separated in the above method. Position sensitive gas detectors with 
small radiation length are effectively transparent to elastic and 
quasi-elastic particles in low mass heavy ion ($<$ 40) induced fission 
experiments and were used in complimentary fission fragment detection method
 to completely eliminate contamination from elastic and quasi-elastic channels.

\indent To get accurate mass and angular distribution, we  applied 
a double arm TOF system using fairly large, low pressure, position sensitive 
Breskin detectors and tuned the measurement of the fissioning systems produced 
in pulsed light heavy ion beam induced fission of heavy targets. The system was 
optimized  for the available flight path and the experimental cross section. 
The details of the method has been extensively discussed in the 
following sections.

\section {Experimental set up} 

\indent To measure the TOF and the folding angle 
between two complementary fission fragments, two large area position 
sensitive detectors were employed. The construction and performance of these 
detectors have been discussed in chapter 2. The detectors were used to 
detect complementary fragments from binary fission.

\indent The experiments were performed using a 15UD Pelletron facility 
of Nuclear Science Centre (NSC), New Delhi, India. The detectors and other 
necessary equipments were  setup  in a large scattering chamber of 1.5 meter 
diameter, popularly referred as General Purpose Scattering Chamber (GPSC) . 
 The two arms in the scattering chamber
 can be rotated in the reaction plane over a wide angular range by means 
of motor driven pulley. The angular positions of the arms, in reference 
to their central line, can be read from outside from a circular scale with 
a vernier, coupled to the rotational movement of the arm. The height of 
the arms from the floor of the scattering chamber was adjustable.
Two detector stands were made in which the detectors can be mounted at
different height to keep the detectors in the reaction plane and electrically 
isolated from the body of the chamber. 
Two stands were fixed generally on the central line of the two arms. 
 Two identical
detectors of active area 24 cm $\times$ 10 cm were mounted on the 
two detector stands with anode plane normal to the particle trajectories 
passing through the center of the detectors. Care was taken 
to position the detectors (detector stands) vertically on the arms and 
secure them rigidly to avoid any accidental movement of the detector in 
course of the 
experiment. These precautions were needed as any small angular deviation 
of the detector in polar or 
azimuthal directions from the normal positions produces systematic errors 
in the calculation of the flight paths for fission fragments hitting 
away from the center of the detector. 

\begin{figure}[h]
\begin{center}
\includegraphics[height=10.0cm]{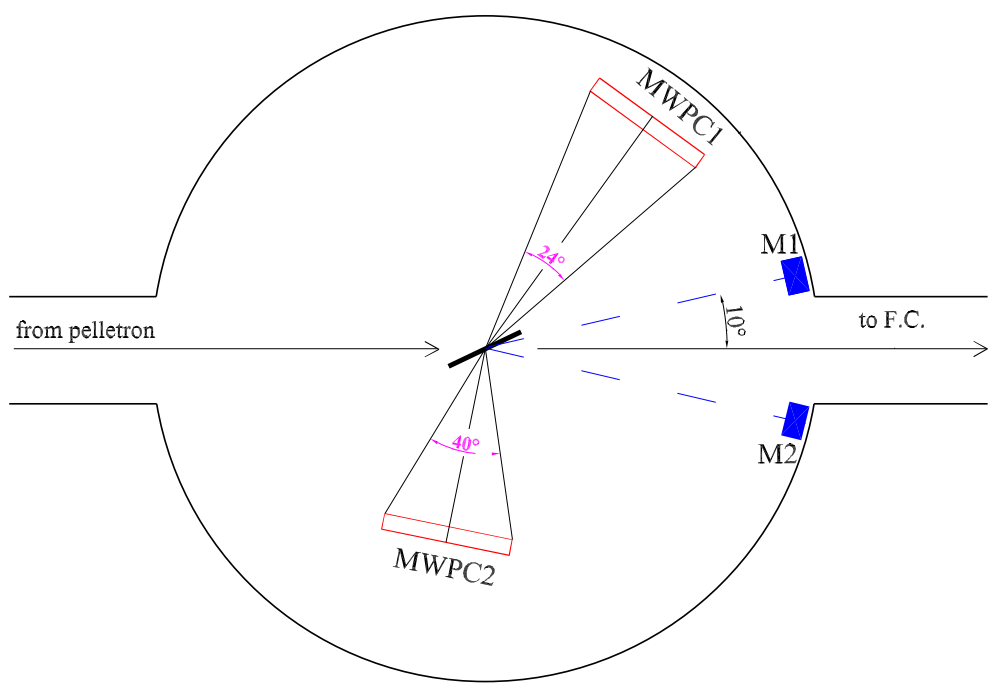}
\end{center}
\caption{\label{exsetup} Schematic diagram of the experimental set-up. }
\end{figure}

\begin{figure}[h]
\begin{center}
\includegraphics[height=10.0cm]{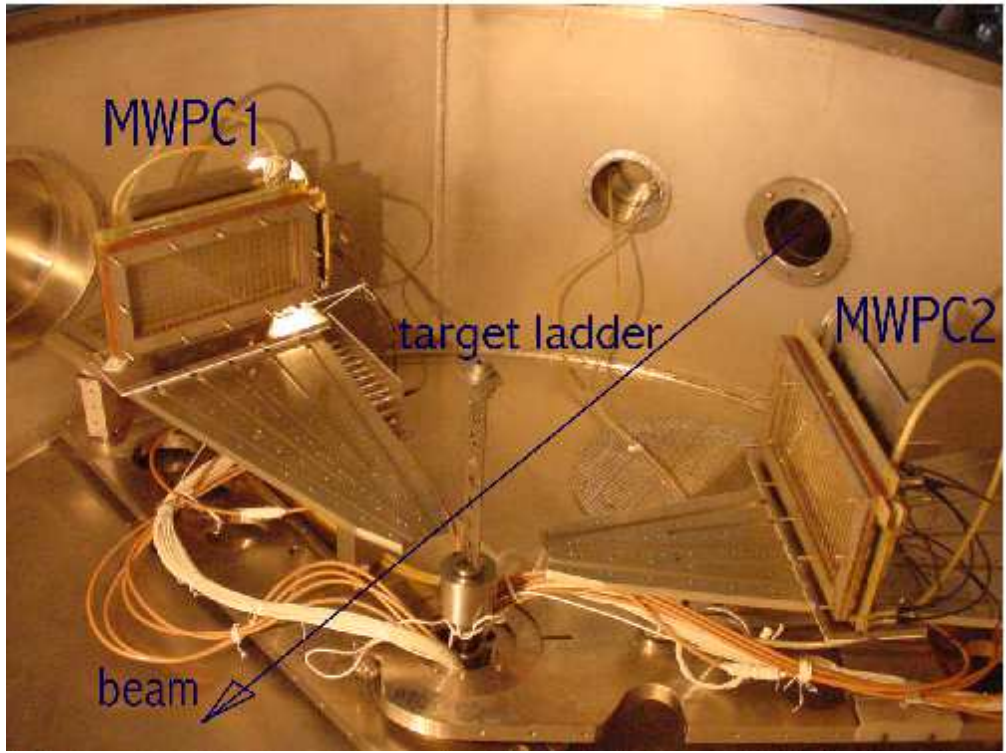}
\end{center}
\caption{\label{tofsetup} TOF set up inside 1.5-meter diameter
scattering chamber at Nuclear Science Centre, New Delhi. }
\end{figure}

\indent The schematic of TOF arrangement for complementary fission 
fragments is shown in Fig. \ref{exsetup}. The detectors labeled MWPC1 and MWPC2 were kept at a 
distance of 52.6 cm and 33.2 cm , respectively from the target.
 The backward detector (MWPC2) 
was kept at smaller distance  to ensure the full folding angle coverage
 for the fission fragments detected in forward (MWPC1) detector. The 
inner view of the scattering chamber is shown in Fig. \ref{tofsetup}.

\indent The target ladder, made of stainless steel is placed at the 
center of the scattering chamber. The ladder can be rotated 
about its own axis and adjusted for different heights vertically from outside 
the chamber. The alignment of the incident 
beam upon the target was adjusted, if necessary, by illuminating a quartz 
target by the beam. During 
the experiment, if necessary,  the target 
ladder can be removed/reintroduced into the chamber, for changing targets, 
with nominal  disturbance to the vacuum of the chamber.

%\clearpage

%\pagebreak

%\clearpage

\indent The beam is dumped on a Faraday cup with suppression for secondary 
electrons. The beam current is measured by a current integrator. Two 
silicon surface barrier detector of thickness 300 $\mu$m, with a slit 
diameter of 2.0 mm in front of it, was kept at angle $\pm 10^{\circ}$ with 
respect to the beam at a distance about 70 cm from the target to monitor the 
yields for elastically scattered particles and were used to normalise the 
fission fragment yields of the PSD's. One of these solid sate detectors was 
also used for on-line monitoring of the time structure of the beam.

\indent The operating pressure in the scattering chamber was required to 
be of the order of 3 $\times$ 10$^{-6}$ mbar. The MWPC's within the chamber 
were connected to a separate gas handling system through two gas feed throughs 
attached to one of the side ports of the scattering chamber. One connecting 
valve was provided between the scattering chamber and the entrance feed-through  of the detector. Cautious handling was required during the process of pumping 
down the system,  operation of the detectors and allowing air into the system  
as the gas windows of the detectors were very thin ($\sim$ 50 $\mu g/ cm^2$). 
A small positive pressure with respect to the chamber needed to be maintained 
inside the detector during slow pumping down ($\sim$ 10 mbar/min) or letting in air into the chamber. During pumping the detectors and the targets were kept 
in such a way that these avoided any direct blast of air. After the rough 
vacuum of the order of 10$^{-3}$ mbar was achieved, the chamber is isolated 
from the detectors and  further pumped down to about 3 $\times$ 10$^{-6}$ mbar. For operation of the gas detectors with a steady flow of gas, the gas flow 
through the detectors were controlled at pressures about 2 mbar by an 
electronic pressure controller (MKS, USA).

\section{Electronic setup}

\begin{figure}[h]
\begin{center}
\includegraphics[height=12.0cm,angle=0]{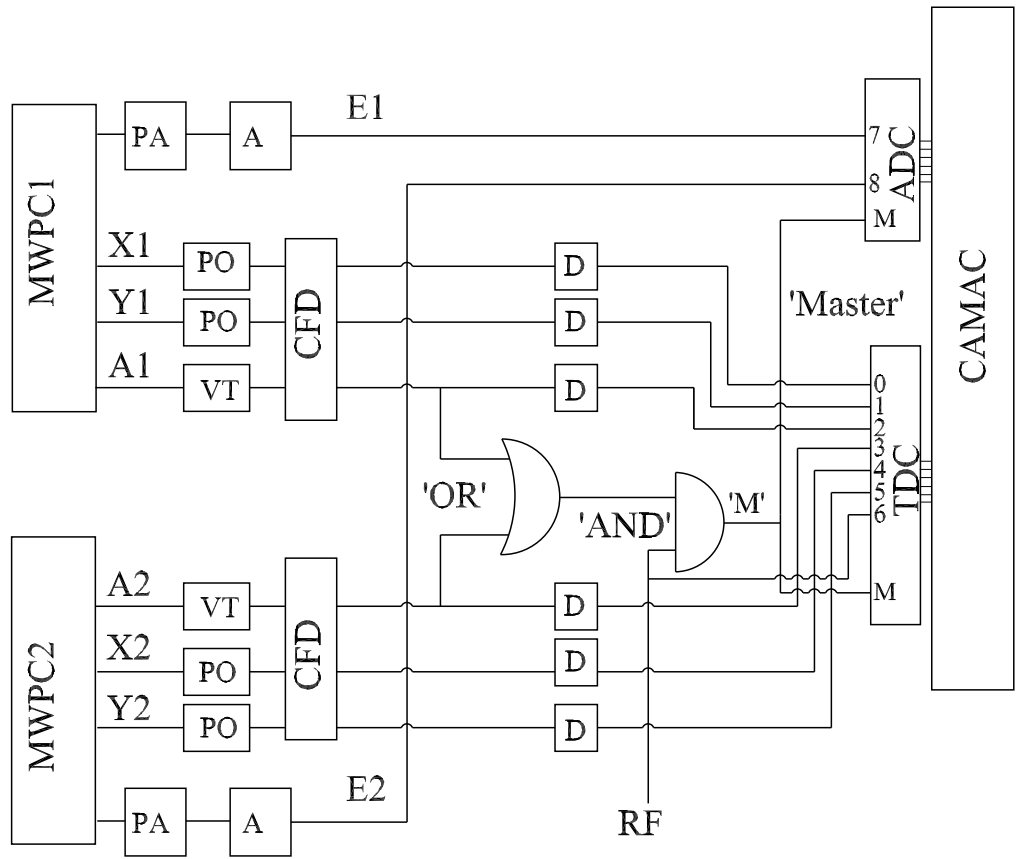}
\end{center}
\caption{\label{circuittof} The schematic of the electronic set up for 
operating the time of flight spectrometer and the acquisition of data.
MWPC(1,2) refer to the detectors. ($A_i, X_i, Y_i$) are the anode, X and 
Y signals. The CFD's refer to the constant fraction discriminator. PO and 
VT refers to the pulse pick-off and wide band voltage sensitive amplifiers, 
while PA and A are charge sensitive preamplifier and shaping amplifiers 
producing energy signals E(1,2) from cathodes. The variable delay generator are labeled as D and the multiplexing/coincidence gates are shown as OR/AND .}
\end{figure}

\begin{figure}[h]
\begin{center}
\includegraphics[height=18.0cm,angle=0]{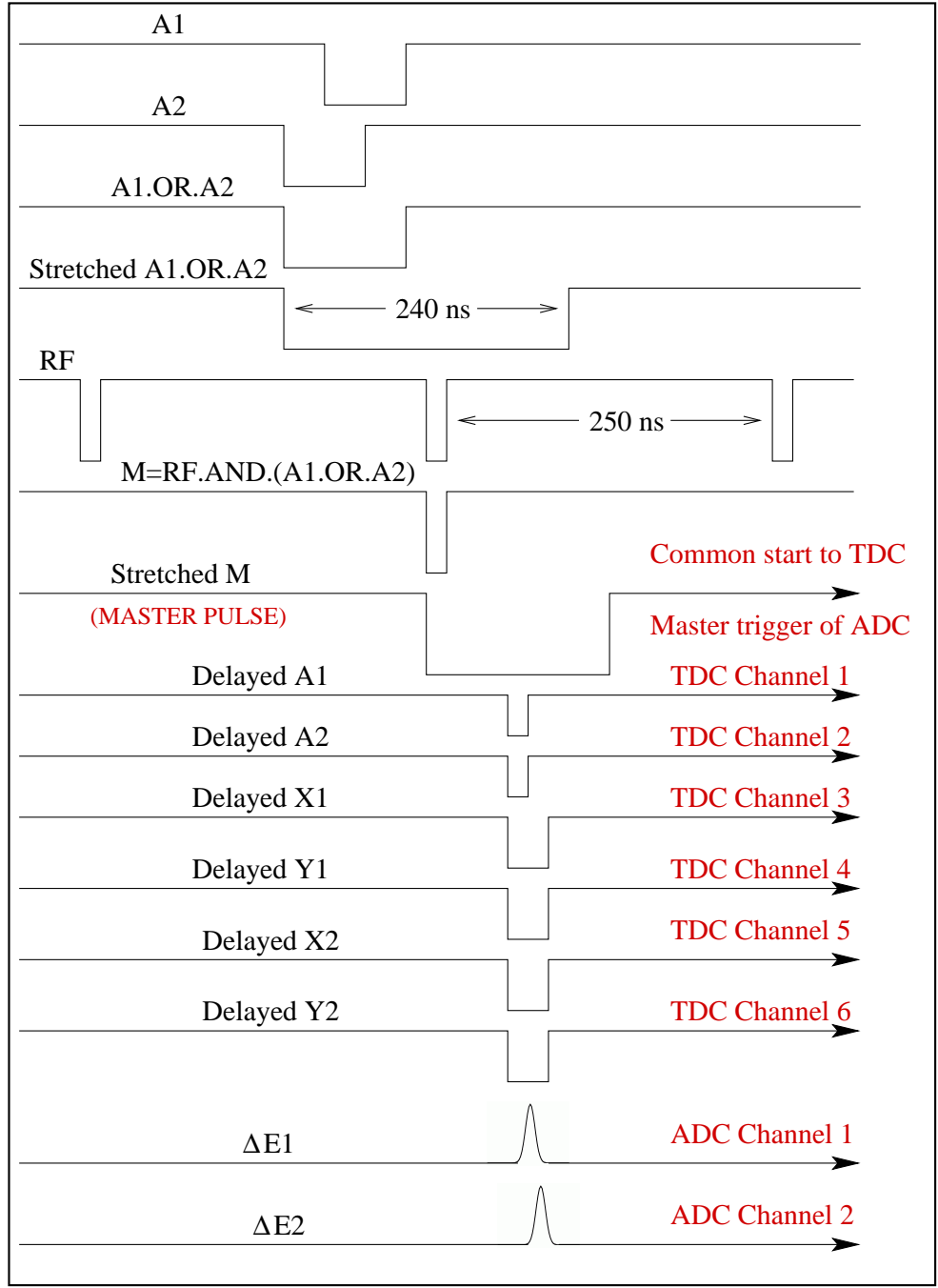}
\end{center}
\caption{\label{master1} The timing diagram for the setup .}
\end{figure}

\indent Fig. \ref{circuittof} shows a simplified block diagram of the 
associated electronics setup during one of the TOF experiment. The electronic 
set-up to obtain the X and Y position signals from a single MWPC has been 
discussed in Chapter 2. Timing pulses from the anodes (A$_1$ and A$_2$) 
were first pre-amplified by VT120A (ORTEC) pre-amplifier. These signals
 were then processed through constant fraction discriminators (CFD) and
 the "OR" of the discriminator pulses corresponding to MWPC1 and MWPC2 
signals were taken in coincidence with the RF pulse from the beam buncher 
of the Pelletron accelerator through the "AND" gate. The coincidence pulse 
is used as master trigger (M) or the master gate  for the computer automated
 measurement and control (CAMAC) 
 base data acquisition system. The X and Y 
signals ($X_1,X_2,Y_1$ \& $Y_2$) were picked off by PHILLIPS 6955B picked off 
amplifier (PO) of gain 100 and  fed to a CFD. The anode signals A$_1$, A$_2$ and
 the X-Y position signal pulses were then time analysed with a 12-bit time 
to digital converter (TDC) which was started by the master trigger. 
The timing diagram for the set up is shown in Fig. \ref{master1}. 
 The energy loss signal in the detector, E$_1$ 
and E$_2$ were pre-amplified by a charge sensitive ORTEC 142IH preamplifier.
The pre-amplifiers were placed as near as possible to the detector, using 
short connecting cable, in order to avoid the degradation of energy resolution. 
The energy signal from the preamplifier was amplified and shaped by 
ORTEC 572 spectroscopic amplifier . The amplified signal was digitised by 
a 12 bit ADC's connected to the CAMAC bus. The energy signal from the 
monitor  detector were also similarly processed ( as the energy pulses of the 
MWPCs) and were recorded as histograms in the CAMAC data acquisition system.
The multi-parameter data acquisition system was controlled by a standard 
control software, called {\it "freedom"}.
%\pagebreak

\indent The timing between one of the monitor detectors (MT$_1$) and 
the RF were measured by a time to amplitude converter (TAC) and was recorded 
by a 13-bit ADC and also stored as histograms.(In some of the 
experiments, the timing of the monitor detector was recorded in TDC as 
list mode data. However, in that case master trigger was [(A$_1$ ."OR". A$_2$)."AND". RF]."OR".(SCA output of the MT$_1$ and RF).)Thus, during experiment
eight parameter list mode data and three histograms were recorded in the 
hard disk of a computer and were latter transfered to CDs for offline 
analysis.

\indent Necessary interconnecting cables with common grounding and adequate 
shielding were provided to transfer all electrical signals from the ground 
floor beam hall to the counting room in the 1st floor where all pulse 
processing and electronic modules were stationed. The NIM modules were 
cooled by circulation of cool air for steady performance over a number of 
days of continuous running. On-line monitoring of the spectra were done 
during the experiment.

\clearpage

%\clearpage

\section{Time calibration}

\indent It has been discussed in the earlier section that the six parameters 
($A_1, X_1, Y_1$ $\&$ $A_2, X_2, Y_2$), out of eight parameters collected in 
the list mode  were basically timing signals. These signals were collected 
in a Time to Digital Converter (TDC), PHILLIPS 7186, in 
 all experiments. Digitization in 12 bits starts following the
 {\it COMMON} start input. The time range was selected to be 400 ns.
The time calibration was done using  a dual pulser. One pulse starts the TDC, 
while another simultaneous output is delayed by calibrated delay cables and stops any particular TDC channel. The delay 
between the start and stop pulses of the TDC were varied and a time spectrum 
was generated.

\begin{figure}[h]
\begin{center}
\includegraphics[height=10.0cm,angle=0]{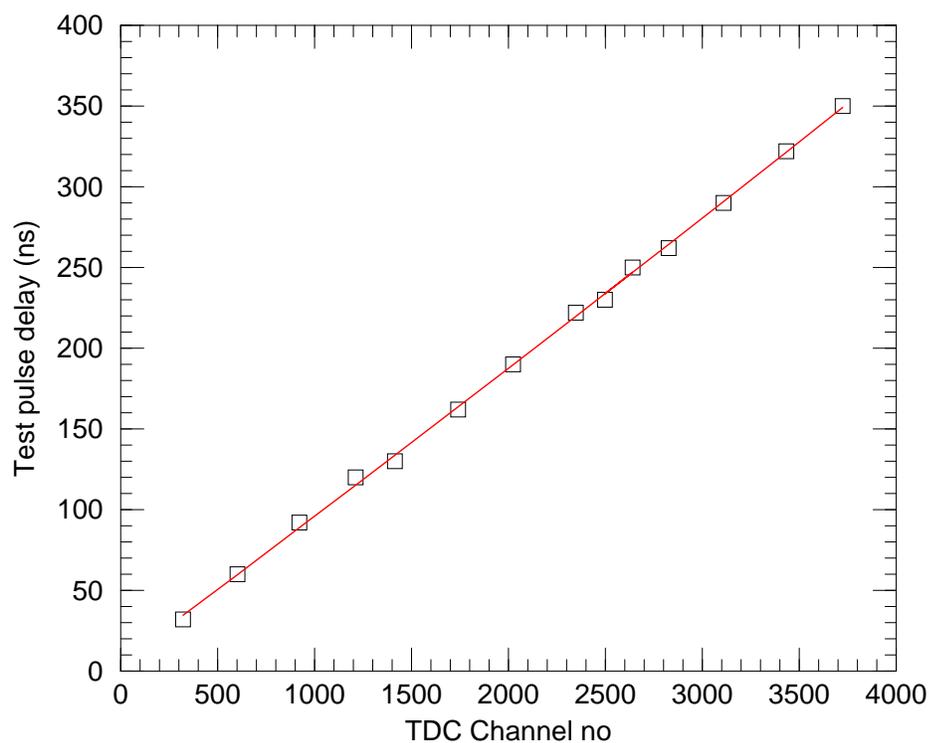}
\end{center}
\caption{\label{tdccalx2} Time calibration of TDC channel for position signal 
$X_2$.}
\end{figure}

\indent Fig. \ref{tdccalx2} shows time calibration of TDC channel 
for $X_2$ spectrum in one of our experiments. The channel was found slightly
 non linear and a second order polynomial was used for the calibration. The 
value of the co-efficients in this particular channel were:\\
$a_0 = $ 5.9360
$a_1 = 893.77 \times 10^{-4}$
$a_2 = 7.4468 \times 10^{-7}$

Same procedure was applied to calibrate all the TDC channels in use. 
 
%\clearpage

\section{Calibration of position}

\indent The position sensitive detector provides the position signal of a 
heavy fission fragment by measuring the delay of the sense wire pulse with 
respect to the anode pulse. After getting the position informations about 
two complementary fission fragments detected in the MWPCs, the corresponding 
angular measures were obtained to know the folding angle of the fission 
fragments. Hence, the conversion from the experimental TDC channels to 
angles is required to find the opening angle of fission fragments. Since 
the position resolution obtainable in these detectors are excellent in terms 
of angular resolution, position calibration procedure with shadows of 
the window support wires can be used for MWPCs.

\begin{figure}[h]
\begin{center}
\includegraphics[height=12.0cm,angle=0]{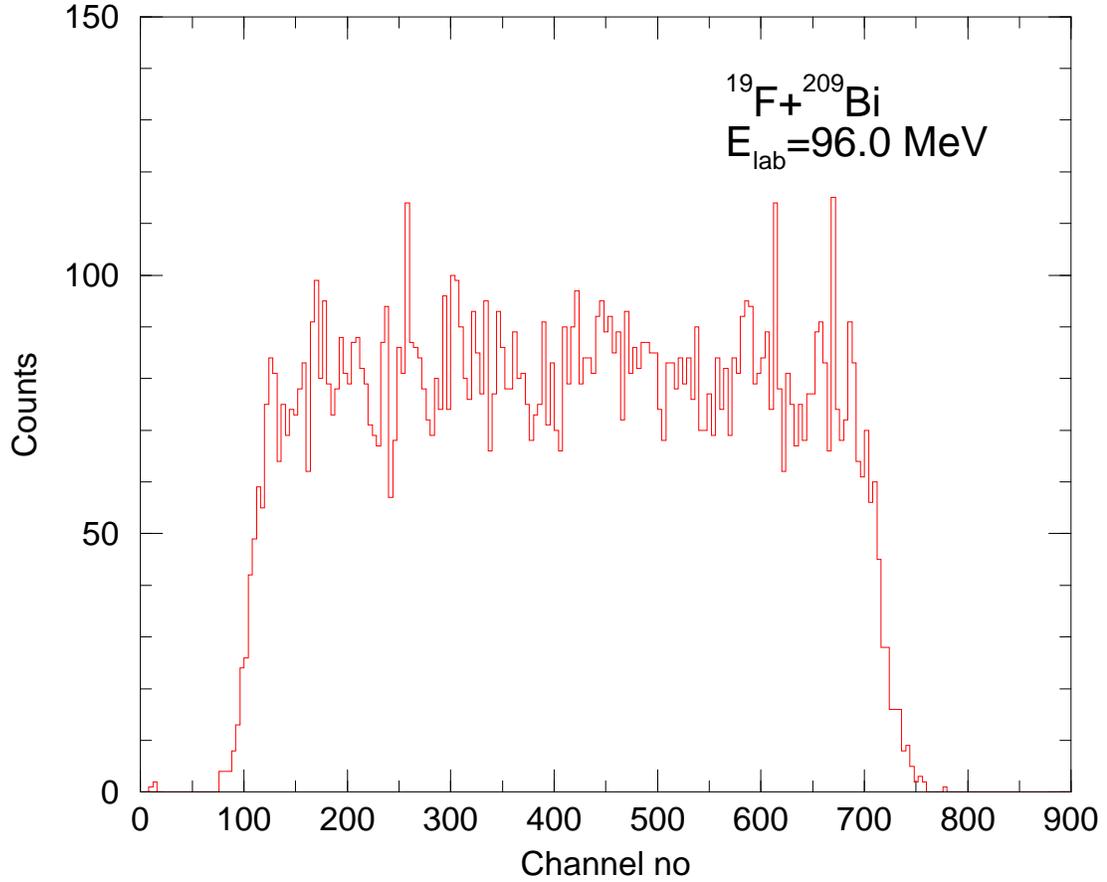}
\end{center}
\caption{\label{spex1_fbi96r10} A typical 1-D spectrum showing the X position 
responses of the MWPC}
\end{figure}

\begin{figure}[h]
\begin{center}
\includegraphics[height=12.0cm,angle=0]{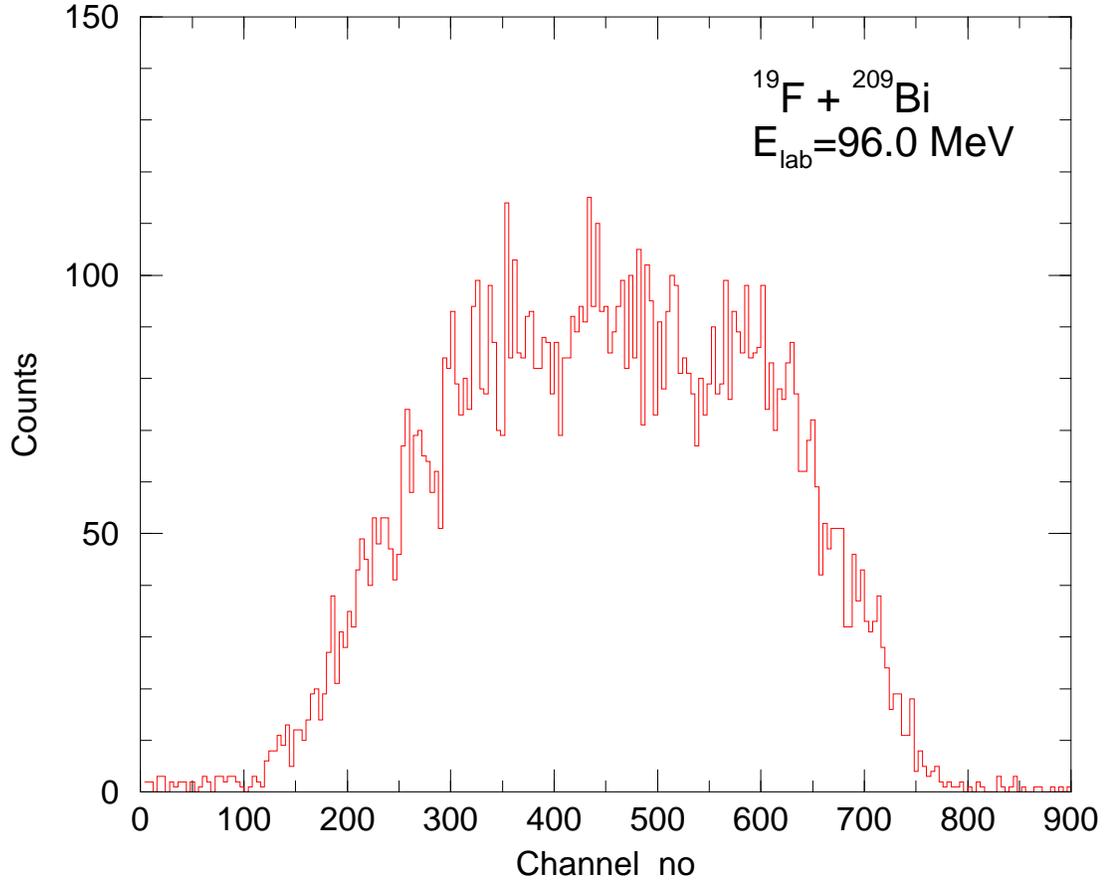}
\end{center}
\caption{\label{spey1_fbi96r10} A typical 1-D spectrum showing the Y position 
responses of the MWPC}
\end{figure}

\indent In the above described calibration procedure, dips in both X and Y 
spectrums, for each of the mesh wire should have to be clearly visible.
 Moreover, because of large distance, very long exposure were needed to get 
clearer shadows of the wires in a two dimensional X-Y spectrum.
 Typical X and Y spectrum in 
one of our experiments for the system $^{19}$F + $^{209}$Bi at laboratory 
energy 96.0 MeV are shown in Fig. \ref{spex1_fbi96r10} and Fig. 
\ref{spey1_fbi96r10}. From previous measurement \cite{NMthesis*3}, it 
is well known that a linear relationship, corrected for solid angle 
effects is adequate for the position calibration.

\indent Here, the following calibration procedure was followed in  our data 
analysis. 

%\clearpage

%\clearpage

\begin{figure}[h]
\begin{center}
\includegraphics[height=10.5cm,angle=0]{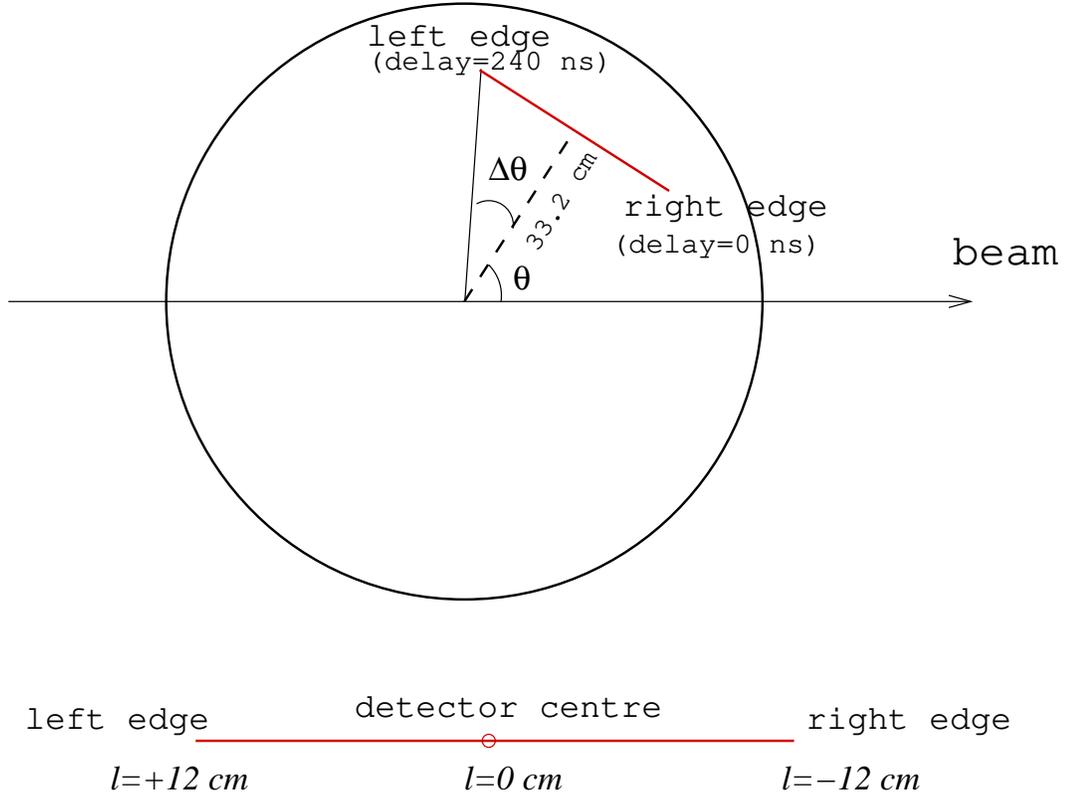}
\end{center}
\caption{\label{angcalib_fig} Schematic of the angular calibration of MWPC's }
\end{figure}

\begin{figure}[h]
\begin{center}
\includegraphics[height=12.0cm,angle=0]{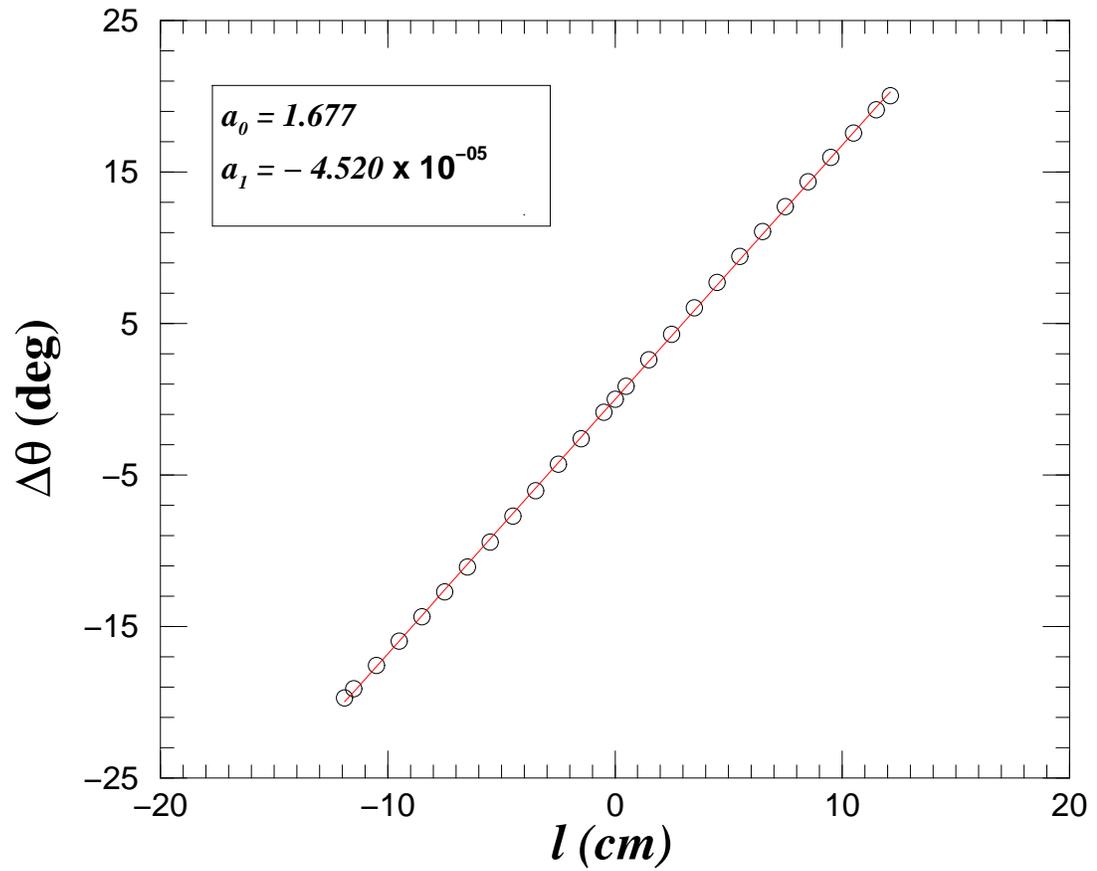}
\end{center}
\caption{\label{lvsdthetacalib} Correlation plot between the detector length 
and off-set angle of the window support wire from the central position of the 
detector.}
\end{figure}

\indent A schematic diagram used for the angular calibration of the 
detector is shown in Fig. \ref{angcalib_fig}. The active area of the 
detector was 24 cm $\times$ 10 cm and was placed 33.2 cm from the 
target. The central position of the detector was assigned as {\it l
$=$ 0 cm} while the right edge and left edge of the detector had 
{\it l $=$ -12 cm} and {\it l $=$ +12 cm} respectively. There were 24 wires 
(diameter 0.4 mm), provided as the support to the window foil of the MWPC, 
along the length of the detector and each were separated by 1 cm . The 
correlation between the length and the off-sets $\Delta\theta$ of the 
wire positions from the central position of the detector is tabulated 
in table \ref{lvsdtheta}. Fig. \ref{lvsdthetacalib} shows the correlation 
plot between {\it l} and $\Delta\theta$. The correlation is given by

%\clearpage

\begin{equation}
\Delta\theta =a_0{\it l} + a_1{\it l^2}
\end{equation}

\indent Value of the coefficients $a_0$ and $a_1$ are also shown 
in the figure. It is noted that the second order term has very small effect 
in calculation of $\Delta\theta$. The position signal was taken from 
the right edge of the detector. It has been discussed in Chapter 2 that 
12 delay line chips, each with 20 ns delay were used in the X-sense wire 
plane. So the delay $tx_2$ at the right edge of the detector was 0 ns while 
at the left edge was 240 ns. Thus we define a relation between length ({\it l}) and delay ($tx_2$) for MWPC2 as follows:

\begin{equation}
{\it l} = \frac{(tx_2 - 120)}{10}
\end{equation}    

\noindent Therefore the angular off-sets $\Delta\theta$ can be found from the
delay by the relation:

\begin{equation}
\Delta\theta = a_0 \Bigg[\frac{tx_{2}-120}{10}\Bigg] + 
a_1 \Bigg[\frac{tx_{2}-120}{10}\Bigg]^2 
\end{equation}

\noindent The value of $tx_2$ for the channel no $x_2$ of any X-spectrum 
can be derived from the time difference between the initial channel  and   
$x_2$ channel, i.e.,

\begin{equation}
tx_2 = tx_{2}^\prime - tx_2^{in} 
\end{equation} 

\noindent $tx_{2}^\prime$ and $tx_2^{in}$ are calculated 
from the channel no ($x_2ch$) of the spectrum using TDC 
calibration which has been discussed in the section 3.4. A second order 
polynomial was fitted to calibrate the TDC channel. The value of 
$tx_{2}^\prime$ and $tx_2^{in}$ are given by,

\begin{equation}
tx_{2}^\prime = c_0 + c_1 \times x_{2ch} + c_2 \times x_{2ch}^2
\end{equation}

\begin{equation}
tx_{2}^{in} = c_0 + c_1 \times x_{2ch}^{in} + c_2 \times {x_{2ch}^{in}}^2
\end{equation}

\noindent with the value of the coefficients

$c_0 =$ 5.9360,   
$c_1 = 893.77 \times 10^{-4},   
c_2=744.68\times 10^{-9}$. 

\indent The actual angular position is found by adding $\Delta\theta$ with 
$\theta$. It is to be noted that to use this calibration procedure the 
calibration data should be taken in singles mode for the full illumination 
of the detector. Using the TDC calibration it had been checked that 
the full $X_2$ spectrum corresponds to 240 ns. 
 
\begin{table}[h]
\begin{center}
\caption{\label{lvsdtheta}~ The angular off-sets of the window support 
wires and the length of MWPC2.}
%\begin{ruledtabular}

\begin{tabular}[t]{||c|c|c||}
\hline\hline
Position & Length (cm) & $\Delta\theta$ \\
\hline
Right edge & -11.9 & -19.7$^{\circ}$ \\
Wire 1 & -11.5 & -19.1$^{\circ}$ \\
Wire 2 & -10.5 & -17.5$^{\circ}$ \\
Wire 3 & -9.5 & -15.9$^{\circ}$ \\
Wire 4 & -8.5 & -14.3$^{\circ}$ \\
Wire 5 & -7.5 & -12.7$^{\circ}$ \\
Wire 6 & -6.5 & -11.0$^{\circ}$ \\
Wire 7 & -5.5 & -9.4$^{\circ}$ \\
Wire 8 & -4.5 & -7.7$^{\circ}$ \\
Wire 9 & -3.5 & -6.0$^{\circ}$ \\
Wire 10 & -2.5 & -4.2$^{\circ}$ \\
Wire 11 & -1.5 & -2.5$^{\circ}$ \\
Wire 12 & -0.5 & -0.8$^{\circ}$ \\
Centre & 0.0 & 0.0$^{\circ}$ \\
Wire 13 & 0.5 & 0.8$^{\circ}$ \\ 
Wire 14 & 1.5 & 2.5$^{\circ}$ \\
Wire 15 & 2.5 & 4.2$^{\circ}$ \\
Wire 16 & 3.5 & 6.0$^{\circ}$ \\
Wire 17 & 4.5 & 7.7$^{\circ}$ \\
Wire 18 & 5.5 & 9.4$^{\circ}$ \\
Wire 19 & 6.5 & 11.0$^{\circ}$ \\
Wire 20 & 7.5 & 12.7$^{\circ}$ \\
Wire 21 & 8.5 & 14.3$^{\circ}$ \\
Wire 22 & 9.5 & 15.9$^{\circ}$ \\
Wire 23 & 10.5 & 17.5$^{\circ}$ \\
Wire 24 & 11.5 & 19.1$^{\circ}$ \\
Left edge & 12.1 & 20.2$^{\circ}$ \\

\hline\hline
\end{tabular}
\end{center}
\end{table}
\clearpage

\section{Measurements of mass distributions}

\begin{figure}[h]
\begin{center}
\includegraphics[height=10.0cm,angle=0]{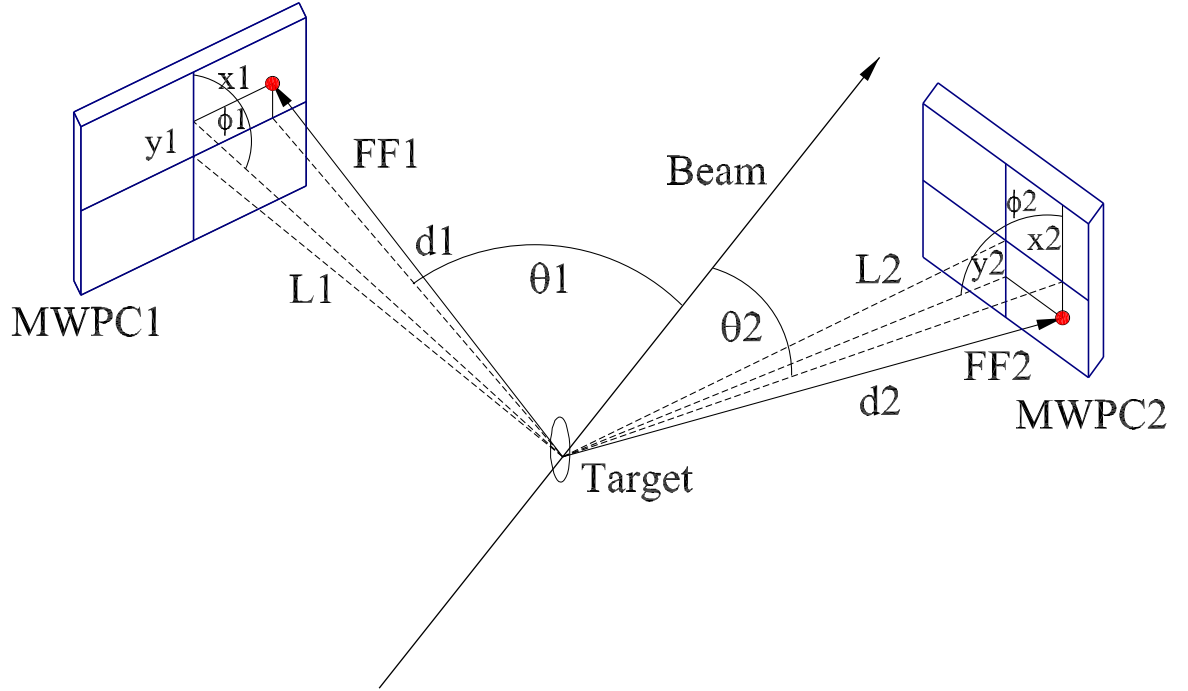}
\end{center}
\caption{\label{layout} The schematic of the set up and the co-ordinates for
the hit positions of complementary fission fragments in two MWPC's.}
\end{figure}

\indent The kinematics of the TOF determination of masses is schematically 
is shown in Fig. \ref{layout}. The masses of the fission fragments were 
determined from the angles $\theta$, $\phi$ and TOF information obtained 
from the experiment using following expressions:

\begin{equation}
m_1=\frac{[(t_1-t_2) + \delta t_0 + m_{CN}(\frac{d_2}{p_2})]}
{(\frac{d_1}{p_1} + \frac{d_2}{p_2})}
\end{equation}

\begin{equation}
m_2=(m_{CN} - m_1)
\end{equation}

\begin{equation}
p_1=\frac{m_{CN}V_{CN}}{(cos\theta_1 + sin\theta_1cot\theta_2)}
\end{equation}

\begin{equation}
p_2=\frac{p_1sin\theta_1}{sin\theta_2}
\end{equation}

\noindent where $t_1$, $t_2$ are the flight times of the fragments FF1 and 
FF2 respectively, over flight paths $d_1$ and $d_2$. These flight paths 
($d_1$ and $d_2$) are determined from the distances $L_1$ and $L_2$ and the 
local coordinates $X_1, Y_1$ and $X_2, Y_2$ of the impact points of fission 
fragments in detectors, MWPC1 and MWPC2. The momentum of the FFs are $p_1$ 
and $p_2$ in the laboratory frame, while the polar and azimuthal angles 
were ($\theta_1, \phi_1$) and ($\theta_2, \phi_2$) respectively.  $\delta t_0$  is the machine delay between the two anode pulses. This was measured from the 
identity of the mass distributions in the two detectors. $m_1$, $m_2$ and 
$m_{CN}$ are the fragment and compound nuclear masses. 

\begin{figure}[h]
\begin{center}
\includegraphics[height=9.0cm,angle=0]{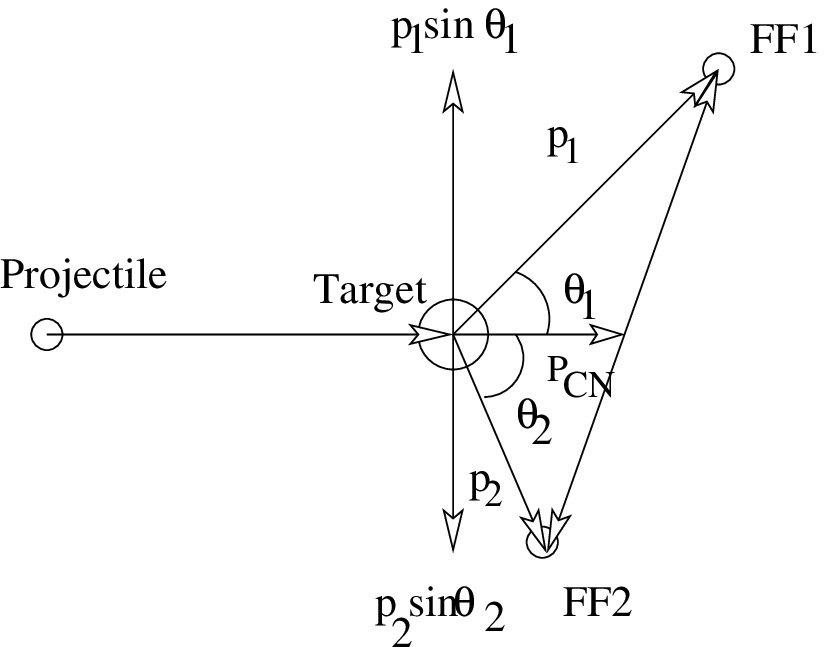}
\end{center}
\caption{\label{kine} Kinematic diagram for fusion-fission with compound 
nucleus recoiling in the beam direction with momentum $P_{CN}$.}
\end{figure}

\indent The above four expression can be derived from simple kinematics 
of mass and momentum conservation. This is illustrated in  Fig. \ref{kine}.
Assumption of mass conservation gives equation 3.23. Equations 3.24 and 
3.25 are derived using momentum conservation

\begin{eqnarray*}
p_{1}cos{\theta_1} + p_{2}cos{\theta_2} = m_{CN}V_{CN}\\
p_{1}sin{\theta_1} = p_{2}sin{\theta_2}
\end{eqnarray*}

\noindent The time of flight difference of the two fission fragments can be 
written as,

\begin{eqnarray*}
t_1 - t_2 = \frac{d_1}{v_1} - \frac{d_2}{v_2}\\
=\frac{d_1 m_1}{p_1} - \frac{d_2 m_2}{p_2}\\
=\frac{d_1 m_1}{p_1} - \frac{d_2}{p_2} (m_{CN} - m_1)\\
=m_1(\frac{d_1}{p_1} + \frac{d_2}{p_2}) - \frac{m_{CN}d_2}{p_2} 
\end{eqnarray*}

\noindent which gives,

\begin{equation}
m_1=\frac{[(t_1-t_2) + m_{CN}(\frac{d_2}{p_2})]}
{(\frac{d_1}{p_1} + \frac{d_2}{p_2})}
\end{equation}

\indent The experimentally measured  time of flight of a fission fragment 
$t_1$ is basically the sum of exact time of flight and  delay ($\delta t_{01}$) 
introduced during the experiment by electronic modules and cables and the 
beam pulse time structure. So, in 
equation 3.26 $t_1$ and $t_2$ should be replaced by $t_1 + \delta t_{01}$ 
and $t_2 + \delta t_{02}$ respectively to get equation  3.22. The difference 
of $\delta t_{01}$ and $\delta t_{02}$ is written as $\delta t_0$, which is 
independent of the beam profile.

\subsection{Mass distributions data analysis}

\begin{figure}[ht]
\begin{center}
\includegraphics[height=10.0cm,angle=0]{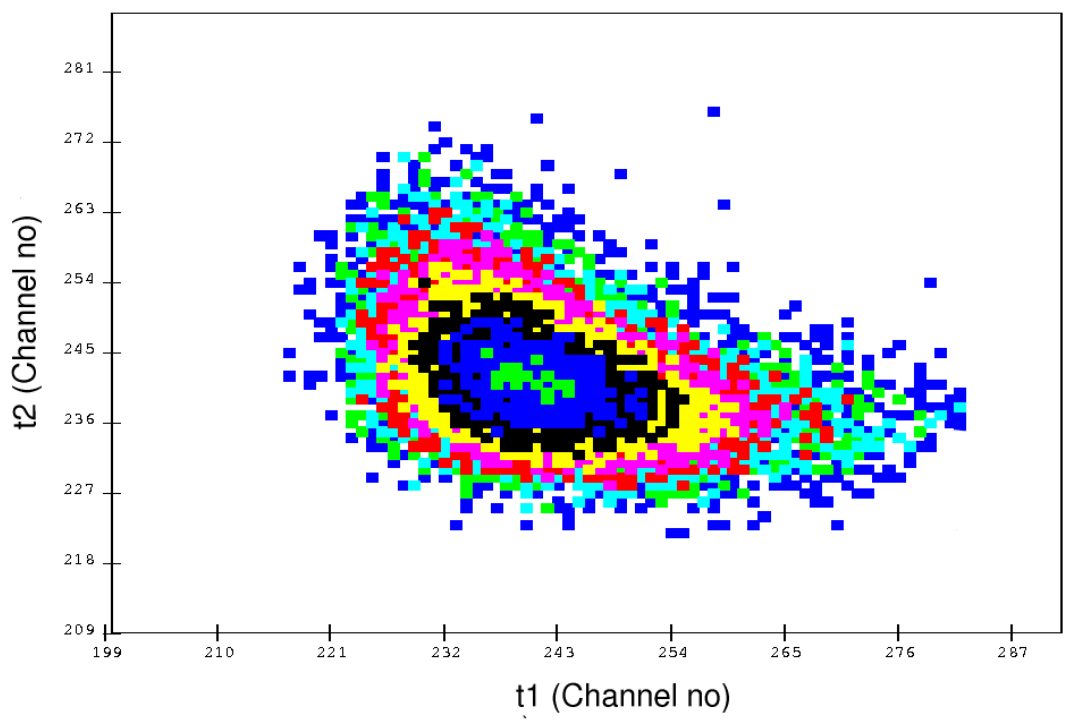}
\end{center}
\caption{\label{a1a2fbi96} The timing correlation between anode pulses of two MWPCs. The observed counts are for complimentary fission fragments}
\end{figure}

\indent A typical 2-D spectrum, of the time correlation between the 
two MWPCs for the system $^{19}$F + $^{209}$Bi measured  at lab 
energy 95.5 MeV is shown in Fig. \ref{a1a2fbi96}.  
 
\begin{figure}[h]
\begin{center}
\includegraphics[height=10.0cm,angle=0]{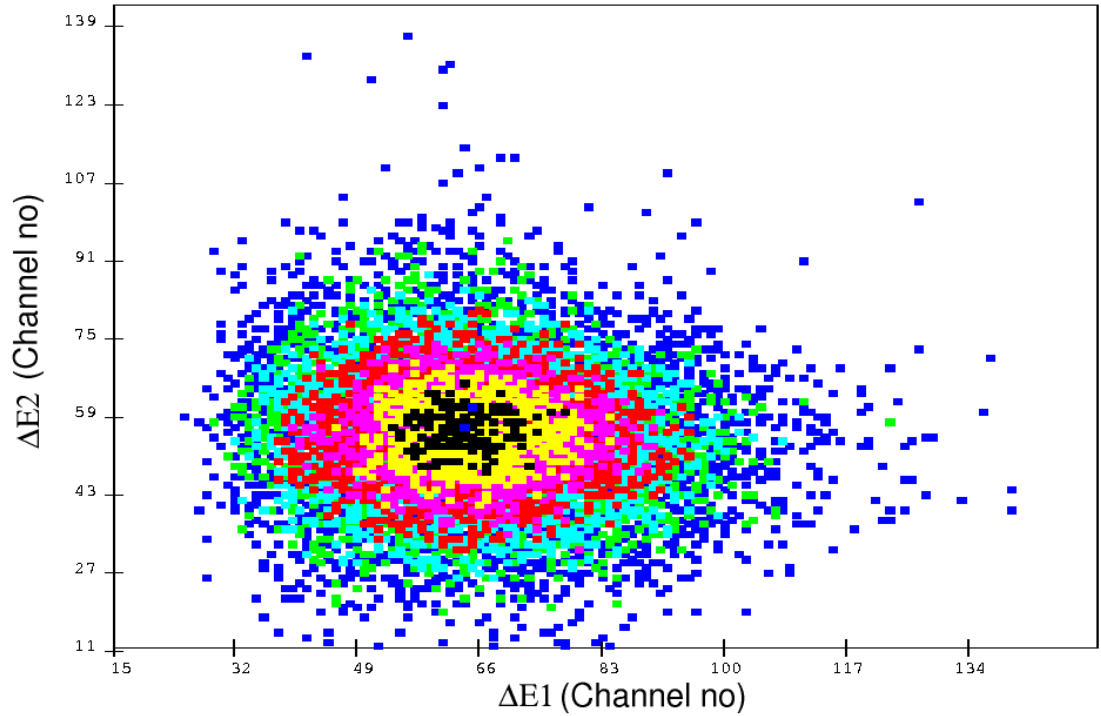}
\end{center}
\caption{\label{e1e2fbi96} A typical 2-D spectrum of the cathode signals 
(energy losses $\Delta E_1$ and $\Delta E_2$) from two MWPCs,
 obtained for $^{19}$F + $^{209}$Bi at bombarding energy 95.5 MeV.}
\end{figure}

\indent It can be easily observed that events from elastic and quasi-elastic 
reactions were effectively eliminated and purely binary FFs were selected.
 Contributions of elastic and quasi-elastic channels were less 
than 0.5 $\%$. Additional elimination of elastic and inelastic 
channels from fission fragments 
were obtained from the correlation of energy deposition signals from cathodes 
of two MWPCs. Fig \ref{e1e2fbi96} shows the correlation between the energy 
depositions in the two MWPCs at the same energy. Since the detectors are 
thin and operated at low pressure, the elastic  and quasi-elastic channels 
have poor response and almost all 
($>$ 99 $\%$) the events in the 2-D plot are from fission fragments.

%\clearpage

%\clearpage

\begin{figure}[ht]
\begin{center}
\includegraphics[height=14.0cm,angle=0]{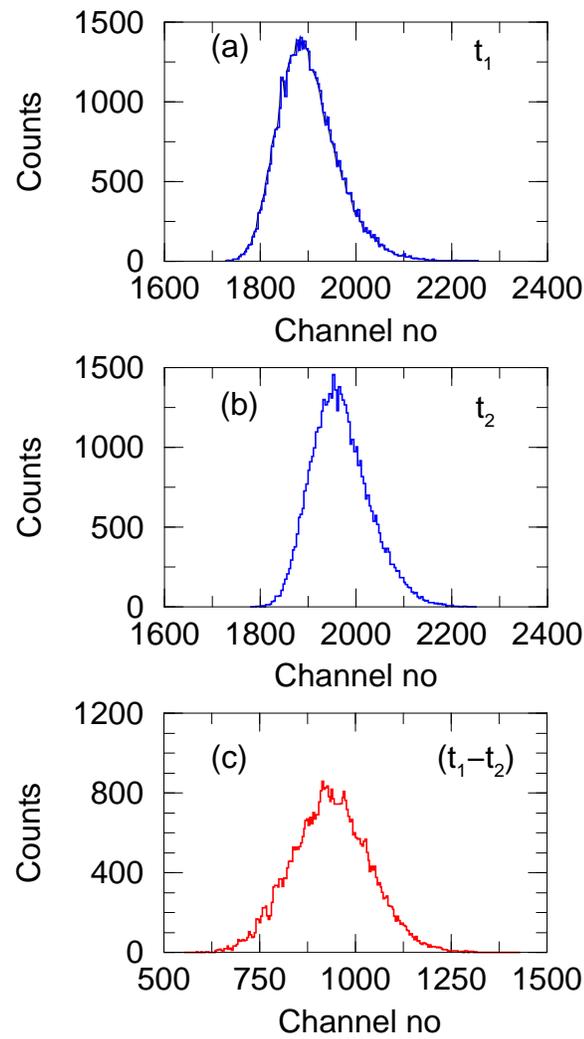}
\end{center}
\caption{\label{spetime} Timing spectrums from  (a) MWPC1 and (b) MWPC2. The 
TOF difference between the coincident fission fragments is shown in (c). The 
spectrums were obtained at the bombarding energy 88.0 MeV for the 
fissioning system of $^{19}$F + $^{232}$Th. }
\end{figure}

\begin{figure}[h]
\begin{center}
\includegraphics[height=10.0cm,angle=0]{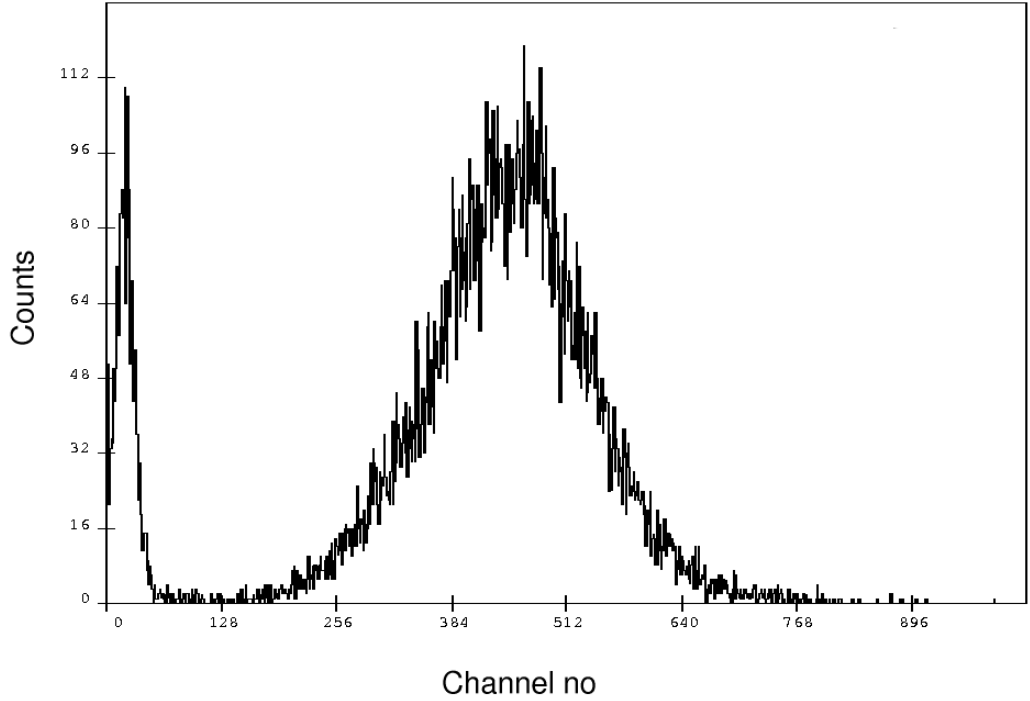}
\end{center}
\caption{\label{e1fbi96} A typical 1-D energy ($\Delta E_1$ ) spectrum
in the backward MWPC obtained for $^{19}$F + $^{209}$Bi at bombarding 
energy 95.5 MeV.}
\end{figure}

The upper two panels of Fig. \ref{spetime} shows the one 
dimensional spectrums 
of the time of flights $t_1$ and $t_2$. The difference of $t_1$ and $t_2$ 
which basically determines the mass distribution is shown in the lower 
panel. A typical 1-D energy spectrum is shown in Fig. \ref{e1fbi96}.

\clearpage

\indent The machine time delays of the anode pulses ($\delta t_{01}$ $\&$ 
$\delta t_{02}$) from the buncher of the Pelletron varies with the change 
in the incident energy of the beam. At low bombarding energies, in some 
specific experimental arrangement (viz., at forward angles) we could 
determine the machine time delays from the timing spectra of elastically 
scattered projectiles from the target. In all cases, however, the machine 
time delays [($\delta t_{01} - \delta t_{02}$) i.e., $\delta t_0$] and 
the difference in the TOF of FFs in two MWPCs could be accurately 
determined from the identity of the mass spectra at the two detectors.

\begin{figure}[h]
\begin{center}
\includegraphics[height=10.0cm,angle=0]{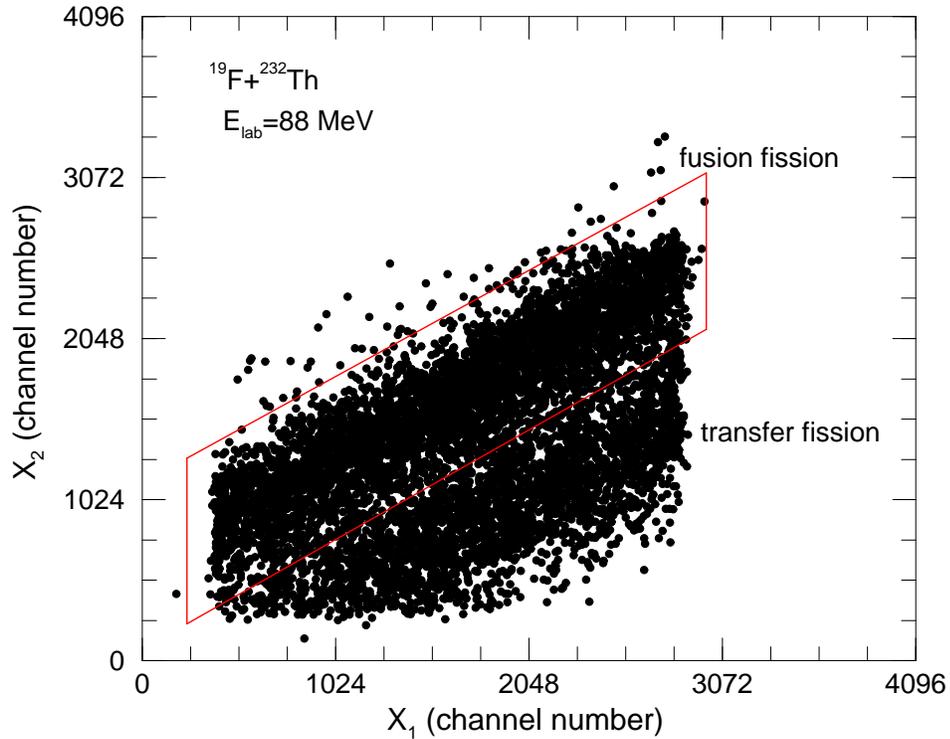}
\end{center}
\caption{\label{x1x2fth88} The correlation between the X-positions of two 
MWPCs. The two bands are for different complimentary FF folding angles - the 
upper band for complete fusion-fission and the lower band is for incomplete 
fusion-fission channels }
\end{figure}

\indent For determining the actual flight paths of the complimentary 
FFs in two MWPCs, the local co-ordinates of the impact point of the FFs
with respect to the centers of the detectors, ($X_1$, $Y_1$) for MWPC1 and 
($X_2$, $Y_2$) for MWPC2, were determined. In addition to the complementarity 
of the FFs, the correlation of the polar and azimuthal angles of FFs 
also carry the imprint of the linear momentum of the projectile transferred 
to the fused system of the projectile and target before fission occurs.
 A typical 2-D spectrum, taking the X position signals of two MWPCs, is shown 
in Fig \ref{x1x2fth88}, in terms of the channel numbers, for the fissioning 
system $^{19}$F + $^{232}$Th, measured at the sub-barrier energy ($E_{lab} =$
88.0 MeV) shows the correlation between the polar angles in two detectors. 
Two bands, representing the folding angle distributions of two different 
mechanism, 

\begin{figure}[h]
\begin{center}
\includegraphics[height=10.0cm,angle=0]{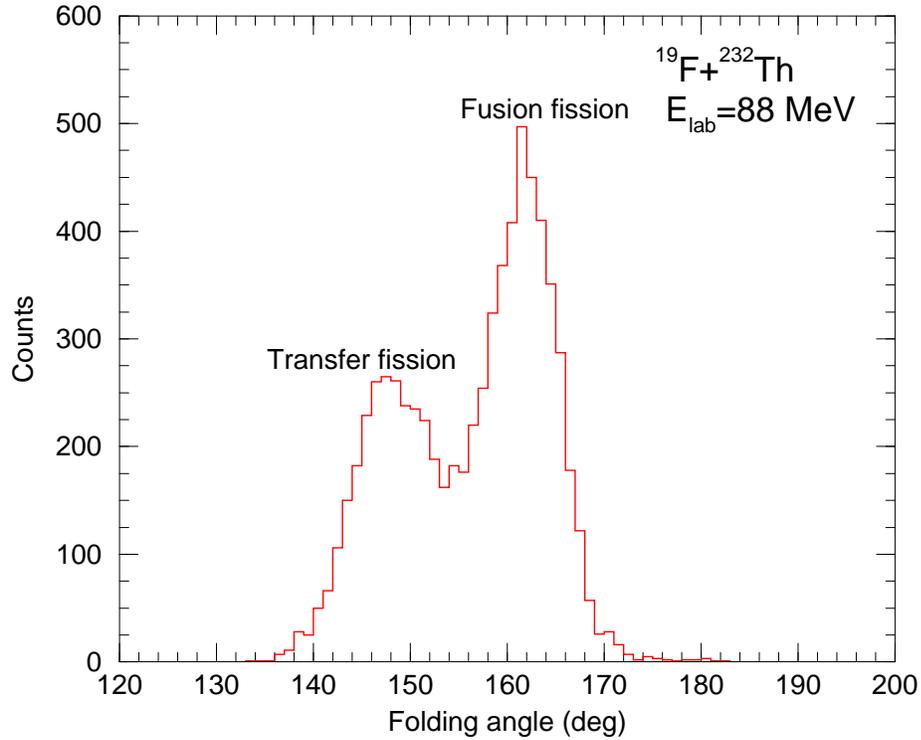}
\end{center}
\caption{\label{fold1d} Folding angle distributions measured for the 
fissioning system $^{19}$F + $^{232}$Th at the laboratory energy 88.0 MeV.}
\end{figure}
%\clearpage

\noindent fusion fission and transfer fission, are clearly visible.

% The one dimensional projection to the slice shown in Fig \ref{x1x2fth88} 
%produces the spectrum  shows the distribution 
%of the positions of coincident fragments corresponding to fragments 
%detected with the first detector in %%%Fig. \ref{fold1dx1}.%introduce this
%%%%figure%%%%%%%

\indent For each event, the 
X positions in the two detectors were transformed to polar angles plugging 
in the geometrical factors. The resulting distributions of counts as a 
function of the folding angle is shown in Fig \ref{fold1d}. In the 1-D 
spectrum, the two distinct peaks 
for the fusion fission (FF) and transfer fission (TF) are observed. 
The position of the 
peak, marked  "fusion fission" at larger folding angle corresponds with the 
expected position calculated with symmetric fragment masses and total 
kinetic energy release in fission according to Viola's systematics 
\cite{Viola85*3} and are due to fission after full momentum transfer from 
the projectile to the target. The broader structure, marked "transfer 
fission", at lower folding angles signifies a larger momentum transfer than 
that occurs in the fusion fission, presumable due to the ejectiles emitted 
in the backward direction after transfer of a few nucleons to the target. 
The distribution for the transfer fission 
 component is wider than that of the fusion fission component at near and 
sub-barrier energies due to widely varying recoil angles and velocities. 

\begin{figure}[h]
\begin{center}
\includegraphics[height=10.0cm,angle=0]{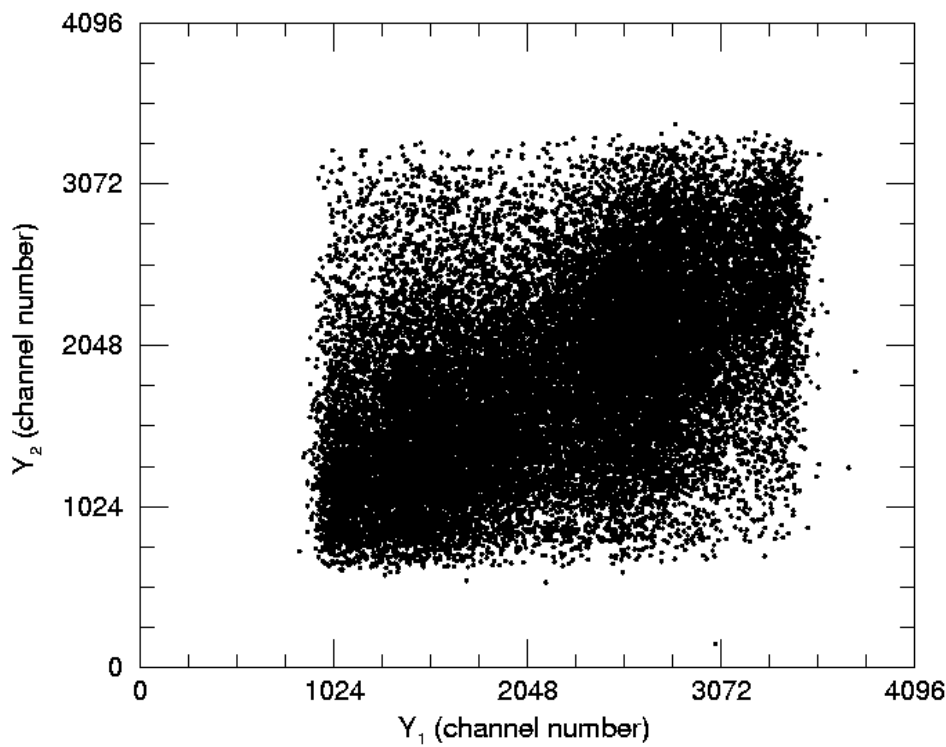}
\end{center}
\caption{\label{y1y2oth82} The correlations of $Y_1$ and $Y_2$ positions of 
fission fragments. The band in the figure shows a complete coverage for 
complementary fragments for fusion-fission reaction channel in the azimuthal 
plane.}
\end{figure}

\indent Similarly correlations in $Y_1 - Y_2$ were also monitored during the 
experiment. Fig. \ref{y1y2oth82} shows the correlation between $Y_1$ and $Y_2$ 
measured for the fissioning system $^{16}$O + $^{232}$Th at laboratory energy 
82 MeV and confirms full acceptance of the two MWPCs for complementary 
FFs in azimuthal plane.

%\clearpage

\begin{figure}[h]
\begin{center}
\includegraphics[height=10.0cm,angle=0]{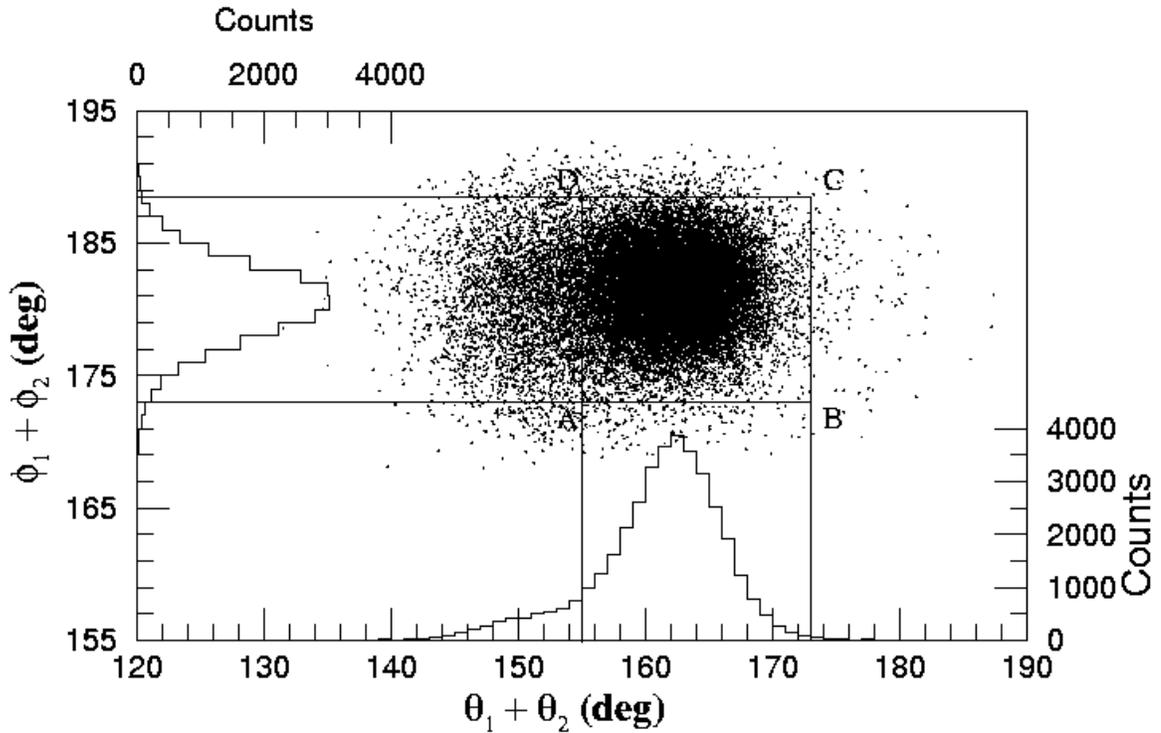}
\end{center}
\caption{\label{thetafioth84} Distributions of complimentary fission fragments 
in ({$\theta$}, {$\phi$}) for the system $^{16}$O + $^{232}$Th at 
$E_cm =$ 77.3 MeV. Rectangle ABCD indicates the gate used to select the 
fusion fission events.}
\end{figure}

\begin{figure}[h]
\begin{center}
\includegraphics[height=10.0cm,angle=0]{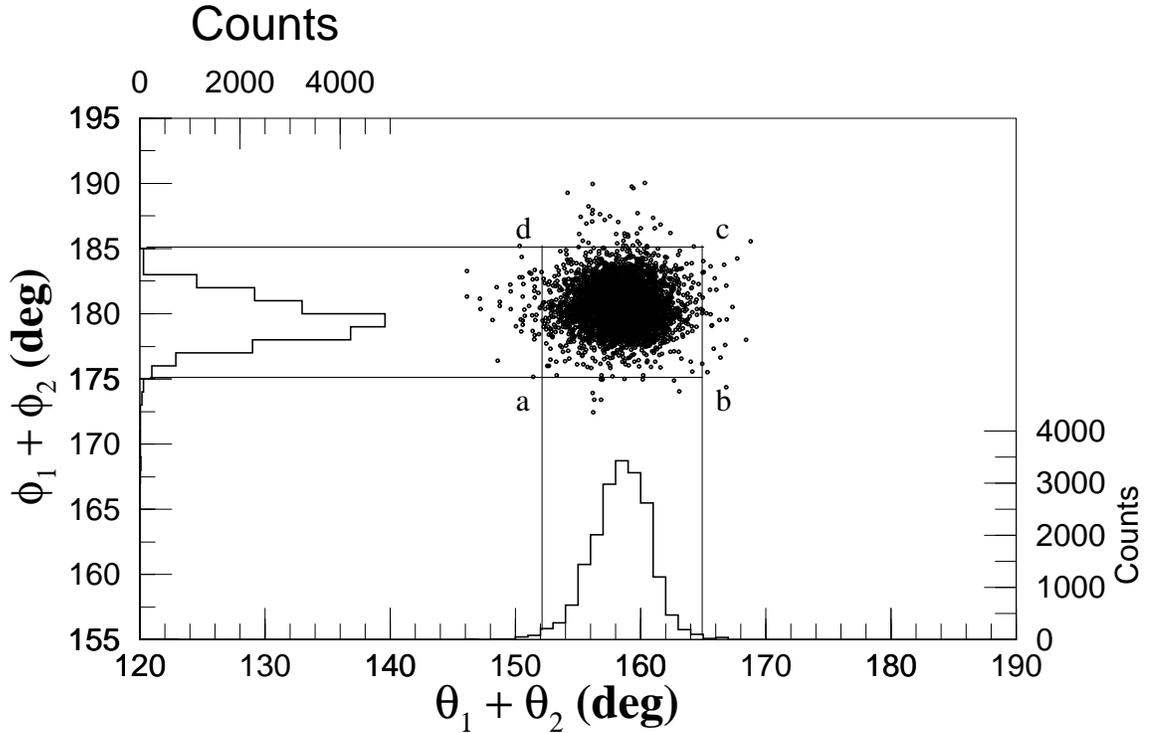}
\end{center}
\caption{\label{thfi_sep_obi} Distributions of complimentary fission fragments 
in ({$\theta$}, {$\phi$}) for the system $^{16}$O + $^{209}$Bi at 
$E_cm =$ 76.2 MeV.}
\end{figure}

\indent It is to be noted that $^{19}$F, $^{16}$O, $^{12}$C + 
$^{232}$Th reactions had a considerable cross sections for transfer 
induced events. The fission fragments for fusion fission events were 
exclusively determined from the distribution of polar and azimuthal angles. 
The measured  polar ({$\theta$}) and azimuthal ({$\phi$}) correlations 
for the fissioning system of $^{16}$O + $^{232}$Th is shown in 
Fig. \ref{thetafioth84}. The events within the rectangle "ABCD" were due 
to fusion fission. However, in  $^{19}$F, $^{16}$O + $^{209}$Bi 
reactions, the transfer fission channel contributes less than 1$\%$. 
This can be observed in Fig. \ref{thfi_sep_obi}, where very few 
events lie outside of the rectangle marked "abcd" in correlated 
{$\theta$} - {$\phi$} distributions of fragments. The projections 
on {$\theta$} and {$\phi$} planes are shown in the insets.   

%\clearpage

\indent In a very similar analysis, the velocities of the recoiling fissioning 
system in the reaction plane and perpendicular to it can be used to 
separate the FF and TF component \cite{HindePRC96*3}. The two components 
of the velocity vector of each fissioning nucleus are determined . The 
component in the beam direction, $v_{par}$ was calculated from the folding angle and the velocities of the two fragments. The other component $v_{perp}$ is 
in the plane perpendicular to the beam and is perpendicular to the 
projection of the scission axis onto this plane. It was determined from the 
azimuthal folding angle and the projection of the measured fragment velocities 
onto this plane.

\begin{figure}[ht]
\begin{center}
\includegraphics[height=10.0cm,angle=0]{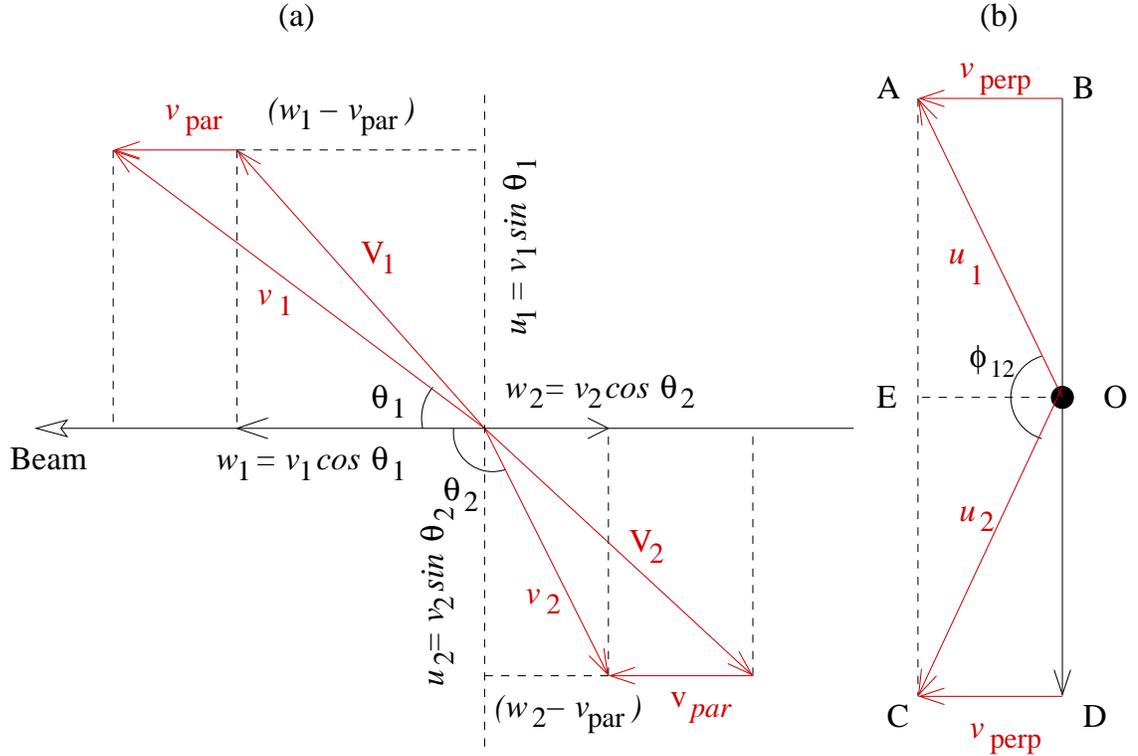}
\end{center}
\caption{\label{velkine1} Diagrams of the fission fragment velocity components.
(a)Plane including the fission fragment velocity vectors and the beam axis 
(b) Plane perpendicular to the beam.}
\end{figure}

\indent In Fig. \ref{velkine1}, the velocities of the fragments are denoted
 by $v_1$ and $v_2$ , 
the scattering angles are $\theta_1$ and $\theta_2$, measured with respect to 
the beam direction. The parallel component of the recoil of the fissioning nucleus is given by $v_{par}$, while the component of the recoil velocity 
perpendicular to the reaction plane is $v_{perp}$.  Initially, it is 
taken that the two velocity vectors and the beam axis are co-planer which 
is equivalent to neglecting $v_{perp}$. In Fig \ref{velkine1}(a), the 
velocities in the centre of mass frame are denoted by $V_1$ and $V_2$ while 
$v_{par}$ represents the component in the beam direction of the center of 
mass velocity of the fissioning system. The measured velocity vectors 
are decomposed into orthogonal components parallel (denoted by $w_1$ 
and $w_2$) and perpendicular (denoted by $u_1$ and $u_2$) to the beam axis.
The velocity vectors $u$'s and $w$'s define the reaction plane for 
compound nuclear reaction. For compound nuclear reaction $<v_{perp}>$ 
is essentially zero while a non-zero value of $v_{perp}$ signifies a 
non-compound reaction. Thus a scatter plot of $v_{par}$ versus $v_{perp}$ 
clearly de-markets the events for compound and non-compound fission events. 

\indent From the figure it is evident, $w_1=v_1cos\theta_1$,
$w_2=v_2cos\theta_2$, $u_1=v_1sin\theta_1$ and $u_2=v_2sin\theta_2$. 
Neglecting the small effects of prescission particle evaporation, the two 
fragments are taken as co-linear and co-planer in the centre-of-mass frame, and the ratio 

\begin{equation}
\frac{u_1}{u_2}= - \frac{w_1-v_{par}}{w_2-v_{par}}
\end{equation}

\noindent can be defined. The minus sign is due to the fact that $u$ values 
(unlike $w$) can only be positive. Thus $v_{par}$ is given in terms of the 
measured velocity component by
 
\begin{equation}
v_{par}=  \frac{u_1w_2+u_2w_1}{u_1+u_2}
\end{equation}

\noindent For fission following complete absorption of the projectile by the 
target, the full momentum of the projectile is transferred and $v_{par}$ should 
be equal to the calculated centre of mass velocity for the collision $v_{c.m.}$.
However, deviations from binary kinematics due to emission of light particles 
perturbs the fission fragment vectors, resulting in a significant spread in 
$v_{par}$.

In Fig \ref{velkine1} the geometry in the plane perpendicular to the beam is 
shown. The measured component $u_1$ and $u_2$ are related to the actual velocities 
of the fragments in the centre of mass frame of the fissioning system by an 
in-plane vector having two components. From the triangle AOB and AOC we find,

\begin{equation}
\frac{v_{perp}}{u_1}= cos\frac{\phi_{12}}{2}
\end{equation}

\begin{equation}
\frac {{\frac{1}{2}}{\sqrt{(u_1^2 + u_2^2 - 2u_1u_2cos\phi_{12})}}}{u_2}
= sin\frac{\phi_{12}}{2}
\end{equation}

\noindent Combining the above two equations velocities of the fissioning system 
perpendicular to the scission axis can be written as, 

\begin{equation}
v_{perp}=\frac {u_1u_2sin\phi_{12}}{\sqrt{(u_1^2 + u_2^2 - 2u_1u_2cos\phi_{12})}}
\end{equation}

\noindent For full momentum transfer fission, only the light particle 
emission causes $v_{perp}$ to deviate from zero which is very small. 

\begin{figure}[ht]
\begin{center}
\includegraphics[height=13.0cm,angle=-90]{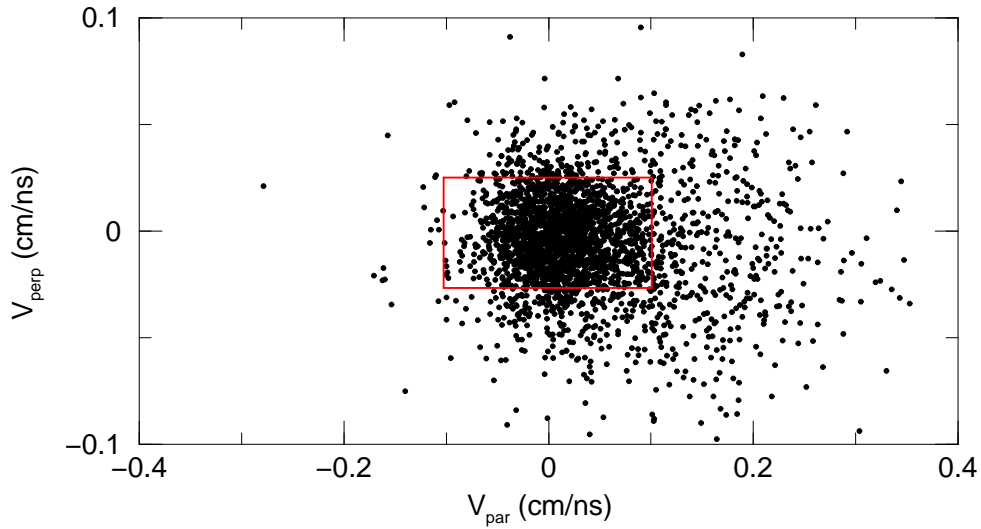}
\end{center}
\caption{\label{vdis} Experimentally determined velocity components of the 
fissioning nuclei $v_{par}$ and $v_{perp}$ at beam energy 92.3 MeV for the 
reaction $^{19}$F + $^{232}$Th. The rectangle indicate the cut used to select 
the FF events.}
\end{figure}

\indent Fig \ref{vdis} shows the  plot of the measured $v_{par}$ relative 
to $v_{cm}$ against $v_{perp}$ for the fissioning system 
of $^{19}$F + $^{232}$Th below the Coulomb barrier energy (92.3 MeV). 
For full momentum transfer fission, the distribution of $v_{perp}$ would 
be expected to be centered around zero and  $v_{par}$ would be expected 
to peak around the centre of mass velocity $v_{c.m.}$. The high concentration 
of events in the middle of the plots within the box corresponds to 
FF events where as the scattered events outside the box are attributed 
to transfer fission (TF) events. 

\begin{figure}[ht]
\begin{center}
\includegraphics[height=12.0cm,angle=-90]{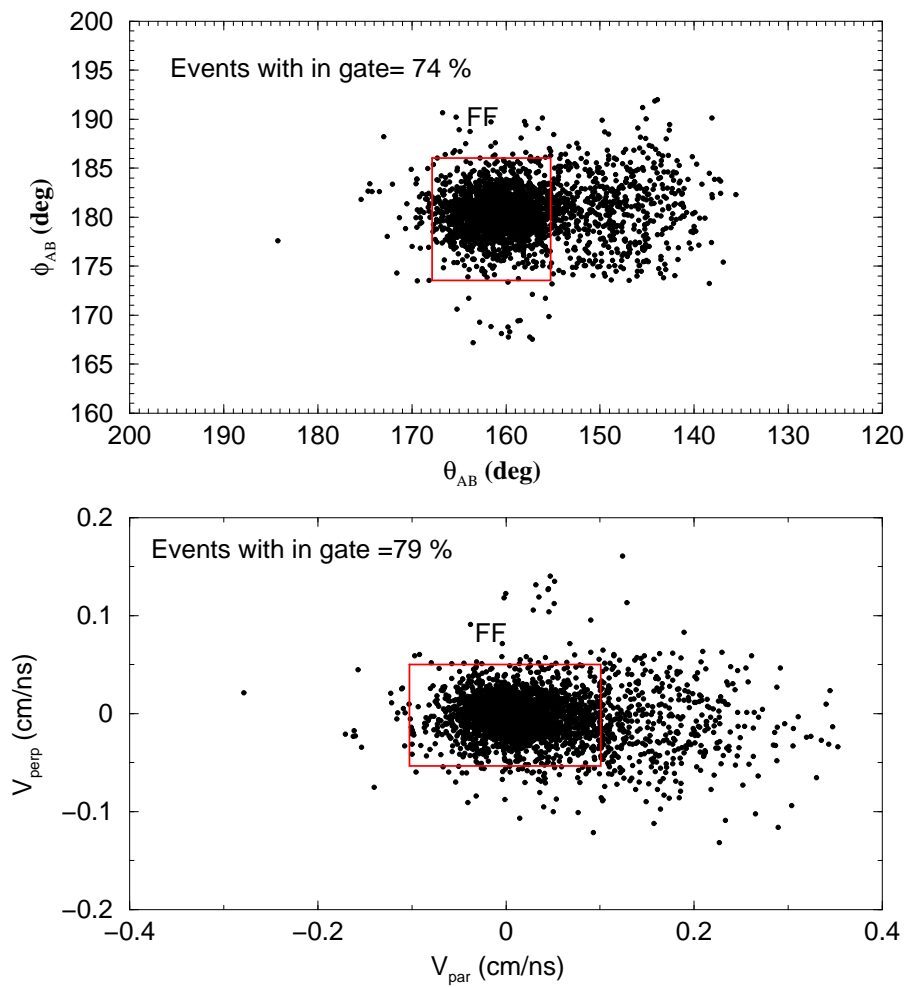}
\end{center}
\caption{\label{compare1} Distributions of complementary fission fragments in 
$\theta-\phi$ plane (upper panel) and $v_{par}-v_{perp}$ (lower panel).
 The rectangle indicate the cut used to select the FF events.}
\end{figure}

\indent Fig \ref{compare1} shows the comparisons of the separation of FF from 
TF events using both the procedures which do not show any significant 
difference of the contours of the FF events. In fact, the coincidence gates 
on $\theta-\phi$ are more compact excluding more transfer fission events than 
in the correlation of velocities of the fissioning system. We have used 
the gates on $\theta-\phi$ in our experimental analysis and cross checked with 
$v_{par}$-$v_{perp}$ distributions.

\subsection{Mass resolution of the spectrometer}

\begin{figure}[h]
\begin{center}
\includegraphics[height=14.0cm,angle=0]{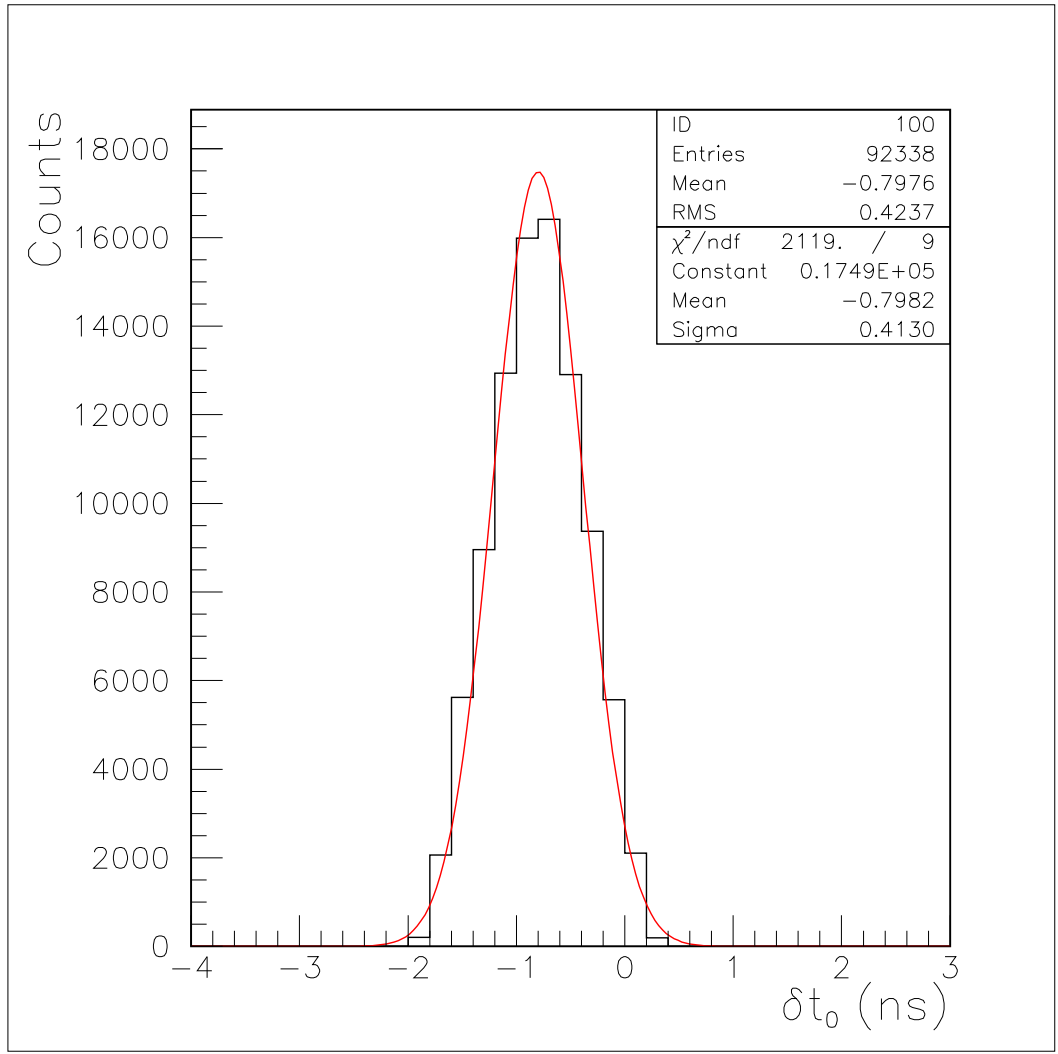}
\end{center}
\caption{\label{delt0} Distribution of machine time}
\end{figure}

\indent The mass resolution depends on errors of time measurements, 
($t_1$ - $t_2$) and $\delta t_0$, and also on the accuracy in measuring the 
angles $\theta$, $\phi$. Contributions of the MWPCs to the timing error 
is small, less than 250 ps because of short rise time ($<$ 1 ns) and 
large signal ($\sim$ 1 Volt for fission fragments) to noise ($\sim$ 10 mV) 
ratio of the anode pulse. In fact, the transit time jitter of the primary 
electrons before striking avalanche is the main contributing factor 
to prompt resolution of the MWPC. The position resolution in the X-direction is 
of the order of 400 $\mu$m and the corresponding polar angular accuracy 
was about 0.07$^\circ$. The accuracy in the azimuthal angle is about 
0.15$^\circ$. Considering this, the variance ($\sigma$) of the delay 
$\delta t_0$ and the pulse width of the beam bunch are the crucial   
determining factor for the mass resolution of the TOF. The distribution 
of $\delta t_0$ is shown in Fig \ref{delt0}. The root mean square (r.m.s) 
value of the distribution was about 420 ps.

\begin{figure}[h]
\begin{center}
\includegraphics[height=12.0cm,angle=0]{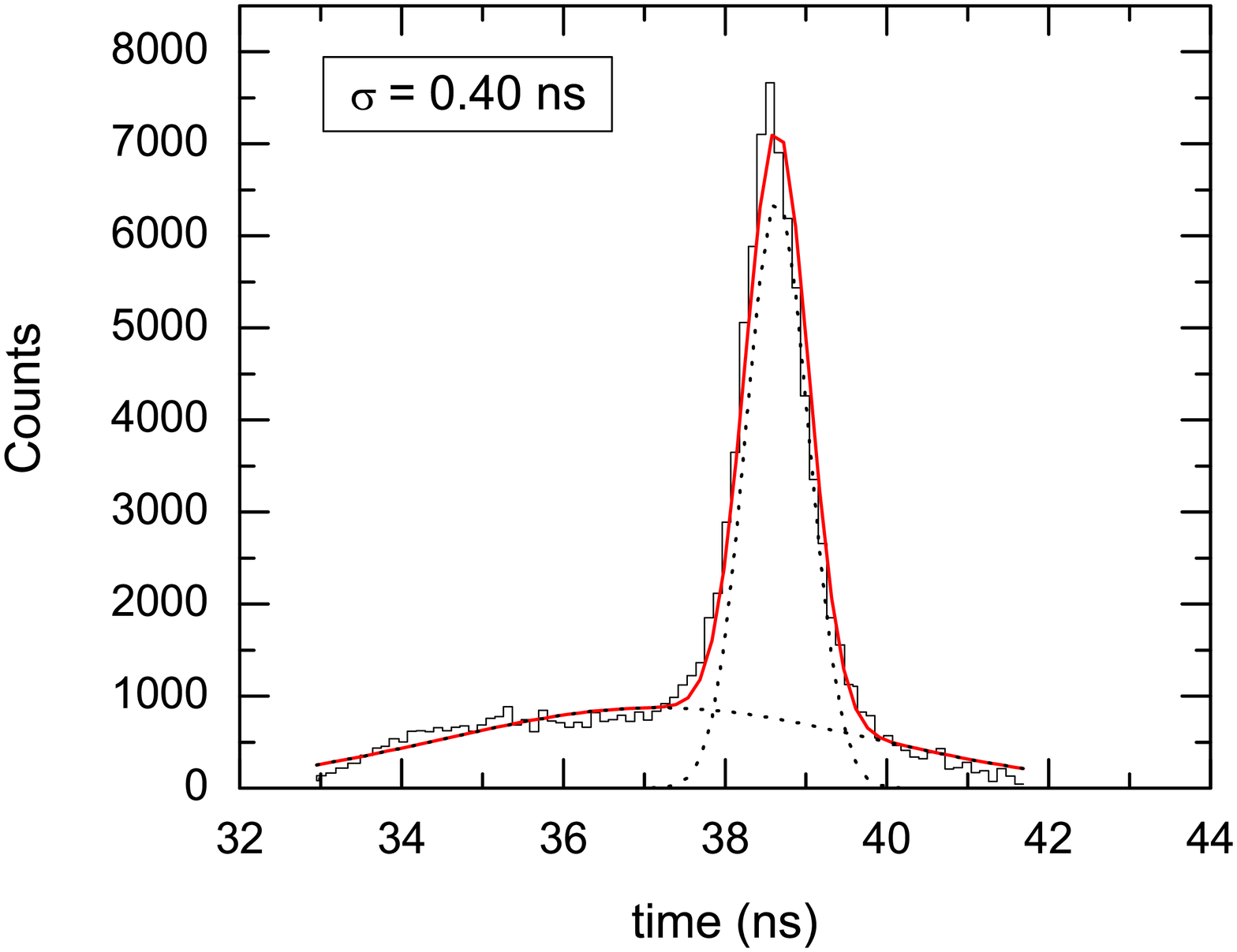}
\end{center}
\caption{\label{beamreso} Time structure of the bunches beam from the 
Pelletron accelerator. The tailing at lower time is due to slower neutrons. 
The spectrum (histograms) is fitted by two Gaussian 
distributions (dotted line). The solid red line represent the overall fitting. }
\end{figure}

\indent The width of the beam bunch was 
measured by the distribution of delays of gamma rays from the target with 
fast plastic scintillator. The timing spectrum of gamma rays shows a variance 
of about 380 ps and is shown in Fig \ref{beamreso}.
 The tailing at higher delay
 was coming from the slow neutrons. The time jitter in MWPC and the 
triggering time jitter of the event on the bunched beam time structure 
together with the uncertainty in measuring the machine delays in each 
TOF arm, defines the ultimate mass resolution of the double arm TOF if the 
masses are measured from the {\it ratio} method 
\cite{HindePRC96*3} [Ratio of the fragment mass $M=u_1/(u_1+u_2)$].
However, the {\it difference} of the TOF's in two arms cancels the 
effect of the beam structure of the bunched beam. The mass resolution for the 
TOF spectrometer after considering the width of the beam bunch and the response of the detectors and the geometrical factors was found to be 4 a.m.u., which 
in most cases is sufficient for fission fragment mass distribution studies.

\clearpage

\begin{figure}[h]
\begin{center}
\includegraphics[height=12.0cm,angle=0]{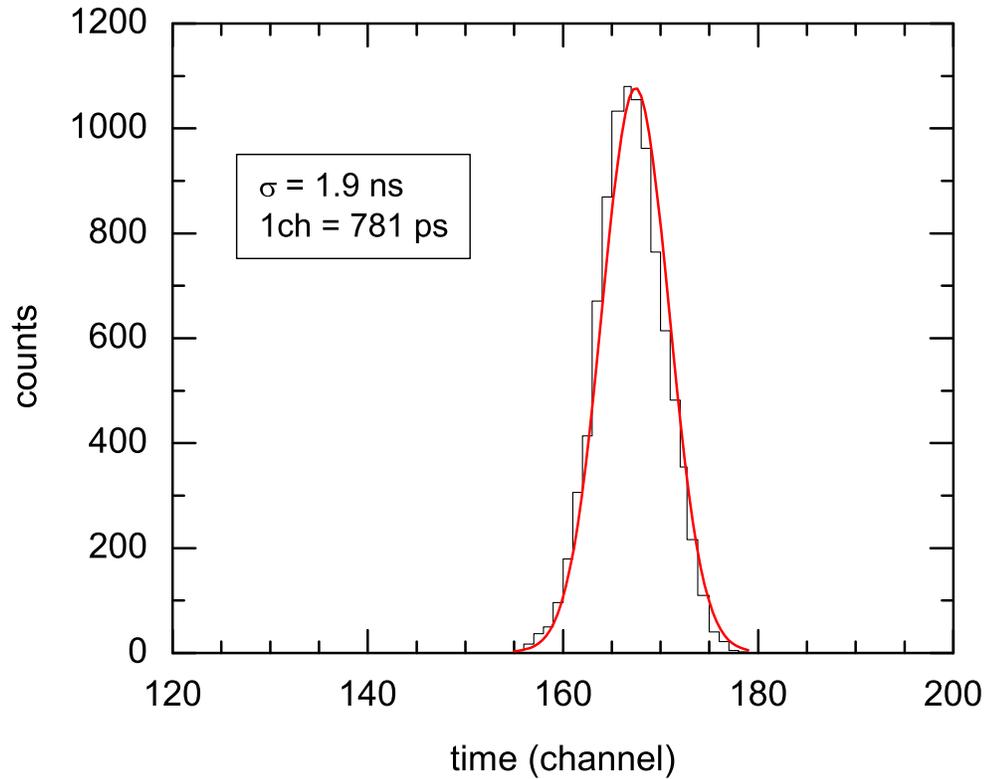}
\end{center}
\caption{\label{mwpcreso} Time of flight data collected using the bunched beam 
and the MWPC to observe elastically scattered 64 MeV $^{12}C$ ions }
\end{figure}

\indent The MWPCs showed very poor response to elastics and only at a larger 
operating gas pressure and at extreme forward-backward geometry of the 
two detectors we could get anode pulses for elastics, but the corresponding 
position signals were poor. For the backward detector at 33.2 cm, the TOF 
spectrum collected using the bunched beam and the MWPC to observe 
elastically scattered 64MeV $^{12}$C ions is shown in Fig \ref{mwpcreso}.
 The variance ($\sigma$) of timing distribution was about 1.9 ns. The 
time distribution widens because of large acceptance angle of the detector 
and the mixture of elastic and quasi-elastic particles with a considerable 
energy spread. Because of the large variance in time, the start time delay 
in the two detectors, if measured separately, to determine the parameter $\delta t_0$ in equation 2.11, resulted in larger variance ($\sigma \sim$ 2.7 ns) in 
 $\delta t_0$ and hence, worse mass resolution of about 11 a.m.u.

\subsection{Systematic error}

\begin{figure}[h]
\begin{center}
\includegraphics[height=12.0cm,angle=-90]{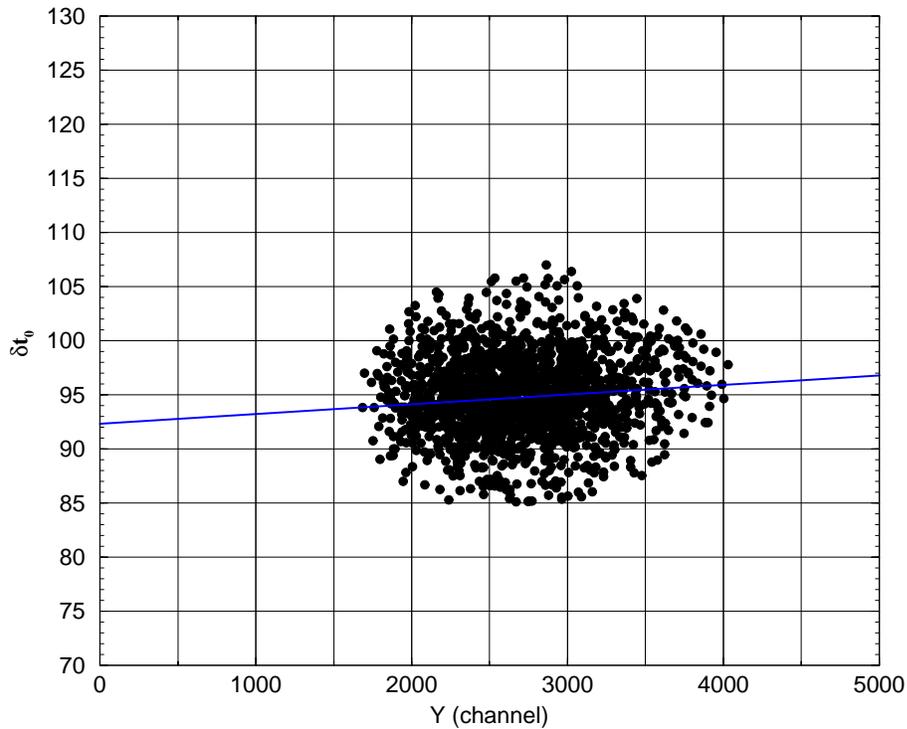}
\end{center}
\caption{\label{errory} Variation of $\delta t_0$ (in arbitrary units) 
with Y-positions of fission fragments in the backward detector in one of
 the experiments.}
\end{figure}

\indent A systematic error on the measured mass depends critically on the 
geometry of the experimental setup. Any rotation or tilt in the anode wire 
plane with respect to the reaction plane renders the machine time delay, 
$\delta t_0$ as a function of X or Y positions ( $X_1$, $Y_1$ and $X_2$, $Y_2$) 
or both, depending upon which of the detector were misaligned. In 
the Fig \ref{errory} a plot of Y positions of backward detector is 
shown against $\delta t_0$. A straight line fit (solid blue line) clearly 
shows that the detector was tilted in azimuthal plane.

\begin{figure}[h]
\begin{center}
\includegraphics[height=16.0cm,angle=-90]{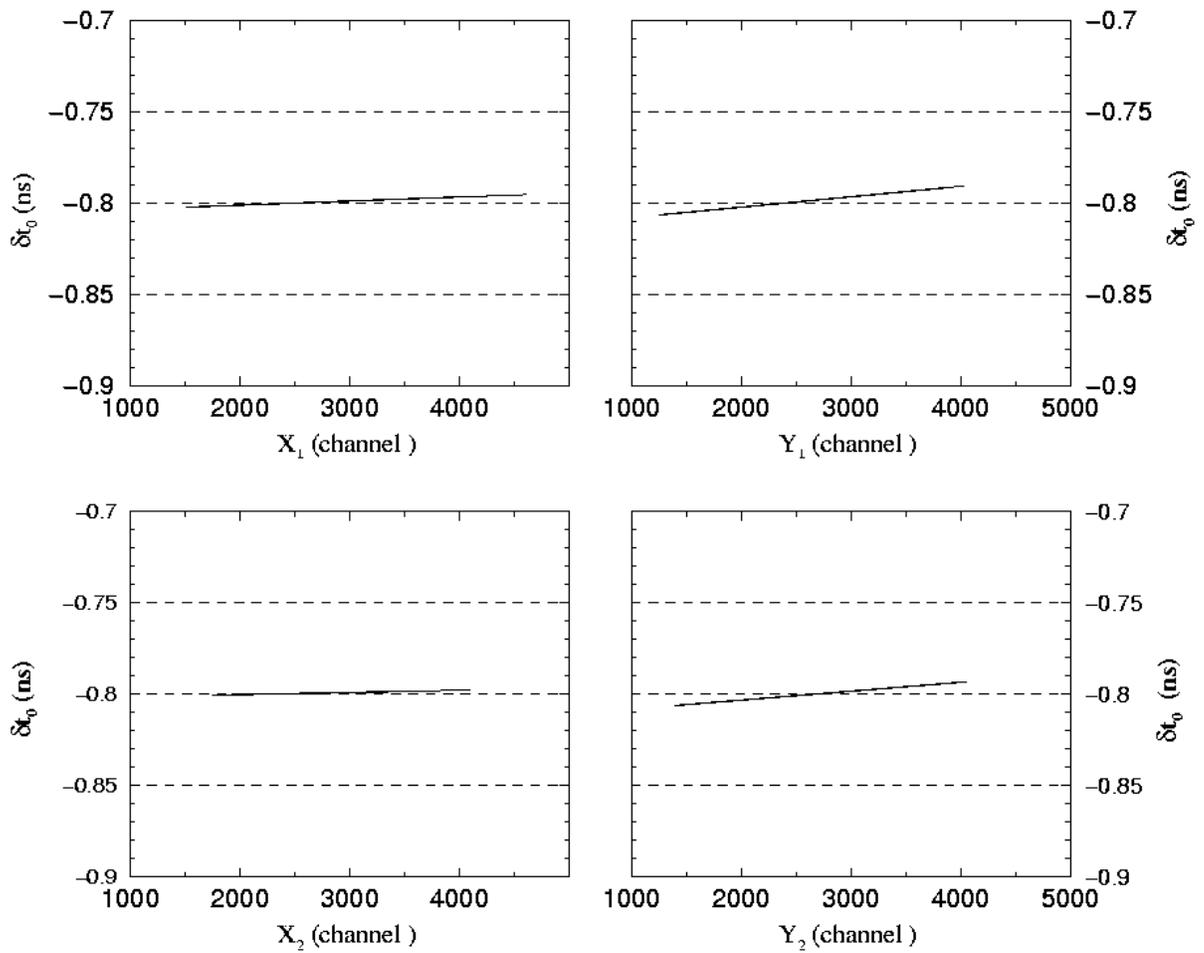}
\end{center}
\caption{\label{syserror} Variation of machine time  
with experimental positions of fission fragments in two detectors.}
\end{figure}

\indent Correlations of the $\delta t_0$ with $X_1$, $Y_1$, $X_2$ and $Y_2$ 
are shown in Fig \ref{syserror} for the system $^{19}$F + $^{232}$Th at 
c.m. energy 92 MeV. Corrections were applied after determining the angle of 
rotation or tilt of the detectors. For example, from the data of $Y_2$ vs
$\delta t_0$, the slope in $\delta t_0$ was found to be 0.01 cm/ns which 
corresponds to a rotation of 0.14$^\circ$ of the MWPC2 about the vertical 
axis. Similarly, the systematic error due to tilt of the MWPC2 was corrected 
from the variation of $X_2$ with $\delta t_0$.

\clearpage

\section{Measurement of fragment angular distributions}

\indent The fission fragment angular distribution is the distribution of 
yields of fission fragments at different c.m. angles. The measure of angular 
anisotropy, which can be related macroscopic properties of the nucleus at 
saddle point, expressed as the ratio of 
the yields at 0$^\circ$ and 90$^\circ$ with respect to the direction of beam 
 or that of yields at 180$^\circ$ and 90$^\circ$.

\begin{equation} 
A=\frac{W(0^\circ)}{W(90^\circ)}=\frac{W(180^\circ)}{W(90^\circ)}
\end{equation}

\begin{figure}[h]
\begin{center}
\includegraphics[height=12.0cm,angle=0]{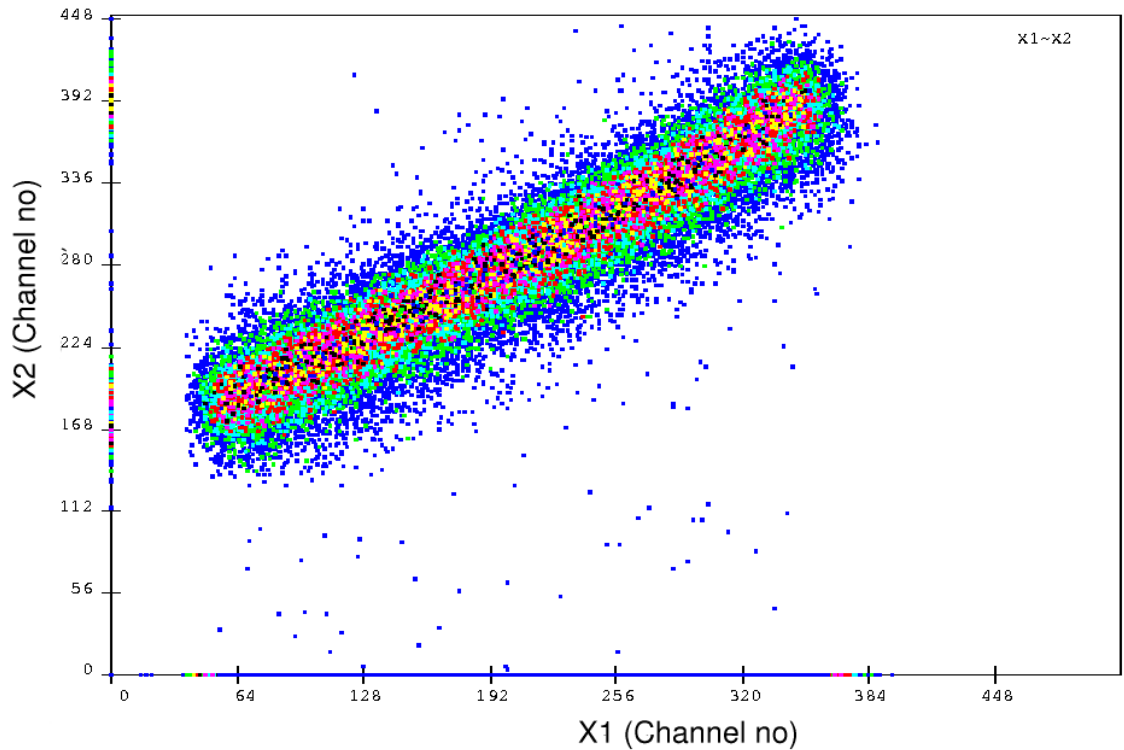}
\end{center}
\caption{\label{x1x2fbi96r10} The correlation between the X-positions of two 
MWPCs measured for the fissioning system $^{19}$F + $^{209}$Bi at the laboratory energy 96.0 MeV. }
\end{figure}

\indent To measure the angular distributions of the fusion fission 
events, transfer fission events should be exclusively separated using
 the folding angle measurement technique. However, in this thesis work we 
have measured the fragment angular distribution for the fissioning system 
$^{19}$F + $^{209}$Bi, where the contribution of transfer induced fission 
events are negligible. The folding angle for the complementary fragments 
were calculated from the event by event information on their positions. 
The polar angular correlation of the fission fragments in two MWPCs is 
shown in Fig. \ref{x1x2fbi96r10} for the fissioning system of 
$^{19}$F + $^{209}$Bi consists of a clear single band contrary to a double
band structure for the fissioning system of $^{19}$F + $^{232}$Th as shown 
in Fig. \ref{x1x2fth88}.The folding angle distribution essentially consisted 
of a single peak, as shown in Fig. \ref{fold_100}, since the transfer 
induced fission channel was populated for this system even at 
lower energies. 

%\begin{figure}[h]
%\begin{center}
%\includegraphics[height=10.0cm,angle=0]{CHAPIII/foldfbi96r10}
%\end{center}
%\caption{\label{foldfbi96r10} Measured  folding angle distributions for the 
%fissioning system $^{19}$F + $^{209}$Bi over a angular range of 
%43$^\circ$ to 67$^\circ$ at 96.0 MeV. The central position of the 
%forward detector was kept at 55$^\circ$. }
%\end{figure}

\subsection{Angular distribution data analysis}

\indent The angular distributions were calculated from the distributions
of the fission yields as a function of the polar angles. The azimuthal angles 
being the same for all the polar angles, the Y position co-ordinates of 
coincident counts (Y$_1$ and Y$_2$) were used only to effectively 
restrict the active area of the detectors about the reaction 
plane. Typical two dimensional plot of X1 and X2 are shown in 
Fig \ref{x1x2fbi96r10}. 
The total counts corresponding slices of spectra for a fixed angular interval 
(e.g. 1$^\circ$ or 3$^\circ$), translated to the corresponding X1 channels, 
are determined from 2D spectrum. For each energy, angular distributions 
were obtained by normalising the fragment counts with geometrical factors and 
beam intensities. The angular distributions were converted to 
the c.m. frame using the fragment velocities 
in the c.m. frame evaluated from the total kinetic energies using Viola's 
systematics \cite{Viola85*3} and recoil of the compound nucleus calculated 
using the two body kinematics . The anisotropy 
of the angular distribution was obtained from the ratio of the yields 
at 0$^\circ$ to that at 90$^\circ$ to from the fitting of the data with 
Legendre polynomial up to P$_6$ terms.

\section{Measurement of excitation function}

The excitation function of the fusion fission events is the variation of cross
 sections of the fragments coming from the fusion fission events as a function 
of incident energy. In this thesis work we have measured the excitation function  for the system $^{19}$F + $^{209}$Bi and compared with coupled channel 
calculations. To measure the excitation function for the fusion fission events, folding angle measurement technique was essentially the basic procedure to 
follow to separate the fusion fission events from the transfer induced fission 
events. However, it has be mentioned earlier that the transfer component in the total cross section is negligible in the reaction for this system. The 
total yields of the fusion fission events were obtained by integrating 
the fragment angular distribution functions over the angular range 
0$^\circ$ to 90$^\circ$ and calculating the fission cross section in 
comparisons to elastic yields which was monitored on two solid state 
detectors at $\pm 10^\circ$ to the beam.

\setcounter{equation}{0}
\setcounter{figure}{0}
\chapter{Results}
\markboth{nothing}{\it Results}

%\newpage
%\vspace*{5cm}
%\raisebox{12cm}{\fbox{\fbox{\parbox[b]{14cm}{
%\indent {\it Precise results of TOF measurements of fission fragment mass 
%distributions have been presented for judiciously chosen five 
%fissioning systems of $^{16}$O, $^{19}$F + $^{209}$Bi and 
%$^{19}$F, $^{16}$O, $^{12}$C + $^{232}$Th. 
%Fission fragment angular distribution measurement have also been presented
%for the reaction $^{19}$F + $^{209}$Bi. Results are compared with earlier 
%measurements.}
%}}}}

\newpage
\indent The mass distributions observed in heavy ion induced fission 
reactions are generally of symmetric shape  because the 
compound nucleus is generally formed with large excitation energy ($E^\star$)  
well above the fission barrier. The fragment shell effects observed in the 
mass distributions in the case of spontaneous and neutron or light heavy 
ion induced reactions at lower bombarding energies are not 
evident in the case of heavy ion induced reactions, due to washing out 
of the shell effects at high excitation energy and angular momenta brought 
into the fissioning composite system by the heavy ions. In general, an 
average increase in the width of the mass distribution is 
observed with the increase in the excitation energy of the  
fissioning nucleus \cite{ShenPRC87*4,ItkisYad*4}. It has also been shown 
in earlier studies that the mass 
distributions of fission fragments in heavy ion induced fission may provide 
information on the reaction mechanism involved in the fission process, due 
to admixture of fully equilibrated compound nuclear events and non-compound 
nuclear reactions such as fast fission \cite{PLBGre81*4}, quasi-fission 
 \cite{Swi81*4} and pre-equilibrium fission etc \cite{PRL85Ram*4,PRL90Ram*4}.
 Mass distributions following such an admixture would be 
expected to be broader than those for normal fission, because non-compound 
fission reactions are expected to have more asymmetric component arising due 
to  incomplete equilibration in mass degree of freedom.

\subsection{Salient features of the different target projectile combinations:}

\indent The nucleus $^{232}$Th is  deformed  with quadrapole 
deformation parameter $\beta_2 =$ 0.217 while $^{209}$Bi is a spherical 
nucleus. Large deviations from the statistical theory 
\cite{HSgeneva58*4} predictions in fragment anisotropy were reported 
\cite{BackPRC90*4,ZhangPRC90*4,ZhangPRC94*4,PRLNM96*4} for deformed $^{232}$Th 
target, while the those for spherical $^{209}$Bi target followed 
\cite{SamEPJ00*4,VulPRC86*4,KaiPR97*4} the statistical prediction. For the 
spherical bismuth target, the entrant system  is compact for any orientation 
whereas the entrance channel compactness in shape changes quite appreciably for the impact point of the projectile changing from equatorial to polar regions of the prolate thorium nuclei. The mass asymmetry parameters $\alpha$ (defined as 
($M_t - M_p$)/($M_t + M_p$) where $M_t$ and $M_p$ are the masses of the 
target and projectile respectively) for the systems $^{19}$F + 
$^{232}$Th (0.85) and $^{16}$O+$^{232}$Th (0.87) are less than the Businaro 
Gallone critical value $\alpha_{BG}$ ($ \sim $0.90 for both systems). Thus 
for these two systems mass flow is from target to projectile. But $\alpha 
(0.90) > \alpha_{BG}$ (0.89) for the $^{12}$C + $^{232}$Th system and 
the flow of mass is from the projectile to the target. Thus the mass flow 
for $^{12}$C  is opposite to that of $^{16}$O and $^{19}$F nuclei. However 
the expected mass flows are from target to projectile for both $^{19}$F + 
$^{209}$Bi ($\alpha=$ 0.83 and $\alpha_{BG} =$ 0.88) and $^{16}$O + $^{209}$Bi 
($\alpha=$ 0.86 and $\alpha_{BG} =$ 0.88) system.

\section{Results for $^{16}$O + $^{209}$Bi}

\begin{figure}[h]
\begin{center}
\includegraphics[height=18.0cm, angle=0]{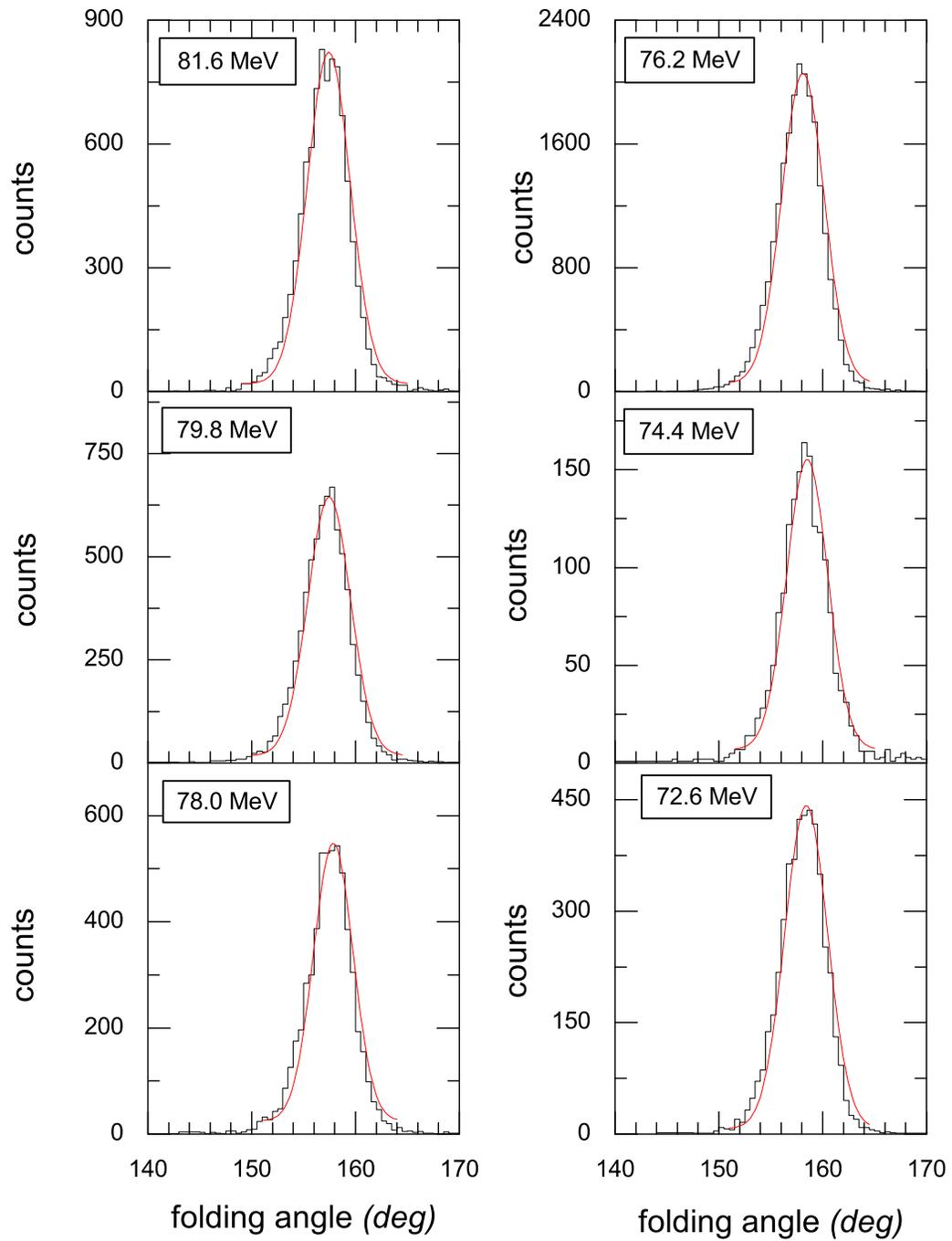}
\end{center}
\caption{\label{foldobi} Folding angle distributions at forward angles for the 
fissioning system of $^{16}$O + $^{209}$Bi at different $E_{c.m.}$.}
\end{figure}

\indent Fission fragment mass distributions were studied for this system 
at six bombarding energies of the projectile near and below the Coulomb barrier. The component of transfer induced fission is expected to be quite small due 
to large fission barrier heights ( $\sim$ 10 to 12 MeV) of the target like 
fissioning nuclei in this mass region. The measured folding angle distributions  and Gaussian fits to the experimental points at all energies are shown in Fig. \ref{foldobi}. It is observed that the 
folding angle distributions are essentially single peaked and  the events are
 almost entirely
from the fission following complete fusion. It is found that
 the peak of the 
measured distribution matches with the calculated folding angle using Viola's 
systematics \cite{Viola85*4} assuming binary fission.  

%\clearpage

\begin{figure}[ht]
\begin{center}
\includegraphics[height=18.0cm, angle=0]{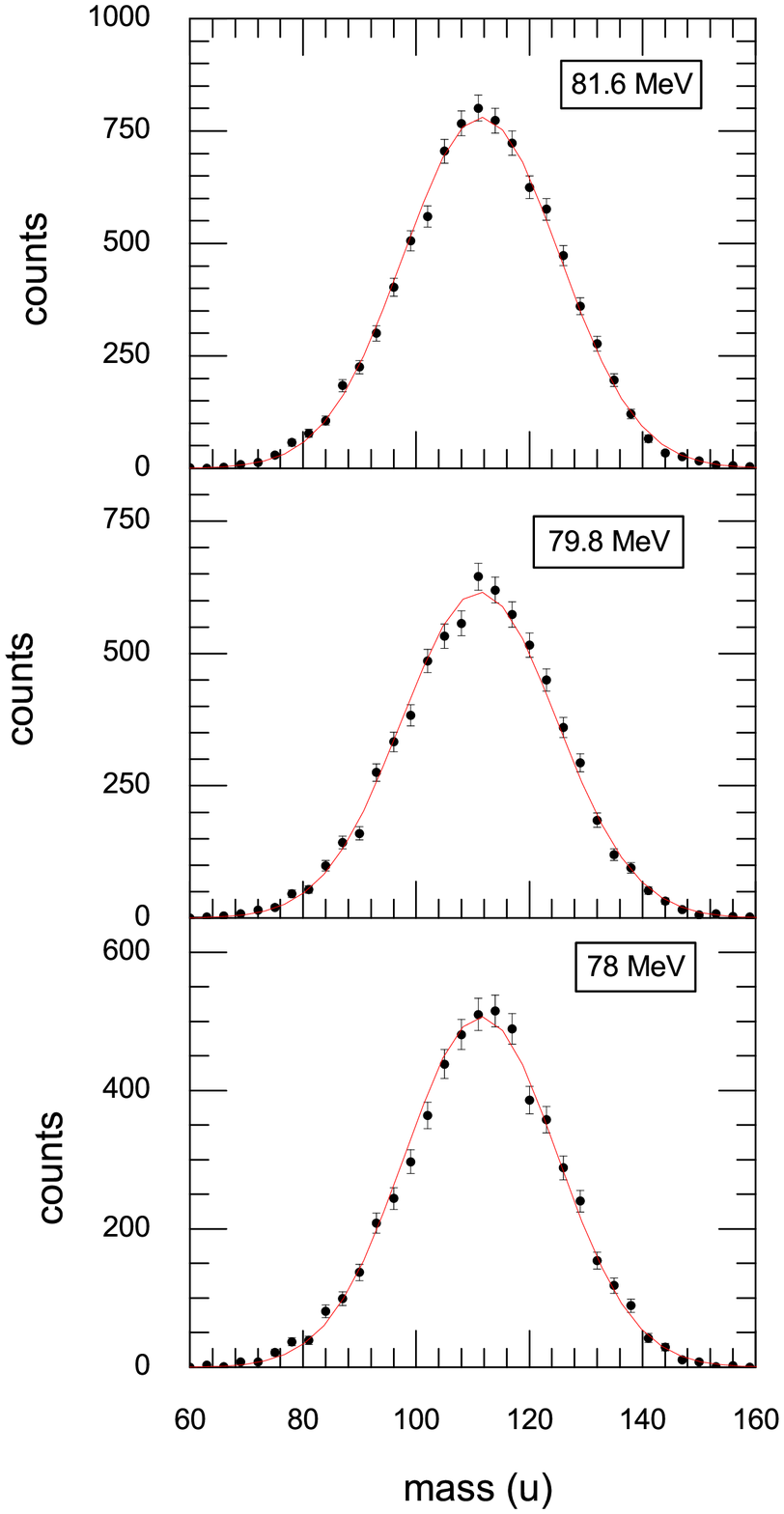}
\end{center}
\caption{\label{massdisobi1} Mass distributions for the fission 
system of $^{16}$O + $^{209}$Bi at different $E_{c.m.}$. The Gaussian fits 
are shown by solid red lines.}
\end{figure}

\begin{figure}[ht]
\begin{center}
\includegraphics[height=18.0cm, angle=0]{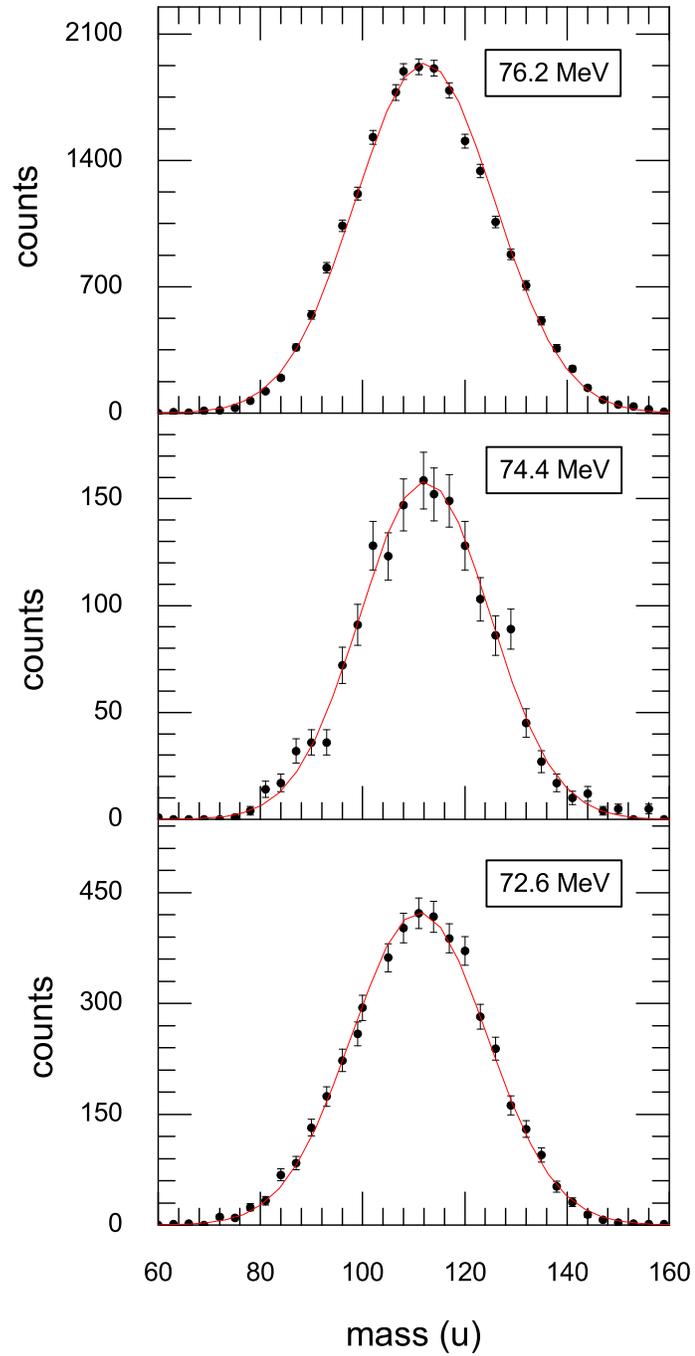}
\end{center}
\caption{\label{massdisobi2} Mass distributions for the fission 
system of $^{16}$O + $^{209}$Bi at different $E_{c.m.}$. The Gaussian fits 
are shown by solid red lines.}
\end{figure}

\indent  The mass distributions  calculated from the experiment following 
the procedure discussed in Chapter 3, at different 
centre of mass energies are shown in Fig. \ref{massdisobi1} at 81.6, 79.8 and 
78.0 MeV, and Fig. \ref{massdisobi2} at 76.2, 74.4 and 72.6 MeV. It is observed 
that the mass distributions are in general symmetric in shape peaking 
around $A_{CN}$/2 where $A_{CN}$ is the mass number of the compound nucleus. 
The width of the mass 

\begin{figure}[h]
\begin{center}
\includegraphics[height=12.0cm, angle=0]{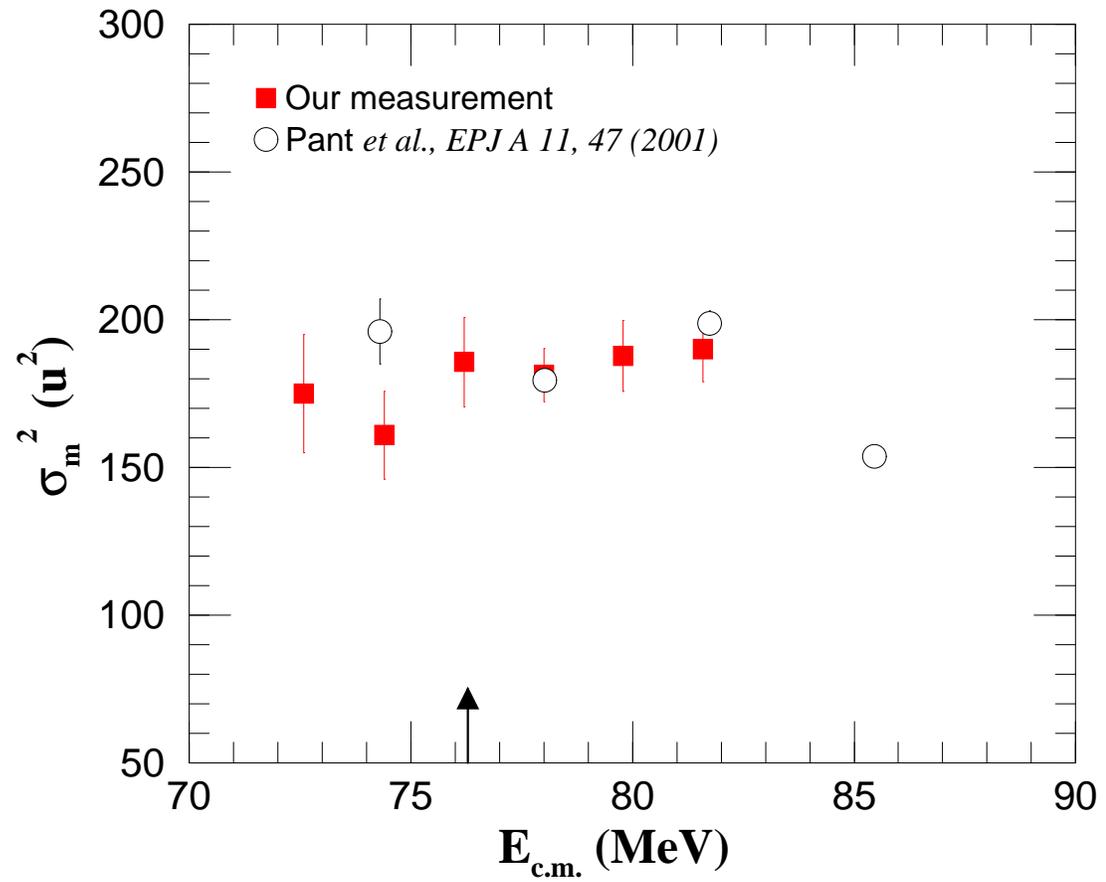}
\end{center}
\caption{\label{varobi_pant_ghosh} The measured variance of mass distributions 
 $\sigma_m^2$, as a function of $E_{cm}$ for the system $^{16}$O + $^{209}$Bi 
(solid square) with other measurement (open symbol). The Coulomb barrier is 
indicated by an arrow.}
\end{figure}

\clearpage

\noindent distributions were determined by fitting the spectra with Gaussian distribution. The values of the measured variance of the mass 
distributions $\sigma_m^2$ at different bombarding energies of the projectile 
is tabulated in table \ref{sigma2obi}. Fig. \ref{varobi_pant_ghosh} shows 
the variance of the fitted Gaussian, $\sigma_m^2$ , to the experimental 
masses as a function of the c.m. energy for the system $^{16}$O + $^{209}$Bi. 
It is seen that the variance of the mass distributions increases very slowly 
with increase  in $E_{cm}$.

\begin{table}[ht]
\begin{center}
\caption{\label{sigma2obi}~ Variance of the mass distributions for the system 
$^{16}$O + $^{209}$Bi. The Coulomb barrier for this system $V_b$ = 76.3 MeV 
in centre of mass frame and the Q value of the reaction is -47.3 MeV.}
%\begin{ruledtabular}
\setlength{\tabcolsep}{0.5cm}
\renewcommand{\arraystretch}{1.1}
\begin{tabular}[t]{|c|c|c|c|c|}
\hline\hline
$E_{lab}$ & $E_{cm}$ & $E_{cm}/V_b$ &$E^{\star}$ & $\sigma_m^2$ \\
(MeV) & (MeV) &  & (MeV) & ($u^2$) \\
\hline
90.0 & 81.6 & 1.09 & 34.3 & 189.8 $\pm$ 10.9\\
88.0 & 79.8 & 1.07 & 32.5 & 187.8 $\pm$ 11.9 \\
86.0 & 78.0 & 1.05 & 30.7 & 181.2 $\pm$ 09.0\\
84.0 & 76.2 & 1.02 & 28.9 & 185.6 $\pm$ 15.1\\
82.0 & 74.4 & 0.99 & 27.1 & 160.9 $\pm$ 14.9\\
80.0 & 72.6 & 0.97 & 25.3 & 175.0 $\pm$ 20.0\\
\hline\hline
\end{tabular}
\end{center}
\end{table}

\indent L.M.Pant {\it et al}., measured \cite{PantEPJ01*4} the mass distributions 
for the system $^{16}$O + $^{209}$Bi system at four bombarding energies 
near the Coulomb barrier. The variation of variances of the mass distributions 
$\sigma_m^2$, measured by Pant {\it et al}. is shown in 
Fig. \ref{varobi_pant_ghosh} 
along with our measurement. It is seen that their measurements match well 
with our measurement at the overlapping energies. However it may be noted that 
Pant {\it et al} measured fragment masses by finding the kinetic energies 
of complementary

\noindent fragments. The present results for 
$\sigma_m^2$ are, however, somewhat systematically higher than the that reported by 
Choudhury {\it et al}., \cite{ChoudhuryPRC99*4} for the $^{16}$O + $^{209}$Bi 
system. From our measurements, we 
conclude that for this system the width of the fragment mass distributions 
varies smoothly with increasing excitation energy.

\clearpage

\section{Results for $^{19}$F + $^{209}$Bi}

Fission fragment mass and angular distributions has been measured for this 
system over a range of bombarding energies from 86 MeV to 100 MeV in 
laboratory frame,simultaneously in the same  experimental setup.

\subsection{Mass distributions}

\begin{figure}[h]
\begin{center}
\includegraphics[height=18.0cm, angle=0]{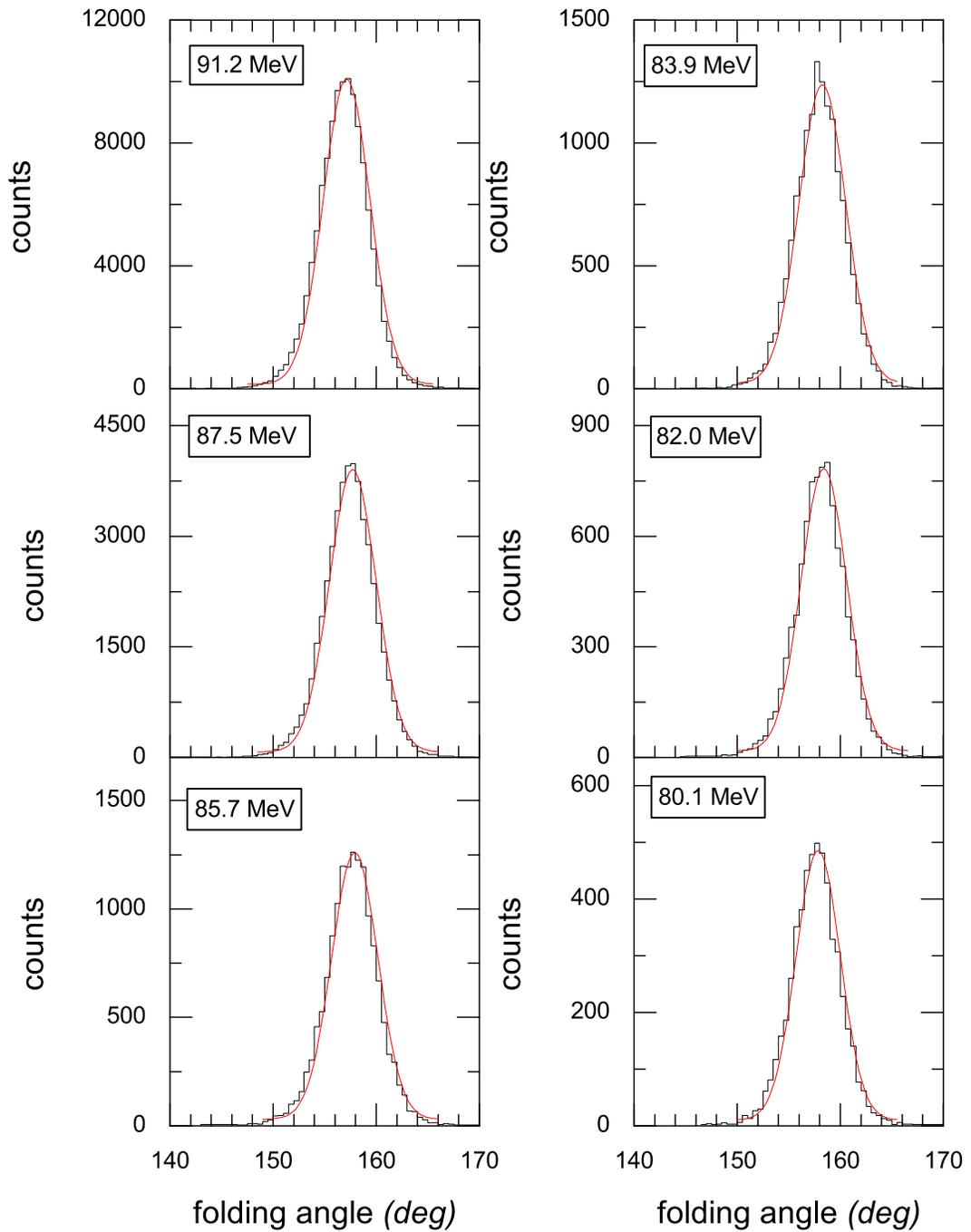}
\end{center}
\caption{\label{foldfbi} Folding angle distributions at forward angles 
for the fissioning system of $^{19}$F + $^{209}$Bi at different $E_{c.m.}$.}
\end{figure}

\indent Fission fragment mass distributions were measured for this system 
at six bombarding energies. For the fissioning system $^{19}$F + $^{209}$Bi, 
the non-compound nuclear fission channel following the transfer of a few 
nucleons is expected to be  quite small . In the present measurement, it 
showed no significant contribution up to 5 MeV below the Coulomb 
barrier. The experimental folding angle distributions at forward positions 
about 55$^\circ$ of the first detector are shown in fig \ref{foldfbi}. 
 It is observed that all the distributions are single peaked and all the 
events are from the fission following complete fusion.

%\clearpage

\begin{figure}[h]
\begin{center}
\includegraphics[height=18.0cm, angle=0]{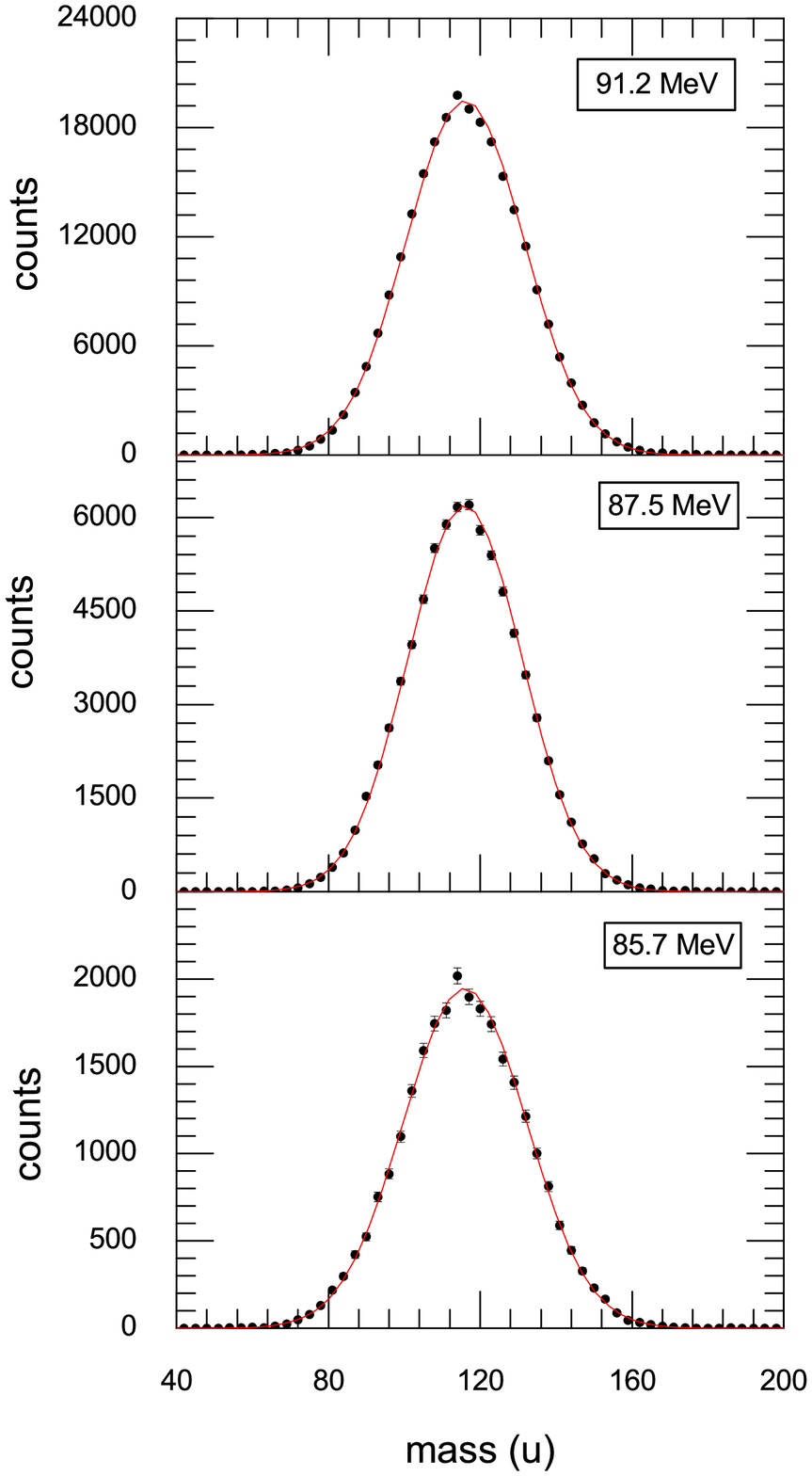}
\end{center}
\caption{\label{massdisfbi1} Mass distributions for the fission 
system of $^{19}$F + $^{209}$Bi at different $E_{c.m.}$. The Gaussian fits 
are shown by solid lines.}
\end{figure}

\begin{figure}[h]
\begin{center}
\includegraphics[height=18.0cm, angle=0]{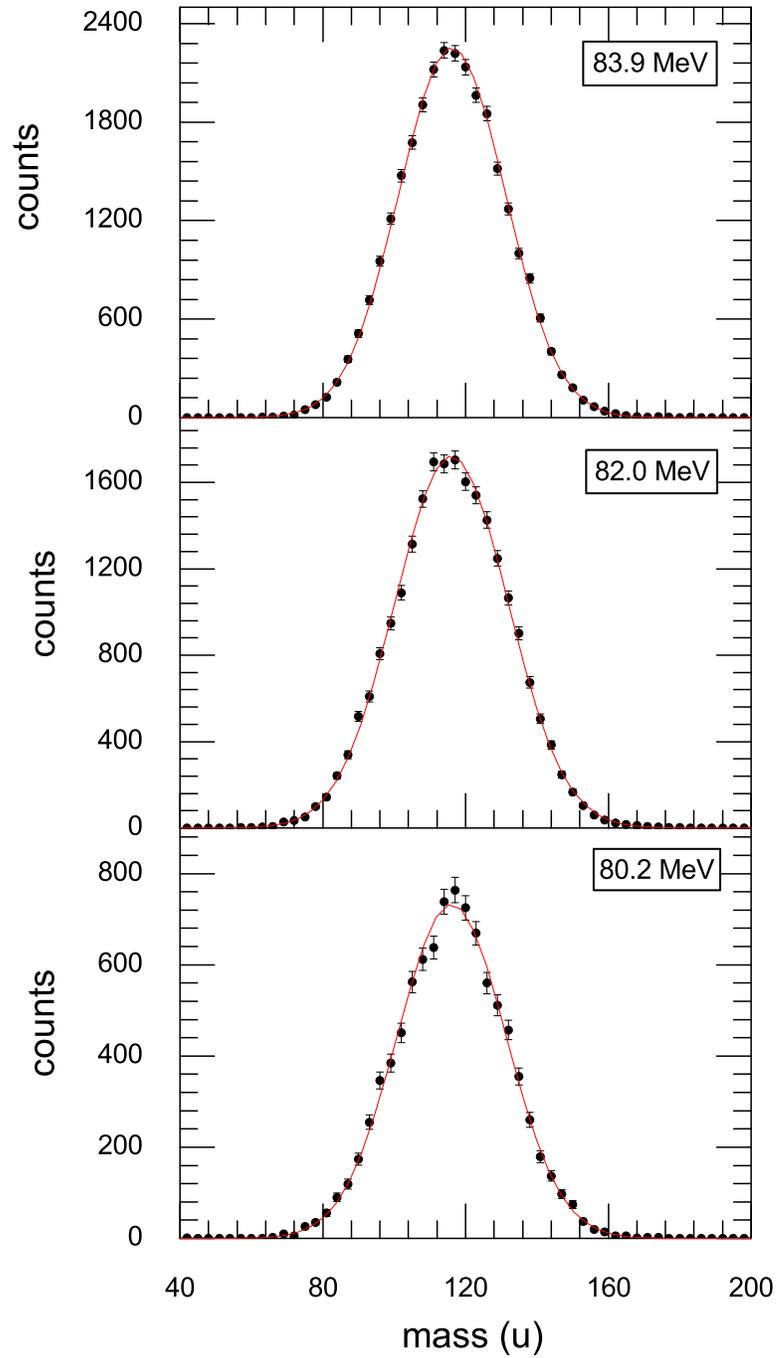}
\end{center}
\caption{\label{massdisfbi2} Mass distributions for the fission 
system of $^{19}$F + $^{209}$Bi at different $E_{c.m.}$. The Gaussian fits 
are shown by solid lines.}
\end{figure}

\indent The measured mass distributions are shown in Fig. 
\ref{massdisfbi1} at 91.2, 87.5 and 85.7 MeV and  in Fig. \ref{massdisfbi2} 
at 83.9, 82.0 and 80.2 MeV . The distributions were fitted with a Gaussian 
function. It is seen that the mass distributions were symmetric in shape with 
centroid around $A_{CN}$/2. We have not observed any significant 
departure from a single Gaussian fit even at lowest energies. The width 
of the mass distributions were derived from the fitted Gaussian distributions 
and the values of $\sigma_m^2$ are tabulated in the table \ref{sigma2fbi}.

\begin{figure}[h]
\begin{center}
\includegraphics[height=11.0cm, angle=0]{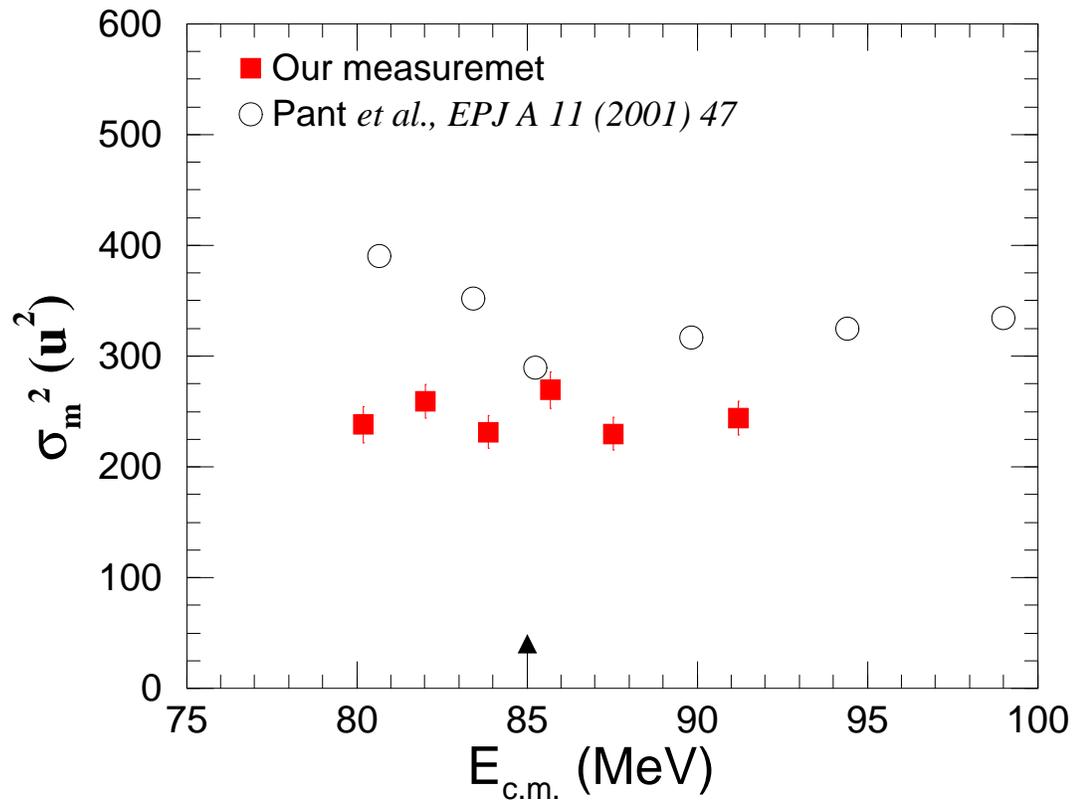}
\end{center}
\caption{\label{varfbi_pant_ghosh} The measured variance of mass distributions 
 $\sigma_m^2$, as a function of $E_{c.m.}$ for the system $^{19}$F + $^{209}$Bi 
(solid square) with other measurement (open symbol). The Coulomb barrier is 
indicated by an arrow.}

\end{figure}

\indent The variation of the variances of the mass distributions is shown 
in Fig. \ref{varfbi_pant_ghosh}. It is observed that the variation in variance 
of the mass distributions with beam energies to be linear with small slope.  

\clearpage

\noindent The measured variation in $\sigma_m^2$ 
by L.M. Pant {\it et al}., is 
shown by open circle in the Fig \ref{varfbi_pant_ghosh}. In their measurement 
an increase in variance of the mass distributions was reported near the 
Coulomb barrier. The increase in $\sigma_m^2$ was shown to correlate with 
the variation of the fragment kinetic energy with the bombarding energy near 
the Coulomb barrier. However, no anomaly in the variance of the mass 
distributions was observed in our measurement.

\indent It is to be  noted that the measurement of Pant {\it et al}., were 
performed using $\Delta E-E$ technique. Thus fission fragment mass 
distributions were derived  form the measured kinetic energies of the 
fission fragments.

\begin{table}[h]
\begin{center}
\caption{\label{sigma2fbi}~ Variance of the mass distributions for the system 
$^{19}$F + $^{209}$Bi. The Coulomb barrier for this system $V_b$ = 85.3 MeV 
in centre of mass frame and the Q value of the reaction is -48.9 MeV.}
%\begin{ruledtabular}
\setlength{\tabcolsep}{0.5cm}
\renewcommand{\arraystretch}{1.1}
\begin{tabular}[t]{|c|c|c|c|c|}
\hline\hline
$E_{lab}$ & $E_{cm}$ & $E_{cm}/V_b$ &$E^{\star}$ & $\sigma_m^2$ \\
(MeV) & (MeV) &  & (MeV) & ($u^2$) \\
\hline
99.5 & 91.2 & 1.07 & 42.2 & 244.3 $\pm$ 09.2\\
95.5 & 87.5 & 1.03 & 38.6 & 229.8 $\pm$ 08.9\\
93.5 & 85.7 & 1.01 & 36.7 & 269.3 $\pm$ 10.7\\
91.5 & 83.9 & 0.98 & 34.9 & 231.6 $\pm$ 09.1\\
89.5 & 82.0 & 0.97 & 33.1 & 259.2 $\pm$ 09.3\\
87.5 & 80.2 & 0.95 & 31.2 & 238.4 $\pm$ 11.4\\
\hline\hline
\end{tabular}
\end{center}
\end{table}

\clearpage

\subsection{Angular distribution}

\begin{figure}[h]
\begin{center}
\includegraphics[height=16.0cm,angle=0]{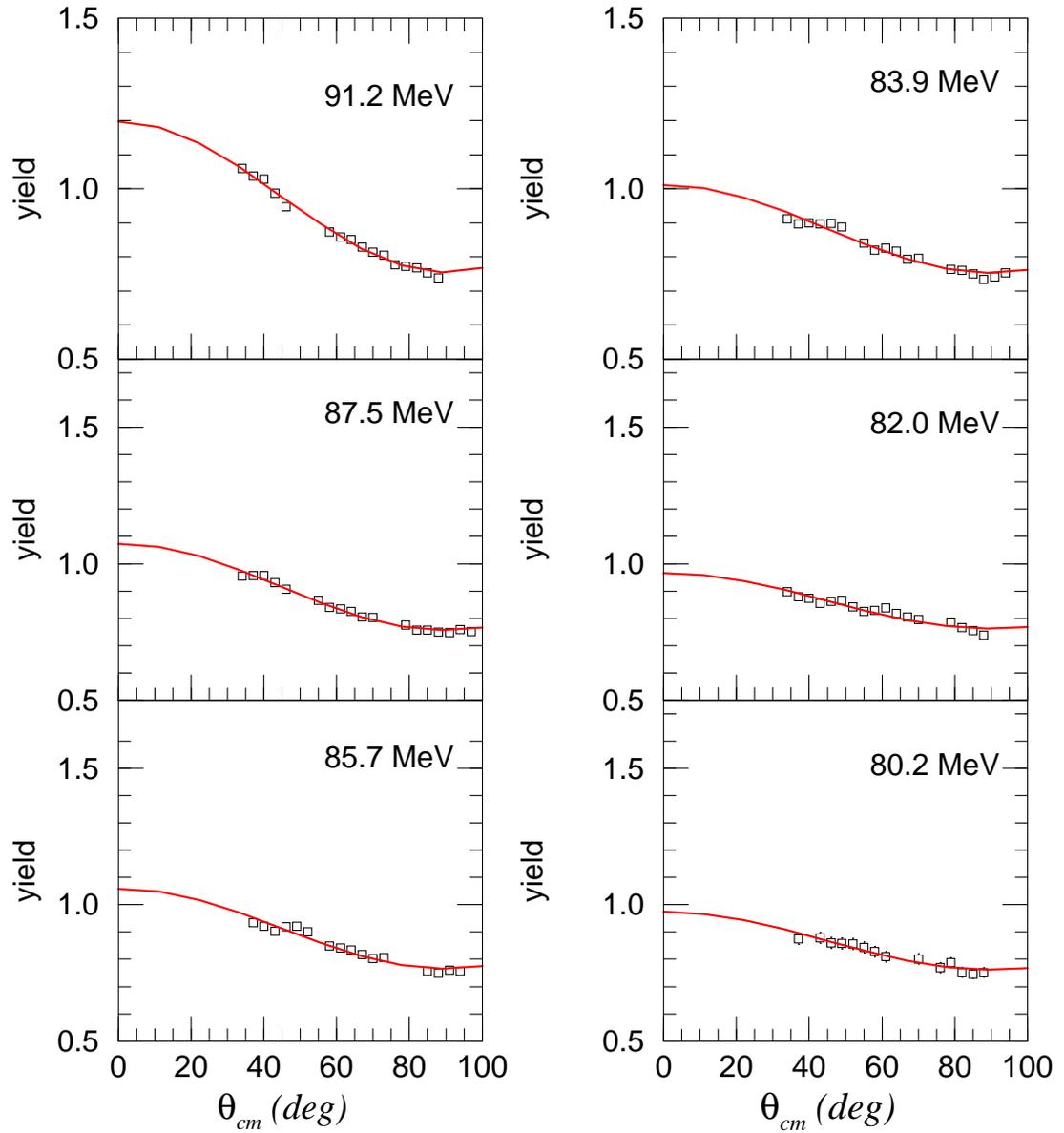}
\end{center}
\caption{\label{angdisfbi} The fragment angular distributions for the 
fissioning system of $^{19}$F + $^{209}$Bi of the FFCF events (open circle) 
in the c.m. frame. The Legendre polynomial fits are shown by solid red lines.}  
\end{figure}

\indent The fragment angular distributions were measured at seven bombarding 
energies. The angular distributions for all the events in the c.m. frame 
are shown in Fig. \ref{angdisfbi} for all the bombarding energies after 
kinematic transformation from the lab. frame. The calculation of the 
transformation factor has been discussed in the previous chapter. The fusion 
fission yields are shown by the open symbols.

\indent To calculate the anisotropies 
of the angular distributions, they were fitted with Legendre polynomials 
up to $P_4$ terms (solid red lines). From the fitted values of the yields 
at 0$^\circ$ and 90$^\circ$, the anisotropies of the angular distributions 
were calculated. The anisotropies obtained from distributions, fitted with 
polynomials up to $P_6$ terms did not change significantly and were within 
statistical error limits.

\indent The values of the calculated anisotropies ($A_{exp}$) of the measured 
angular distributions at all bombarding energies are given in table 
\ref{aniso_tab_fbi}. The theoretical anisotropies ($A_{SSPM}$), calculated 
following the SSPM are also shown in the table. The values were corrected for 
pre-saddle neutron multiplicities. The values of pre-saddle neutrons which 
lead to cooling at the saddle-point, have been taken from the works of 
Rossner {\it et al}., \cite{RossPRC92*4} who have determined these from actual 
measurements for $^{16}$O + $^{208}$Pb system. The calculated values of 
the corrected excitation energies, corrected temperatures and corrected 
variances of the K-distribution, along with some of more physical 
quantities are given in table \ref{fbi_tab_K02}. The method of calculation 
of the relevant parameters are described in Appendix.

\begin{figure}[h]
\begin{center}
\includegraphics[height=12.0cm,angle=0]{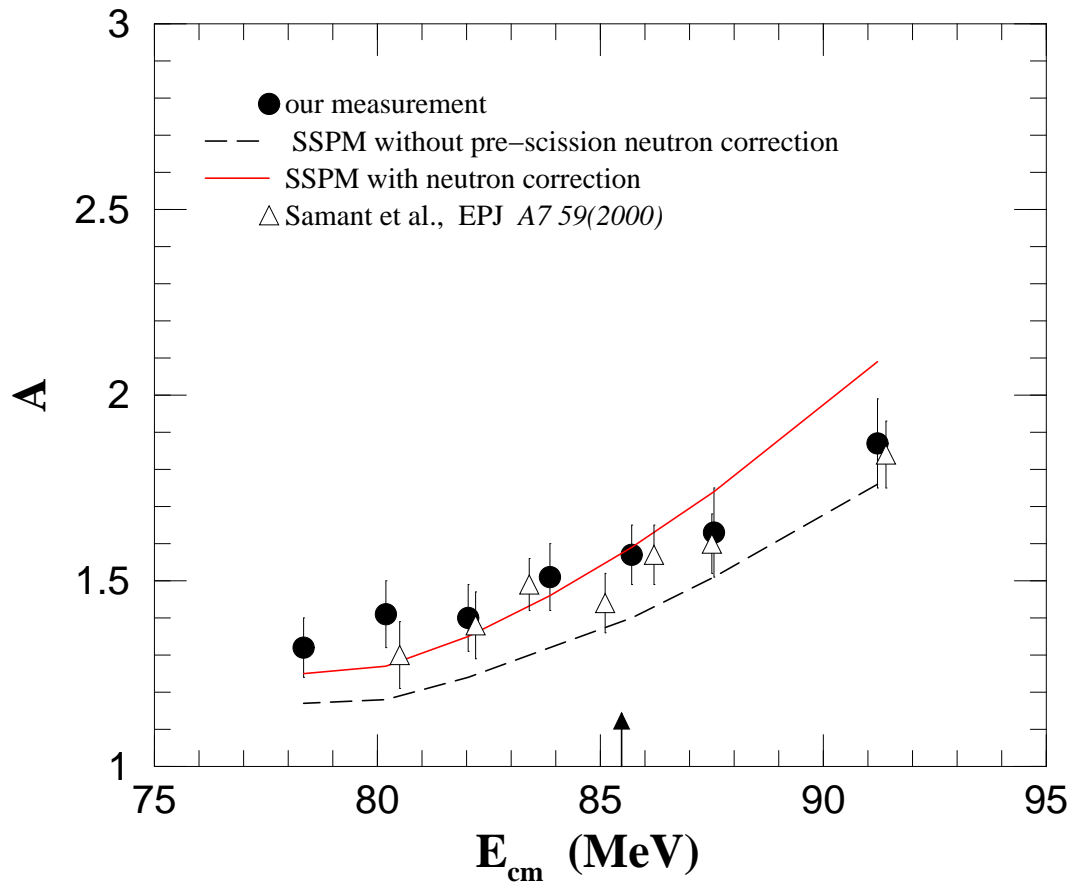}
\end{center}
\caption{\label{aniso_fbi} The measured fragment anisotropies  for the 
fissioning system of $^{19}$F + $^{209}$Bi of the FFCF events (solid circles). 
The neutron emission corrected SSPM calculation is shown by solid line.}  
\end{figure}

\indent The values of the anisotropies for the system $^{19}$F + $^{209}$Bi
 as a function of c.m. energy are shown in Fig. \ref{aniso_fbi}, illustrated 
by solid circles. The predicted values of anisotropies according to SSPM, 
incorporating the correction for pre-scission neutron emission are shown by 
solid line. The trend of the anisotropies obtained in the present 
measurement for the system $^{19}$F + $^{209}$Bi was found to be similar to 
that observed by Samant {\it et al}., \cite{SamEPJ00*4}. 
The values of the measured 
anisotropies by Samant {\it et al}. are  

\clearpage

\noindent shown by open triangles in 
Fig \ref{aniso_fbi}. The dashed line represents the calculation which does 
not include the neutron emission correction. It is found that the calculation 
which included the pre-scission neutron emission provides a better description 
of the anisotropy data.

%\clearpage

\begin{table}[h]
\begin{center}
\caption{\label{fbi_tab_K02}~ Several physical quantities calculated for 
calculation of anisotropy for $^{19}$F + $^{209}$Bi, incorporating the 
correction for pre-scission neutron emission .}
%\begin{ruledtabular}
%\setlength{\tabcolsep}{0.5cm}
\renewcommand{\arraystretch}{1.1}
\begin{tabular}[t]{|c|c|c|c|c|c|c|c|c|c|}
\hline\hline
$E_{cm}$ & $E^\star$ & $\nu_{pre}$ & $T_{n}$ & $E^\star_{corr}$ & $B_f(l)$
&$T_{corr}$ & $J_{eff}$ & $K_0^2$ \\
(MeV) & (MeV) &  & (MeV) & (MeV) & (MeV) & (MeV) & ($J_0^{-1}$) &  \\
\hline
91.2 & 42.23 & 2.23 & 1.25 & 22.16 & 3.67 & 0.90 & 1.145 & 122.14 \\
87.5 & 38.56 & 2.05 & 1.19 & 20.36 & 3.93 & 0.85 & 1.134 & 114.25 \\
85.7 & 36.72 & 1.97 & 1.15 & 19.38 & 4.06 & 0.82 & 1.131 & 109.92 \\
83.9 & 34.88 & 1.88 & 1.12 & 18.45 & 4.15 & 0.79 & 1.130 & 105.81 \\
82.0 & 33.05 & 1.77 & 1.08 & 17.72 & 4.21 & 0.77 & 1.127 & 102.86 \\
80.2 & 31.21 & 1.70 & 1.04 & 16.62 & 4.28 & 0.74 & 1.125 & 98.67  \\
78.3 & 29.36 & 1.59 & 1.00 & 15.85 & 4.30 & 0.71 & 1.125 & 94.68  \\

\hline\hline
\end{tabular}
\end{center}
\end{table}

\begin{table}[h]
\begin{center}
\caption{\label{aniso_tab_fbi}~ The experimental and the theoretical 
anisotropies for $^{19}$F + $^{209}$Bi at all bombarding energies .}
%\begin{ruledtabular}
\setlength{\tabcolsep}{0.4cm}
\renewcommand{\arraystretch}{1.1}
\begin{tabular}[t]{||c|c|c|c|c|c|c||}
\hline\hline
$E_{lab}$ & $E_{cm}$ & $E_{cm}/V_b$ & $<l^2>$ & $K_0^2$ & $A_{SSPM}$ 
& $A_{exp}$ \\
(MeV) & (MeV) &  & (coup) & (corr) & &  \\
\hline
99.5 & 91.2 & 1.07 & 534.19 & 122.14 & 2.09 & 1.87 \\
95.5 & 87.5 & 1.03 & 339.05 & 114.25 & 1.74 & 1.63 \\
93.5 & 85.7 & 1.01 & 257.40 & 109.92 & 1.59 & 1.57 \\
91.5 & 83.9 & 0.98 & 195.93 & 105.81 & 1.46 & 1.51 \\
89.5 & 82.0 & 0.97 & 143.80 & 102.86 & 1.35 & 1.40 \\
87.5 & 80.2 & 0.95 & 105.80 & 98.67  & 1.27 & 1.41 \\
85.5 & 78.3 & 0.93 &  96.69 & 94.68  & 1.25 & 1.32 \\

\hline\hline
\end{tabular}
\end{center}
\end{table}

\clearpage

\section{ Result for $^{19}$F + $^{232}$Th}

\indent Fission fragment mass distributions were precisely measured for the
fissioning system of $^{19}$F + $^{232}$Th at thirteen energies from above 
to below Coulomb barrier in the energy range 105.4 - 84.2 MeV in lab. frame. 
A large contribution from the target-like fragment fission (TLFF) 
events for the system, $^{19}$F + $^{232}$Th, was reported 
by Leigh {\it et al}.,
\cite{LeighPRC79*4}. Measurement of N. Majumdar {\it et al}., \cite{NMthesis*4}
showed contribution of TLFF channel of approximately 20 $\%$ of the 
total fission yield at energies around the Coulomb barrier and it amounted 
to almost 85 $\%$ of the total yield at deep sub-barrier regime. For this 
system, the folding angle measurement technique was of utmost interest 
to separate the FFCF events from all non-compound fission channels. This 
technique was extensively used \cite{NMPRC95*4,ZhangPRC94*4} in the measurements 
for this system where the chance of admixture of non-compound nuclear 
fission with the FFCF events were eliminated .

\begin{figure}[h]
\begin{center}
\includegraphics[height=18.0cm,angle=-0]{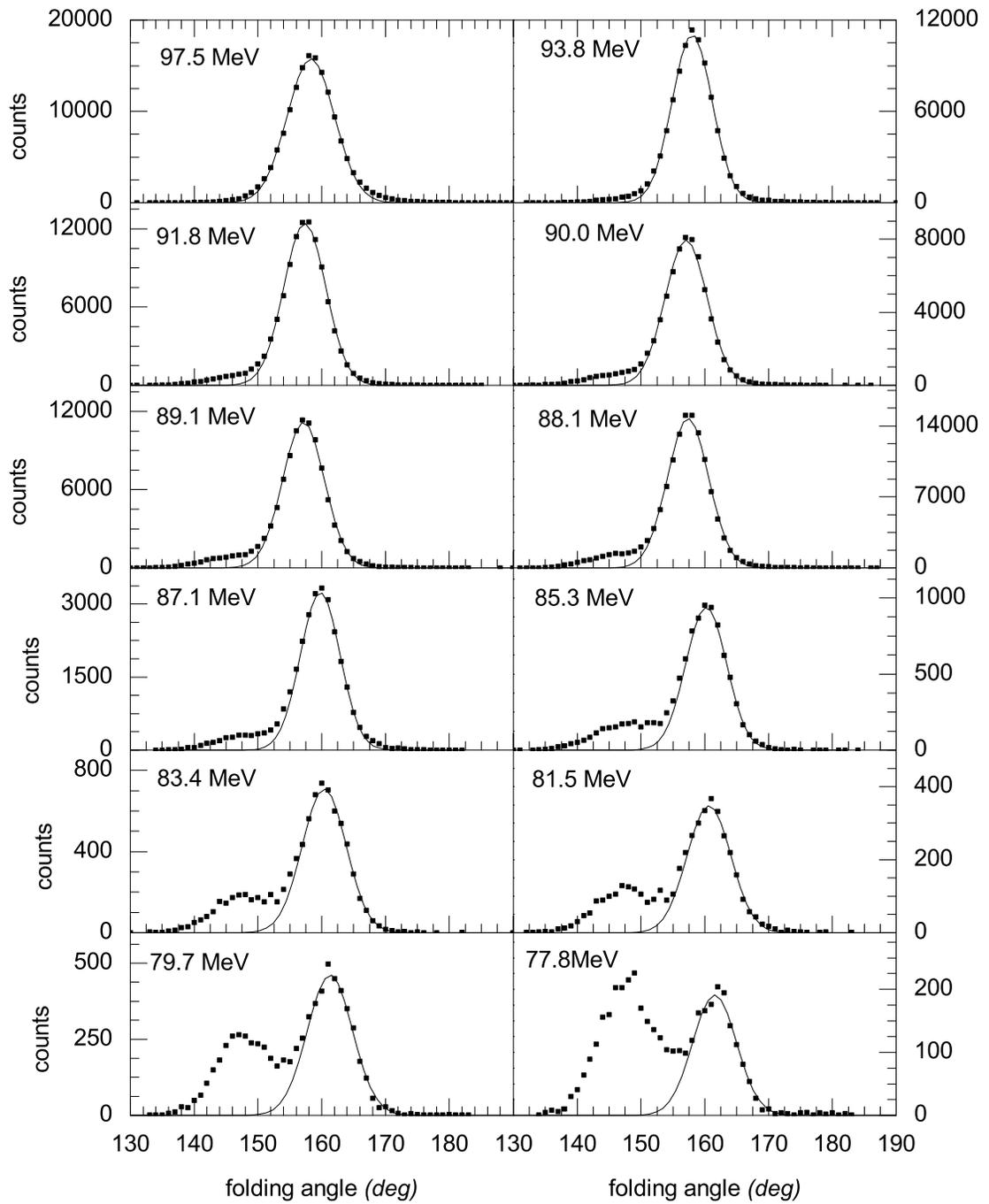}
\end{center}
\caption{\label{foldfth} The measured folding angle distributions for the 
fissioning system of $^{19}$F + $^{232}$Th at different energies in c.m. frame.}  
\end{figure}

\indent The folding angle distributions of the fission fragments for forward 
position around 65$^\circ$ of the forward detector are shown in 
Fig. \ref{foldfth}. The experimental distributions show the distinction 
between FFCF and TLFF yields. It is seen that the position of the 
experimentally observed peaks matches with the simulated peaks of the  
folding angle distribution of the FFCF events. Below 
Coulomb barrier, the distributions at lower folding angles represent the 
folding angle distributions of the TLFF events.  The distribution for the 
TLFF component is wider than that of the FFCF at near and sub-barrier energies
 due  to widely varying recoil angles and velocities.

\begin{figure}[h]
\begin{center}
\includegraphics[height=10.0cm,angle=0]{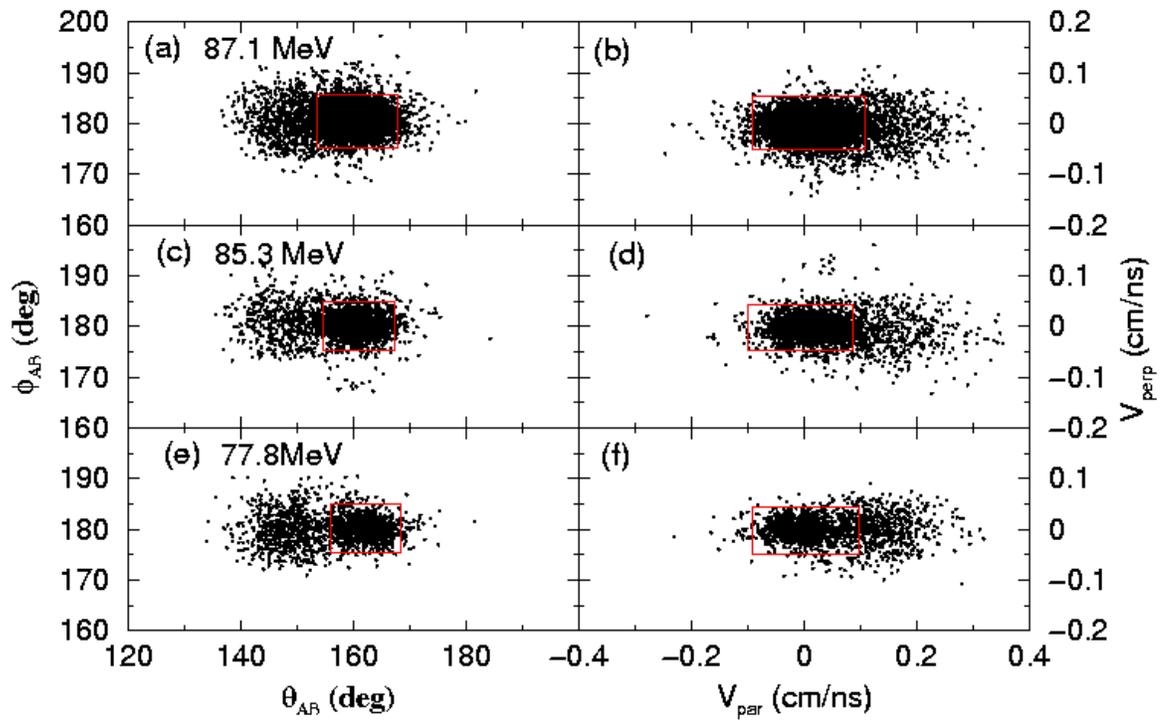}
\end{center}
\caption{\label{v_thfi1} Distributions of the complementary fission 
fragments in $\theta - \phi$ (left panels (a),(c) and (e) ) and 
$V_{par}-V_{perp}$ (right panels (b), (d),(e) ). The contour represents 
the gate used to select the fusion fission events.} 
\end{figure}

\indent As discussed in Chapter 3, transfer fission fragments can be 
separated from the FFCF from the distributions  of the parallel component 
and perpendicular velocity component of the fissioning nucleus. In 
Fig. \ref{v_thfi1} we have shown the separation of FFCF and TLFF 
events in both $\theta - \phi$ and $V_{par} - V_{perp}$ plane at 
three representative energies.

\clearpage

Table \ref{tab:vthfitab} shows the events 
used to calculate the variance of the mass distributions using both 
the gates in angles and velocities of the fissioning system. It is 
observed that the coincidence gates on  $\theta - \phi$ are more compact, 
excluding more TF events than in the correlation of velocities of the 
fissioning system.

\begin{table}[h]
\begin{center}
\caption{\label{tab:vthfitab}~ Gated events used to calculate the mass variance}
%\begin{ruledtabular}
%\setlength{\tabcolsep}{0.4cm}
\renewcommand{\arraystretch}{1.1}
\begin{tabular}[t]{|c|c|c|}
\hline\hline

$E_{cm}$ & Events within gate $\theta,\phi$ & Events within gate $V_{par},V_{perp}$ \\
(MeV) & ($\%$) & ($\%$)  \\
\hline
%97.5 & 87 & 88 \\
%93.8 & 81 & 86 \\
%89.1 & 77 & 85 \\
87.1 & 75 & 84 \\
85.3 & 74 & 79 \\
77.8 & 42 & 49 \\
\hline\hline
\end{tabular}
\end{center}
%\end{ruledtabular}
\end{table}

\begin{figure}[ht]
\begin{center}
\includegraphics[height=17.0cm,angle=0]{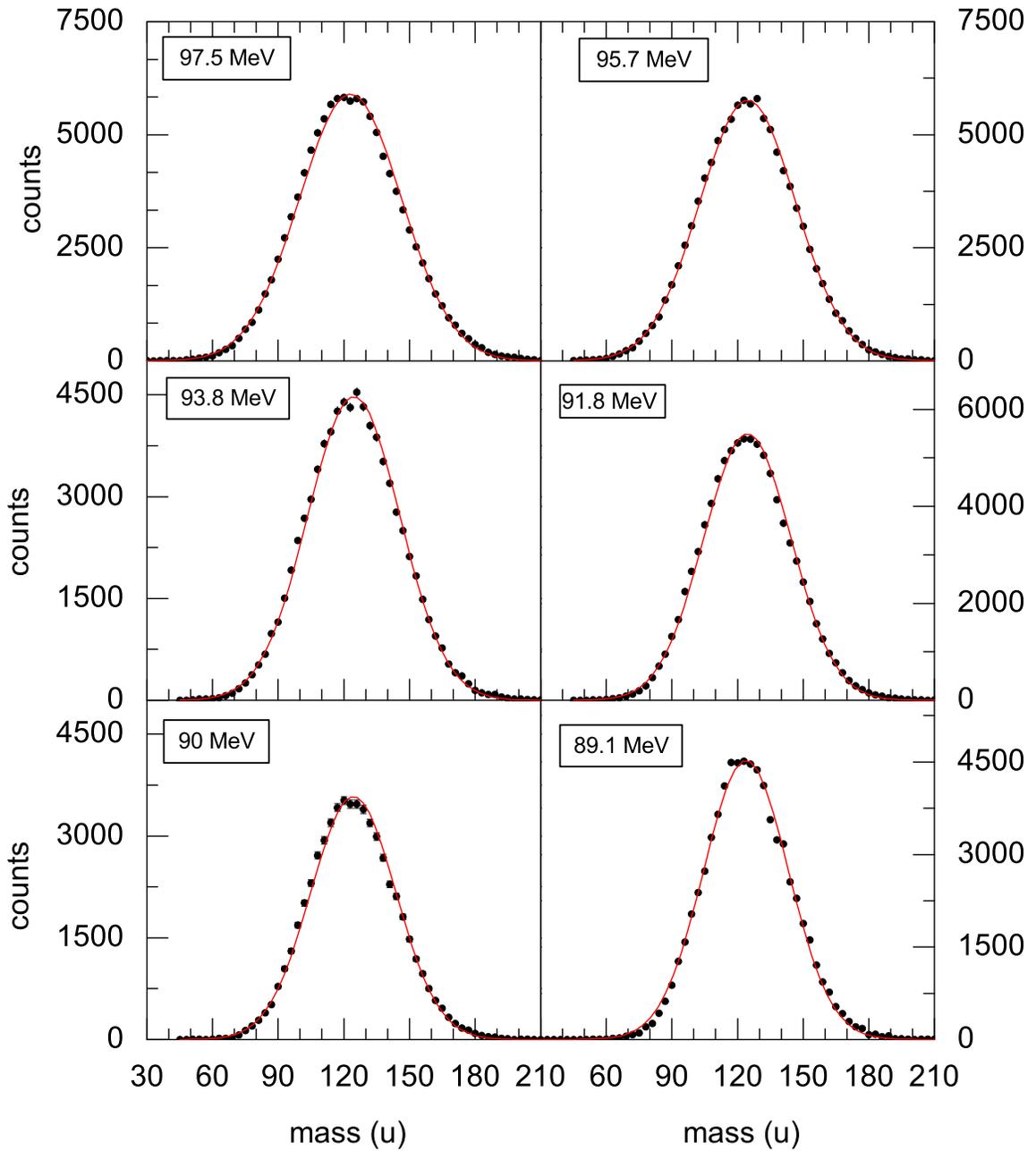}
\end{center}
\caption{\label{massdisfth1} The measured mass distributions for the 
fissioning system of $^{19}$F + $^{232}$Th at different energies in the 
c.m. frame.}  
\end{figure}

\begin{figure}[ht]
\begin{center}
\includegraphics[height=17.0cm,angle=0]{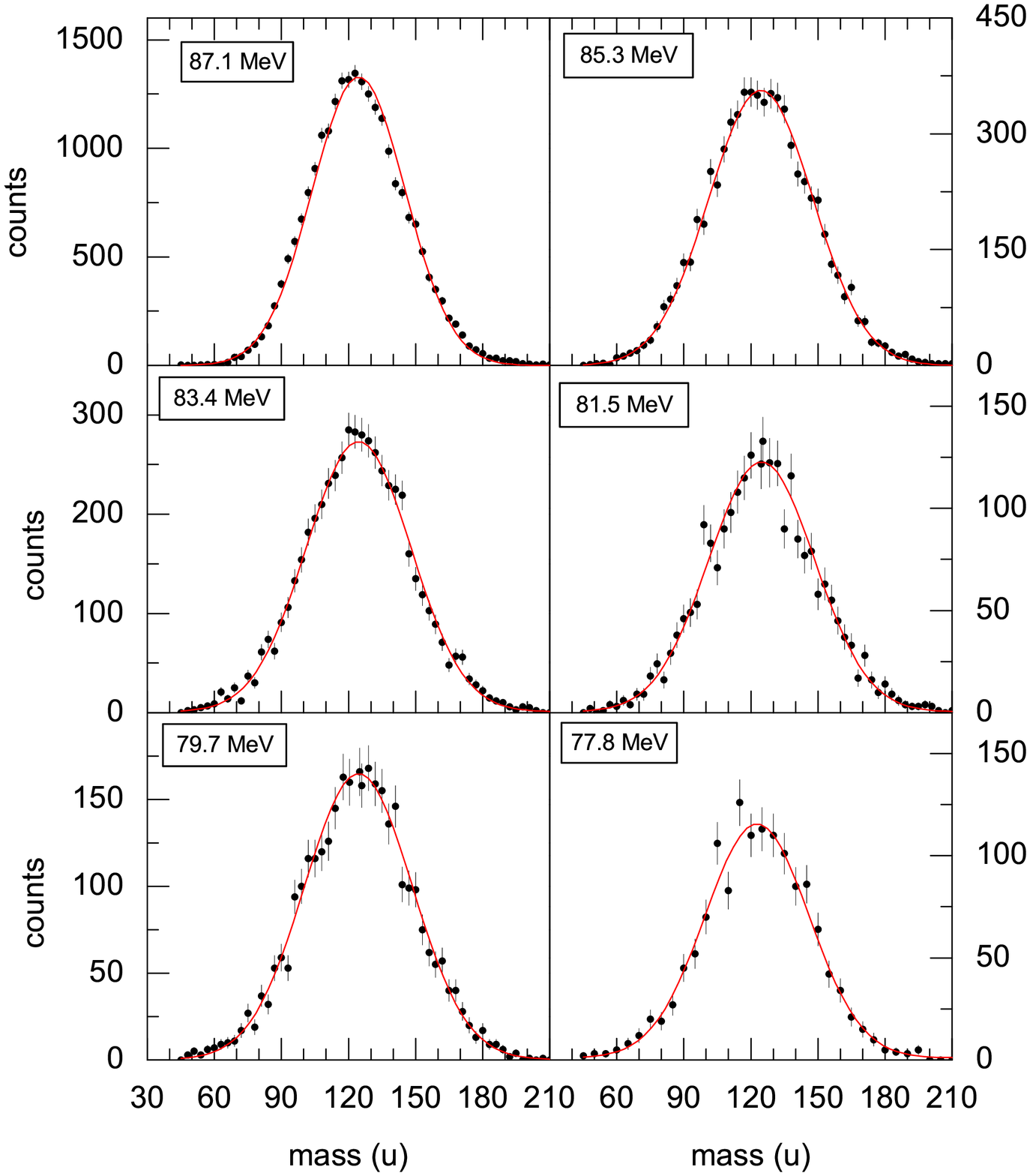}
\end{center}
\caption{\label{massdisfth2} The measured mass distributions for the 
fissioning system of $^{19}$F + $^{232}$Th at different energies in the 
c.m. frame.}  
\end{figure}

\indent The mass distributions of the FFCF reaction  are shown in 
Fig. \ref{massdisfth1} and Fig. \ref{massdisfth2} for the different 
incident energies in the c.m. frame.
The yields shown by the solid circles, are shown at a mass bin of 3 {\it amu}. 
To obtain the variance of the mass distributions, the distributions were fitted
 with Gaussian distributions. The solid line show the fits to the data. It is 
observed the distributions at all energies can be well fitted with a single 
Gaussian with peak close to the half of the combined target plus projectile 
mass. We have not observed any significant admixture of an asymmetric mass 
distribution in the measured mass distributions.

\begin{figure}[ht]
\begin{center}
\includegraphics[height=12.0cm,angle=0]{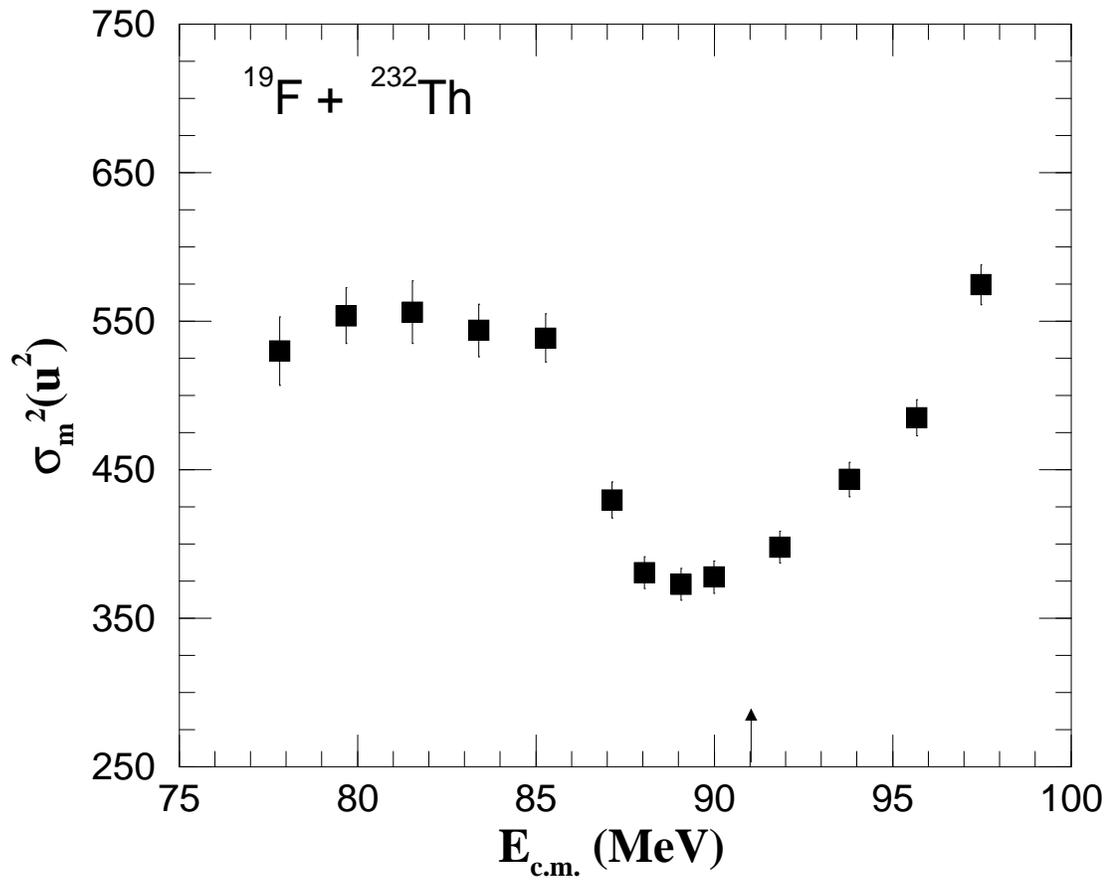}
\end{center}
\caption{\label{fthvaronly} The measured mass variance $\sigma_m^2$ for the 
fissioning system of $^{19}$F + $^{232}$Th as a function of projectile 
energies in the c.m. frame. Coulomb barrier is indicated by an arrow.}  
\end{figure}

\indent The variation of the variance of the fitted Gaussians $\sigma_m^2$ to 
the experimental masses as a function of c.m. energies are shown on Fig. 
\ref{fthvaronly} and the values are tabulated in table \ref{fthvar_tab}. 
Above the fusion barrier, $\sigma_m^2$ decreases with 
decrease in energy. However as the energy is decreased below the barrier, 
a sudden, almost 50$\%$ increase in the value of $\sigma_m^2$ is observed.
With further decrease to sub-barrier energies, $\sigma_m^2$ remains nearly 
constant with a small decreasing trend. However these values are substantially 
larger than the value at the barrier.

\clearpage

\begin{figure}[ht]
\begin{center}
\includegraphics[height=10.0cm,angle=0]{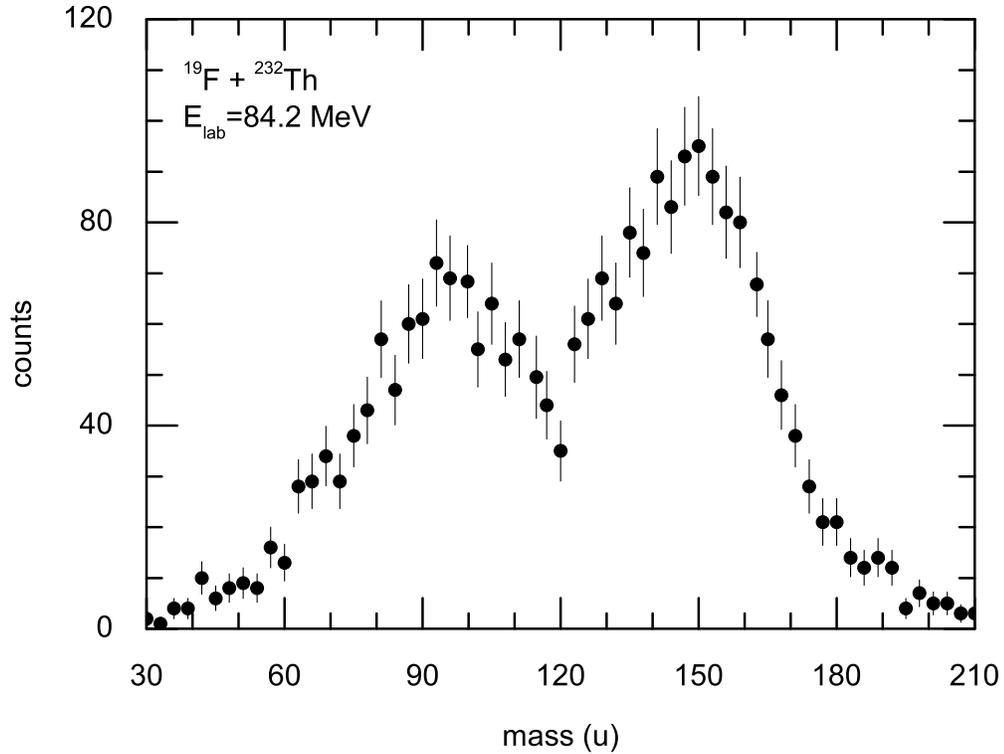}
\end{center}
\caption{\label{massdistf} The measured mass distributions for the TF components only for the  fissioning system of $^{19}$F + $^{232}$Th at 84.2 MeV in 
lab. energy.}  
\end{figure}

\indent The mass distributions for transfer fission is double humped
(shown in Fig. \ref{massdistf}. So it is important to ensure that the 
admixture of TF component was not going to affect the $\sigma_m^2$ values 
for FFCF reactions. Hence the effect of the admixture of TF was also 
investigated for the $^{19}$F + $^{232}$Th reaction.

%\clearpage

\begin{table}[h]
\begin{center}
\caption{\label{fthvar_tab}~ Variance of the mass distributions for the system 
$^{19}$F + $^{232}$Th. The Coulomb barrier for this system $V_b$ = 91.2 MeV 
in centre of mass frame and the Q value of the reaction is -40.55 MeV.}
%\begin{ruledtabular}
\setlength{\tabcolsep}{0.5cm}
\renewcommand{\arraystretch}{1.1}
\begin{tabular}[t]{|c|c|c|c|c|}
\hline\hline
$E_{lab}$ & $E_{cm}$ & $E_{cm}/V_b$ & $E^\star$ & $\sigma_m^2$   \\
(MeV) & (MeV) &  & (MeV) & $(u^2)$   \\
\hline
105.4 & 97.5 & 1.07 & 56.95 & 574.56 $\pm$ 13.42\\
103.4 & 95.7 & 1.05 & 55.15 & 484.88 $\pm$ 12.11\\
101.4 & 93.8 & 1.03 & 53.25 & 443.52 $\pm$ 11.58\\
 99.3 & 91.8 & 1.01 & 51.25 & 398.00 $\pm$ 10.77\\
 97.3 & 90.0 & 0.99 & 49.45 & 377.91 $\pm$ 10.88\\
 96.3 & 89.1 & 0.98 & 48.55 & 372.87 $\pm$ 10.62\\
 95.2 & 88.1 & 0.97 & 47.55 & 380.64 $\pm$ 10.73\\
 94.3 & 87.1 & 0.96 & 46.55 & 429.73 $\pm$ 12.23\\
 92.3 & 85.3 & 0.94 & 44.75 & 538.70 $\pm$ 16.24\\
 90.2 & 83.4 & 0.92 & 42.85 & 543.82 $\pm$ 17.72\\
 88.2 & 81.5 & 0.89 & 40.95 & 556.01 $\pm$ 20.98\\
 86.2 & 79.7 & 0.87 & 39.15 & 553.66 $\pm$ 18.82\\
 84.2 & 77.8 & 0.85 & 37.25 & 529.92 $\pm$ 23.01\\
\hline\hline
\end{tabular}
\end{center}
\end{table}

\begin{figure}[h]
\begin{center}
\includegraphics[height=12.0cm, angle=0]{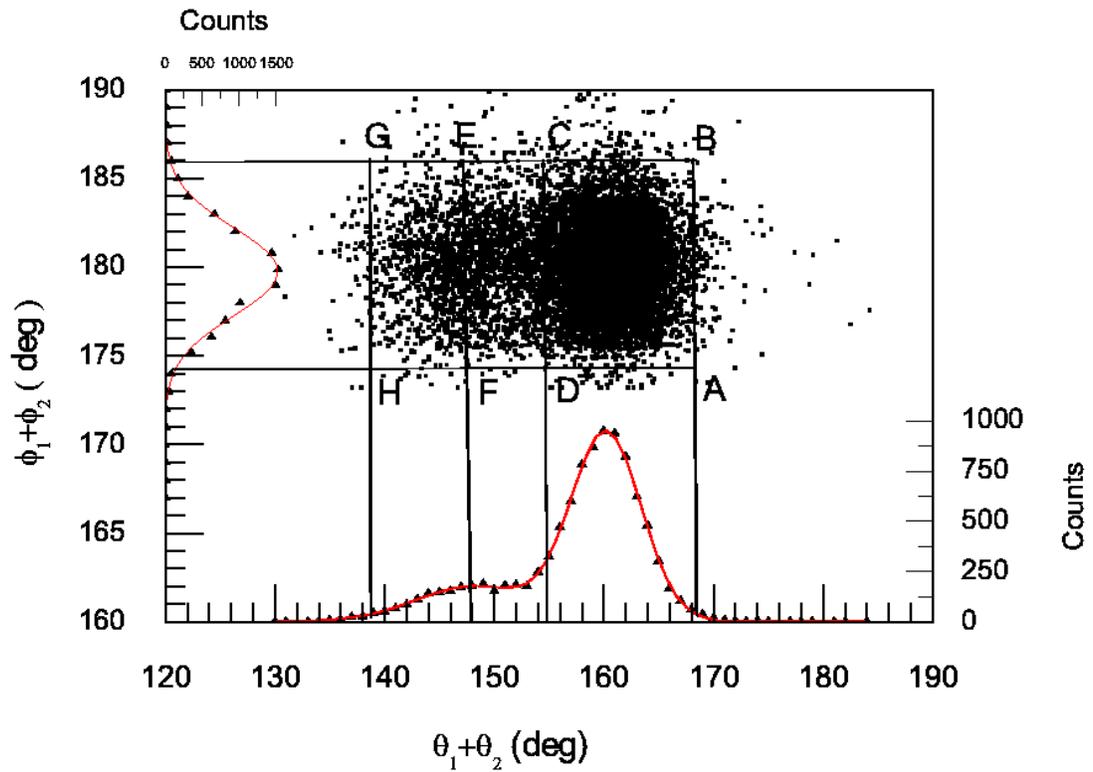}
\end{center}
\caption{\label{thfi_fth_gate} Distributions of complimentary fission 
fragments in $\theta, \phi$  at $E_{cm}=$ 85.3 MeV for the fissioning 
system of $^{19}$F + $^{232}$Th. Rectangle ABCD indicates the gate used to 
select the fusion-fission events for mass determination. Rectangles ABEF and 
ABGH indicate the gate used to add $50\%$ and $100\%$ of TF events, 
respectively}
\end{figure}

\begin{figure}[h]
\begin{center}
\includegraphics[height=10.0cm, angle=0]{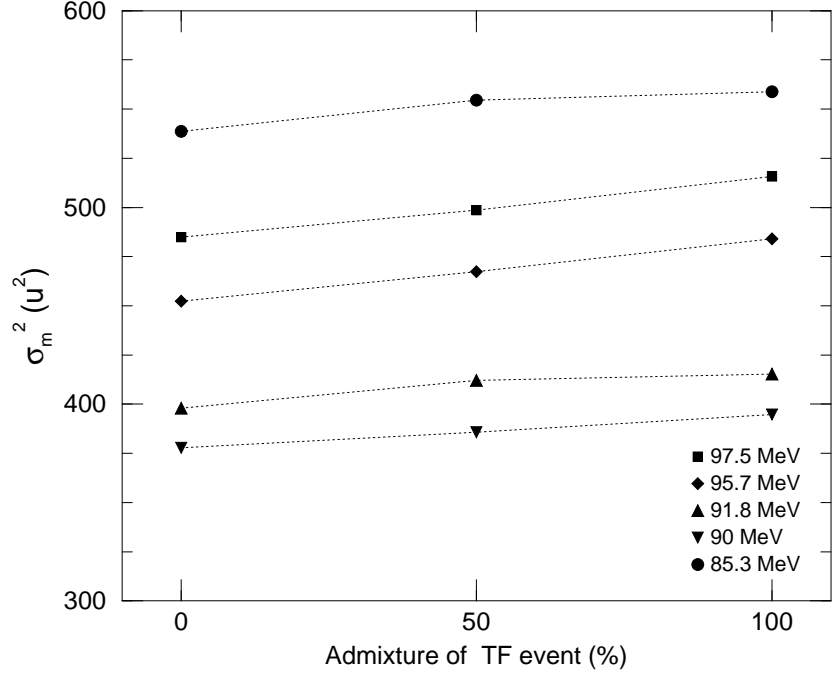}
\end{center}
\caption{\label{errortf} Variance of mass distributions at different projectile energies (c.m.) as a function of admixture of transfer fission (TF) events for 
the fissioning system $^{19}$F + $^{232}$Th at different $E_{cm}$.}
\end{figure}

\noindent It has been discussed in chapter 2 that fission fragments from 
compound nuclear fission events can be exclusively determined from the 
distributions of polar ($\theta$) and azimuthal ($\phi$) angles. The observed 
distributions of the complementary fission events in ($\theta$,$\phi$) plane 
are shown in Fig \ref{thfi_fth_gate} for the $^{19}$F + $^{232}$Th system at
 $E_{c.m.} =$ 85.3 MeV . The events enclosed by rectangle ABCD in the 
figure are the fragments from fusion fission reaction. The projections on 
 $\theta$ and $\phi$ planes are shown in the insets. At different energies, 
the window on the folding angles of the fission fragments were varied to 
estimate the effect of any admixture of non compound fission channels. In 
Fig \ref{errortf} the measured variances of the fission mass distributions 
are shown as a function of the admixture of transfer followed by fission (TF) 
events at different c.m. energies. The width of the distribution for any 
energy shows a slow increase (less than $10\%$) with admixture of TF events. 
Even at lower energies the contribution of transfer fission events does not affect
 the mass distributions significantly. The slow and linear increase in 
$\sigma_m^2$ values with increasing admixture of TF events as shown in Fig \ref{errortf} 
clearly indicates that the observed anomalous variation in $\sigma_m^2$ with 
energy for the deformed target can not be due to admixture of TF events.    

\clearpage

\section{ Result for $^{16}$O + $^{232}$Th}

\indent Fission fragment mass distributions has been measured for the 
fissioning system of $^{16}$O + $^{232}$Th at fourteen energies in the 
energy range 102.8 MeV to 78.6 MeV in the laboratory frame. For this 
fissioning system, it was observed earlier \cite{LestoneNPA90*4} that the 
non-compound nuclear fission channel following the transfer of a few nucleons 
is significantly populated to contribute to the total fission cross section.
The measurement of Majumdar {\it et al.} \cite{NMthesis*4} showed that the 
contribution of TF to the yield at deep sub-barrier energies was about 
30 $\%$. Folding angle technique was used to separate the TF events from the 
FF events.

\begin{figure}[h]
\begin{center}
\includegraphics[height=17.0cm,angle=0]{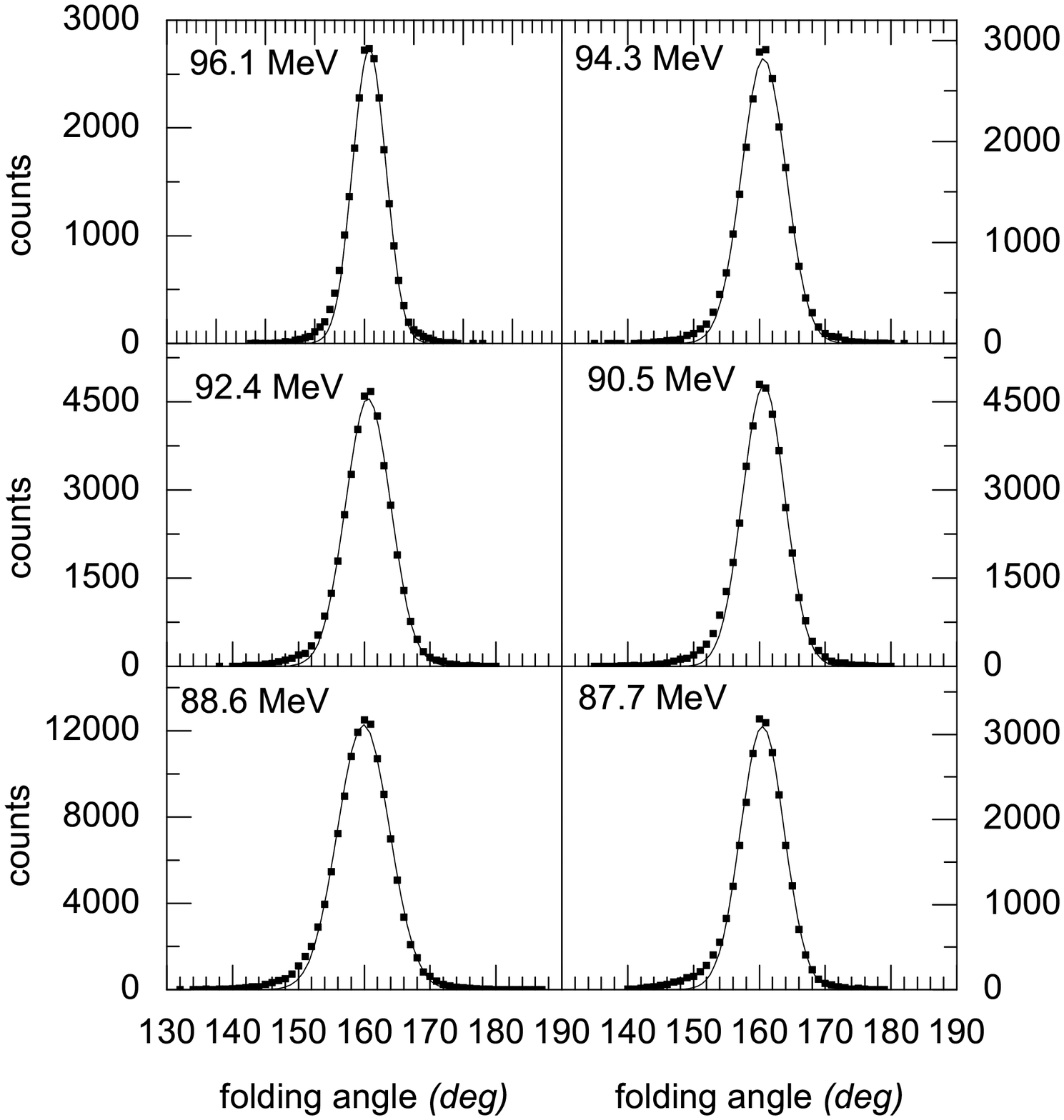}
\end{center}
\caption{\label{foldoth1} The measured folding angle distributions for the 
fissioning system of $^{16}$O + $^{232}$Th at different energies in c.m. frame.}  
\end{figure}

\begin{figure}[h]
\begin{center}
\includegraphics[height=17.0cm,angle=0]{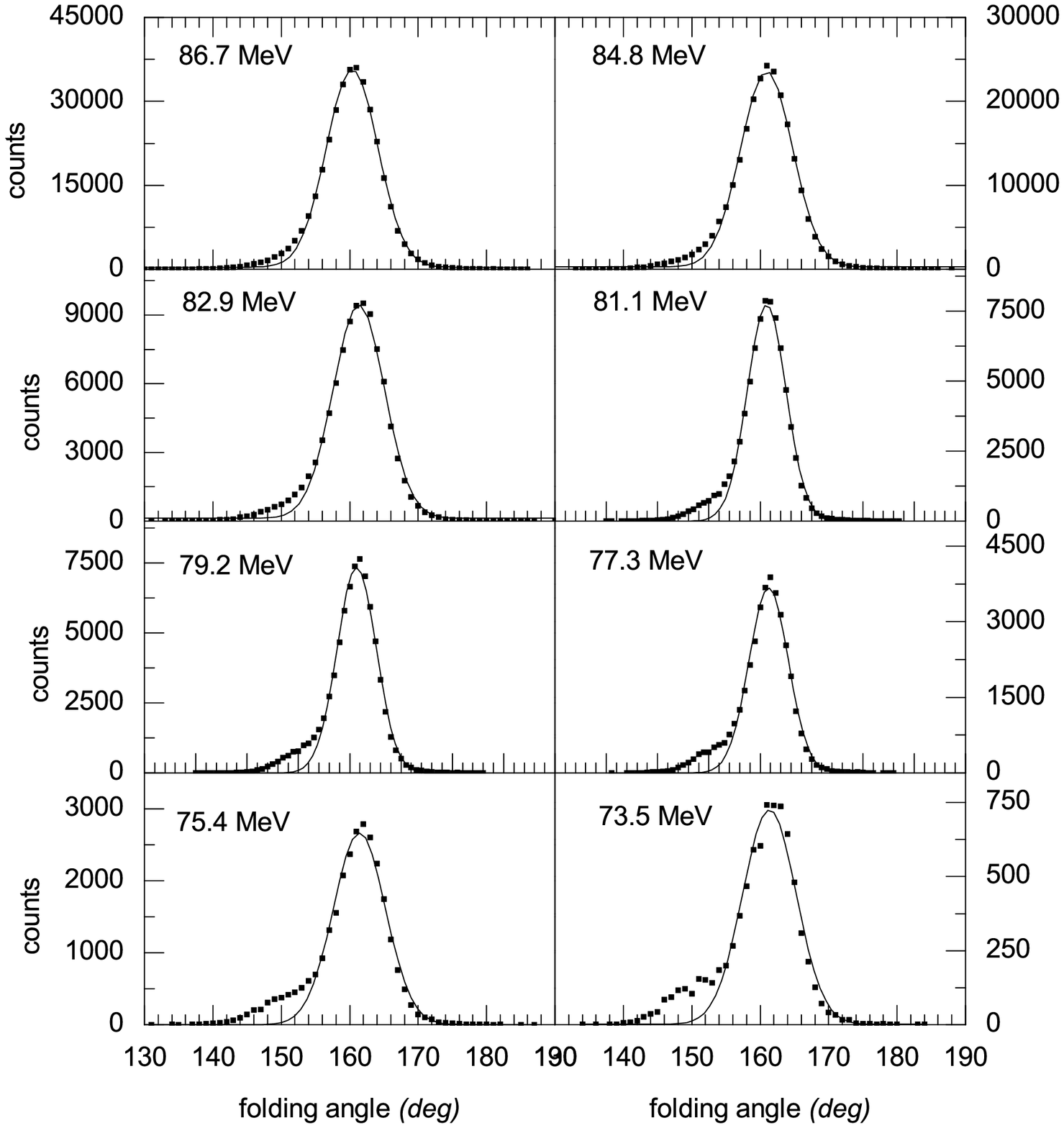}
\end{center}
\caption{\label{foldoth2} The measured folding angle distributions for the 
fissioning system of $^{16}$O + $^{232}$Th at different energies in c.m. frame.}  
\end{figure}

\indent  The experimental folding angle distributions at 
about 60$^\circ$ to the beam direction  are shown in
 Fig \ref{foldoth1} and \ref{foldoth2}. The distributions at lower 
folding  angles, noticeable clearly  below the Coulomb barrier, represent the 
folding angle distributions of the TLFF. Measured peak of the folding angle 
distributions of the FF events matches with the simulated value. Here also, 
like the fissioning system of $^{19}$F + $^{232}$Th, it was found that the 
distribution of TLFF component is wider than that of FFCF at near and 
sub-barrier energies due to widely varying recoiling angles  and velocities.
The FFCF events are separated from the distribution of events within specific 
$\theta-\phi$ values, as described in the case of $^{19}$F + $^{232}$Th 
reaction.

\begin{figure}[ht]
\begin{center}
\includegraphics[height=17.0cm,angle=0]{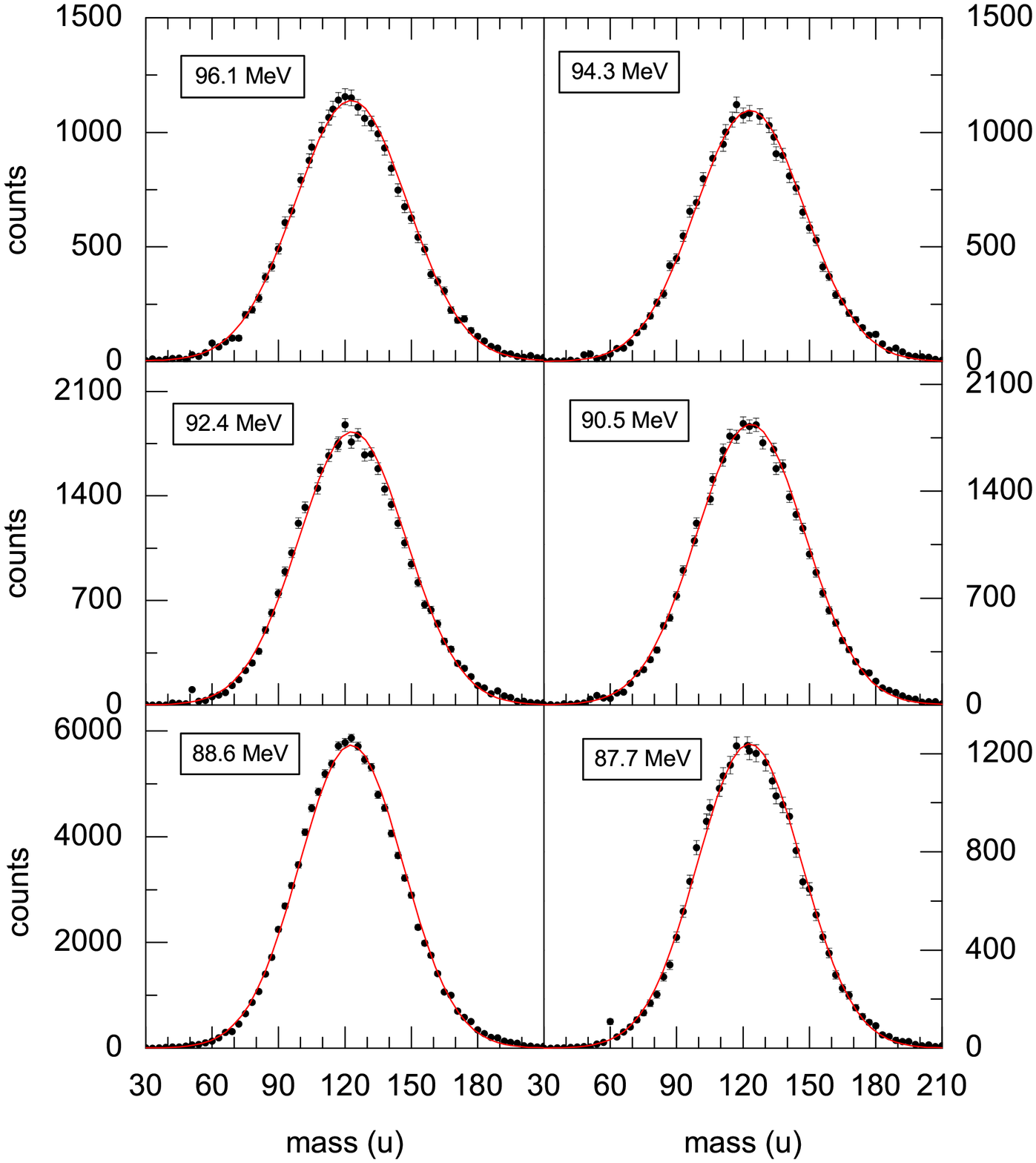}
\end{center}
\caption{\label{massdisoth1} The measured  mass distributions for the 
fissioning system of $^{16}$O + $^{232}$Th at different energies in the 
c.m. frame.}  
\end{figure}

\begin{figure}[ht]
\begin{center}
\includegraphics[height=17.0cm,angle=0]{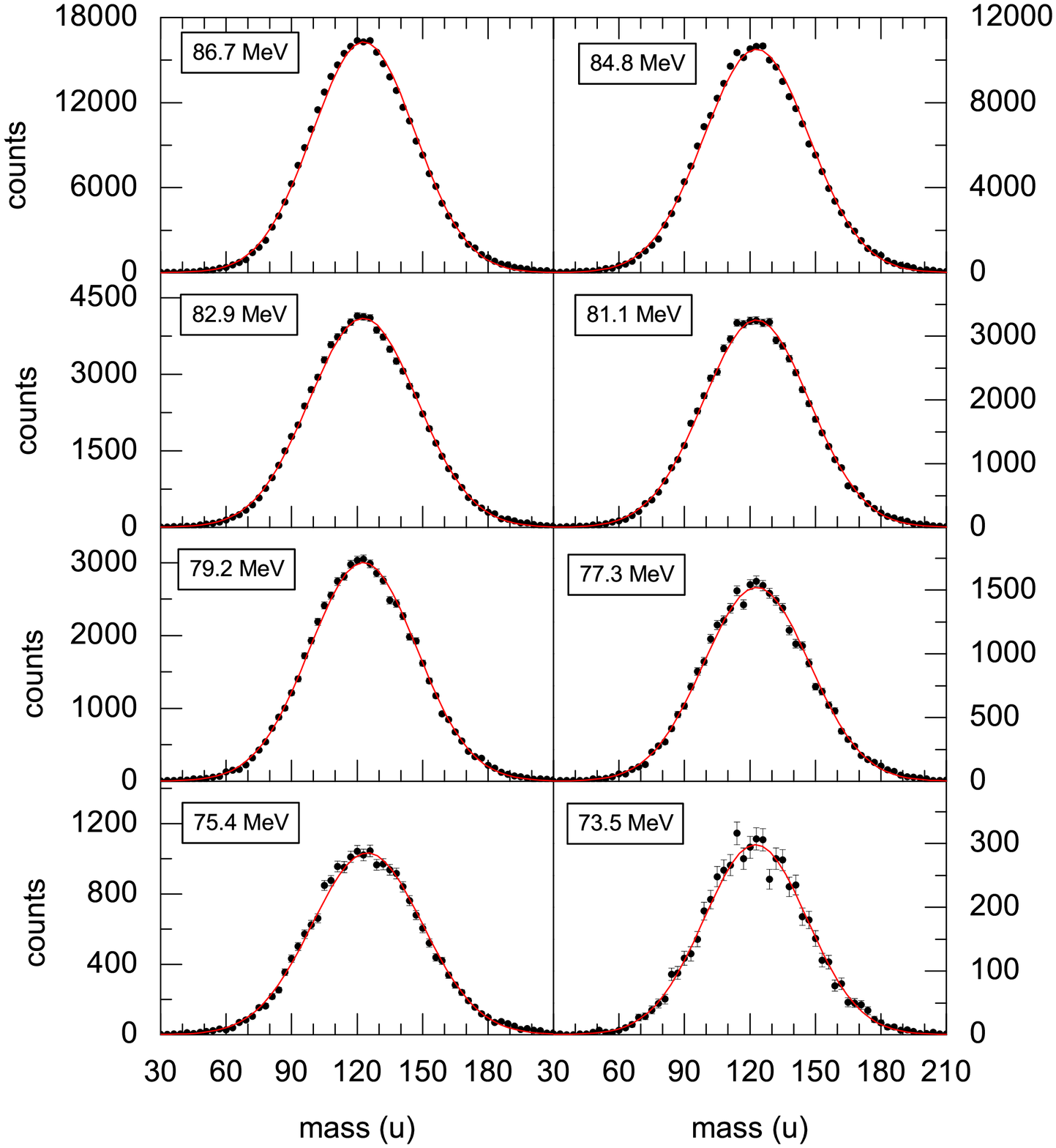}
\end{center}
\caption{\label{massdisoth2} The measured  mass distributions for the 
fissioning system of $^{16}$O + $^{232}$Th at different energies in the 
c.m. frame.}  
\end{figure}

\indent The mass distributions for the FFCF events are shown in Fig 
\ref{massdisoth1} and Fig \ref{massdisoth2} for all bombarding energies 
in the c.m. frame. The FFCF yields are shown by solid circle for 3 {\it amu} 
mass bin. To calculate the variance of the mass distributions, they were 
fitted with Gaussian distributions. The solid lines show the fit to the 
data. It can be observed that mass distributions are well fitted with Gaussian distributions even at lowest energies.

\begin{figure}[ht]
\begin{center}
\includegraphics[height=12.0cm,angle=0]{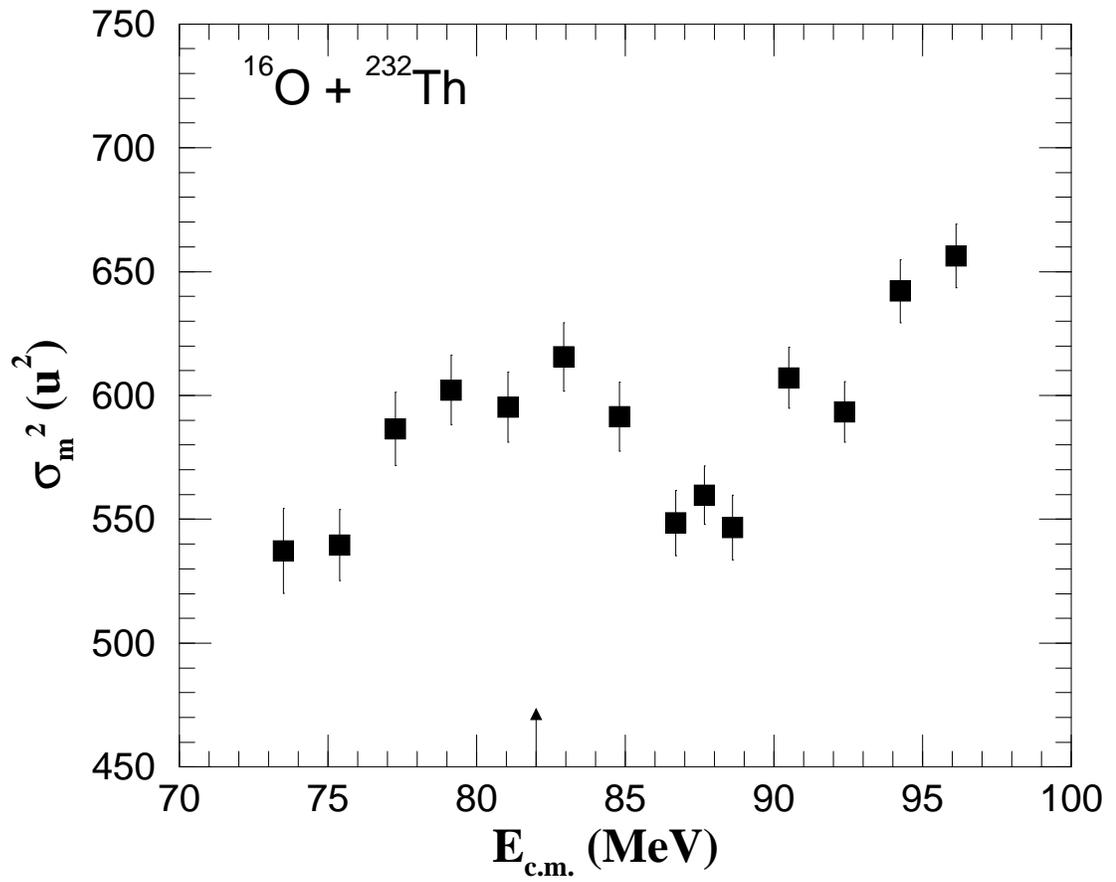}
\end{center}
\caption{\label{othvaronly} The measured mass variance $\sigma_m^2$ for the 
fissioning system of $^{16}$O + $^{232}$Th as a function of projectile 
energies in the c.m. frame. Coulomb barrier is indicated by an arrow.}  
\end{figure}

\indent The bombarding energy after correction for average energy degradation 
in target, $E_{lab}$, the corresponding energy in c.m. system, $E_{c.m.}$, 
the fraction of the energy relative to Coulomb barrier, $E_{c.m.}/V_b$, 
the excitation energy of the fissioning system $E^*$ and the variance 
($\sigma_m^2$) of the mass distributions are tabulate in table \ref{othvar_tab}.
To show the variation of $\sigma_m^2$ versus beam energy graphically, 
Fig \ref{othvaronly} is drawn with experimental $\sigma_m^2$ values 
depicted with solid squares. We observed a very similar trend of the 
variation of $\sigma_m^2$ with decreasing energy, as was observed for 
the $^{19}$F + $^{232}$Th system. In $^{16}$O + $^{232}$Th, above the fusion
 barrier, $\sigma_m^2$ decreases 
smoothly with the energy, but near the barrier at about 87 MeV, $\sigma_m^2$ 
starts to rise and reaches a peak around 81 MeV. At still lower energies, 
it again starts to fall smoothly with decrease in energy. However the 
rise in $\sigma_m^2$ is about 15$\%$ compared to as almost 50$\%$ 
rise observed in $^{19}$F + $^{232}$Th  \cite{myrapid1*4}.

%\clearpage

%\begin{figure}[h]
%\begin{center}
%\includegraphics[height=8.0cm,angle=0]{CHAPIV/foldoth1}
%\end{center}
%\caption{\label{foldoth1} The measured folding angle distributions for the 
%fissioning system of $^{16}$O + $^{232}$Th at different energies in c.m. frame.}  
%\end{figure}

\begin{table}[h]
\begin{center}
\caption{\label{othvar_tab}~ Variance of the mass distributions for the system 
$^{16}$O + $^{232}$Th. The Coulomb barrier for this system $V_b$ = 81.9 MeV 
in centre of mass frame and the Q value of the reaction is -36.53 MeV.}
%\begin{ruledtabular}
\setlength{\tabcolsep}{0.4cm}
\renewcommand{\arraystretch}{1.2}
\begin{tabular}[t]{|c|c|c|c|c|}
\hline\hline
$E_{lab}$ & $E_{cm}$ & $E_{cm}/V_b$ & $E^\star$ & $\sigma_m^2$   \\
(MeV) & (MeV) &  & (MeV) & $(u^2)$   \\
\hline
102.8 & 96.1 & 1.17  & 59.60 & 656.38 $\pm$ 12.80\\
100.8 & 94.3 & 1.15  & 57.73 & 642.11 $\pm$ 12.67\\
98.8  & 92.4 & 1.13  & 55.85 & 593.40 $\pm$ 12.17\\
96.7  & 90.5 & 1.10  & 53.97 & 607.12 $\pm$ 12.32\\
94.7  & 88.6 & 1.08  & 52.08 & 546.62 $\pm$ 13.09\\
93.7  & 87.7 & 1.07  & 51.14 & 559.79 $\pm$ 11.79\\
92.7  & 86.7 & 1.06  & 50.18 & 548.50 $\pm$ 13.11\\
91.7  & 85.8 & 1.047 & 49.25 & 621.50 $\pm$ 12.46\\
90.7  & 84.8 & 1.035 & 48.29 & 591.46 $\pm$ 13.86\\
88.7  & 82.9 & 1.012 & 46.41 & 615.53 $\pm$ 13.89\\
86.7  & 81.1 & 0.99  & 44.53 & 595.36 $\pm$ 14.15\\
84.6  & 79.2 & 0.97  & 42.63 & 602.21 $\pm$ 13.98\\
82.6  & 77.3 & 0.943 & 40.75 & 586.60 $\pm$ 14.77\\
80.6  & 75.4 & 0.920 & 38.87 & 539.63 $\pm$ 14.40\\
78.6  & 73.5 & 0.897 & 36.97 & 537.31 $\pm$ 17.15\\
\hline\hline
\end{tabular}
\end{center}
\end{table}

\clearpage

\section{ Result for $^{12}$C + $^{232}$Th}

\begin{figure}[ht]
\begin{center}
\includegraphics[height=16.0cm,angle=0]{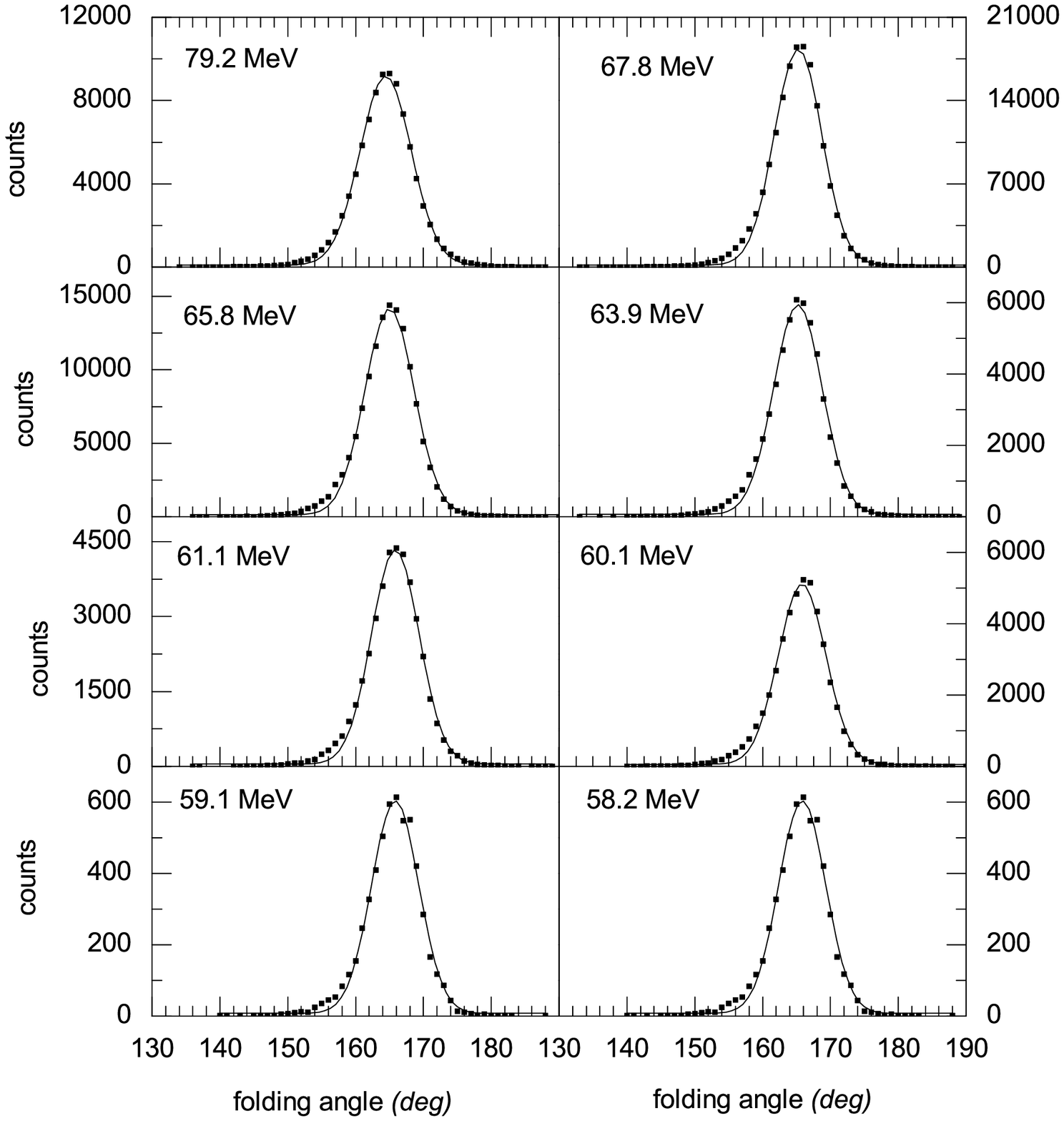}
\end{center}
\caption{\label{foldcth} The measured folding angle distributions for the 
fissioning system of $^{12}$C + $^{232}$Th at different energies in c.m. frame.}\end{figure}

\indent The fragment mass distributions of the FFCF events for the system 
$^{12}$C + $^{232}$Th were measured at ten bombarding energies. For this system, TLFF channel is weakly populated, compared to that observed in two fissioning 
system of $^{19}$F + $^{232}$Th and $^{16}$O + $^{232}$Th. The folding angle 
distributions in the first detector at forward 
positions around 60$^\circ$ are shown in Fig \ref{foldcth}. Measured peak of the folding angle distributions of the FF events matches with the simulated value. On;y at lower energies below coulomb barrier, a contribution ($\sim$ 5 $\%$)
 of TLFF could be observed, and separated by gates on $\theta$ and $\phi$ as in the earlier cases.

\begin{figure}[h]
\begin{center}
\includegraphics[height=13.0cm,angle=0]{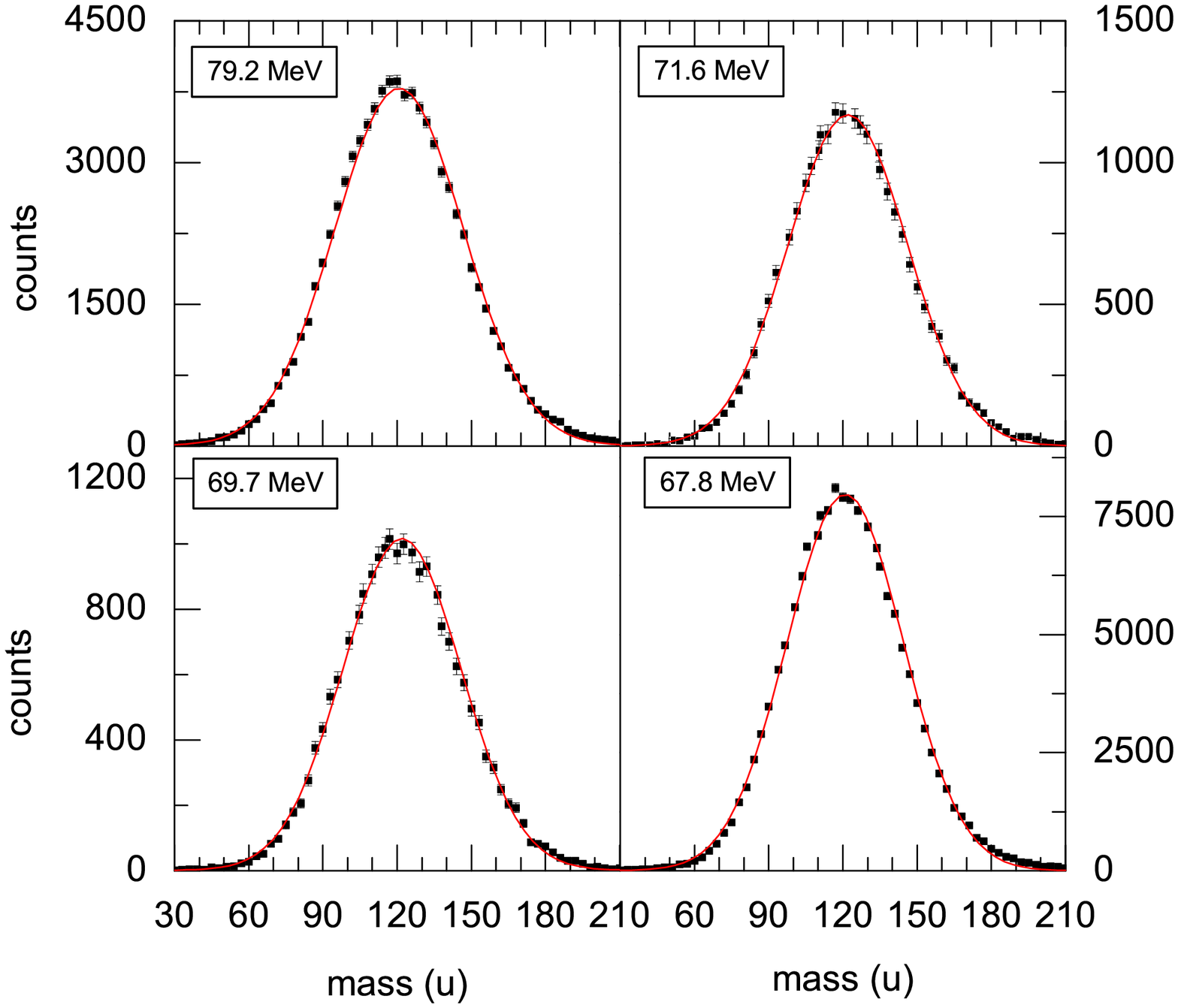}
\end{center}
\caption{\label{massdiscth1} The measured  mass distributions for the 
fissioning system of $^{12}$C + $^{232}$Th at different energies in the 
c.m. frame.}  
\end{figure}

\begin{figure}[h]
\begin{center}
\includegraphics[height=17.0cm,angle=0]{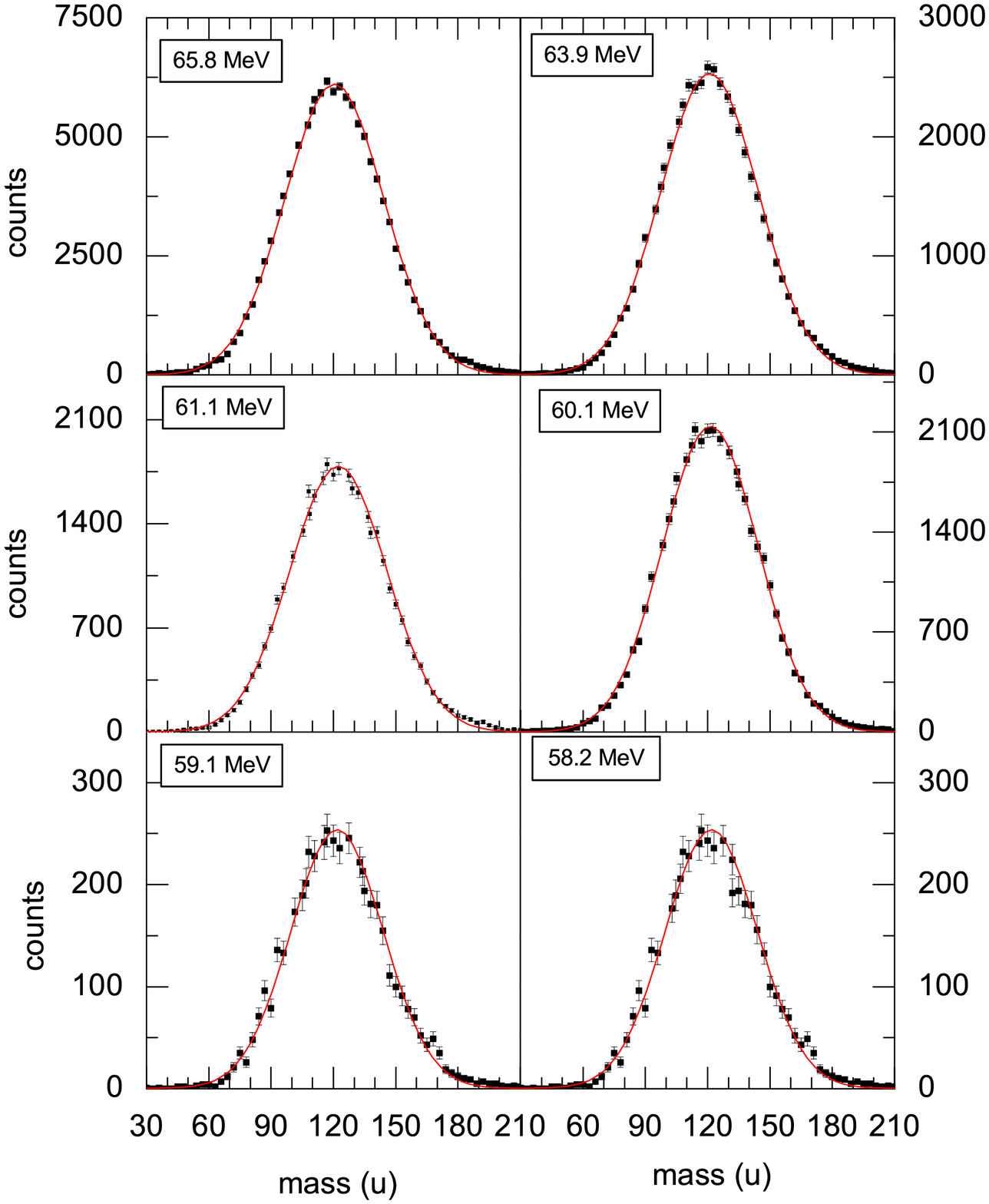}
\end{center}
\caption{\label{massdiscth2} The measured  mass distributions for the 
fissioning system of $^{12}$C + $^{232}$Th at different energies in the 
c.m. frame.}  
\end{figure}

\indent The fragment mass distributions at all bombarding energies in c.m.
frame are shown in Fig \ref{massdiscth1} and Fig \ref{massdiscth2} in 3 
{\it amu} mass bin. It is observed that the mass distributions can be well 
fitted with Gaussian distributions even to the lowest energies. 
The fittings are shown by solid lines in the figures. Variances of the mass 
distributions were calculated from the fitted Gaussians.

\begin{figure}[ht]
\begin{center}
\includegraphics[height=12.0cm,angle=0]{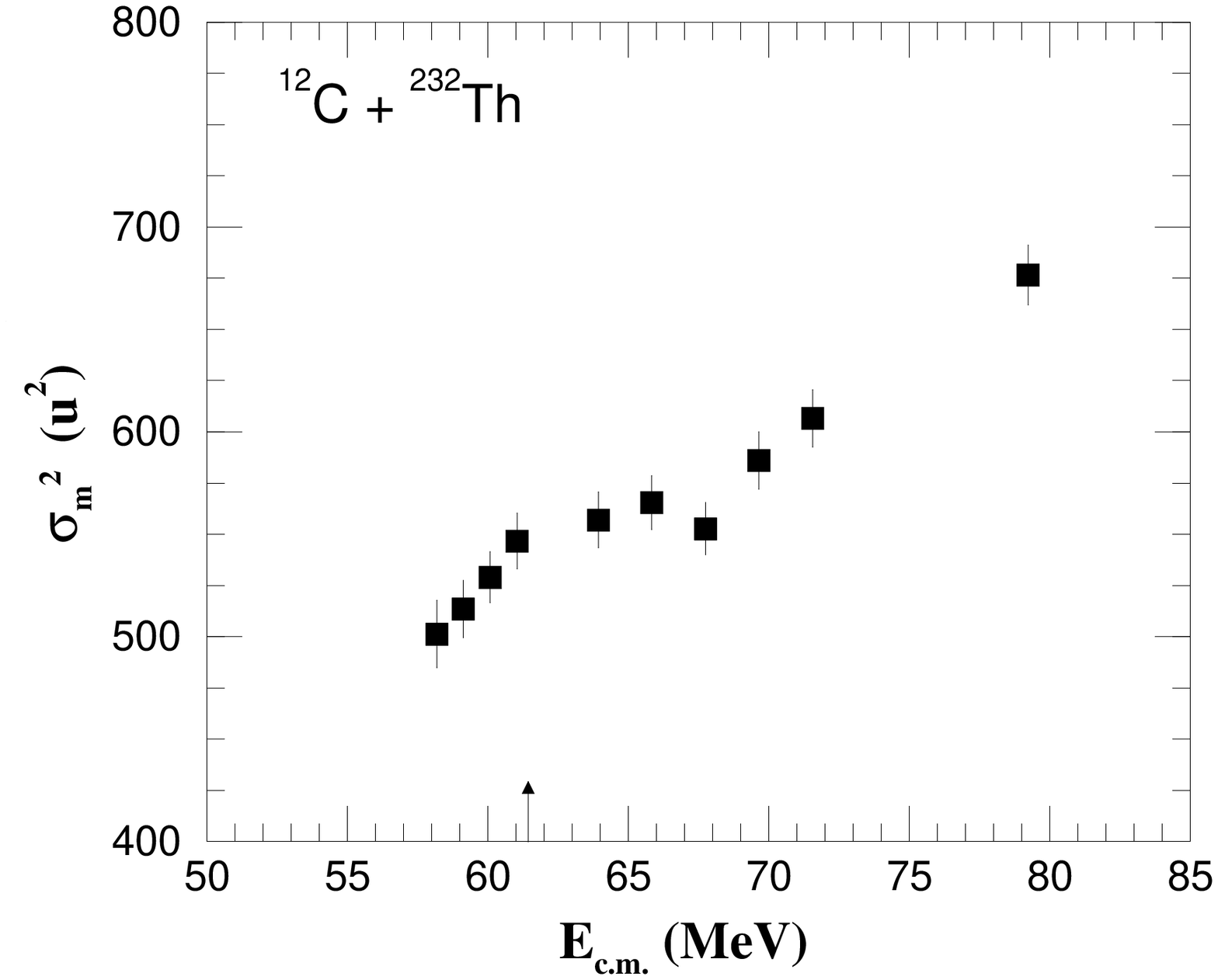}
\end{center}
\caption{\label{cthvaronly} The measured mass variance $\sigma_m^2$ for the 
fissioning system of $^{12}$C + $^{232}$Th as a function of projectile 
energies in the c.m. frame. Coulomb barrier is indicated by an arrow.}  
\end{figure}

\indent The values of the calculated variance ($\sigma_m^2$) of the 
measured mass distributions of the FFCF events at all energies are given in 
table \ref{cthvar_tab} along with the target loss corrected beam energies in different frames and scales and excitation energies. The calculated values of 
excitation energy are 
listed in the column four. The values of the variances, $\sigma_m^2$,  of the 
FFCF events for the system of  $^{12}$C + $^{232}$Th as function of the 
c.m. energies are shown in Fig \ref{cthvaronly}. Here also, we observed 
trend of the variation of $\sigma_m^2$ with decreasing energy 
similar to that observed for the $^{19}$F + $^{232}$Th and $^{16}$O + $^{232}$Th
systems. In $^{12}$C + $^{232}$Th system, above the fusion barrier, 
$\sigma_m^2$ decreases smoothly with the energy, but near the barrier, 
$\sigma_m^2$ starts to increase . At still lower energies, it again starts to 
decrease smoothly with decrease in energy. However the rise in $\sigma_m^2$ is 
about 10$\%$ compared to as almost 50$\%$ rise observed in 
$^{19}$F + $^{232}$Th  \cite{myrapid1*4} and 15$\%$  in $^{16}$O + $^{232}$Th  
\cite{myrapid2*4}.

\begin{table}[h]
\begin{center}
\caption{\label{cthvar_tab}~ Variance of the mass distributions for the system 
$^{12}$C + $^{232}$Th. The Coulomb barrier for this system $V_b$ = 62.6 MeV 
in centre of mass frame and the Q value of the reaction is -23.0 MeV.}
%\begin{ruledtabular}
\setlength{\tabcolsep}{0.5cm}
\renewcommand{\arraystretch}{1.3}
\begin{tabular}[t]{|c|c|c|c|c|}
\hline\hline
$E_{lab}$ & $E_{cm}$ & $E_{cm}/V_b$ & $E^\star$ & $\sigma_m^2$   \\
(MeV) & (MeV) &  & (MeV) & $(u^2)$   \\
\hline
83.3 & 79.2  &   1.27  & 56.23 & 676.52 $\pm$ 14.47 \\
75.3 & 71.6  &   1.14  & 48.58 & 606.63 $\pm$ 14.00\\
73.3 & 69.7  &   1.11  & 46.66 & 586.12 $\pm$ 14.00\\
71.3 & 67.8  &   1.08  & 44.75 & 552.72 $\pm$ 12.62\\
69.2 & 65.8  &   1.05  & 42.84 & 565.48 $\pm$ 13.13\\
67.2 & 63.9  &   1.02  & 40.94 & 556.96 $\pm$ 13.52\\
64.2 & 61.1  &   0.975 & 38.06 & 546.62 $\pm$ 13.56\\
63.2 & 60.1  &   0.96  & 37.10 & 529.00 $\pm$ 12.51\\
62.2 & 59.1  &   1.945 & 36.14 & 513.47 $\pm$ 14.02\\
61.2 & 58.2  &   0.93  & 35.19 & 501.31 $\pm$ 16.48\\
\hline\hline
\end{tabular}
\end{center}
\end{table}

%\clearpage

\setcounter{equation}{0}
\setcounter{figure}{0}
\setcounter{table}{0}
\chapter{Discussions $\&$ Conclusions}
\markboth{nothing}{\it Discussions $\&$ Conclusions }

\newpage

\indent In the present work the fragment mass distributions in fusion-fission
 (FF) reactions induced by light  (almost) spherical projectiles on 
spherical and deformed targets are systematically studied near and below 
Coulomb barrier energies. The method and experiments for precise measurements 
with excellent mass resolution were described earlier along with the measured 
values of the width (variances of mass distributions) of the distributions.

\begin{figure}[h]
\begin{center}
\includegraphics[height=10.0cm, angle=90]{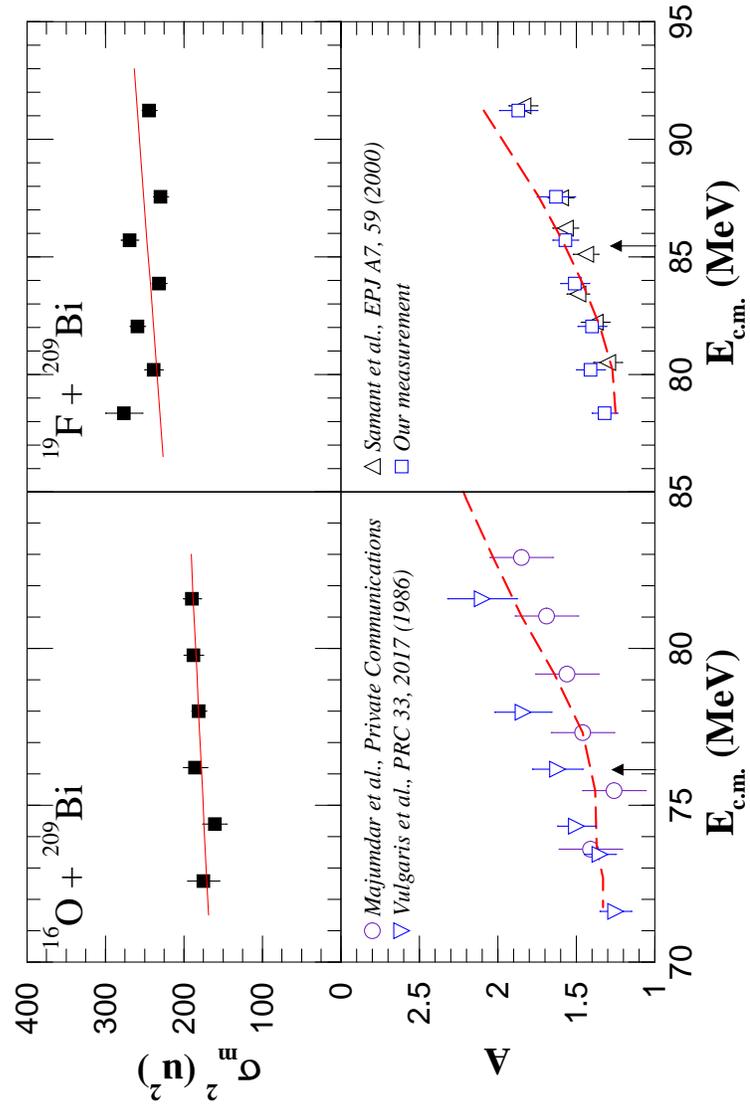}
\end{center}
\caption{\label{spherical_target} Mass variance $\sigma_m^2$ (upper panels)
and  anisotropy A (lower panels), as a function of $E_{c.m.}$ for the fissioning systems  with {\it spherical target}. The solid red line in the upper panels 
is the calculations from the statistical model. The dashed red line in the 
lower panels represent the SSPM calculation with correction for prescission 
neutron emission. The Coulomb barrier is indicated by an arrow.  }
\end{figure}

\indent The variations of $\sigma_m^2$ with energy for $^{16}O$ and $^{19}F$ +  
$^{209}Bi$ are  reproduced in the upper panel of \ref{spherical_target}. 
In the lower panel, the reported results on measurements of angular anisotropy 
(A) of the fission fragments are shown. For  $^{19}F$ +  $^{209}Bi$, the 
angular distribution measured in the same experimental set up by us are also 
shown. It is interesting to note that the $\sigma_m^2$ varies smoothly with 
$E_{c.m.}$ while the trend in the angular anisotropy also follows closely the 
predictions of the microscopic theory (SSPM).

\begin{figure}[h]
\begin{center}
\includegraphics[height=12.0cm, angle=90]{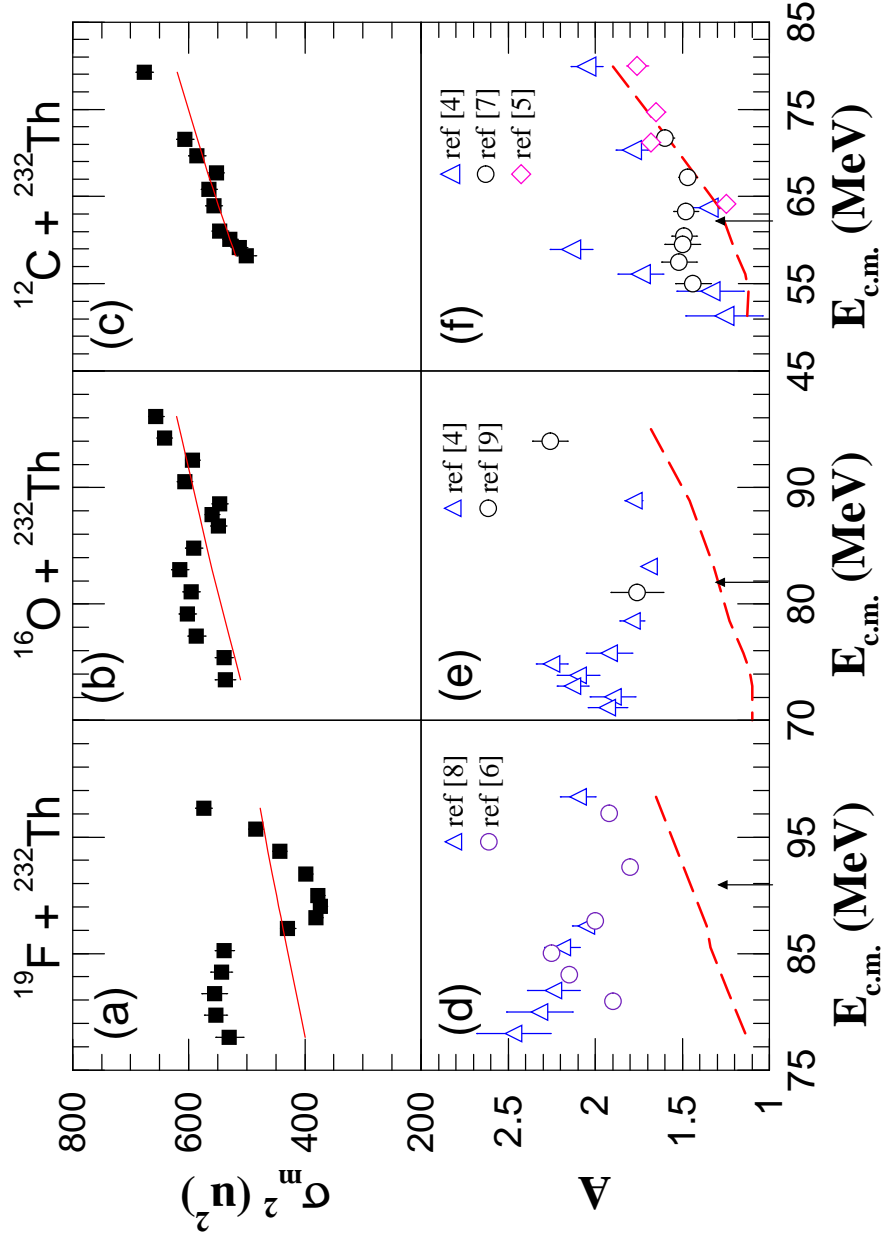}
\end{center}
\caption{\label{deformed_target} Mass variance $\sigma_m^2$ (upper panels)
and  anisotropy A (lower panels), as a function of $E_{c.m.}$ for the fissioning systems  with {\it deformed target}. The dashed red line in the 
lower panels represent the SSPM calculation with correction for prescission 
neutron emission. The Coulomb barrier is indicated by an arrow.  }
\end{figure}

\indent In Fig \ref{deformed_target}, the upper panel shows the variations of 
the $\sigma_m^2$ with energy for the deformed target $^{232}$Th for  $^{12}$C,
 $^{16}$O and $^{19}$F projectiles, while the lower panel shows in variations 
in reported values of angular anisotropy A. It is again interesting to note 
that variations in $\sigma_m^2$ and A with energy is significantly different
 from the observed trend for the $\sigma_m^2$ and A  for spherical target 
projectile combinations. Variations in both the parameters show that those 
are decreasing with energy, but around Coulomb barrier a significant increase is observed with decreasing energy.

\indent The close resemblance of the trends in the two cases, i.e., the 
departure of the observables  for fusion-fission reaction, immediately points 
to a common origin of the observed anomalies in the angular and mass 
distributions. A common explanation of both the "anomalies" would, therefore, 
give an insight into the 
fusion-fission dynamics; or in other words, a pointer 
to the path followed by the system of the two separated nuclei to a mononuclear system and re-separation into fragments with altered  nucleon numbers, 
excitation, temperature, spin etc.

\clearpage

\noindent 

\indent The process of fusion and subsequent evolution of the system, damping 
of the radial motion and excitation followed by cooling and thermal 
equilibration of the fused system had been a topic of investigation for last 
few decades. With tremendous advance in knowledge and computational powers, 
it has been possible to theoretically follow  the transit of the system from 
the initial to final configurations by the minimum potential path through the 
landscape of potential energies in a multi-dimensional space governed by the 
residual strong interactions among the nucleons and the coulomb interactions. 
The paths  can be calculated in microscopic theories of  transport through a 
dissipative system. However, these calculations are complicated 
and   still in the infancy. Hence knowledge about the fusion-fission reactions
 mostly comes from the conclusions drawn from the experimental observations.

\section{Mass distributions with spherical target $^{209}$Bi}

\indent The observation of the smooth variations of the fragment anisotropy 
and the close agreement with the predictions based on the macroscopic pictures 
of the  equilibration  of the projection of the total angular 
momentum on the symmetry axis of  a finite, rotating  liquid drop in the 
SSPM model points to the path followed by such systems in the multidimensional 
energy landscape. The path leads over a fusion barrier to fuse the two 
projectiles with damping of the radial motion and increase in the internal 
excitation of the fused system. The fused system can cool down by emission of 
nucleons and photons and reach the fusion meadow  as a evaporation 
residue. If the fissility parameter ($Z^2/A$) is large and the Coulomb force 
tends to overcome the surface energy of the equilibrated compound nucleus, the 
internal excitation induces multi polar shape oscillations and the system 
reaches a saddle shape, a point of no return - and the system slides down 
the fission barrier to fission valley and reaches a scission configuration 
where two fission fragments fly off in opposite directions.. In macroscopic 
theories, the spin quantum numbers $(J,K)$ of the fissioning systems are 
assumed to be frozen at the saddle point  and the agreement of the calculated 
angular anisotropy with the experimental values in $^{16}O$ and $^{19}F$ fusing 
with spherical bismuth nuclei directs to a possible scenario of the 
fusion-fission reaction as described above.   

\noindent In the standard statistical saddle point model (SSPM) 
\cite{HSGeneva58*5}, the fission fragments are assumed to 
scission along the symmetry axis at the saddle point where the constants of 
motion are frozen. For a compound nuclear state of total angular momentum  
$J$ and projections of it, M and K along the z-axis and symmetry 
axis, respectively, the angular distribution of fission fragments are given 
by the symmetric top wave function as

\begin{equation} 
W_{MK}^J(\theta) ={\frac{2J+1}{4\pi }}|D_{MK}^J(\phi,\theta,\psi)|^2 
={\frac{2J+1}{4\pi }}|d_{MK}^J(\phi,\theta,\psi)|^2
\end{equation}

\noindent Here $\theta $ is the c.m. angle with respect to beam of the
 fragment. For spin less target and projectile, M is zero and the angular
 distribution of fragments are given by 

\begin{equation}
W_{0,K}^J (\theta)={\frac{2J+1}{4\pi}}|d_{0,K}^J|^2
\end{equation}

\noindent In the standard saddle-point statistical theory, the distribution 
of K values are estimated by using a constant temperature level density argument at the fission saddle point and this approach leads to the expression

\begin{equation}
\rho(K) = \frac{e^{-K^2/2K_0^2}}{\sum_{K=_J}^J e^{-K^2/2K_0^2}}
\end{equation}

\noindent where $K_0^2$= $I_{eff}T/\hbar^2$ and 1/$I_{eff}$ = 1/$I_\|$ - 
1/$I_\perp$. The distribution of K values is therefore gaussian with variation
 $K_0$, which is determined by the effective moment of inertia 
$I_{eff}$ and nuclear temperature T, both calculated at the saddle point for 
fission. The nuclear temperature at the saddle point T is obtained from the 
relation

\begin{equation}
T^2= a(E^*-B_f-E_{rot})
\end{equation}

\noindent where $E^*$ is the excitation energy of the compound nucleus, 
$B_f$ is the fission barrier, $E_{rot}$ is the rotational energy and 
$a$ [=8/A (MeV/u), A being the mass number] is the level density parameter. 
We thus obtain the following expression for the distribution of fission
 fragments following fusion:

\begin{equation}
W(\theta)={\frac{\lambda^2}{4\pi}} {\sum_{K=J}(2J+1)T_J }\sum{\frac{2J+1}{4\pi}} |d_{0,K}^J(\theta)|^2 \rho(K)
\end{equation}

\noindent where $T_J$ is the transmission co-efficient for the  $J$th partial 
wave for fusion. The statistical saddle point model clearly emphasizes the
 macroscopic co-ordinates, viz, moment of inertia and temperature and the 
level density parameter, governing the direction of the emission of the 
fragments. The SSPM predictions of angular anisotropies are shown in 
lower panel of Fig. \ref{spherical_target} by dashed lines. The close 
agreement of the SSPM predictions with the experimental data shows that 
macroscopic liquid drop model describes the fusion-fission path adequately.

\indent The above discussed fusion-fission path is also supported by the 
observed shape and variation of the width of the fragment mass distribution 
with bombarding energy. The variations with beam energy in the center of mass 
frame, the width ({$\sigma _m^2$}) of the mass distributions of an equilibrated compound nucleus will be governed by the macroscopic forces, i.e., will have a 
weak and statistical dependence on  energy. The observed smooth variation in 
$\sigma_m^2$ in the $^{16}$O and $^{19}$F  on $^{209}$ Bi is in general 
agreement with the trend expected for fusion-fission reactions. The 
solid red line in the Fig \ref{spherical_target} is the expected dependence 
of the width of the mass distribution on the incident energy as derived from 
the statistical model treatment consistent with the fusion-fission mechanism. 
Here it is assumed that the mass asymmetry potential can be approximated 
by a parabolic shape,

\begin{equation}
U(m)= \frac{1}{2}k(m-m_s)^2
\end{equation}

\noindent where $m$ is the fragment mass, $m_s$ is the mass for symmetric 
fragmentation and $k$ is the stiffness parameter for the mass asymmetry degree 
of freedom. A statistical model treatment leads to a variance of the fragment 
mass distribution given by,

\begin{eqnarray}
\sigma_m^2
=\frac{T}{k}
={\frac{1}{k}}\sqrt {\frac{E_{sc}^*}{a}} 
\end{eqnarray}

\noindent where $T$ is the scission point temperature, $a$ is the level 
density parameter, $E_{sc}^*$ is the excitation energy at the scission point 
which is calculated as follows:

\begin{equation}
E_{sc}^*=E^* + Q_{symm} - E_K - E_{def} - E_{rot}
\end{equation}

\noindent Here $E^*$ is the excitation energy of the compound system, 
$Q_{symm}$ is the $Q$ value for symmetric fission, $E_K$ is the total 
kinetic energy estimated from Viola's systematics \cite{Viola85*5}. $E_{def}$ 
accounts for the fragment deformation energy and $E_{rot}$ is the rotational 
energy at the scission. Detail calculation of $\sigma_m^2$ is given in the 
Appendix. The theoretical fits to the mass distributions are shown by thick 
continuous lines in upper panel of Fig \ref{deformed_target} using equation
5.7 and 5.8. The stiffness parameter $k$ is taken as free parameter.

\indent The linear variation of the $\sigma_m^2$ with the beam energy 
$E_{c.m.}$ (which also relates the excitation energy and equilibrium 
temperature of the system) shows that macroscopic forces in the liquid 
drop model of nucleus is adequate to describe both the angular and 
mass distributions of fragments in fusion-fission reaction. The 
experimental observation fully support a picture of the dynamics of the fusion 
of the two independent ions with initial kinetic energy to overcome 
the Coulomb and coriolis forces to "fuse" over a fusion barrier and cool 
down by few particle emission and reach equilibrium to the so called 
compound nucleus; and then undergoing shape oscillations over a saddle shape 
to roll down the fission ridge to scission at the fission valley. Such a 
picture is nicely described in Fig \ref{moller} as  discussed later.   
  
\section{Mass distributions with deformed target $^{232}$Th}

\indent The above simple picture of the fusion-fission path and evolution 
of a compound nuclear fission, as described in the earlier section, runs 
into difficulty in the case of deformed 
target in the energies close to the Coulomb barrier and had been subject of 
intense research in recent years in multitude of experimental observables. 
The anomalous effect near and below the Coulomb barrier energies in the 
angular anisotropies of the fission fragments were observed in many systems 
\cite{KaiPR97*5}. In the systems studied in the 
present case, the angular anisotropies showed remarkable anomalous increase 
\cite{NMPRL96*5,RamPRL90*5,ZhangPRC94*5,LestonePRC97*5,NMPRC95*5,BackPRC90*5} as shown in 
Fig. \ref{deformed_target} in the lower panels.  The SSPM predictions for the 
angular anisotropy and present calculations of $\sigma_m^2$ from statistical 
model are shown by blue dotted lines in Fig \ref{deformed_target}. The 
experimental data for both the observables shows significant departure from
 the macroscopic model predictions. It is interesting to note that 
for the first time, in precise measurements of $\sigma _m^ 2$, the measured 
values also show anomalous departure from a smooth variation observed for the 
systems with spherical target-projectile systems. Hence a common explanation 
of both angular and mass distributions in term of a dynamical model of the 
macroscopic and microscopic effects are needed for a satisfactory explanation. 

\indent The angular anisotropies in the deformed target and projectile systems
 have been explained in different models.  
 The predictions of the SSPM was found to be quite accurate, 
specially for spherical target and projectile systems and even for deformed
 targets. The departure from the SSPM predictions for deformed targets were 
found to be nominal at energies, or temperatures, above that corresponding to
 the fusion barrier. 
Hence the anomalous angular anisotropies observed, particularly at near and
 below barrier energies were guessed to originate due to microscopic effects, 
viz, binding energies, effects of entrance channel 
mass asymmetry or dynamical effects due to dissipation of the nucleons in the 
forming the mononuclear system in fusion of the two nuclei.  

\indent It had been observed that the anomalous angular anisotropies are also 
accompanied by an order of magnitude increase in the fusion cross sections 
over that of the one-dimensional barrier penetration models. Hence for a few 
years it had been subject of intense debate over two factors -whether the 
average $<l^2>$ values are underestimated or the width of the equilibrated
 K distributions, $K_0^2$ are overestimated from the experimental data, as the 
approximate relation of the angular anisotropy to the above factors are given 
by the approximate relation $A = 1+  <l^2>/4K_0^2$. It had been finally 
resolved by simultaneous agreement of the fusion excitation functions and the 
distributions of barriers in a coupled channel analysis that indeed it had been the overestimation of the $K_0^2$ , which is responsible for the 
increase in the angular anisotropy in the near and sub-Coulomb barrier 
energies in the deformed target-projectile systems \cite{NMPRL96*5,HindePRL95*5}.

\subsection{Reduction of the width of the K distributions: Pre-equilibrium 
fission model}

\indent There has been different ideas and models of the reduction of the 
width of the spread in the orientation of the nuclear symmetry axis of the 
fissioning nucleus at the saddle configuration with respect to the space fixed 
axes. The earliest ideas are based on the non-equilibration of the K quantum 
number before the system reaches saddle shape \cite{RamPRL85*5}. 
The pre-equilibrium 
model of Kapoor and Ramamurthy (KR) assumed that for heavy systems, the 
equilibration times are large compared to the time taken to reach the saddle 
shapes and the resultant width of the K distribution, $K_0^2$ 
is on average smaller than that predicted for a finite rotating liquid drop
 model prediction. They proposed that all fission events taking place in a 
time scale of 8$\times$ 10$^{-21}$ sec or less have a K distribution 
represented by a  
narrow Gaussian  around K=0, with a constant variance $\sigma_K = 
J \sigma_{\theta}$, where $\sigma _\theta =$ 0.06. This smears the angular 
distributions and the average angular anisotropy increases. However, according
 to the model prescription, it was expected that the effect of 
K-non-equilibration will increase with energy, and in fact, it had not 
been expected that at energies close to and below barrier, the 
pre-equilibration of K quantum number would be a major factor in determining 
the fragment angular anisotropy. In addition, it was not clear why the effect 
predominates in the deformed target-projectile systems.

\indent Vorkapic and Ivanisevic \cite{VorPRC95*5} modified the pre-equilibrium 
model 
of KR by calculating the fission probability for a series of time interval 
from the liquid drop model of Bohr and Wheeler \cite{BohrPR39*5}, for specified 
orientation of the symmetry axis with respect to the projectile trajectory. 
The initial K distribution was assumed to be around most probable projection 
of the total spin on the symmetry axis. Considering that at sub-barrier energies, the barrier distribution functions shows that the fusion is almost confined to the region of the tip or polar region of the prolate targets and the 
equatorial regions do not contribute to the fusion, the orientation averaged 
fission times showed pre-equilibrium fission time scales at near and below 
energies. The calculated angular distributions for $^{16}$O+ $^{232}$Th systems could be quite well reproduced for a speed of equilibration 
C $\sim$ 0.75 $\times 10^{20}$ $ s^{-1}$. The equilibration time is related
 to the width of the K distribution as $K_0 = J C t_m $ , 
where $t_m$ is the mean of some time interval. It is worthy to note that as 
energy increased above the Coulomb barrier, the  probability of fusion did 
not depend on the relative orientation, i.e., the dependence on polar 
orientation of the deformed target was washed out, and the calculated width 
of the K distribution converged to the SSPM value.

\indent Both the pre-equilibrium model of KR and its modification assumes that 
a compound nucleus is formed with full damping of radial motion and mass 
relaxation before it reaches the saddle configuration by shape oscillations 
in a time faster than the equilibration time for statistical equilibration of 
the projection of the spin on to the symmetry axis. Thus the angular distributions are not affected by the entrance channel variables, if any, in the primary 
fusion of the binary system. However, the extent, or the quantitative 
estimation of the angular anisotropy shows a strong correlation with 
the entrance channel masses of the target and projectile. The mass asymmetry of 
target-projectile combination, $\alpha$ = $(M_t - M_p)/(M_t + M_p)$, 
largely determines the initial flow of nucleons in binary fusion. The 
macroscopic forces due to 
Coulomb and surface energy combines to put the flow of nucleons in 
opposite directions depending on the $\alpha$ being smaller or larger than 
the critical Businaro mass asymmetry ( $\alpha_{BG} $ $\approx $ 0.90). 
For systems of $^{19}$F + $^{232}$Th or $^{16}$O+ $^{232}$Th ,
{$\alpha < \alpha_{BG}$, the flow of nucleons 
are from heavier to lighter fragments, and reverse is the direction of flow
 of matter in $^{12}$C+ $^{232}$Th systems with $\alpha > \alpha_{BG}$ . 
It is interesting to note that on $^{232}$Th targets, the anomalous rise 
in angular anisotropy gets progressively reduced as entrance channel $\alpha$ is increased.

\indent The experimental evidences that the total fusion cross sections are 
enhanced for the larger initial separation of the projectile fusing with polar
 region of the deformed target and the observation of the influence of the 
entrance channel mass asymmetry on the degree of angular anomaly of the fission fragments points to possible microscopic effects in defining the fusion path 
itself of the binary system to a composite mononuclear system. If the path in 
a multi-dimension potential energy landscape deviates from that over a normal 
fusion barrier governed by the macroscopic properties of a hot, finite 
rotating liquid drop, it can be reasonably assumed that a finite probability 
exists for the system to go over a conditional fission barrier before ever
 reaching a mononuclear configuration. Such a fission event, in all 
experimental situations would be counted as a fusion-fission event following 
full transfer of initial momentum, although statistical theories will not be 
applicable to predict the experimental observables, viz, angular or mass 
distributions of such non-equilibrated fission modes.

\indent It is quite well known that such non-compound fission modes exist 
in heavy ion induced fission reactions. If the excitations  and transfer of 
the orbital momentum is large enough to reduce the $l$ dependent fission 
barrier height to zero, the nascent mononuclear system spontaneously under 
goes fission with memory of entrance channel mass and orientation. Such an
 event is known as {\it fast fission} and the angular 
distribution shows an almost 1/sin$\theta $ behaviour with large anisotropy. 
The mass distribution also shows large asymmetry and  may be double peaked
\cite{Vandenbook*5}. In certain systems with near-symmetric target-projectile 
combinations and at high excitations, 
the normal fusion-fission path may not also follow the minimum energy 
configurations and a finite non-zero mass asymmetric fission barrier may be the 
preferential route for the binary system to follow. Such reactions, termed 
as {\it quasi-fission} \cite{ShenPRC87*5} , also yields high angular anisotropy 
and a 
symmetric mass distribution, In analogy to such reactions, a quasi fission 
reaction mechanism can be postulated for the reactions with deformed targets 
at low energies where the reactions take place preferentially through the polar regions of the target.  

\indent Hinde {\it et al.,} \cite{HindePRL95*5} conjectured a mechanism of
 orientation dependent quasi fission reaction to explain  the observed fragment angular anisotropies in $^{16}$O + $^{238}$U systems. N.Majumdar {\it et al.,}
\cite{NMPRL96*5} extended the model to include systems with entrance channel mass asymmetry on either side of the  Businaro-Gallone ridge. The salient features
 of the orientation dependent quasi fission reaction are discussed below.

\subsection{Nuclear orientation dependent quasi fission model}

\indent Hinde {\it et al.,} \cite{HindePRL95*5} measured the fragment angular 
distribution of the system $^{16}$O + $^{238}$U and observed a rise in the 
anisotropies of fission fragments in FFCF reaction at sub-barrier energies. 
They proposed a orientation dependent quasi-fission model to explain the 
rising trend in angular anisotropy. It was conjectured that the 
admixture of a QF mechanism, characterized by high anisotropy, with the 
statistical compound nuclear  events consistent with SSPM prediction of fragment angular anisotropy, boosted the resultant anisotropies. However, the 
occurrence of the QF mechanism is governed by the relative orientation of the 
projectile and the target.  

\indent Hinde {\it et al.,} \cite{HindePRL95*5} measured the fusion barrier 
distribution for the system $^{16}$O + $^{238}$U and the rise in anisotropies 
with decreasing energy. They pointed out that the higher fusion barrier at 
energies above Coulomb barrier refers to the  contact of the projectile 
irrespective of the equatorial or polar region of the prolate target,  but 
the contact with equatorial region leads to a more compact di-nuclear system.
 Since in quantum mechanical systems, similarity in entrance and exit channel 
shapes are favoured, it is expected that the compound nuclear formation is more likely for contact with equatorial region than that for contact with prolate
 region. On the contrary, the lower fusion barriers at below Coulomb barrier 
energies refers to contact with the polar region, resulting an 
elongated configuration in the entrance channel; consequently a much elongated 
saddle shape leading to quasi-fission is favoured than formation of a compact, 
almost spherical compound nucleus. The two different cases are shown in 
Fig \ref{quasi} where in Case I, with larger fusion barrier,  formation of 
compound nucleus  is most probable.  In Case II,  smaller fusion barrier for 
elongated configuration is likely to generate  quasi-fission .

\begin{figure}[h]
\begin{center}
\includegraphics[height=6.0cm, angle=0]{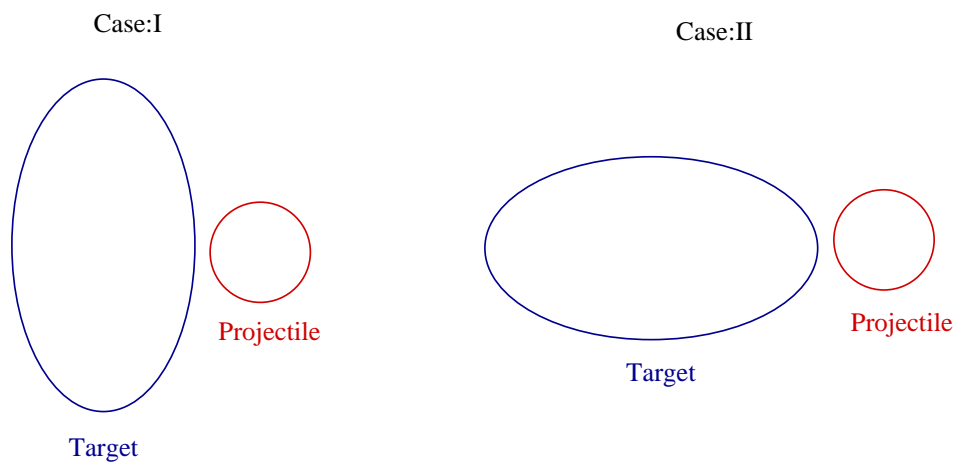}
\end{center}
\caption{\label{quasi} Two different cases for the injection of the projectile 
at the tip and flattened side of the deformed target.}
\end{figure}

\begin{figure}[h]
\begin{center}
\includegraphics[height=5.0cm, angle=0]{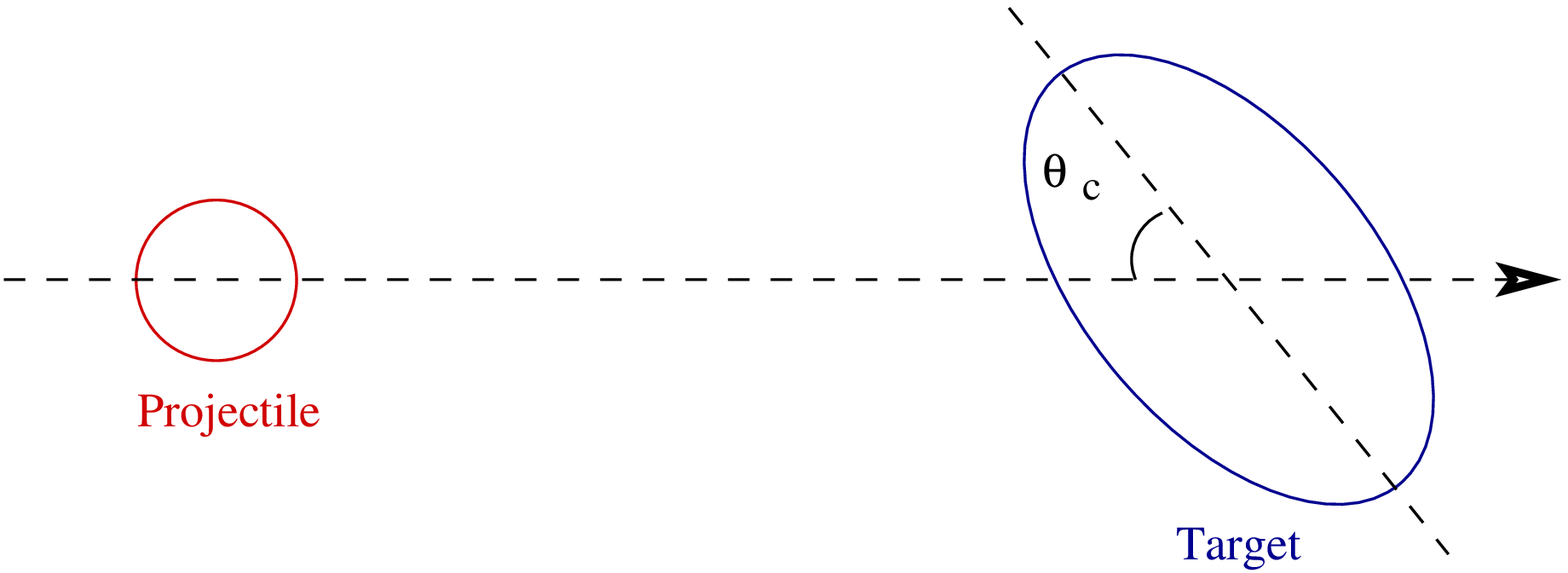}
\end{center}
\caption{\label{quasi2} Injection of a spherical projectile over the surface 
of a deformed target at an relative angle $\theta_c$ .}
\end{figure}

\indent In  the  hypothesis of QF at sub-barrier energies it is assumed  that 
at near and below barrier energies,  the reaction mechanism is  dominated by 
QF mode for relative orientation angle  of the projectile trajectories up to 
a maximum, or critical angle of  $\theta_c$, as shown in Fig \ref{quasi2}. 
Above the critical angle, reaction mechanism is assumed to be dominated by 
compound nuclear fission. Hence an admixture of the two modes changes the 
anisotropies to higher values compared
 to SSPM predictions.  For the QF mode, the contribution to angular anisotropy 
was assumed to have  a  high, constant value. The angular anisotropies for 
compound nuclear fission  was assumed to follow SSPM predictions.  A cross 
section weighted admixture of angular anisotropies fits with experimental 
trend of the variation of A with energy for the systems $^{12}$C, $^{16}$O 
and $^{19}F$+$^{ 232}$Th \cite{NMPRL96*5}. Details of the calculation are 
discussed in reference \cite{NMthesis*5}.

\indent If we compare the two hypotheses of the origin of the non-compound 
effects in fusion of two heavy nuclei leading to anomalous angular anisotropies and extend the expected effects in other reaction channels, viz, cross sections for production of evaporation residues (ER) or the mass distributions, further 
inputs for a clearer understanding of the fusion mechanism may be obtained. 
If the two ions coalesce and produce a compound nucleus equilibrated in 
all degrees of motion, the compound nucleus cools down with emission of particles and 
photons to a evaporation residue. Hence production of ER is pre-conditional 
on the formation of a compound nucleus. Thus observation of ER's in a nuclear
 reaction can be used as a cursor of the reaction path following a fusion 
over the fusion barrier and cooling down to a fusion meadow.

\indent The pre-equilibrium model of fusion-fission reaction (PEF) points to a 
delay in the equilibration of the spin projection on the symmetry axis. Hence
 if we integrate over all orientations and allow for the considerable time for
 cooling of the ER's, it is easy to follow that K-equilibration time will not 
effect the probability of formation of ER's just in the same way that the total fission cross sections are not effected by the pre-equilibrium process.  However, in the quasi fission (QF), the intermediate compound nuclear state is missing and the system directly proceeds from a initial bi-nuclear configuration to 
final bi-nuclear configuration with altered mass, charge and excitation energy
 ratio. Hence the systems do not go through a stage whereby the ER can be
 produced. Hence the cross section for production of  evaporation residue
 is hindered.

\indent Hinde {\it et al.,}  reported \cite{HindePRL02*5} severe inhibition 
of fusion 
in heavy ion induced reactions with different entrance channel masses to 
produce the final compound nucleus $^{220}$Th. The entrance channel mass 
asymmetry differed from 0.85 ($^{16}$O + $^{204}$Pb), 
0.64 ($^{40}$Ar + $^{180}$Hf), 0.56 ($^{48}$Ca + $^{172}$Yb),
0.25 ($^{82}$Se + $^{138}$Ba) and 0.12 ($^{124}$Sn + $^{96}$Zr) as shown 
in Fig \ref{HindePRL02_fig}. The cross sections for evaporation residue 
by (HI,xn) reactions were progressively inhibited as the entrance channel mass 
asymmetry increases. However, $^{40}$Ar with large binding energy inhibited 
the fusion cross sections. This observation was claimed to point toward a 
quasi fission phenomena.

%fig........
\begin{figure}[h]
\begin{center}
\includegraphics[height=11.0cm, angle=0]{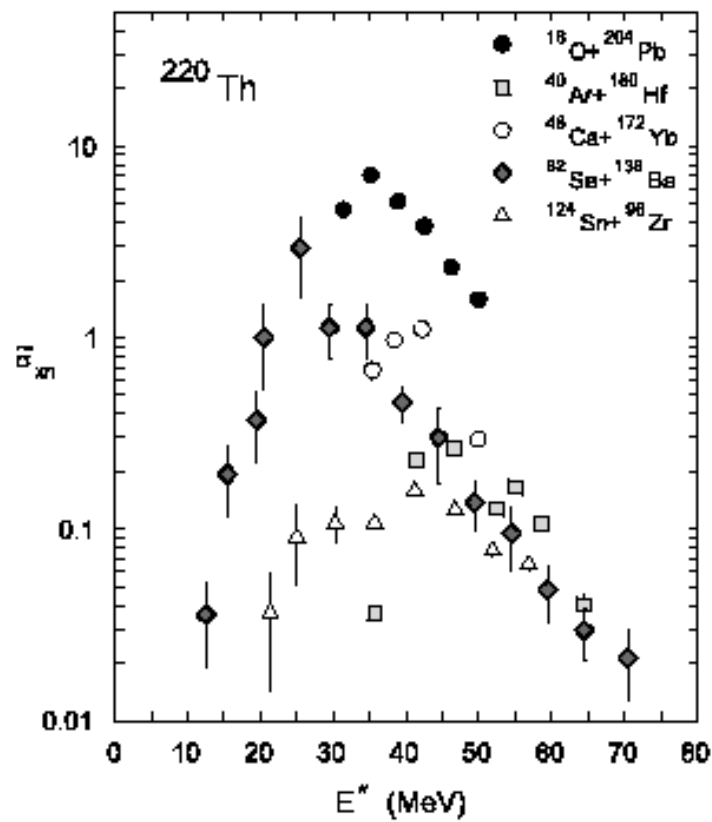}
\end{center}
\caption{\label{HindePRL02_fig} Reduced xn ER cross sections giving Thorium 
residues reported by Hinde {\it et al.,} \cite{HindePRL02*5}. Above the saturation 
energy (35 MeV for $^{16}$O) the yields for the $^{16}$O-induced reaction 
are typically a factor of 10 higher than for the more symmetric reactions.}
\end{figure}

\indent However, Nishio {\it et. al.,}  measured \cite{NishioPRL04*5} the fusion 
of $^{16}$O with $^{238}$U and reported that no inhibition in the prediction of 
evaporation residue even in below Coulomb barrier energies where the measured 
fusion barrier distribution function clearly showed reaction taking place 
through the polar region of Thorium nucleus preferring quasi fission reactions.   These conflicting picture prompted us to use fragment mass distribution 
as a probe for the reaction mechanism.

%\fig...

\indent If we consider the effect of the PEF and QF on the mass 
distributions, it is probable the effects will be different  in two cases. 
Mass relaxation is always considered to be faster than the equilibration of
 spin projections. The mass distribution is also dominated by  macroscopic 
forces. The statistical parameters governing the mass distributions are the 
formation of  the two nascent cores at the saddle and the subsequent parameters of the neck region and its snapping. In other words, the excitations or the 
temperature along with liquid drop properties are the determining factor. The 
liquid drop fission barrier is mass symmetric and at the large excitation 
energies, the shell effects are washed out. The expected mass distribution is 
around zero mass asymmetry and experimentally it has been found that for 
compound nuclear reactions, the mass distributions is peaked around the mean of target and projectile mass (or very close to it, allowing for emission of 
particles before scission).  This can also be observed in case of the present 
study of the mass distributions on bismuth target with oxygen and fluorine 
projectiles. The width of the mass distribution is a signature of the 
statistical or macroscopic processes. Hence we do not expect any large 
deviations in the  average or the width of the mass distributions for 
the pre-equilibrium model. 

\indent In the case of QF, the picture is entirely different. The system  
passes over a  ridge with even a non-zero mass asymmetry. Hence the ratio 
of the masses of the nascent cores may vary from symmetric cores of the 
compound nuclear fission. The mass split may become asymmetric. The width 
of the distribution can be different for two reasons - the asymmetry may 
not be large to separate the light and heavy fragments decisively, and the 
statistical parameters may be different from that of the compound nuclear 
fission.

\indent Hinde {\it et al.,} measured the distribution of fragment masses in 
$^{16}$O + $^{238}$U in near and below energies. The authors concentrated on 
fitting the observed masses with a symmetric (for compound nuclear fission) 
 and a asymmetric (for quasi-fission) distributions and showed that the 
asymmetric component was getting prominent as the energy decreases below the 
Coulomb barrier \cite{HindePRC96*5}. However, it may be noted that  the average 
mass asymmetry of the QF was not large enough for a clear separation of the 
symmetric and asymmetric components and the mixing ratio of the symmetric 
and asymmetric components obtained in the analysis of the experimental data 
may be doubtful. It appears that a more precise measurements as done in the
 present investigations may clearly separate the QF and CNF modes and a 
mixing of the two modes can be attempted to explain the experimental widths 
of the mass distributions.

\noindent Observation of a sudden rise in $\sigma_m^2$ values for the systems 
$^{19}$F, $^{16}$O , $^{12}$C + $^{232}$Th as the 
excitation energy is lowered may signify a mixture of two fission modes, one 
following the normal statistical prediction of fusion-fission path along 
zero lift-right mass asymmetry ($\alpha$) and another following a different 
path in the energy landscape with zero or small mass asymmetry. The mixture of 
the two modes could give rise to wider mass distributions. Similar to the 
postulation of the orientation dependent quasi fission discussed in the 
above section, we postulate that for fusion-fission paths corresponding to the 
projectile orientations up to a critical angle $\theta_c$ of impact on 
the polar region of the prolate thorium, the width and energy slope of the 
symmetric mass distributions are different as shown by dot-dashed curves in 
Fig. \ref{var_fth_fit}, Fig. \ref{var_oth_fit} and Fig. \ref{var_cth_fit}, 
compared to those for the statistical fusion fission path (shown by dotted 
curves). The mass widths weighted by the fission cross sections which are 
assumed to be very close to fusion cross sections as the composite systems are 
of high fissility, are mixed for the two modes using following relation:

\begin{equation}
{\sigma_m^2}=\frac{{\sigma_m^2}_{FF} \times X_{FF} + {\sigma_m^2}_{QF} \times X_{QF}}{X_{FF} + X_{QF}}
\end{equation}

\noindent where, $X_{FF}$ and $X_{QF}$ represent the cross sections for 
fusion-fission and quasi-fission respectively. The values of 
${\sigma_m^2}_{FF}$ and ${\sigma_m^2}_{QF}$ are taken from the fitted lines 
(dotted and dot-dashed respectively) as shown in Fig. \ref{var_fth_fit}, 
Fig. \ref{var_oth_fit} and Fig. \ref{var_cth_fit} for the systems 
$^{19}$F, $^{16}$O and $^{12}$C + $^{232}$Th respectively.

\indent  The fission cross sections 
were taken from the measurement of Majumdar {\it et al.,} \cite{NMPRL96*5}. The
calculated mass widths for the system $^{19}$F + $^{232}$Th is shown in 
Fig. \ref{var_fth_fit}. The solid red line is the calculation for the 
critical angle $\theta_c = 20^\circ$ and the blue line represent the 
calculation with critical angle $\theta_c = 18^\circ$. We note that the 
combination of the two fission modes with differences in the trend of 
variation of $\sigma_m^2$ with energy reproduces the experimentally 
observed $\sigma_m^2$ values, even if a very sharp region on the 
nuclear surface is taken as a demarkation zone for the two fission 
modes. It can also be noted that in explaining the similar trend of 
anisotropy data Majumdar {\it et al.,} also used values of same critical
 angle to explain the observed angular anisotropy. Additionally, a 
cut on critical $l$-values of 11-13$\hbar$ were assumed to separate the 
two modes of fission. However we needed no cuts on critical $l$-values, 
signifying that the spinning of the nucleus was slow enough to significantly 
affect the much faster mass transports.

\indent The calculated mass widths for the system $^{16}$O + $^{232}$Th is 
shown in Fig. \ref{var_oth_fit}. The solid red line is the calculation for the 
critical angle $\theta_c = 20^\circ$ and the blue line represent the 
calculation with critical angle $\theta_c = 25^\circ$. It is interesting to
 note that the relative anomalous rise in $\sigma_m^2$ in $^{16}$O + $^{232}$Th 
is lesser than that in $^{19}$F + $^{232}$Th, although, the same critical 
region 
on the thorium nuclear surface defines the demarkation of the two fission 
modes. We conclude that the quantitative difference in the fissioning systems 
is possibly due to the difference in the feeding point of the system along 
the mass-asymmetry and the elongation axis of the multi-dimensional
 
\begin{figure}[h]
\begin{center}
\includegraphics[height=11.0cm, angle=0]{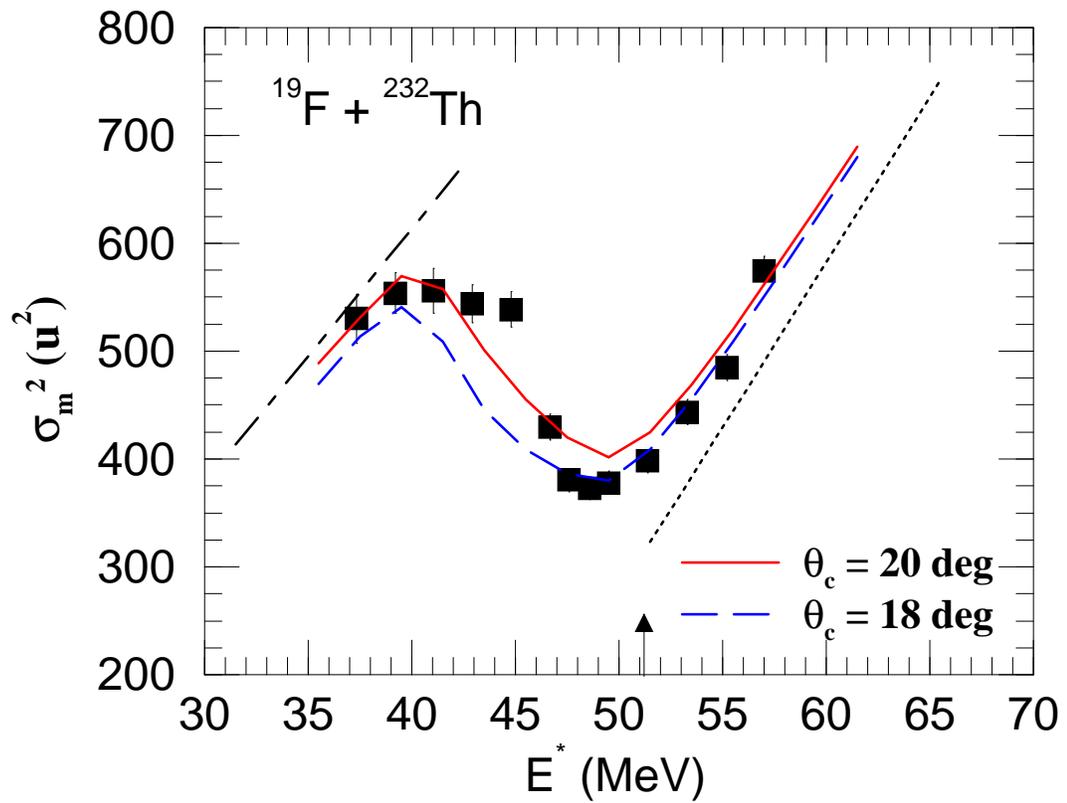}
\end{center}
\caption{\label{var_fth_fit} Variation of $\sigma_m^2$ with excitation energy 
for the system $^{19}$F+ $^{232}$Th. The dotted and dot-dashed curves are 
variation for normal and postulated quasi-fission modes, respectively. 
Calculated $\sigma_m^2$ (red and blue lines) are shown for two critical 
angles ($\theta_c$).}
\end{figure}

\begin{figure}[h]
\begin{center}
\includegraphics[height=11.0cm, angle=0]{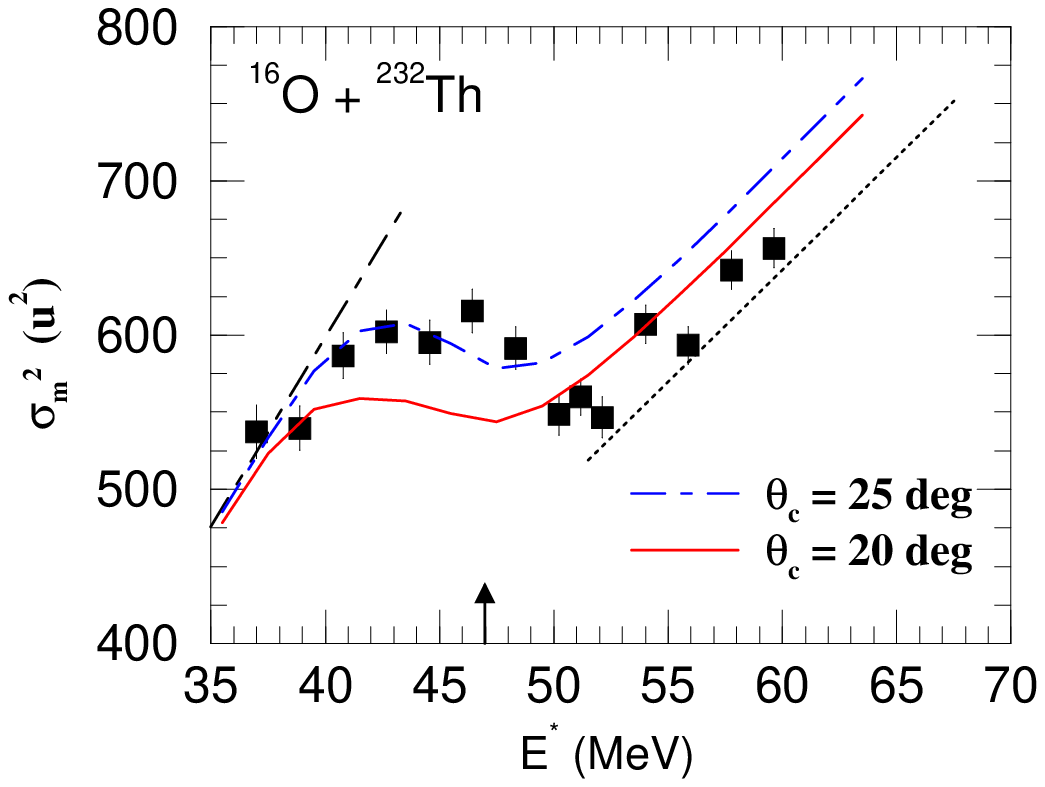}
\end{center}
\caption{\label{var_oth_fit} Variation of $\sigma_m^2$ with excitation energy 
for the system $^{16}$O+ $^{232}$Th. The dotted and dot-dashed curves are 
variation for normal and postulated quasi-fission modes, respectively. 
Calculated $\sigma_m^2$ (red and blue lines) are shown for two critical 
angles ($\theta_c$).}
\end{figure}

\begin{figure}[h]
\begin{center}
\includegraphics[height=11.0cm, angle=0]{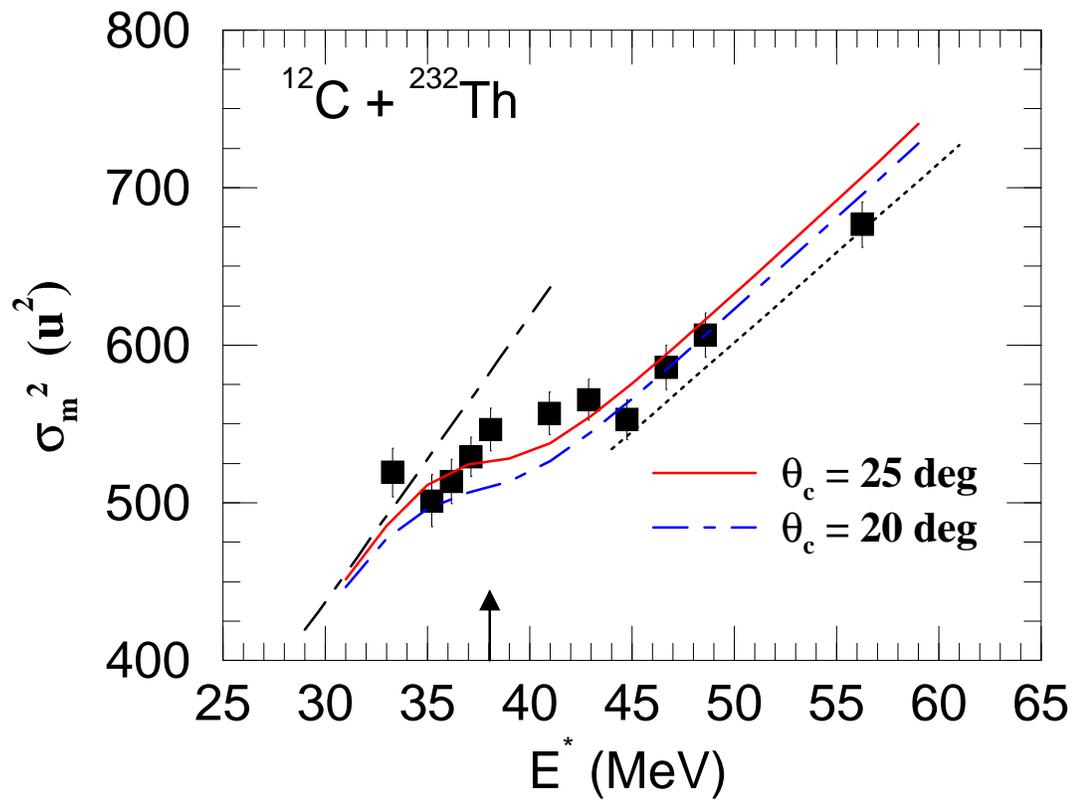}
\end{center}
\caption{\label{var_cth_fit} Variation of $\sigma_m^2$ with excitation energy 
for the system $^{12}$C+ $^{232}$Th. The dotted and dot-dashed curves are 
variation for normal and postulated quasi-fission modes, respectively. 
Calculated $\sigma_m^2$ (red and blue lines) are shown for two critical 
angles ($\theta_c$). }
\end{figure}
\clearpage

\noindent potential energy landscape.It is to be noted that the  similar value 
of 
$\theta_c$ ($20^\circ$) was used \cite{NMPRL96*5} to explain the anomalous 
anisotropies for this system also where QF events contributes.

\indent The above conjecture is supported by till lesser anomaly in case of 
$^{12}$C + $^{232}$Th system where  the calculated mass widths is 
shown in Fig. \ref{var_cth_fit}. The solid blue line is the calculation for 
the critical angle $\theta_c = 20^\circ$ and the red line represent the 
calculation with critical angle $\theta_c = 25^\circ$.  As can 
be seen from the reasonable agreement of the mixed $\sigma_m^2$ values with 
the observed fission fragment mass widths, we can phenomenologically explain 
the observed increase in widths of the mass distributions when energy is 
decreased. It is interesting to note that the fusion-fission process 
is clearly dominated by the normal process at above Coulomb barriers and the 
"anomalous" fission process is dominant at lower energies. However, 
experimental evidence suggests that the variations of mass distributions 
with excitation energies are similar for both the processes, probably 
dominated by macroscopic forces, but differing quantitatively due to 
microscopic effects.

\indent Extensive calculations of the multidimensional potential energy 
surface have successfully explained spontaneous and low energy fission 
phenomena \cite{NatureMoller01*5,PRLMoller04*5}. In Fig. \ref{moller}, 
reproduced 
from reference  \cite{NatureMollerNV03*5}, the normal fusion-fission path 
and the production of evaporation residues are pictorially depicted. 
The system of two 
ions in the entrance have to overcome the fusion barrier and reach the top 
of the fusion hill to slide down to the {\it fusion meadow}, and cool down 
to a ER. The path is shown by shaded (magenta colour) area in picture in 
left. However, if the combined system has a large fissility, {\it i.e.,}
 large Coulomb to surface energy ratio, the saddle point, indicated by a 
"$\times$" may be a more possible path and is reached by the system 
rather than the {\it fusion meadow}, to roll down hill to the 
{\it fission valley}. 

\begin{figure}[h]
\begin{center}
\includegraphics[height=7.0cm, angle=0]{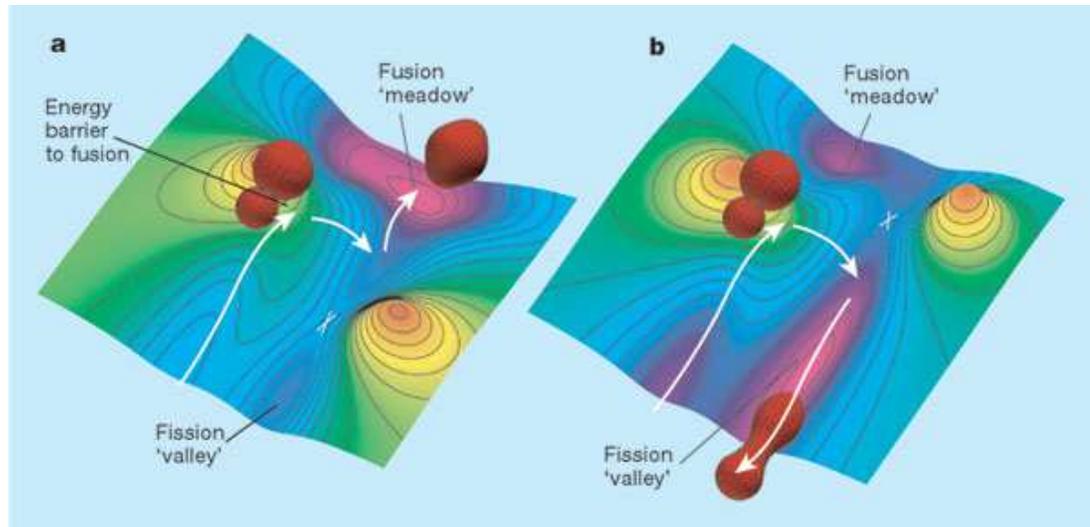}
\end{center}
\caption{\label{moller} Schematic energy landscape for fusing nuclei reported 
by Moller {\it et al.,} \cite{NatureMollerNV03*5}. }
\end{figure}

%figure moleer nature

\indent With advent of numerical processing powers of parallel computations, 
realistic calculations of the energy surfaces became possible. Calculated 
paths through the minimum energy valleys and over ridges in the potential 
surface showed that apart from the deformations and necking of the two nascent 
fragments the left-right mass asymmetry also plays a crucial role. The 
accuracy in the numerical calculations depends on the number of grid 
points on the nuclear surface of the two nascent fusing ions. Moller 
{\it et al.,} \cite{NatureMoller01*5} calculated the shapes of two fusing heavy
 ions 
in their ground state and calculated the heights of fusion and fission 
barriers. The authors could successfully explain the features of symmetric 
and asymmetric fragment mass distributions in the spontaneous or low energy 
fission of $^{228}$Ra, as shown in Fig \ref{moller_nature_fig}. All the 
heavier than actinide nuclei show mass symmetric ($\alpha = 0$) and 
mass asymmetric ($\alpha \not= 0$) saddle shapes with a ridge separating 
the two down the scission path. It can be observed the system preferentially 
follows the asymmetric mass saddle point (shown by blue line) and the 
symmetric mass saddle (red dotted line) lies higher in energy. The path 
down the saddle points for the two cases are separated by ridge, the highest 
of which is shown by triangle. It is the relative heights of the symmetric
 and asymmetric saddle points and ridge which separates the two paths to 
scission, critically determines the fragment mass distribution.  

\begin{figure}[h]
\begin{center}
\includegraphics[height=7.0cm, angle=0]{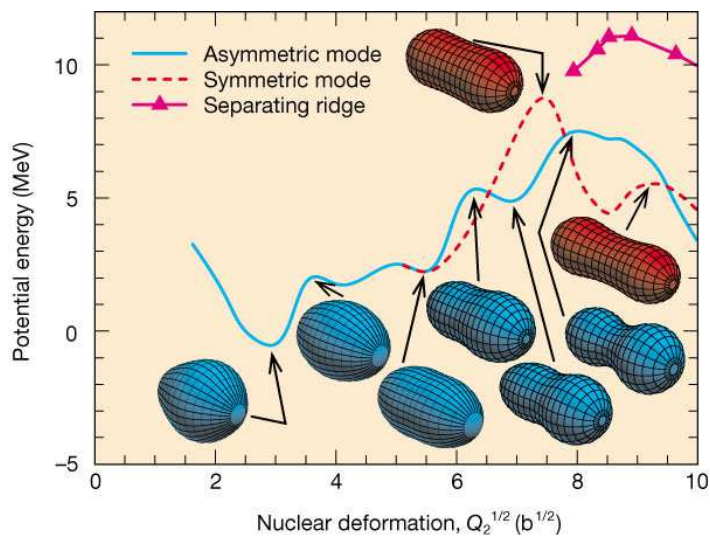}
\end{center}
\caption{\label{moller_nature_fig} Calculated potential energy valleys and 
ridges and corresponding nuclear shapes for $^{228}$Ra as reported by 
Moller {\it et al.,} \cite{NatureMollerNV03*5}. }
\end{figure}

\begin{figure}[h]
\begin{center}
\includegraphics[height=9.0cm, angle=0]{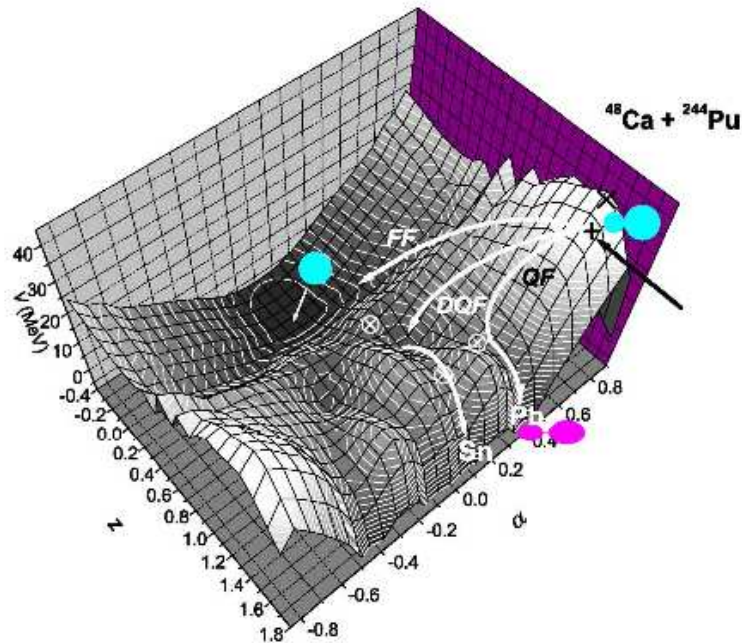}
\end{center}
\caption{\label{aritomo} Potential energy surface of liquid drop model 
calculated by Aritomo {\it et al.,} \cite{NPAAritomo04*5} 
with shell correction. Black arrow shows the injection point of the reaction 
$^{48}$Ca + $^{244}$Pu. }
\end{figure}

\indent Recent extensions \cite{NPAIwamoto04*5,NPAAritomo04*5} of 
the five dimensional energy landscapes for fusion of $^{48}$Ca with  
$^{244}$Pu  have been carried out. The results are shown in the Fig 
\ref{aritomo}. In addition to the calculated minimum energy path  to 
reach the fusion meadow and the subsequent descent 
to the fission valley over a mass symmetric unconditional saddle corresponding 
to the fusion-fission (FF) path, at higher excitations, most of the paths may 
deviate through a mass symmetric saddle shape before fusion to re-separate in 
a quasi-fission (QF) reaction mode.

\indent In a very similar situation, in case of fusion of spherical 
projectiles with deformed $^{232}$Th nuclei, above the Coulomb barrier, the 
system follows a fusion-fission path over the mass symmetric unconditional 
saddle. But as the energy is decreased, these paths are progressively blocked 
and then the microscopic effects come into play. For the polar region of the 
deformed target, the system starts from an initial condition with varying 
deformation, separation and damping of radial motion. This results in the 
system finding a minimum energy path skirting the fusion meadow and over an 
almost mass symmetric saddle. 

\noindent In analogy to  skiers coming down a mountain 
slope from different heights (initial energy), go over a peak( fusion barrier) 
to a meadow (fusion) and continuing to slide over a small hillock 
(unconditional fission barrier) to reach the valley below (scission) in the 
established route (FF), those who start just below or at the peak, the normal
 route is blocked. However, if mountainsides are different (microscopic effect
 due to deformation) and a ridge exists near the peak, some of the skiers can 
reach the ridge and follow it  over a hilltop (conditional mass symmetric 
saddle) and reach almost the same spot at the valley in different route (QF).
 However, for a spherical target, the mountain sides are all similar and no 
ridges exist. The current experimental results strongly indicates the 
likely scenario described above and calls for detailed calculations of the
 energy diagrams for the motion of the nucleons through the dissipative system 
with different initial conditions.

\section{Conclusions}

\indent The present experimental investigations on the heavy ion induced fission reactions were done with a double arm time-of-flight spectrometer with 
difference of T.O.F.'s for precise measurements of widths of mass distributions 
with a mass resolution of 3-4 {\it a.m.u}. The experiments clearly established 
 that width of the mass distribution  is a sensible tool to observe departure 
from the normal fusion-fission path in the fusion of heavy nuclei. It has been
 observed that in case of relative orientation-symmetric target-projectile 
combinations, such as spherical target ($^{209}$Bi) and 
projectiles ($^{19}$F and $^{16}$O), the system at all beam energies follows 
a fusion path over a mass symmetric fission barrier to coalesce to an 
equilibrated compound nucleus (CN), which subsequently undergoes shape 
transitions over a mass symmetric saddle shape to scission into binary 
fission fragments. The angular and mass distributions in such compound 
nuclear fission can be predicted quite accurately in terms of the macroscopic 
properties of a rotating finite drop of liquid equilibrated in spin ($J$), 
projection of spin ($K$) and excitation energies.

\indent However, this picture does not hold in case of systems with 
orientation-asymmetric target-projectile systems, such as in the case of 
deformed target ($^{232}$Th) and spherical projectiles ($^{12}$C, $^{16}$O 
and $^{19}$F) systems, particularly at near and below Coulomb energies, 
where the reaction cross-sections are crucially dependent on the relative 
orientations of the projectile with deformed target. The observed angular 
anisotropies and the present measurement of width of mass distributions 
differ significantly from the predictions based on the macroscopic theories.
This shows the possibility of following of the system in fusion-fission paths 
in alternate routes to that of the normal fusion-fission paths.

\indent  The exact mechanisms for the departure  from normal fusion-fission 
paths are not known 
accurately. However, macroscopic effects such as the direction of mass flow or 
the mass relaxation time being too prolonged may not be the cause. It has been 
established earlier from the experimental barrier distributions, the reaction
 cross sections in $^{19}$F, $^{16}$O, $^{12}$C + $^{232}$Th in near and below 
Coulomb barrier energies are mostly for impact of the projectiles on the polar 
regions of the thorium nuclei. Following the quantum mechanical effects 
favouring  similar shapes in entrance and exit channels 
\cite{NatureMollerNV03*5}, we modify the simple postulation of the microscopic 
effects of the relative  orientation of the projectile to the nuclear symmetry 
axes of the deformed target \cite{HindePRL95*5}. We assume that for the 
non-compact entrance channel shape, the impact of the projectile in the polar
 region of $^{232}$Th target drives the system to an almost mass symmetric 
saddle shape, rather than a compact equilibrated fused system. The observed 
fragment mass widths can be quantitatively explained under such assumptions. 
The above postulation is supported by the observation that for the 
spherical target $^{209}$Bi, where entrance channel compactness of shape is 
same for all relative target-projectile orientations, only normal 
fusion-fission paths, as characterized by the smooth variation of fragment 
mass widths with excitation energy, are observed. It is also worthwhile to 
note that effect of the anomalous mass widths increases with left-right mass 
symmetry in the entrance channel in case of $^{19}$F, $^{16}$O, $^{12}$C + 
$^{232}$Th system in consonance with our description. The present string of 
 measurements indicate that higher entrance channel mass asymmetry and 
energies close to the Coulomb barrier are preferable to increase the 
probability of reaching the fusion meadow in synthesis of super-heavy 
elements in heavy ion reactions.

\setcounter{equation}{0}
\setcounter{figure}{0}
\chapter{Appendix}
\markboth{nothing}{\it Appendix}

\newpage

\section{Statistical model of angular distribution}

\indent The Statistical Saddle Point Model (SSPM) was developed by 
Halpern and Strutinsky \cite{HSgeneva58*6} 
to predict the angular distribution of the fission fragments yielded from 
the fission of a compound nucleus formed by the complete fusion of target 
and the projectile. The theoretical framework of the fragment angular 
distribution involves complete statistical equilibration of the tilting mode 
of the separating fragments in the intermediate composite system.

\indent According to SSPM, the angular distribution can be described in 
terms of quantum numbers $J\hbar$, the magnitude of the total angular 
momentum, $\vec{J}$ of the fissioning nucleus; $M\hbar$, the projection of 
$\vec{J}$ along the space axis which can be identified with the direction of 
projectile, and $K\hbar$, the projection along the symmetry axis of the 
system.

\begin{figure}[ht]
\begin{center}
\includegraphics[height=8.0cm,angle=0]{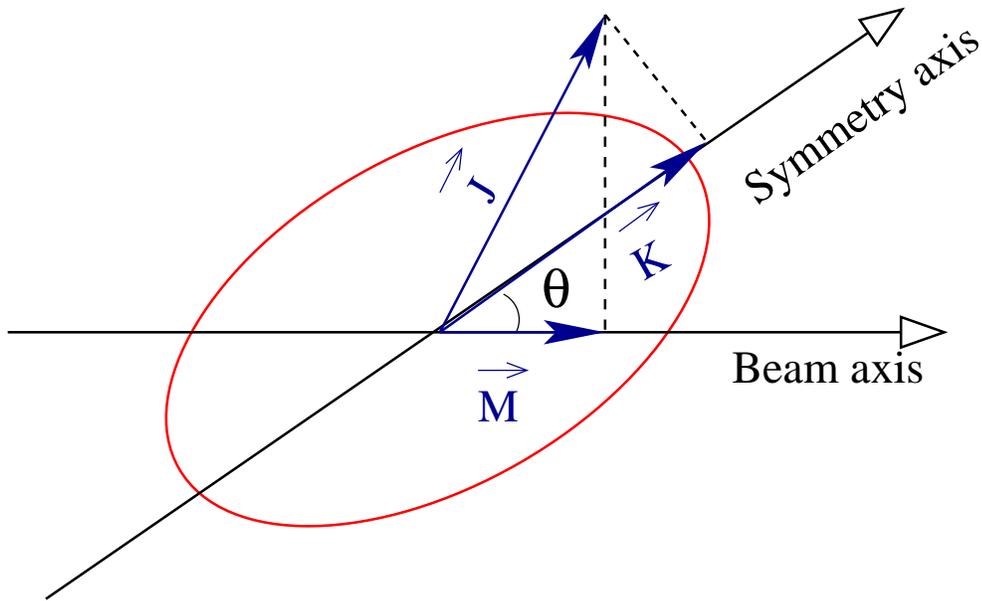}
\end{center}
\caption{\label{sspm} Quantum numbers of a deformed nucleus}  
\end{figure}

\indent The quantum numbers $K\hbar$ describes the {\it tilting mode} of the 
fragments with respect to the space axis, designed by the angle $\theta$. The 
quantum numbers $J$ and $M$ are conserved throughout various extended shapes 
during the passage of the fissioning nucleus from an initial state of 
formation of compound nucleus to final state of saddle point configuration, 
but there is no such quantum number $K$. According to the assumption made, 
the final $K$ distribution is determined by distributions of the $K$ values 
which characterizes the initial states of the system at the saddle point and 
it is good quantum number beyond this point of fission process.

\indent In the statistical limit, when the level density of the internal 
states of transition state is high, $K$ distribution may be obtained by the 
Boltzmann factor $e^{-E_{rot}/T}$ where $E_{rot}$ is the rotational energy 
expended during the passage of the fissioning nucleus from the initial 
excited composite state to highly deformed transition state. The rotational
energy can be written as,

\begin{eqnarray}
E_{rot}^{J,K}={\frac{\hbar^2}{2I_{\perp}}}(J^2-K^2) + 
{\frac{\hbar^2}{2I_{\|}}}K^2\nonumber\\
={\frac{\hbar^2}{2I_{\perp}}}J^2 + {\frac{\hbar^2K^2}{2}}
(\frac{1}{I_\|} - \frac{1}{I_\perp})\nonumber\\
={\frac{\hbar^2}{2I_{\perp}}}J^2 + {\frac{\hbar^2}{2I_{eff}}}K^2
\end{eqnarray}

where, 
\begin{eqnarray*}
\frac{1}{I_{eff}}= \frac{1}{I_{\|}}-\frac{1}{I_{\perp}}\nonumber
\end{eqnarray*}

\noindent Here $I_{\|}$ is the moment of inertia parallel to the symmetry 
axis and $I_\perp$ is the moment of inertia  perpendicular to the symmetry 
axis. 

\noindent The density of levels in the transition state is dependent on the 
thermodynamic energy $(E-E_{rot}^{J,K})$ available to the nucleus,

\begin{equation}
\rho(J,K) \propto e^{\frac{(E-E_{rot}^{J,K})}{T}}
\end{equation}

\noindent where E is the total energy and T is the temperature at the saddle 
point. For fixed E and T and J,
\begin{eqnarray*}
\rho(K) \propto e^{-\frac{E_{rot}^K}{T}}\nonumber\\
\propto e^{-\frac{\hbar^2K^2}{2J_{eff}T}}
\end{eqnarray*}

\noindent This is equivalent to a truncated Gaussian K distributions,

\begin{eqnarray*}
\rho(J,K) \propto e^{-\frac{K^2}{2K_0^2}}: when K \le J\\
=0 : when K>0
\end{eqnarray*}
 
\noindent which is characterized by a variance 

\begin{eqnarray*}
K_0^2=\frac{I_{eff}T}{\hbar^2}
\end{eqnarray*}

\indent The probability of emitting fission fragments from a transition state 
with quantum numbers J,M and K at angle $\theta$ is given by 

\begin{eqnarray}
P_{M,K}^J(\theta)= (2J+1) {\frac{2\pi sin\theta Rd\theta}{4\pi R^2}}
                   |d_{M,K}^J (\theta)|^2\nonumber\\ 
                 =\frac{2J+1}{2}|d_{M,K}^J (\theta)|^2 sin\theta d\theta 
\end{eqnarray}

\noindent where $P_{M,K}^J(\theta)$ represents the probability of emitting 
fission fragments at angle $\theta$ into the conical volume defined by 
the angular increment $d\theta$. The normalization is such that the 
probability integrates to unity for limits 0 to $\pi$. The area of the 
angular ring on a sphere of radius R through which the fission fragments 
are passing is given by the width of the strip $Rd\theta$ times the 
circumference of the ring $2\pi R sin\theta$. The annular ring area must be 
divided by the total area of the sphere $4\pi R^2$ in order to give the 
probability as given by $P_{M,K}^J(\theta)$. 

\indent The foregoing probability depends on $d_{M,K}^J(\theta)$ function 
and is universal in the sense that it is independent of the polar angle, the 
angle of rotation about the symmetry axis and the moment of inertia. Hence, 
the probability distribution depends only on the angle $\theta$ between the 
space fixed and body fixed axis.

\indent The $d_{M,K}^J(\theta)$ functions are defined by the following 
relation:

\begin{equation} 
 d_K^J(\theta)=[J!J!(J+K)!(J-K)!]^{\frac{1}{2}}
\sum_{x}\frac{(-1)^x(sin{\frac{\theta}{2}})^{K+2x}(cos{\frac{\theta}{2}})
^{2J-K-2x}}{(J-K-x)!(J-x)!(x+K)!x!}
\end{equation}

\noindent where the sum is over x=0,1,2,..... and contains all terms in which 
no negative values appears in the denominator of the sum for any one of the 
quantities in parentheses. The angular distribution $W_{M,K}^J(\theta)$ is 
obtained by dividing the probability for emitting fission fragments at angle 
$\theta$ by $sin\theta$,

\begin{equation}
W_{M,K}^J(\theta) \propto {\frac{2J+1}{2}}|d_{M,K}^J|^2
\end{equation}  

\indent The angular distribution of the fission fragments, produced due to the 
fission of a completely fused nucleus mainly depends upon three factors:\\
(i) Transmission coefficient, $T_J$ for passage through the transition state\\
(ii) Level density $\rho(E,K)$\\
(iii) Probability distribution $P_{M,K}^J(\theta)$.  

\noindent In the limit when the target and projectile spins are zero and no 
particle emission from the initial compound nucleus occurs before fission, 
{\it i.e.,} M=0, the angular distributions for many J values and a Gaussian 
K distribution is given by,

\begin{equation}
W(\theta) \propto \sum_{0}^J(2J+1) T_J \sum_{K=-J}^J (2J+1)|d_{M,K}^J(\theta)|^2
\frac{e^{-\frac{K^2}{2K_0^2}}}{\sum_{K=-J}^Je^{-\frac{K^2}{2K_0^2}}} 
\end{equation}

\noindent The symmetric top wave function can be approximated as,

\begin{equation}
|d_{0,K}^J(\theta)|^2 \cong {\frac{1}{\pi}}[(J+ \frac{1}{2})^2 sin^2\theta 
- K^2] ^ {-\frac{1}{2}}
\end{equation}

\noindent and by integrating over all J and K states
 
\begin{equation}
W(\theta) \propto 1 + {\frac{<J^2>}{4K_0^2}}cos^2\theta
\end{equation}

\indent Thus, according to the SSPM model \cite{Vandenbook*6}, the simplified 
expression for fission anisotropy is given as 

\begin{equation}
A=\frac{W(0^\circ)}{W(90^\circ)}= 1+ \frac { < J^2 >}{4K_0^2}
\end{equation}

\noindent Here $K_0^2$ is the variance of the K distribution and $< J^2 >$ 
is the second moment of the compound nuclear spin distribution. $K_0^2$ is 
given as 

\begin{equation}
K_0^2= \frac {I_{eff}T}{\hbar^2}
\end{equation}
\noindent where $I_{eff}$ is the effective moment of inertia  and T is the 
temperature at the saddle. The temperature T is given as 

\begin{equation}
T= \sqrt{\frac {E^\star}{a}}
\end{equation}
 
\noindent where $a$ is the level density parameter taken to be equal to 
$A_{CN}$/10 and $E^\star$ is the effective excitation energy at the saddle.

\subsection{Excitation function for $^{19}$F + $^{209}$Bi }

\indent The cross sections for the fissioning system of $^{19}$F + $^{209}$Bi 
were measured at five energies. The cross sections were obtained from the 
angular distribution data and were normalised to the elastic cross sections 
produced in the monitor detector. Further, the excitation function was 
normalised with respect to the data measured by Majumdar {\it et al}., 
\cite{NMprivate*6} at higher energies. The experimental normalised 
fission cross-sections for various bombarding energies are listed in 
Table \ref{xsec_tab_fbi}.

\begin{table}[h]
\begin{center}
\caption{\label{xsec_tab_fbi}~ Fission cross sections for the system 
$^{19}$F + $^{209}$Bi at all bombarding energies .}
%\begin{ruledtabular}
\setlength{\tabcolsep}{0.5cm}
\renewcommand{\arraystretch}{1.1}
\begin{tabular}[t]{|c|c|c|}
\hline\hline
$E_{lab}$ & $E_{cm}$ & $\sigma_{fiss}$  \\
(MeV) & (MeV) & (mb)   \\
\hline
99.5 & 91.2 & 318.0   \\
95.5 & 87.5 & 151.4 \\
89.5 & 82.0 & 12.5  \\
87.5 & 80.2 &  3.0  \\
85.5 & 78.3 &  1.1  \\

\hline\hline
\end{tabular}
\end{center}
\end{table}

\indent The cross sections in unit of millibarn are illustrated in 
Fig. \ref{excitation_fbi} as function of bombarding energies in the c.m. frame, 
by solid circles. The cross sections reported by Samant {\it et al}., 
\cite{SamEPJ00*6} 
and Majumdar {\it et al}.,\cite{NMprivate*6} are shown by open square and open 
triangles respectively. It is noted that earlier measurements agreed with 
present measurement.

\indent The excitation function was fitted with the couple channel model. 
Using a modified computer code CCDEF \cite{CCDEF*6} the mean square spin values 
$< l^2 >$ were obtained. We have used four  states of the projectile 
$^{19}$F: {\it (i)} $ E^\star=$ 0.198 with $\beta_2=$ 0.55, 
{\it (ii)} $ E^\star=$ 1.346 with $\beta_3=$ 0.33, {\it (iii)} $ E^\star=
$ 1.554 with $\beta_2=$ 0.58 and {\it (iv)} $ E^\star=$ 2.780 with $\beta_4=
$ 0.22. Taking the levels of the target $^{209}$Bi to be the same as   
$^{208}$Pb we have considered one vibrational states with $ E^\star=$ 2.62
 with $\beta_3=$ 0.12 for the best fit of the data. The coupled channel 
calculation is shown by the solid line in Fig. \ref{excitation_fbi}. The mean 
square spin values at each energy was deduced from the above calculation. 

\begin{figure}[ht]
\begin{center}
\includegraphics[height=12.0cm,angle=0]{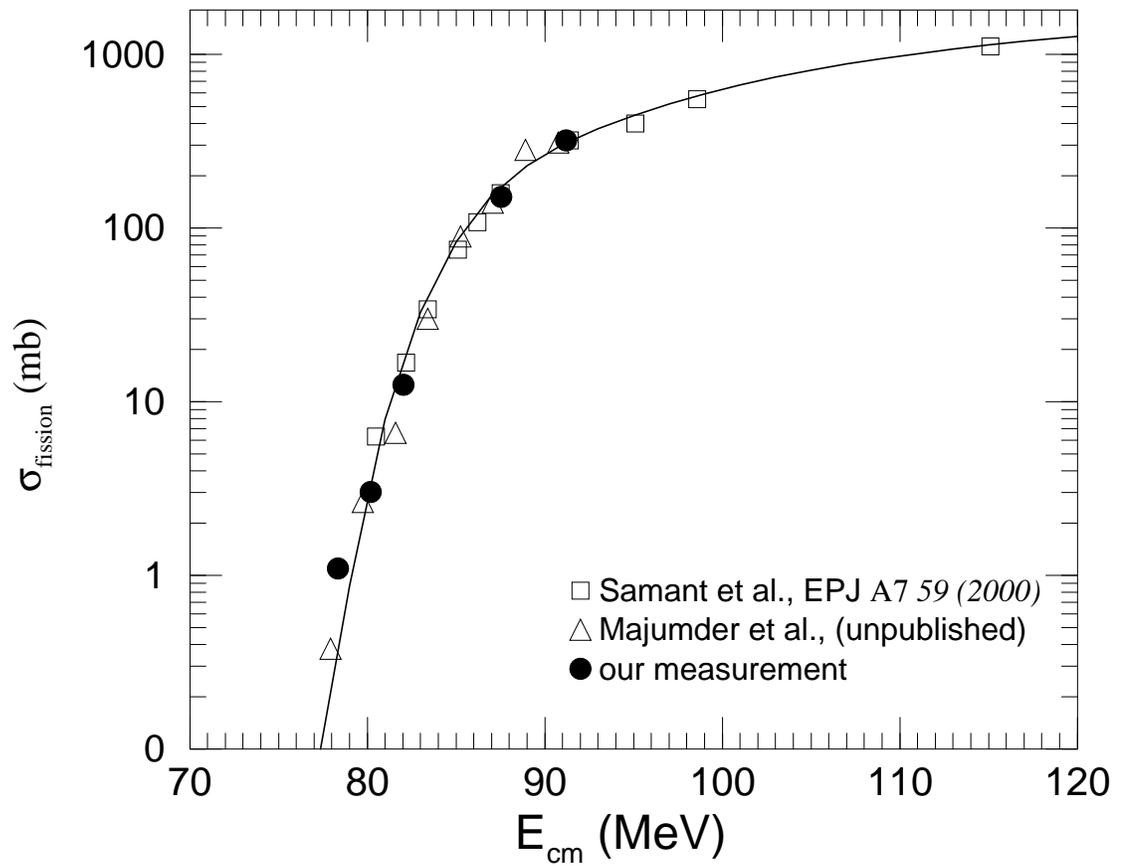}
\end{center}
\caption{\label{excitation_fbi} The measured excitation function for the 
fissioning system of $^{19}$F + $^{209}$Bi (solid circles). 
The coupled channel calculation is shown by solid line. Other measurement are 
indicated by open symbols}  
\end{figure}

\clearpage

\section{Statistical and dynamical aspects of mass 
distribution} 

\indent From the microscopic point of view the phenomenon of fission is 
extremely complicated. As the fissioning nucleus is torn apart in a very 
short time, the motion of each individual nucleon is radically altered. The 
entire interval from the instant that some energy is imparted to the nucleus 
({\it e.g.,} by a neutron capture) to the time at which it is torn apart 
(scission) may be divided into two parts.

\indent During the first interval of time, the imparted energy is 
redistributed among the nucleons over a comparatively large ($\sim 10^{-15}s$) 
period involving many nucleonic collisions. During this time, the energy 
that goes into the collective degrees of freedom causes increased deformation 
of the nucleus and this change in deformation takes place very slowly compared 
to the time of individual nucleonic motion. If none of the individual 
nucleons attains enough energy to escape the nucleus ({\it e.g.,} by  neutron 
emission), the imparted energy may be nearly all spent in causing a large 
critical deformation, beyond which the system is unstable. This metastable 
state is reached at the saddle point at the end of the first interval of time. 
Once the nucleus is over the saddle point, shape instability takes place 
and the nucleus can distort very easily to reach the point of scission.

\indent The second time interval, from the saddle point to scission, is very 
short ($\sim 10^{-21}s$) and the separation of collective and intrinsic 
coordinates may not be meaningful at this stage, although there is evidence 
for the persistence of shell effects even here. Here we shall concentrate on 
the collective motion of the nucleus up to the saddle point, where we can still 
meaningfully describe the motion in terms of collective variables. There are 
two aspects of the problem:{\it (i) the statics} i.e., the potential energy 
surface of the nucleus with increasing deformation and {\it (ii) the dynamics} 
- it's equation of motion. In the absence of any complete theory, it is 
instructive to construct intuitive models of fission to see whether the 
salient experimentally observed characteristics can be extracted and the 
liquid drop model is an attempt in this direction.

\indent From the static approach in the frame work of the liquid drop model 
described above it is not possible to say how the energy is realised in the 
fission process is divided between the different degrees of freedom in the 
fissioning nucleus or what is the distribution of the corresponding variables. 
These are the problems whose solution ultimately leads to description of the 
distributions of the observed fragment characteristics: the masses, charges, 
kinetic energy and excitation energy. An important role here is placed by the
 stage of descent of the fissioning nucleus from the top of the barrier, which 
serves as the reference point for the process of release of the energy 
concentrated on the fissioning degree of freedom. This problem is solved in the frame work of statistical and dynamical models of the fission process.  

\indent The assumption of the statistical model are the simplest. This model 
is based on the assumption that in a considered distinguished state of the 
nucleus (for example, at the scission point or saddle point) the condition 
of statistical equilibrium with respect to all degrees of freedom is ensured. 
This is an assumption of the transition state model, justified by the argument 
that in the neighborhood of the saddle point the fission is the slowest. It 
is obvious that if the condition of statistical equilibration is to be 
realised at the scission point one must require that the exchange between the 
various degrees of freedom take place sufficiently rapidly compared with the 
descent time $t_d$. In Fong's model the exchange time is taken to be the 
characteristic nucleon time $\tau_n \sim 3 \times 10^{-22}$ sec and this ensures 
the inequality:

\begin{eqnarray*}
t_d >> \tau_n\nonumber
\end{eqnarray*}

\noindent The transition state model leads to the Bohr-Wheeler formula for the 
barrier penetrability:

\begin{equation}
 T_f (E) = \int_0^{E-E_f} \rho(U) dU 
\end{equation}

\noindent where $\rho(U)$ is the density of the transition states. In the 
Fermi gas model:

\begin{eqnarray}
\rho(U) \sim e^{[2\sqrt{aU}]}\\
T= \sqrt{\frac{(E-E_f)}{a}}
\end{eqnarray}

\noindent where $A$ is the level density parameter. 

\indent For the conditional barriers $E_f=V(\eta)$ for fixed mass symmetric 
deformations $\eta$ and also an expansion of $V(\eta)$ in the neighborhood 
of $\eta=0$ :

\begin{eqnarray}
V(\eta)= V(0) + ({\frac{dV}{d\eta}})_{\eta=0}\eta + 
{\frac{1}{2}}({\frac{d^2V}{d\eta^2}})_{\eta=0}\eta^2 + ......\\
= V(0) + {\frac{1}{2}}({\frac{d^2V}{d\eta^2}})_{\eta=0}\eta^2 + ......
\end{eqnarray} 

\noindent since $dV/{d\eta}$ at $\eta = 0$ vanishes. As a mass-symmetric 
deformation, Strutinsky \cite{StruNPA67*6} took a quantity related to the 
ration of the volumes of the right ($v_r$) and left ($v_l$) parts, namely, 

\begin{equation}
\eta = 2 \frac{v_r-v_l}{v_r + v_l}
\end{equation}

\noindent The dimensionless drop rigidity parameter for such variation of the 
shape 

\begin{equation}
K= \frac{1}{2E_s^0}({\frac{d^2V}{d\eta^2}})_{\eta=0}\eta^2 
\end{equation}

\noindent Thus,

\begin{equation}
V(\eta) = V(0) + E_s^0 K \eta^2 + .......
\end{equation}

\noindent So we can obtain the distribution of $\eta$, the first approximation 
to which corresponds to the Gaussian distribution

\begin{equation}
Y(\eta) \sim e^{-E_s^0K(\frac{\eta^2}{T})}
\end{equation}

\noindent We assume that the volumes of the parts of the saddle figure are 
proportional to the masses M and (A-M) of the future fragments, i.e.,

\begin{eqnarray*}
v_r \propto M\nonumber\\
v_l \propto (A-M)
\end{eqnarray*}

\noindent Therefore,
  
\begin{eqnarray*}
\eta= 2 \frac{M-A+M}{A}={\frac{2}{A}}(2M-A)={\frac{4}{A}}(M- \frac{A}{2})
\end{eqnarray*}

\noindent Using above equations we get,

\begin{eqnarray} 
Y(M) \sim e^{-{\frac{E_s^0K}{T}}{\frac{16}{A^2}}(M- \frac{A}{2})^2}\nonumber\\
\sim e^{-\frac{(M-\frac{A}{2})^2}{2\sigma_m^2}}
\end{eqnarray}

\noindent where, 

\begin{eqnarray}
\sigma_m^2=\frac{A^2T}{32E_s^0K}=\frac{T}{k}
\end{eqnarray}

\noindent and $k$ is given by,

\begin{equation}
k=\frac{32E_s^0K}{A^2}
\end{equation}

\section{Solid angle correction to yields of a position sensitive detector}

\indent Let us assume that a small central segment of a MWPC, {\it dx}
is located at a perpendicular distance, R, from the centre of a fission 
source. The solid angle subtended at the source by the segment is,

\begin{equation}
d\Omega=\frac{dx}{R^2}
\end{equation}

\begin{figure}[h]
\begin{center}
\includegraphics[height=7.0cm, angle=0]{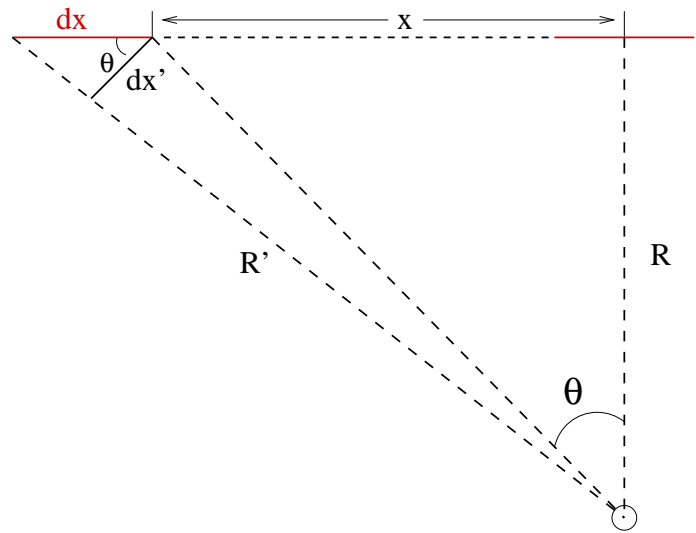}
\end{center}
\caption{\label{solidang} Schematic representation of the effect of solid 
angle of a MWPC, situated at a distance from a point source.}
\end{figure}
%\clearpage

\noindent Any segment at the side of the PSD is similarly located at a distance of $R^\prime$ from the source. But the yield produced at this segment should 
be a function of angle it makes with the central segment. It is shown in 
Fig. \ref{solidang}, the angle is $\theta$. The solid angle subtended by this 
segment at the source is,

\begin{equation}
d\Omega^\prime=\frac{dx^\prime}{R^{\prime2}}
\end{equation}

\noindent The distance $R^\prime$ can be written as,
\begin{equation}
R^\prime=\frac{R}{cos\theta}
\end{equation}

\noindent The length of the second segment is,
\begin{equation}
dx^\prime={dx}{cos\theta}
\end{equation}

\noindent Combining the above equations one gets,

\begin{equation}
d\Omega^\prime=({\frac{dx}{R^2}})cos^3\theta
\end{equation}

\noindent Then the correction factor for the yield at a side segment is 
$cos^3\theta$ which should be multiplied with the observed yields to 
obtain corrected yield due to solid angle effect.

\section{Gas handling system}

\indent The operating pressure in the scattering chamber was required to be of 
the order of 2 $\times 10^{-6}$ torr. The MWPCs within the chamber were 
connected to a separate gas handling system, shown schematically in 
Fig \ref{gashandling}, through two gas feed throughs attached to one of the 
side ports of the scattering chamber. One connecting valve was provided 
between the scattering chamber and the entrance-feed throughs of the 
detector. For pumping down the chamber, the bypass valve at the port of the 
chamber was required to be kept open to draw the air from the detectors and 
the gas handling system also, keeping all the valves open in the gas flow 
controlling system, except inlet control valve. After the rough vacuum of the 
order of  $ 10^{-3}$ torr was achieved, the connecting valve was closed 
to disconnect the chamber from the detectors and the chamber was further pumped down to 2 $\times 10^{-6}$ torr. The shunt and bypass valves in the flow 
handling system were then closed. For the operation of the MWPCs with a
 steady flow of gas, the gas flow through the detectors were controlled at 
pressure in the range of 1.5-3.0 torr by an electronic pressure controller 
(Make: MKS,USA). The flow rate of the gas through the detector was controlled 
by a metering valve.

\begin{figure}[ht]
\begin{center}
\includegraphics[height=12.0cm,angle=0]{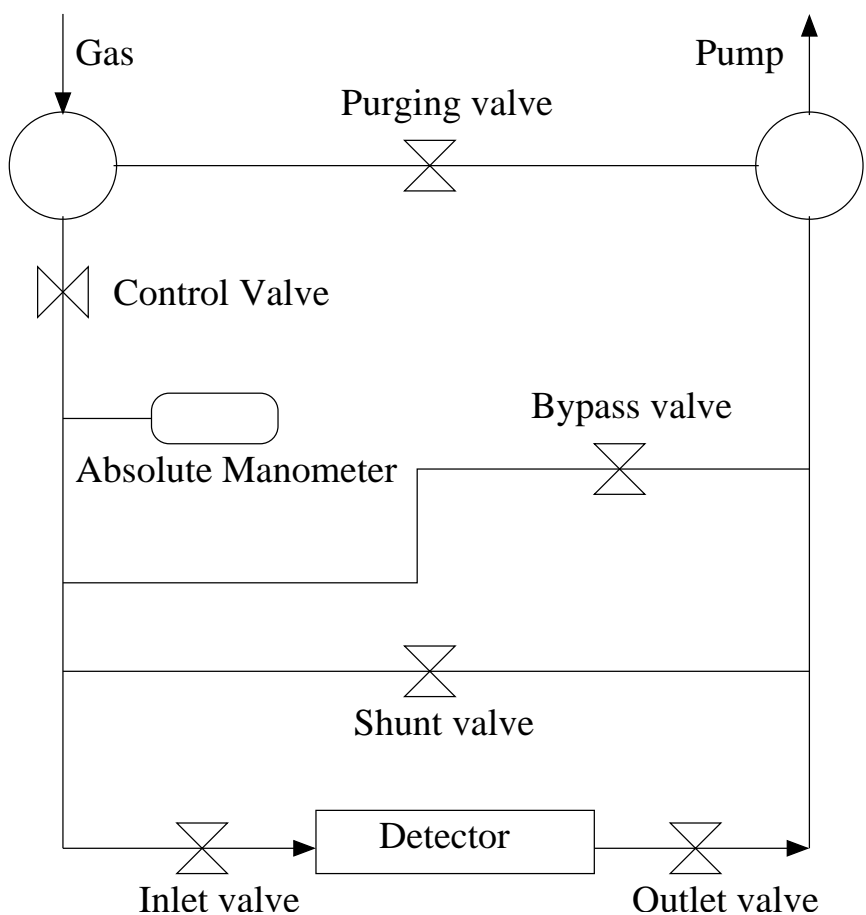}
\end{center}
\caption{\label{gashandling} Flow chart of the gas handling system with 
necessary valves, used to operate gas flows through MWPC.}  
\end{figure}

\end{document}